%% file: main.tex
\documentclass[11pt,a4paper]{article}

\usepackage{jheppub}

\usepackage{bm}
\usepackage{amsthm}
\usepackage{mathtools}

\usepackage{xcolor}
\usepackage{booktabs}
\usepackage{longtable}
\usepackage{enumitem}

\usepackage{orcidlink}

\usepackage{csquotes}

\usepackage{tikz}
\usetikzlibrary{arrows.meta, positioning, calc}
\usepackage{pgfplots}
\pgfplotsset{compat=1.18}

\theoremstyle{definition}
\newtheorem{proposition}{Proposition}[section]
\newtheorem{definition}{Definition}[section]

\input{macros}

\input{frontmatter-jhep}

\hypersetup{
  pdftitle={The Spectrum of a Compact Internal Space II: Effective Couplings and Mass Scales},
  pdfauthor={Jinku Guo},
  pdfkeywords={Three fermion generations; Fermion mass hierarchies; Fermion mass relations; Gauge coupling constants; Higgs boson mass; Neutrino masses; Cosmological parameters},
}

\makeatletter
\gdef\@fpheader{}
\makeatother

\begin{document}

\let\acknowledgmentsOrig\acknowledgments
\renewenvironment{acknowledgments}{\acknowledgmentsOrig}{}

\maketitle\flushbottom

% --- Sections ---
\input{sections-en/sec01_introduction}
\input{sections-en/sec02_computation}

\input{sections-en/sec04_content_symmetries}
\input{sections-en/sec03_internal_scale}

\input{sections-en/sec05_gauge_couplings}
\input{sections-en/sec07_ewsb_cp}
\input{sections-en/sec06_flavour}
\input{sections-en/sec08_cosmology}
\input{sections-en/sec09_qcd_bbn}
\input{sections-en/sec10_results}

\input{sections-en/sec11_conclusion}

% --- Appendices ---
\appendix
\input{sections-en/appendix_a}

\input{sections-en/appendix_b}
\input{sections-en/appendix_c}
\input{sections-en/appendix_d}
\input{sections-en/appendix_e}
\input{sections-en/appendix_f}
\input{sections-en/appendix_g}

% --- Acknowledgements ---
\begin{acknowledgments}
AI-assisted tools were used for language editing and for drafting and
checking the numerical verification and formalization code. They also
assisted selected numerical consistency checks. The author is responsible
for the scientific content and conclusions.
\end{acknowledgments}

% --- Declarations (conflict of interest + data availability, before references) ---
\input{sections-en/declarations}

% --- Bibliography ---
\bibliographystyle{JHEP}
\bibliography{references}

\end{document}

%% file: macros.tex
\newcommand{\eff}{\text{eff}}
\newcommand{\SM}{\text{SM}}

\newcommand{\Tr}{\operatorname{Tr}}
\newcommand{\SU}{\mathrm{SU}}
\newcommand{\U}{\mathrm{U}}

\newcommand{\order}[1]{\mathcal{O}(#1)}

\newcommand{\GN}{G_{\mathrm{N}}}
\newcommand{\MP}{M_{\mathrm{P}}}
\newcommand{\MG}{M_{\mathrm{G}}}
\newcommand{\kL}{kL}

%% file: frontmatter-jhep.tex
\title{The Spectrum of a Compact Internal Space II: Effective Couplings and Mass Scales}

\author[a]{Jinku Guo\,\orcidlink{0009-0000-6600-6171}}
\affiliation[a]{Northwestern Polytechnical University, Xi'an 710072, China}
\emailAdd{guojk@nwpu.edu.cn}

\abstract{%
Minimal unbiased compositional change generates the Gaussian source
of the compact internal construction. We derive joint constraints on
effective couplings and mass scales from its gauge and matter content,
scalar sector, spectral endpoint and kinetic normalizations.
Compatibility of source and colour screening selects three matter
copies. Within the response family indexed by integers, shell support
and position retention select the level \(R=5\), whose transverse
traceless endpoint equation has a unique positive root. Eliminating shared transfer
parameters gives a relation among six geometric fermion masses.
Bandwidth variation correlates the electron and electroweak scales, while
two cosmological ratios are independent of the common neutrino mass
normalization. Finite matching through one loop implements the generated
Fermi, top and Higgs conditions in a common action. The absorptive top
self energy gives the Born decay width as an additional output.
The compact source covariance and positive RMS amplitude determine a
local nonlinear flux that couples matter motion to source feedback.
A model with an auxiliary radius admits autonomous source evolution,
positive reduced propagation forms and a matter phase clock
on a specified finite cosmological interval.
Physical comparisons use the stated mass definitions, matching and thermal
approximations. The fixed DESI
DR2 distance comparison gives \(\chi^2=29.5087\) for thirteen components
and \(p=0.00554\), indicating tension in the adopted cosmological mapping.
Spectral tail and quadrature bounds control the finite response calculation.
}

\keywords{Three fermion generations; Fermion mass hierarchies;
Fermion mass relations; Gauge coupling constants;
Higgs boson mass; Neutrino masses;
Cosmological parameters}

%% file: sections-en/sec01_introduction.tex
\section{Introduction}\label{sec:intro}

A common internal origin for particle interactions must account for
their different strengths and the large separation of mass scales.
The scalar and Dirac spectra of a compact space specify mode positions
and multiplicities. Source operations couple the modes, and kinetic
normalization converts the resulting matrix elements into interaction strengths.
We study how these ingredients determine effective couplings and masses
and what relations follow when different sectors share one source and
one internal scale. Their shared endpoint constrains the sectors jointly.
Dimensionless combinations
eliminate common transfer parameters or mass normalizations and test
the relative response laws with the matching appropriate to each
observable.

In the Standard Model the Yukawa matrices of charged fermions are
independent inputs. At tree level, electroweak symmetry breaking converts
their singular values into masses~\cite[Electroweak and Higgs reviews]{PDG2024}.
Here the allowed operations of an internal response space determine
mass relations. Each physical quantity is associated with a source
variation and a normalized field or observable response. The same source
data constrain particle parameters and the collective response of the
occupied system.

Minimal unbiased compositional change is the common starting point of
the mathematical and physical constructions. The source law is isotropic
and centered. Weak continuity and independent convolution composition
together with self similarity under square root scaling and finite normalized
covariance determine its Gaussian form. The correlation window and reflected spectrum supply the common
carrier~\cite{Guo2026SpectralArithmetic,Guo2026NumberDynamics}.
The mathematical continuation uses native mode operations to generate
arithmetic. The physical construction in
Ref.~\cite{Guo2026CompactInternalSpaceI} generates the compact projective
geometry. Relative frame localization fixes the faithfully acting gauge group.
Recursive minimal realization of faithful matter preserving adjoints fixes
the sixteen dimensional matter module. An elementary reversal real chiral
variation fixes the single complex scalar doublet.
The corresponding uniqueness statements hold within the generated class;
the constructions introduce no fitted numerical parameter.

Mode content determines particle identity, normalized source matrix
elements determine interaction strengths, and repeated amplitude transfer
produces mass hierarchies. Effective fields represent these quantities at the scale of
observation and are varied through their actual dependence on the
generated variables. Clock readings compare changes within the same
history. The macroscopic metric represents relations between matter
motion and source response together with clock rates in the inherited Einstein--matter
description. This interpretation determines which variables and
variations enter each calculation.

Spectral geometry provides precedents for connecting internal data to
particle parameters. A spectral triple encodes metric and differential
structure~\cite[Chapter~VI]{Connes1994}. The spectral action constructs a
bosonic action from a Dirac spectrum~\cite{ChamseddineConnes1997}.
For a product of a spin manifold and a finite noncommutative geometry
the Chamseddine--Connes--Marcolli construction~\cite[Sections~3.7 and 5.1--5.4]{ChamseddineConnesMarcolli2007}
obtains high-scale relations among gauge and Higgs coefficients together with Yukawa couplings.
These relations are evolved to low energy. Ref.~\cite{vanSuijlekom2015} develops the construction further.
The response evaluated here is a projected Gaussian spectral sum with
discrete multiplicities and finite source-frequency bandwidth retained.
Its endpoint depends on the resolved modes and is evaluated directly.
Heat-kernel methods~\cite{Gilkey1995} supply the spectral-geometric
background.

The scalar analysis inherited from
Ref.~\cite[Section~8 and Appendix~E]{Guo2026CompactInternalSpaceI}
establishes a unique condensed branch of the relative spectral potential
and a positive gap for its assigned scalar mass operator. It also gives uniform
truncation bounds for the stationary amplitude and normalized mass.
We use the same feedback, subtraction, weights, drive and kinetic
normalization. The resulting positive restoring response supplies the
scalar branch used in the mass and electroweak calculations.

The first result is a joint determination of content and endpoint.
Extensive source screening and the colour-screening condition select
three matter copies. Their charge traces fix the torsion response
while the shell-support and three-position conditions select the
integer response level \(R=5\). At fixed content and bandwidth
homogeneity reduces the tensor endpoint equation to
\(A_R/c=4/27\). Here \(A_R>0\) is independent of \(c=kL\) and hence gives one
positive root. That endpoint retains three positive Dirac positions.
The three-position plateau persists for every continuously extended
bandwidth at least five. The count is stable under this continuation,
whereas the dimensional scales retain a bandwidth dependence.

The second result is a set of cross-sector constraints. Eliminating
common transfer parameters gives a relation among six geometric fermion
masses. Correlated running-mass data~\cite{HuangZhou2021} constrain its
finite matching at a common scale. The common neutrino mass scale cancels
from the vacuum-density fraction and a dimensionless temperature--Hubble
ratio. Bandwidth variation also correlates the electron and electroweak
scales. These relations test relative transfer laws even when a common
dimensional normalization is eliminated. A canonically normalized action
realizes the family transfer and relative mixing rotations. Finite
one-loop matching implements the generated Fermi condition and the top and Higgs
conditions in that action. Its absorptive top self-energy supplies the
Born decay width as an additional output.

The macroscopic applications use the Einstein--matter response of
Ref.~\cite{Guo2026SpectralGravity}. The homogeneous branch is tested against
the DESI distance vector with its full covariance. The variational construction of Bekenstein and
Milgrom~\cite{BekensteinMilgrom1984} supplies the nonrelativistic field
equation for local nonlinear gravity. The endpoint covariance and RMS response fix its acceleration
normalization and interpolation function here. The static problem has a
unique weak solution for the stated boundary data. Coupling it to
pressureless trajectories gives source feedback and an energy balance
during smooth evolution. A covariant auxiliary-radius realization
further admits an occupied FLRW background with autonomous source
feedback and positive reduced propagation forms together with a matter-phase clock on
a finite interval. Source work and total stress conservation follow
from the common action. Its conserved phase charge makes the
matter phase monotone on that interval.
A separate independent-foliation realization of the
type introduced by Blanchet and Marsat~\cite{BlanchetMarsat2011} is
analysed by the constrained slow-motion reduction of
Ref.~\cite{Flanagan2023}. Its kinetic signs are determined for the
backgrounds specified in that analysis.

The accompanying implementation~\cite{Guo2026GenesisFramework} evaluates
the spectral response mode by mode. It solves the endpoint and propagates
shared inputs through the coupling and mass relations. Newton's constant
sets the dimensional normalization. The calculation distinguishes gauge
evolution with fixed geometric Yukawa data from joint five-coupling evolution
and from finite one-loop matching. Functional-renormalization
methods~\cite{Berges2002,Dupuis2021} provide the setting for scale-dependent
composite responses. The finite response and transfer equations used here
are specified explicitly.
Physical comparisons retain the mass definition and perturbative order and
observational covariance appropriate to each quantity.

The archived records document numerical reproduction of the spectral,
coupling and mass calculations and interval certificates for the endpoint
and its specified propagated readouts. The accompanying Lean
development~\cite{GuoLeanFoundations} provides machine-checked proofs of
the indexed source and basic physical-interface statements in their
declared comparison classes. Appendix~\ref{app:repro} identifies the
formal correspondence and numerical coverage and distinguishes
floating-point values from rigorous enclosures.

The appendices contain the detailed source-response and transfer
calculations, error bounds and numerical records.

%% file: sections-en/sec02_computation.tex
\section{Spectral inputs and physical normalization}\label{sec:comp}

Spectral positions and multiplicities specify the channels in the
internal response sums. Source matrix elements couple these channels;
division by the kinetic norms gives the physical interaction strengths.
Every sector uses the same endpoint, content data and dimensional
normalization.

Three dimensionless quantities organize the calculation. The charged content
trace gives the torsion response \(\tau\). Shell support and position retention
select the integer response level \(R\). The response balance determines
the tensor endpoint \(\kL\). Newton's constant \(\GN\) sets the dimensional
unit. These common data determine the electroweak and fermion scales
together with the cosmological scales.

\subsection{Dimensional normalization}\label{sec:anchor}

Newton's constant sets the dimensional normalization. In the natural unit
convention of the PDG physical-constants table~\cite{PDG2024} its value is
\[
\GN = 6.70883 \times 10^{-39}\ \mathrm{GeV}^{-2}.
\]
It defines the reduced Planck mass
\[
\MP = \frac{1}{\sqrt{8\pi\GN}} = 2.43532\times10^{18}\ \mathrm{GeV}.
\]
The particle mass scales are dimensionless multiples of $\MP$ determined by the spectral data and the stated hierarchy relations. The unreduced Planck mass $M_{\mathrm{Pl}}=\sqrt{8\pi}\,\MP$ is used in the conventional radiation-era expansion formulas. Appendix~\ref{sec:derivbbn} specifies the additional effective inputs and thermal thresholds in the abundance calculation.

The torsion modulus $\tau$ and the jointly selected content and endpoint $\kL$ are evaluated from the response relations. The root equation for $c=kL$ depends only on dimensionless spectral and content data.

\subsection{The spectral-sum representation}\label{sec:specsum}

The compact carrier and its response algebra are constructed in Sections~3--4 of Ref.~\cite{Guo2026CompactInternalSpaceI}.  The same source generates the projective geometry and global gauge group together with the charged module and elementary scalar connector. Its curvature and volume enter the internal response sums. The Gaussian factor represents the inherited unbiased composition law. Appendix~\ref{sec:p5-inherited-gaussian} states the scale convention used here and cites the uniqueness proof.

The scalar Laplacian on $RP^3$ of radius $L$ has spectrum and multiplicities
\[
\lambda_l=\frac{l(l+2)}{L^2},\qquad d_l=(l+1)^2,\qquad l=0,2,4,\ldots.
\]
The spin structure $\tau_-$ has positive Dirac eigenvalues $(n+3/2)/L$ for even $n$ and negative eigenvalues $-(n+3/2)/L$ for odd $n$ with multiplicity $(n+1)(n+2)$. The numerical channels resolve these signed branches with their respective projectors. The physical radius enters through dimensionless scale ratios. Converting back to masses uses the common anchor.

In the inherited field realization, expansion in the internal mode
basis gives macroscopic fields as coefficients. The internal geometry
determines their mode labels, representations and allowed products.
The coefficient fields describe the corresponding amplitudes in the
macroscopic spacetime representation.
A source or scalar connector acts between modes through an
overlap matrix element. Dividing out the kinetic norms of the fields
gives the canonically normalized interaction vertex. Mode multiplicities
count the contributions; matrix elements and kinetic forms determine
their strength.
The generated source identifies the physical variation being
measured. A charge insertion probes its current response whereas a
deformation of the internal quadratic form probes its restoring
response. The associated projection and canonical normalization determine how that
response enters the observable.

The scalar sector inherits three further results from
Section~8 and Appendix~E of Ref.~\cite{Guo2026CompactInternalSpaceI}.
These are the renormalized conformal spectral response, a positive
gap for the assigned scalar mass operator and uniform truncation
bounds on a condensed scale interval. They determine the positive stationary
amplitude and colour-singlet scalar mass $m_\delta$. A common cutoff controls
the mass error along a compact condensed scale path. Their use retains the
spectral feedback and subtraction with the original weights and drive
and the same kinetic normalization. Appendix~\ref{sec:p5-complete-condensate}
states the complete condensate used here and its relation to the
electroweak window readout. These results determine the scalar branch on which the mass and
electroweak responses are evaluated.

\subsection{Response definitions and scale-transfer relations}\label{sec:struct}

The response relations comprise the $R=5$ TT endpoint with normalization
$4/27$ and the content screening and position-retention rules. Their scale
transfers determine the gauge normalizations, sector mass ladder,
VEV proportional to the squared window and QCD catalogue. Cosmological
readouts and the RMS-to-flux map use the same source data. Each generated
source operation fixes the admissible variation and channel. Its projection
and normalization determine the amplitude, coupling or density being
read. Representation and charge moments fix the algebraic factors. The
scale convention identifies the physical regime of the response.
In the local spatial channel the occupation covariance fixes the
acceleration variance. The quadratic entry represents squared field
strength, and the positive RMS readout supplies the response amplitude.
Appendix~\ref{app:deriv} derives these responses and transfer operators
with their compatibility relations and error bounds.

Physical readouts are functions of this fixed generated source and
its allowed operations with the inherited comparison maps and
normalizations retained. The generating arithmetic and the closure
of admissible spectral operations are developed in
Refs.~\cite{Guo2026SpectralArithmetic,Guo2026NumberDynamics}.
Here the response and transfer relations identify which of those
operations realize each physical quantity. A change of coordinates
must transform the source and kinetic form together with the observable.
Such transformations preserve the physical content of a realization.
Composition determines the kernel. Joint source compatibility selects
the content. The response balance fixes the endpoint. The corresponding
uniqueness results retain the source operations and projectors and
normalizations entering each construction.

For a generated history $H$ denote a clock readout by $T_{\mathcal C}[H]$
and another observable by $O[H]$. On an interval where the clock is
monotone their relation defines $O$ as a function of $T_{\mathcal C}$.
The order of construction is determined by the dependencies of the
generating operations~\cite{Guo2026NumberDynamics}. The Gaussian
composition parameter measures accumulated change. The
renormalization-group parameter labels the scale of a response.
Physical rates use the selected clock normalization. In the macroscopic
metric representation a nonzero matter amplitude with phase $\vartheta$ and
$X=-(\nabla\vartheta)^2>0$ gives the proper-time readout
$ds=d\vartheta/\sqrt X$ along its future-directed phase flow
(Proposition~\ref{prop:p5-native-phase-congruence}).
The clock counts phase change and calibrates each
increment by its instantaneous frequency. This material readout
supplies the interval used to express propagation and reaction rates.

For effective fields $\Phi=\Gamma(q)$ the generated action is
$S_{\rm gen}[q]=S[\Gamma(q)]$ and its equations follow from variations
of the admissible $q$. Its source exchange, constraints and quadratic
response are calculated from this same action. An invariant-subsystem realization in an independently specified field
theory additionally satisfies its embedding and evolution-preservation
conditions.

The physical state space used here is the generated image
\[
 \mathcal P_{\rm phys}=\Gamma(\mathcal Q_{\rm gen}),
\]
with its source and admissible variations retained. The relative-frame and
least-matter continuations together with the connector continuation act on this image with the
universal properties established in the companion construction. Their
finite responses and sector readouts define the observables evaluated
here. Enlarged field spaces serve to compare effective realizations.
Physical variations follow the stated generating map.

\subsection{Theoretical inputs and comparison conventions}\label{sec:status}

Table~\ref{tab:status} distinguishes the generated source structure and
its physical continuation from the finite response readouts evaluated
here. It also identifies the dimensional and effective inputs and the
quantities derived from them. Observations enter the comparison stage.

\begin{table}[t]
\centering
\caption{Logical status of the quantities entering the calculation.  The
generated structural and physical-continuation results are inherited from
Ref.~\cite{Guo2026CompactInternalSpaceI}. The source operations and response
rules define the calculation. Their joint conditions select the content
and finite response level and endpoint.}
\label{tab:status}
\small
\begin{tabular}{p{2.7cm} p{5.0cm} p{4.1cm}}
\toprule
Logical status & Objects & Role in the calculation \\
\midrule
Generated source structure & Minimal unbiased composition and its Gaussian source.
Projective carrier and signed spectra with their multiplicities and source algebra. &
Defines the admissible modes and products and source operations inherited from
the common generating construction. \\
Generated physical continuation & Relative-frame localization and recursive
minimal matter realization with the elementary reversal-real connector.
The resulting gauge group and matter module with the scalar connector and
the condensed scalar branch for its stated spectral feedback and mass
operator. & Defines the physical representation class and the scalar
background inherited from the companion construction. \\
Response structure & Channel projectors with kinetic and response
normalizations. Integer response-level family and endpoint normalization.
Scale-transfer and readout maps. & Defines the finite response problem
whose joint consequences are calculated throughout the paper. \\
Dimensional and effective inputs & Newton's constant for dimensional
reconstruction together with the separately stated thermal thresholds and
effective inputs used in the abundance calculation. & Supplies the physical
unit and the declared closures of the extended sectors. \\
Derived quantities & Torsion response and replicated content. Selected level
$R=5$ and endpoint. Couplings and masses with widths and densities and macroscopic responses. & Obtained
after the generated structure and physical realization have been fixed. \\
Comparison data & Reference measurements and Standard-Model extrapolations with
observational covariance matrices. & Enter residuals and likelihoods and other
tests after the corresponding theoretical quantities have been evaluated. \\
\bottomrule
\end{tabular}
\end{table}

The comparison tiers A--D classify approximation level and evidentiary
status. Table~\ref{tab:status} instead identifies the logical origin of
the inputs and outputs. Appendix~\ref{app:table} assigns a comparison
tier to each numerical entry.

The response conditions fix the dimensionless endpoint before Newton's
constant supplies the unit for dimensional reconstruction. Gauge,
flavour, electroweak and cosmological quantities use these common data
with their respective matching conventions. Reference measurements and
Standard-Model extrapolations enter the comparison after the generated
response laws have been evaluated.
An explicitly calibrated effective extension identifies its calibration
at the defining relation and tests additional quantities independently.
Spectral truncation, quadrature error and physical sensitivity are
treated separately in Appendix~\ref{app:sensitivity}.

\subsection{Common endpoint data}\label{sec:repro}

The content and endpoint first satisfy the joint conditions of Appendix~\ref{sec:p5-joint-content}. On the selected branch the unrounded endpoint and source responses are propagated through the coupling and mass relations. The dimensionless root equation is independent of the dimensional anchor. Appendix~\ref{app:repro} specifies the spectral traces and numerical precision together with the electroweak corrections used in the tables.

\subsection{Renormalization-group evolution}\label{sec:rguse}

The spectral responses supply boundary couplings at their designated scales. The retained two-loop RG equations~\cite{LuoWangXiao2003,LuoXiao2003} evolve the emergence-scale data from $\MG$ to the internally determined $Z$-boson mass $M_Z$. The representation content fixes their group factors. The production gauge evolution holds the geometric Yukawa input $y_0=1$ fixed. Joint five-coupling evolution and finite one-loop matching are separate calculations. Appendix~\ref{sec:rgflow} gives the transfer convention. The high-scale reference couplings come from a separate Standard-Model extrapolation with the boundary values listed in Appendix~\ref{app:repro}. Figure~\ref{fig:dependency} shows the dependence of the computed outputs.

\input{figures/fig_dependency}

%% file: figures/fig_dependency.tex
%% Dependence of the observables on the common spectral data
\begin{figure}[t]
\centering
\begin{tikzpicture}[
  font=\scriptsize,
  >=Stealth,
  b/.style={draw, rounded corners=1pt, align=center, inner sep=3.5pt},
  a/.style={b, fill=black!7},
  h/.style={b, very thick},
  c/.style={b, dashed, fill=black!3},
  arr/.style={-{Stealth[length=1.3mm]}, thick},
  carr/.style={dashed, -{Stealth[length=1.3mm]}, black!55},
]
\node[a] (GN) at (-3.8,0) {observed anchor\\ $\GN$};
\node[a] (SC) at (2.4,0) {minimal-change source and Gaussian law\\
 and generated spectrum};
\node[b] (cont) at (2.4,-1.2) {generated physical continuation\\
 relative frames and matter module\\ scalar connector};
\node[b] (real) at (2.4,-2.4) {evaluated response readouts\\
 projectors and $R=5$\\ normalizations and transfer maps};
\node[h] (end) at (2.4,-3.6) {derived content and endpoint\\
 $\tau=\tfrac1{50}$\quad $(n_g,\kL)=(3,2.4935\ldots)$};
\node[b] (data) at (0,-4.9) {common dimensionless data and physical scales\\
 $\kL,\tau,\text{content};\quad \MP,\MG,k_{\rm GUT}$};
\node[b] (g) at (-3,-6.5) {gauge normalization and running\\ $g_i(\mu)$};
\node[b] (v) at (3,-6.5) {window and family transfers\\ $v$, masses and mixing};
\node[b] (obs) at (0,-8.4) {electroweak and weak observables\\
 QCD scales and masses\\ cosmological densities and temperatures\\ distances};
\node[c] (cmp) at (0,-10.0) {comparison references\\ measurements and estimates};

\draw[arr] (GN.west) -- (-5.4,0) -- (-5.4,-4.9) -- (data.west)
 node[pos=0.75,above,font=\tiny] {$\MP$};
\draw[arr] (SC) -- (cont);
\draw[arr] (cont) -- (real);
\draw[arr] (real) -- (end);
\draw[arr] (end) -- (data);
\draw[arr] (data) -- (g);
\draw[arr] (data) -- (v);
\draw[arr] (g) -- (obs);
\draw[arr] (v) -- (obs);
\draw[carr] (obs) -- (cmp) node[midway, right, font=\tiny, text=black!55] {comparison};
\end{tikzpicture}
\caption{Dependence of the observables on the common generated structure.
The upper branch shows the minimal-change source and spectrum followed by the
generated physical continuation and the finite response readouts evaluated
in this paper. Screening selects the content branch. The resulting response
problem then gives the highlighted joint content and endpoint solution.
Gauge normalization and mass transfer use the common source data in
parallel. Their outputs combine in the lower-scale calculations.
Arrows show the principal dependencies of these grouped calculations.
The observed anchor supplies $\MP$ for dimensional reconstruction.
Comparison values enter at the last stage.}
\label{fig:dependency}
\end{figure}

%% file: sections-en/sec04_content_symmetries.tex
% =============================================================================
% Section 4 - Matter content and torsion response
% The Spectrum of a Compact Internal Space II: Effective Couplings and Mass Scales
% =============================================================================

\section{Matter content and torsion response}\label{sec:content}

The charged representations specify the response of the matter content
to chiral and weak sources and to hypercharge. Their finite traces
determine both the normalized chiral response used for torsion and the
charge susceptibilities entering gauge matching. We use the minimal
joint module, its charged response quotient, the primitive charge
lattice and the pointwise vacuum stabilizer developed in the section on
joint response module and charge in
Ref.~\cite{Guo2026CompactInternalSpaceI}. Anomaly cancellation and Yukawa
covariance check the resulting assignments. The chiral derivative and
charge Hessians refer to the same normalized source and give
\(\tau=1/50\). Its charge moments also enter screening and hypercharge
matching.

\subsection{Complete matter and the charged response}\label{sec:count}

The full matter carrier has charge decomposition
$8_L+(7_R+1_{\rm neutral})$.  Exchange preserves its total capacity.
A common response on the two complete chiral copies has zero centre
trace.  The charge operator has a one-dimensional neutral
right kernel.  The charged quotient has $8_L+7_R$ components. These quotient counts enter the torsion trace. The
neutral line remains in the full matter carrier for interactions whose
operators act on it.

The five representations per generation are $Q_L = (3,2)$ and $L_L = (1,2)$
together with the singlets $u_R = (3,1)$ and $d_R = (3,1)$ and $e_R = (1,1)$.
The left-handed doublet $Q_L$ has $3 \times 2 = 6$ components and $L_L$
has $1 \times 2 = 2$ components. Hence
\[
N_L = 6 + 2 = 8 .
\]
The right-handed singlets give
\[
N_R = 3 + 3 + 1 = 7 .
\]
The number of Weyl components per generation in the charged response
quotient is
\[
N_f = N_L + N_R = 15 ,
\]
and the total over three generations is $45$.
The left-handed count equals the number of colour generators
\[
N_L = N_g = N_c^2 - 1 = 8 ,
\]
and the right-handed count is one less. The chiral excess in the charged response is
\[
N_L - N_R = 1 .
\]
The spin-structure convention on
$\pi_1(RP^3)=\mathbb Z_2$ is specified by $\tau_-$
in the representative convention of
Ref.~\cite{Guo2026CompactInternalSpaceI}.

Writing all fields as left-handed fields includes the conjugates of
the right-handed species. In this convention the hypercharge assignment satisfies
\[
\sum Y = 0 ,
\]
the vanishing of the mixed gravitational anomaly. The quadratic hypercharge moment is
\[
\sum Y^2 = \frac{10}{3} ,
\]
the sum $6 \cdot (1/6)^2 + 3 \cdot (2/3)^2 + 3 \cdot (1/3)^2 + 2 \cdot (1/2)^2 + 1 \cdot (1)^2$. The fermion content also obeys the identity
\[
N_f = 2 \sum Y^2 \Delta_f^2 = 15 ,
\]
with fermion conformal weight $\Delta_f=3/2$. These per-generation
identities determine the normalized chiral-content ratio used for torsion.

\subsection{The conformal--colour content identity}\label{sec:duality}

The full quadratic response algebra and local colour-frame
propositions of Ref.~\cite{Guo2026CompactInternalSpaceI} generate
$SU(W)\cong SU(3)$ on $W=\operatorname{Sym}^2S$ under the
frame-localization premise. Here $S$ is its complex two-dimensional spinor carrier. The associated
content identity uses the conformal coupling of a scalar in dimension $d$
\[
\xi = \frac{d-2}{4(d-1)} ,
\]
and with internal dimension $d = 3$ it is $\xi = 1/8$. The resulting identity is
\[
N_g \xi = 1 ,
\]
equivalently $N_g\Delta=2(d-1)$ with scalar conformal weight
$\Delta=(d-2)/2=1/2$. The generated value $N_g=8$ and the conformal
coefficient $\xi=1/8$ satisfy the identity. Since $N_c^2-1=8$ has
the single positive solution $N_c=3$ the relation $d=N_c=3$
compares the independently fixed internal dimension and colour number.

\subsection{The torsion parameter}\label{sec:torsion}

The torsion readout compares the net chiral response with the charge
capacity of the same one-generation source. The first derivative with
respect to the chiral source measures its left--right imbalance. The
connected second derivative with respect to hypercharge measures the
quadratic charge response. We evaluate both on the equal-weight charged
carrier with the inherited normalization of the hypercharge operator.
On the oriented three-dimensional carrier the isotropic and totally
antisymmetric tensor channel is spanned by its volume form. The
homogeneous torsion response therefore has a single amplitude
parametrised by
\[
T^a{}_{bc} = \frac{\tau}{L} \varepsilon^a{}_{bc} ,
\]
where the dimensionless modulus $\tau$ is the normalised chiral-content invariant. Its numerator is the chiral excess per fermion component
\[
\langle \chi \rangle = \frac{N_L - N_R}{N_f} = \frac{1}{15} ,
\]
and its denominator is the quadratic hypercharge moment
\[
\Pi_{\mathrm{ren}} = \sum Y^2 = \frac{10}{3} .
\]
Thus
\[
\tau = \frac{\langle \chi \rangle}{\Pi_{\mathrm{ren}}} = \frac{N_L - N_R}{N_f \sum Y^2} = \frac{1}{15 \cdot 10/3} = \frac{1}{50} .
\]
Per-component normalization of the chiral response and per-generation
normalization of the charge capacity give the source balance
$\Pi_{\mathrm{ren}}\tau=\langle\chi\rangle$. The chiral excess supplies
the oriented bias. The positive charge susceptibility sets the
restoring response against which that bias is measured. This balance
fixes the constitutive normalization of the single isotropic torsion
amplitude. The inherited normalization of $Y$ fixes the charge unit.
The per-component and per-generation traces fix the relative units of
the two responses. Both are evaluated on the equal-weight finite
content carrier before applying the spatial projectors
$P_\pm$. The resulting $\tau$ then enters the spectral mass shifts of
each assigned spatial branch. Replicating the one-generation carrier
multiplies the extensive bias and capacity by the same factor and preserves
their ratio. Relative chiral weights change the source itself and hence
its normalized bias. The charge operator has the normalization of
Ref.~\cite{Guo2026CompactInternalSpaceI}.
Appendix~\ref{sec:derivtau} evaluates this source and gives the
unequal-branch-weight expression \eqref{eq:p5-tau-weighted}.

\subsection{Torsion and geometric response}\label{sec:geodyn}

The content parameter enters geometry through torsion and the $J=2$
squash, which correct the tensor and gauge responses and the electroweak
readouts. The $J=2$ response uses the following dimensionless geometric
coefficients. Their dimensional form
carries an additional $L^{-2}$.
The Einstein--Cartan coefficient is
\[
\lambda_{\mathrm{EC}} = N_g \Bigl(1 + \frac{\tau}{2}\Bigr)^2 + 6 = 14 + 8\tau + 2\tau^2 ,
\]
and the effective tensor-response coefficient is
\[
\lambda_{\mathrm{TT}} = 14 = 2 N_R .
\]
Here $\mathrm{TT}$ labels the effective tensor response. The rough TT Laplacian eigenvalues are defined
separately in Ref.~\cite{Guo2026CompactInternalSpaceI}.
The squash amplitude is
\[
s_0 = n_{\mathrm{broken}} \, \tau = 2\tau ,
\]
with $n_{\mathrm{broken}} = 2$ the number of broken generators of $\mathfrak{su}(2)_R$. Its content ratio is
\[
\frac{s_0}{N_R} = \frac{1}{175} .
\]
The curvature relation of the Einstein-Cartan geometry is
\[
R(\omega) = R_{\mathrm{LC}} - \frac32 \Bigl(\frac{\tau}{L}\Bigr)^2 ,
\]
with $R_{\mathrm{LC}} = 6/L^2$. Torsion corrects the curvature at order $\tau^2 \sim 10^{-4}$. The gauge matching and electroweak order parameters use these same curvature and squash normalizations.

%% file: sections-en/sec03_internal_scale.tex
% =============================================================================
% Section 3 - The internal scale and the window capacity
% The Spectrum of a Compact Internal Space II: Effective Couplings and Mass Scales
% =============================================================================

\section{The internal scale and the window capacity}\label{sec:intscale}

For fixed torsion $\tau$ the remaining dimensionless data are the
response level $R$ and the tensor endpoint $\kL$. The ratio $c=\kL$
compares the internal radius with the resolution length $k^{-1}$ and
enters both the generation window and the amplitude transfers that set
the mass hierarchy. The common screening relation selects the
three-copy content branch. At fixed content and bandwidth the tensor
source response determines a unique endpoint, whose spectral window
must retain the same number of family positions. These requirements
give a unique endpoint on the selected branch and a finite bandwidth
range with unchanged active shells. The same screening response enters
the weak coupling.

\subsection{The common scale parameter}\label{sec:closurescope}

The window capacity $\kL$ relates the emergence scale, gauge couplings,
fermion masses and electroweak scale through a common internal resolution.
The response problem is specified by the field content, species projectors,
torsion and bandwidth. The Gaussian profile and transverse-traceless
readout with endpoint normalisation $4/27$ determine the endpoint. The
Planck anchor then supplies the mass unit.
The resulting hierarchy is exponentially sensitive to $\kL$. At fixed
other inputs $v\propto(\kL)^{-1}e^{-4\pi\kL}$ and
$m_e\propto e^{-20\kL}$. The logarithmic derivatives and exact finite changes are given in
Appendix~\ref{app:sensitivity}. The content ratio $\Omega_\Lambda$
is independent of this common scale.

\subsection{Spectral data}\label{sec:datum}

The internal space is $RP^3 = S^3/\mathbb{Z}_2$ with round radius $L$.
We use the spectral data and spin conventions collected in
Appendix~A of Ref.~\cite{Guo2026CompactInternalSpaceI}. The scalar Laplacian has the spectrum
\[
\lambda_l = \frac{l(l+2)}{L^2}, \qquad d_l = (l+1)^2, \qquad l = 0, 2, 4, \ldots ,
\]
where the restriction to even $l$ is the antipodal projection and $d_l$ is
the multiplicity. Fix the spin structure $\tau_-$ and the orientation and
deck-lift convention supplied by the companion paper. Its signed
Dirac spectrum is
\begin{equation}
\begin{aligned}
 \operatorname{Spec}D_{\tau_-}
 &=\left\{\frac{n+3/2}{L}:n\in2\mathbb N_0\right\}
   \cup\left\{-\frac{n+3/2}{L}:n\in2\mathbb N_0+1\right\},\\
 d_n^D&=(n+1)(n+2).
\end{aligned}
\label{eq:p5-signed-dirac-datum}
\end{equation}
The other spin structure interchanges the two parity assignments. Thus
$|D_{\tau_-}|$ has eigenvalues $(n+3/2)/L$ for all
$n\in\mathbb N_0$ and $D_{\tau_-}^2$ has eigenvalues
$(n+3/2)^2/L^2$ for all $n$ with multiplicity $d_n^D$ in each case.
Each $n$ contributes one signed eigenspace with this multiplicity.

Write $P_\pm=\mathbf1_{\mathbb R_\pm}(D_{\tau_-})$. The three quantities entering the response and generation count are
\begin{equation}
\begin{aligned}
 \text{generation positions: }&
 \operatorname{Spec}(L D_{\tau_-}|_{P_+})
       =\{n+3/2:n\in2\mathbb N_0\},\\
 \text{positive-branch squared tower: }&
 D_{\tau_-}^2|_{P_+},\qquad n\in2\mathbb N_0,\\
 \text{negative-branch squared tower: }&
 D_{\tau_-}^2|_{P_-},\qquad n\in2\mathbb N_0+1.
\end{aligned}
\label{eq:p5-dirac-consumer-types}
\end{equation}
The position-retention rule counts distinct positive positions
without weighting by $d_n^D$. The fermionic spectral trace on the charged
response carrier $\mathcal M_{\rm ch}$ uses squared eigenvalues and their
multiplicities. The chiral assignment
uses $P_-$ for $Q_L,L_L$ and $P_+$ for $u_R,d_R,e_R$.  Its uniform-even
comparison uses $P_+$ for every species.  These are specified restrictions
of $D_{\tau_-}^2$. The association of intrinsic sign projectors with
the indicated chiral species is part of the retained matter assignment.
Appendix~\ref{sec:p5-spectral-implementation} specifies these traces
and their numerical evaluation.

The volume is $\mathrm{Vol}(RP^3) = \pi^2 L^3$ and is half the round-sphere volume.
The six Killing vectors span the isometry algebra
$\mathfrak{so}(4)$. The vanishing first cohomology
$H^1(RP^3,\mathbb R)=0$ excludes harmonic one-forms. Connectedness
likewise makes the constants the only scalar Laplacian zero modes.

\subsection{Internal radius and Kaluza--Klein masses}\label{sec:kk}

The remaining radius is fixed by the dimensionless endpoint together
with the Planck anchor. We use the Kaluza--Klein mass readout of the
inherited internal spectrum~\cite{Guo2026CompactInternalSpaceI}
with the usual Kaluza--Klein interpretation~\cite{Kaluza1921,Klein1926}.
The scale relation is
\[
\GN \ \to\ \MP = \frac{1}{\sqrt{8\pi\GN}} \ \to\ M_{\mathrm{G}} = \frac{\MP\sqrt{\pi}}{\kL} ,
\]
where $M_{\mathrm{G}}$ is the emergence scale. The Kaluza--Klein masses are
\[
m_n^{\rm KK} = \frac{n + 3/2}{\kL} \, M_{\mathrm{G}} ,
\]
and the window edge is
\[
\kL \, M_{\mathrm{G}} = \MP \sqrt{\pi} .
\]
Put $\varkappa:=\kL$. At the emergence scale the physical radius and dimensionless Dirac operator
and generation cutoff are
\begin{equation}
\begin{aligned}
 L_{\mathrm{phys}}&=\frac{\varkappa}{M_{\mathrm G}},
 &\widehat D&=L_{\mathrm{phys}}D_{\tau_-},\\
 c_{\mathrm{gen}}&=\varkappa^2,
 &\Lambda_{\mathrm{gen}}&=\frac{c_{\mathrm{gen}}}{L_{\mathrm{phys}}}
   =\varkappa M_{\mathrm G}=\MP\sqrt\pi.
\end{aligned}
\label{eq:p5-paper4-window-map}
\end{equation}
On the positive branch $\widehat\lambda_n=n+3/2$ so the condition
$\widehat\lambda_n<c_{\mathrm{gen}}$ is equivalent to
$m_n^{\rm KK}<\Lambda_{\mathrm{gen}}$. The generation cutoff selects the family
positions. The Gaussian spectral trace evaluates each assigned spatial
tower with its own mode multiplicities and response bandwidth.
The Kaluza--Klein scale measures the internal spectral cost of resolving
each mode. The family masses measure the chirality-changing response
between retained matter states. They are read from the mass vertex in
the canonical kinetic normalization with the hierarchy set by accumulated
transfer along the retained positions.

\subsection{The fixed-branch spectral endpoint}\label{sec:closure}

The endpoint is defined by the transverse-traceless (TT) response at
zero external momentum. Each internal mode contributes its tensor
response with its field weight, multiplicity and Gaussian suppression.
Summing these contributions fixes the compact-source scale used by
the different sectors.  The tensor-survival maximum fixes the normalization of the summed response.
Before restriction to the source bandwidth the Gaussian sum on $RP^3$ is
\[
\Pi^2_{\mu\nu}(L,\Lambda^2) = \frac{1}{V_3} \sum_{\text{fields}}\sum_{\text{modes}} d_n \, w \int_0^\infty \frac{d\omega}{\pi}\; \frac{(\omega^2+\lambda_n)^2}{(\omega^2+\lambda_n+m^2_{\eff})^2}\, e^{-(\omega^2+\lambda_n)/\Lambda^2} ,
\]
Here $V_3=\pi^2L^3$. The mode multiplicity is $d_n$ and $w=4$
for each gauge polarisation or Weyl fermion and $w=1$ for each real
scalar. The second argument is the squared Gaussian cutoff. The
field content and $\tau$ are suppressed. With $q^2=\omega^2+\lambda_n$
the kernel satisfies $K_{\rm TT}(q^2,0)=1$ for $q^2>0$.
The Gaussian factor encodes the inherited composition law for unbiased
change. The rational factor is the nonlinear response of each mode to
its effective mass. The trace combines this response with the Gaussian
weight on the same generated spectrum.
The flat-space comparison uses the Ward-normalised subtracted amplitude.
Appendix~\ref{sec:derivkl} specifies the curvature and torsion shifts
and the assigned spectral branches.

The endpoint normalization is the mass-response extremum
\[
\frac{4}{27} = \max_{0 \le y \le 1} y(1-y)^2 \qquad \text{at} \qquad y = \frac13 ,
\]
where $y=m^2/(q^2+m^2)$ is the mass fraction of
$K_{\rm TT}(q^2,m^2)=(1-y)^2$. The variable $q^2$ is the internal
spectral argument whereas $k$ is the running scale.
The single-leg survival fraction $1-y$ multiplies under tensor
composition and gives the two-leg response $K_{\rm TT}=(1-y)^2$.
For $0<y<1$ its response to the logarithmic mass--momentum ratio
$\zeta=\log[y/(1-y)]=\log(m^2/q^2)$ is normalized per tensor leg as
\[
 -\frac12\frac{dK_{\rm TT}}{d\zeta}=y(1-y)^2.
\]
Thus the extremum measures the largest change of the surviving tensor
response under a relative change of mass and momentum.
The endpoint rule uses the maximum $4/27$ of this two-leg response
derivative to normalize the summed
compact-source response.
The macroscopic propagation criterion uses the metric response described
in Appendix~\ref{sec:p5-endpoint-boundary}.
The emergence endpoint is
\[
\frac{V_3 \, \widehat\Pi^2_{\mu\nu}(\kL, k^2/M_{\mathrm{P}}^2)}{32\pi^2} = \frac{4}{27}
\qquad \text{at} \qquad k = M_{\mathrm{G}} .
\]
The endpoint calculation uses the dimensionless
arguments $\ell_{\mathrm{end}}(k)=\varkappa M_{\mathrm G}/k=C/k$ and
$\widehat\Lambda_{\mathrm{tr}}(k)=k/\MP$ where
$C=\MP\sqrt\pi$. Its amplitude $\widehat\Pi^2$ is the displayed spectral
sum restricted to the specified response domain and evaluated with
$L=\ell_{\mathrm{end}}$ and squared cutoff
$\widehat\Lambda_{\mathrm{tr}}^2$.
The physical radius is instead $L_{\mathrm{phys}}$ in
\eqref{eq:p5-paper4-window-map}. At $k=M_{\mathrm G}=C/\varkappa$ the
arguments are $(\ell_{\mathrm{end}},\widehat\Lambda_{\mathrm{tr}})
=(\varkappa,\sqrt\pi/\varkappa)$ and leave one equation in one unknown.
The numerical calculation uses bisection on
$[0.3\,M_{\mathrm{G}},3\,M_{\mathrm{G}}]$ and self-consistency iteration.
The frequency response in the numerical tables has support
$\omega^2+\lambda_n<R\Lambda^2$ with $R=5$. In operator terms the
common collective source is read through the spectral projection
$P_R=\mathbf1_{[0,R\Lambda^2)}(H)$ with $H=\omega^2+A$.
Thus the retained Gaussian multiplier has relative weight greater than
$e^{-5}$ throughout the response domain. The inherited Gaussian law
fixes the envelope while $P_R$ specifies the bandwidth accepted by the
collective source. The $R=5$ response defines one common endpoint
for the downstream sectors. Numerical quadrature evaluates the integral
on this domain. The endpoint and mass relations test the physical
consequences of the common bandwidth.
Here TT denotes the tensor projection used to read the compact-source
susceptibility and fix this internal endpoint. Macroscopic tensor
propagation is described by the Einstein--matter realization of
Section~\ref{sec:gravitational}. The common Newton normalization links
that description to the dimensional scale used here.
The shell support and the three-position condition select the finite
response level used in the numerical evaluation.

\begin{definition}[Finite response level]\label{def:p5-response-level}
The finite response domain is parametrized by a real cutoff $R$ in
$P_R=\mathbf 1_{[0,R\Lambda^2)}(H)$. The physical level family consists
of the projections $P_N$ indexed by positive integers $N$ with
$\Lambda^2$ fixing the unit of the cutoff. A real cutoff
$\mathcal R$ interpolates this family in the sensitivity analysis.
\end{definition}

The shell-support condition preserves the species projectors entering
the compact-source trace, while position retention fixes the family
coefficient space. Both conditions must hold on the same endpoint
branch. Their inequalities select one member of the integer response
family.

\begin{proposition}[Finite response level]\label{prop:p5-response-level}
For the three-copy branch with $\tau=1/50$ and the stated positive shell
support the unique integer response level is $R=5$.
\end{proposition}

\begin{proof}
The endpoint $c_R=27A_R/4$ is strictly increasing in $R$ by
Proposition~\ref{prop:p5-response-bandwidth}.  The outward interval evaluation
at $R=4$ gives
\begin{equation}
 c_4\in\left[\frac{187}{80},\frac{23939}{10240}\right]
 =[2.3375,\,2.33779296875],
 \label{eq:p5-c4-enclosure}
\end{equation}
which lies below $\sqrt{11/2}=2.3452078799$.  The interval also separates the
endpoint response from $4/27$ with one derivative sign on the bracket. It
certifies a unique root at $R=4$.  Since $c_5=2.49353$ lies above
$\sqrt{11/2}$ the crossing $c_{R_*}=\sqrt{11/2}$ satisfies $4<R_*<5$.  The
shell support of Eq.~\eqref{eq:p5-shell-support-interval} gives
$R\leq16/\pi<6$ and the three-position condition requires $R>R_*$.  The only
integer in this interval is $R=5$.
\end{proof}

For $R=5$ the endpoint is
\[
\kL = 2.49353 .
\]

\begin{proposition}[Unique endpoint at fixed content and bandwidth]\label{prop:klunique}
The dimensionless endpoint has fixed-point function
\[
F(\varkappa) = \frac{\pi^2\varkappa^3\,
 \widehat\Pi^2_{\mu\nu}(\varkappa,\pi/\varkappa^2)}{32\pi^2}
 - \frac{4}{27}.
\]
Fix the field content and $\tau$ together with the species projectors and
dimensionless integration domain $R>0$. With $c=\varkappa=\kL$ the coefficient $A_R$ defined in
\eqref{eq:p5-homogeneous-coefficient} is independent of $c$ and $\MP$ and satisfies
\[
 F_R(c)=\frac{A_R}{c}-\frac4{27},\qquad
 F_R'(c)=-\frac{A_R}{c^2}<0,\qquad
 c_R=\frac{27}{4}A_R.
\]
Thus each fixed content branch has exactly one positive endpoint.
The reference level is $R=5$. The value $R=\infty$ denotes the full Gaussian integral
with the full specified towers. These roots have separate error bounds. This
specializes the TT scaling law derived in
Ref.~\cite{Guo2026CompactInternalSpaceI}.
Appendix~\ref{sec:p5-homogeneity} checks the present inputs and the
infinite-domain extension. The finite- and full-domain values are compared in
Appendix~\ref{sec:p5-gaussian-tail}.
\end{proposition}
The domain comparison is quantitatively significant. Holding the content and
projectors fixed together with the torsion modulus, Planck anchor and loop factor,
Eq.~\eqref{eq:p5-domain-hierarchy-enclosures} gives
\[
 v_\infty/v_5\in[0.10,0.13],\qquad
 m_{e,\infty}/m_{e,5}\in[0.03,0.04].
\]
The position count is discrete whereas the endpoint coefficient
integrates continuously varying mode weights over the response domain.
The displayed ratios measure the dependence of absolute masses on that
bandwidth. The integration error at a fixed bandwidth is controlled
separately. Proposition~\ref{prop:p5-response-bandwidth} also shows that the three-position plateau persists for
every $R\geq5$ whereas the two mass scales decrease together according to
\[
 \frac{v_R}{v_5}
 =\frac{c_5}{c_R}
   \left(\frac{m_{e,R}}{m_{e,5}}\right)^{\pi/5}.
\]
The common bandwidth therefore produces a correlated displacement of
the electroweak and electron scales, which can be tested across the two
sectors.

The shell support provides a further constraint. For the positively
weighted endpoint trace used here retaining the same active shells as
at $R=5$ and imposing three generation positions gives
\[
 R_*<R\leq\frac{16}{\pi},\qquad
 c_{R_*}=\sqrt{11/2},\qquad R_*\simeq4.0187119842.
\]
Equation~\eqref{eq:p5-joint-bandwidth-interval} proves existence and
uniqueness of this lower edge using monotonicity and outward finite-shell
bounds. The ratio of the two mass scales then removes both the Planck
anchor and their common squash factor
\[
 \frac{v}{m_e}
 =\frac{3}{16\pi^{5/2}}\frac{e^{(20-4\pi)c}}{c}.
\]
With both masses in common units this ratio is strictly increasing on
the three-position branch. An independently matched value therefore
determines at most one compatible endpoint. At fixed source data it also
determines at most one bandwidth. Equation~\eqref{eq:p5-scale-ratio-identification}
gives the inverse response. The charged-lepton invariant in
\eqref{eq:p5-lepton-transfer-invariant} tests the transfer rule after
eliminating the bandwidth as well.

Under the stated torsion mass shifts the three-copy capacity plateau
is stable throughout $0\leq\tau\leq0.1$ with the other response data
fixed. The finite endpoint bound is derived in
Appendix~\ref{sec:derivtau}. The central calculation uses the content
value $\tau=1/50$.

\paragraph{Weak-coupling inversion.}
For fixed $N_c>0$ and real $\tau$ the screened law
$4\pi/g_2^2=c^5/2+1/N_c-\pi\tau^2/2$ is strictly monotone in
$c>0$. A positive reference coupling therefore determines the unique
capacity
\begin{equation}
 c_{\rm ref}=\left[2\left(\frac{4\pi}{g_{2,\rm ref}^{2}}
 -\frac1{N_c}+\frac{\pi\tau^2}{2}\right)\right]^{1/5},
 \label{eq:p5-weak-window-inverse}
\end{equation}
whenever the quantity in parentheses is positive.

The independently extrapolated Standard-Model reference coupling gives
\(c_{\rm ref}\) while the spectral coefficient \(A_R\) determines the
endpoint of Proposition~\ref{prop:klunique} before that comparison.
Under the common content and torsion response and matching convention their
unrounded central values satisfy
\begin{equation}
 \left|\frac{c_{\rm ref}}{c}-1\right|<3\times10^{-6}.
 \label{eq:p5-endpoint-consistency}
\end{equation}
Both give \(2.49353\) at the displayed precision with signed percentage
residual \(100(c_{\rm ref}/c-1)\%\simeq0.00022\%\).
This is a central-value consistency test of the spectral endpoint and
screening law. The certified endpoint enclosure and reference-coupling
rounding retain the separate accuracy stated in
Appendix~\ref{app:sensitivity}.

The Gaussian width of the scale field is $L_{\mathrm{Cg}}=\sqrt\pi$
which fixes the critical curvature
\[
R_{\mathrm{c}} = \frac{6}{\pi} .
\]
The window capacity is then
\[
(\kL)^2 = 6.2177 ,
\]
which sits below the Euclidean-period value $2\pi = 6.2832$ by $1.04\%$.

\subsection{The generation count}\label{sec:gencount}

The generation rule assigns a family position to each retained positive
Dirac level with unit weight for each position. The retained-position
count must agree with the copy number entering the same source trace.
With the map \eqref{eq:p5-paper4-window-map} the criterion is
\[
n + \frac32 < (\kL)^2 .
\]
With $(\kL)^2 = 6.2177$ the retained positions are
\[
n = 0 \ (1.5), \qquad n = 2 \ (3.5), \qquad n = 4 \ (5.5) ,
\]
The position $n = 6$ gives $7.5$ and lies above the edge. The
selected window therefore contains exactly three positions and the
generation count under this retention rule is
\[
n_g = 3 .
\]
The family coefficients have a concrete representation on these shells.
Write $F_j^+$ for the positive shell with $n=2j$ and
$S\simeq\mathbb C^2$ for the spin-$1/2$ carrier of the isotropy group
$H_{\rm iso}\simeq SU(2)$. Each of the three shells contains one copy
of $S$. Its normalized radial intertwiner supplies one family
coefficient. Their coefficient space is
\[
 K=\operatorname{Hom}_{H_{\rm iso}}
       (S,F_0^+\oplus F_1^+\oplus F_2^+),\qquad \dim K=3.
\]
The products of these intertwiners act as matrix units on $K$ and identify
it with the minimal three-copy family carrier. The kinetic form and mass
vertices transform together under this identification
(Proposition~\ref{prop:p5-replicated-quadratic-response}). The spin
structure and species projectors fix the spatial modes in each copy.
The family matrices act on their common coefficient space.
The full $|D_{\tau_-}|$ window at the same edge contains five distinct
positions $n=0,1,2,3,4$ and the retained positive-branch eigenspaces
have total dimension $2+12+30=44$. These are the spatial counts.
The three-dimensional intertwiner space carries the family count.

The mass ladder assigns family amplitudes $m_n\propto e^{-\alpha_f n}$ to these positions
with extrusion orders $n=\{0,2,4\}$ and sector-dependent rates $\alpha_f$.
The capacity check fixes the retained positions. The weighted traces
remain those assigned to each species.
The step structure of the criterion is shown in Figure~\ref{fig:plateau}.

The step boundaries give a direct interval certificate for this count.
Every positive capacity satisfying $11/2<c^2\leq15/2$ selects exactly
$n=0,2,4$. The third position is $4+3/2=11/2$ whereas the next is
$6+3/2=15/2$. Hence an enclosure $[c_-,c_+]$ certifies the
three-position plateau whenever $c_-^2>11/2$ and
$c_+^2\leq15/2$.
The content used in the trace must agree with the count read from its
endpoint. For $g$ copies of the inherited one-generation carrier the
extensive adjoint-to-matter response ratio is $1/g$. Requiring the same
inverse weak-source coefficient to obey the colour-normalised
screening closure $N_c\delta_{\rm col}=1$ gives
$1/g=1/N_c$ and hence $g=N_c=3$
(Proposition~\ref{prop:p5-screening-compatibility}). The endpoint of this
three-copy source then passes the three-position plateau test. The
screening conditions select the content branch and the window verifies
that its retained family positions reproduce the same count. These
selections are preserved by simultaneous unitary changes of mode basis.
The source and kinetic form transform together with the mass vertices and leave
the response traces and canonically normalized masses invariant. The endpoint
branches are enumerated in Appendix~\ref{sec:p5-joint-content}.

\input{figures/fig_plateau}

%% file: figures/fig_plateau.tex
\begin{figure}[t]
\centering
\begin{tikzpicture}[font=\scriptsize, >=Stealth]
\draw[->] (0,0) -- (9.6,0) node[below] {$c_{\mathrm{gen}}=(kL)^2$};
\draw[->] (0,0) -- (0,4.9) node[left] {$N_{\mathrm{win}}(c_{\mathrm{gen}})$};
\draw[fill=black!7] (5.5,0) rectangle (7.5,4.5);
\foreach \x/\lab in {1.5/$\frac32$,3.5/$\frac72$,5.5/$\frac{11}{2}$,7.5/$\frac{15}{2}$} {
  \draw[dashed, black!35] (\x,0) -- (\x,4.5);
  \node[below] at (\x,0) {\lab};
}
\draw[very thick] (0,0) -- (1.5,0);
\draw[very thick] (1.5,1) -- (3.5,1);
\draw[very thick] (3.5,2) -- (5.5,2);
\draw[very thick] (5.5,3) -- (7.5,3);
\draw[very thick] (7.5,4) -- (9.0,4);
\foreach \x/\y in {1.5/0,3.5/1,5.5/2,7.5/3} {
  \fill (\x,\y) circle (1.5pt);
  \draw[fill=white] (\x,\y+1) circle (1.5pt);
}
\foreach \y in {1,2,3,4} \node[left] at (0,\y) {$\y$};
\node[above] at (6.5,4.5) {three-position plateau $N_{\mathrm{win}}=3$};
\draw[red!70!black, thick] (6.2177,0) -- (6.2177,4.5);
\node[red!70!black, below] at (6.2177,0) [yshift=-8pt] {$(\kL)^2=6.2177$};
\draw[blue!60!black, dashed] (6.2832,0) -- (6.2832,4.5);
\node[blue!60!black, right] at (6.2832,2.0) {$2\pi$};
\end{tikzpicture}
\caption{Window count for the selected positive Dirac positions. Each position has unit weight. The thresholds are $n+\tfrac32$ for $n=0,2,4,6$ with the strict retention rule giving $N_{\mathrm{win}}=3$ on $\tfrac{11}{2}<c_{\mathrm{gen}}\leq\tfrac{15}{2}$. Both the recorded endpoint $(\kL)^2=6.2177$ and the Gaussian benchmark $2\pi=6.2832$ lie inside this interval.}
\label{fig:plateau}
\end{figure}

%% file: sections-en/sec05_gauge_couplings.tex
% =============================================================================
% Section 5 - Gauge couplings
% The Spectrum of a Compact Internal Space II: Effective Couplings and Mass Scales
% =============================================================================

\section{Gauge couplings}\label{sec:gauge}

Ref.~\cite{Guo2026CompactInternalSpaceI} fixes the common connection and
its action on each matter representation. Canonical normalization of
the quadratic response determines the interaction strengths. The
endpoint of Section~\ref{sec:intscale} fixes the common scale, and the
representations of Section~\ref{sec:content} determine the matter
response. Channel normalization, screening and finite matching give
the boundary couplings for scale evolution. Experimental values serve
as comparison data.
The isometry-breaking matching scale is
$k_{\rm GUT}=\MP\sqrt\pi\,\tau/\sqrt3$ set by the $J=2$
isometry breaking at $L_{\rm GUT}=\sqrt3/\tau$.
Here $k_{\rm GUT}$ denotes this geometric matching scale. Each channel
has its own generator normalization and finite matching coefficient.
The three boundary couplings are normalized channel responses of one
connection. Their evolution with the retained matter content correlates
the weak, electromagnetic and strong interactions across scales.

\subsection{Weak coupling and screening}\label{sec:g2}

For the $\mathfrak{su}(2)_L$ isometry modes on $RP^3$ the integrated
Killing-mode norm fixes the quadratic response. Canonical normalization
therefore makes the squared coupling inversely proportional to this norm.
The Killing-response convention with scale factor $(M_{\mathrm G}/\MP)^2$ gives the bare coupling
\[
g_2^2 = 8 \, \Bigl(\frac{M_{\mathrm{G}}}{\MP}\Bigr)^2 (\kL)^{-3} .
\]
The window-edge identity $\kL M_{\mathrm{G}} = \MP \sqrt{\pi}$ of Section~\ref{sec:kk} gives
\[
g_2^2 = \frac{8\pi}{(\kL)^5}, \qquad \alpha_W(M_{\mathrm{G}}) = \frac{g_2^2}{4\pi} = \frac{2}{(\kL)^5} ,
\]
so the bare weak coupling at the emergence scale depends only on the
window capacity in this response normalization. The common action
fixes the strength by normalizing the gauge kinetic term together with
the matter current. A constant rescaling of a mode changes both
coefficients and leaves the canonical coupling invariant.
Appendix~\ref{sec:derivg2} gives this reduction and identifies the
absolute response normalization. In the common connection convention screening changes
the coefficient of the quadratic gauge term. The matter and torsion
contributions are organized additively in the inverse coupling through
the relation
\[
\frac{1}{\alpha_{\mathrm{SM}}} = \frac{1}{\alpha_W} + \frac{1}{N_c} - \frac{\tau^2 \pi}{2},
\qquad
N_c\Bigl(\frac{1}{\alpha_{\mathrm{SM}}} - \frac{1}{\alpha_W} + \frac{\tau^2 \pi}{2}\Bigr) = 1 ,
\]
where $\alpha_W=2/(\kL)^5$. The colour term is the ratio of extensive
adjoint and matter responses equal to $1/N_c$ on three generations.
The torsion term combines the quadratic Einstein--Cartan coefficient with
the Euclidean period and $N_g\xi=1$ in this additive screening relation.
Appendix~\ref{sec:derivg2} evaluates these factors. 
Proposition~\ref{prop:p5-screening-compatibility} selects the common
copy number $g=3$ under the two screening rules. The resulting value is
\[
g_2(M_{\mathrm{G}}) = \sqrt{\frac{4\pi}{1/\alpha_W + 1/N_c - \tau^2 \pi/2}} = 0.508848 ,
\]
The unrounded value differs from the reference $0.508845$ by
$+0.0005\%$. The two screening terms reduce the bare Killing value.
Its difference from the same reference is $+0.34\%$.

\subsection{The hypercharge coupling}\label{sec:g1}

The $\mathrm{U}(1)_Y$ matching rule accompanies the
$\mathfrak{su}(2)_R \to \mathrm{U}(1)_R$ breaking by the $J = 2$ squash.
The surviving axial mode acquires a different curvature norm from its
round-space value. With the axial Killing generator held fixed its
metric-dual one-form changes with the squash. For an amplitude constant
over the internal space, preservation of the curvature quadratic norm
of the same Abelian connection rescales its amplitude by $\kappa(s)$. The
inverse squared norm ratio is
\[
\kappa^2(s) = \frac{1 + s}{(1 - 2s)^{5/2}} ,
\]
which at $s_0=2\tau$ gives $\kappa(2\tau)=1.13183$.
Appendix~\ref{sec:p5-axial-normalisation} computes the norm ratio and
the corresponding matter-mode profile. The matching is imposed at
$k_{\rm GUT}=\MP\sqrt\pi\,\tau/\sqrt3$ for which
$k_{\rm GUT}/M_{\rm G}=\kL\tau/\sqrt3\simeq0.02879$.
The coupling is then transported to $M_{\rm G}$ by the running equations.
The multiplicative matching correction uses
\[
\delta_{g_1} = - \tau \, r_{23} \, \sum Y^2 \Delta_f \xi ,
\]
where $r_{23}=1/(\sqrt3\,\sum Y^2)=3/(10\sqrt3)$ is the
hypercharge-trace hierarchy ratio. The content product is
$\sum Y^2\Delta_f\xi=5/8$ with $\Delta_f=3/2$ and $\xi=1/8$ from
Section~\ref{sec:content}. The complete boundary relation is
\[
 g_1(k_{\rm GUT})
 =g_2(k_{\rm GUT})\,\kappa(2\tau)\,(1+\delta_{g_1}).
\]
It combines the axial curvature factor with the finite content
correction in the stated external-mode normalization before scale
evolution. The curvature and derivative terms of the internal axial
profile have distinct Gram factors; their calculation is given in
Appendix~\ref{sec:p5-axial-normalisation}. Transport to the emergence scale gives
\[
g_1(M_{\mathrm{G}}) = 0.60499 .
\]
Here and in the running equations $g_1=\sqrt{5/3}\,g_Y$.
Together with $g_2$ it determines the fine-structure constant through
$e^{-2}=g_Y^{-2}+g_2^{-2}$ and $\alpha_{\rm em}=e^2/(4\pi)$.
The common two-loop transport gives $\alpha_{\rm em}^{-1}(M_Z)=128.206$.
The reference value $127.952$ gives a deviation of $+0.20\%$.

\subsection{The colour coupling}\label{sec:g3}

The colour coupling is matched to the $\mathfrak{su}(2)$ Killing normalization at leading order. The long-root correction at the breaking scale is
\[
g_3(k_{\mathrm{GUT}}) = g_2(k_{\mathrm{GUT}}) \, \Bigl(1 + \frac{\alpha_{\mathrm{GUT}}^2}{K}\Bigr) ,
\]
where $\alpha_{\mathrm{GUT}} = g_2(k_{\mathrm{GUT}})^2/(4\pi)$ and
\[
K = \frac{J(J+2)}{d} = \frac{8}{3}
\]
is the coefficient assigned by the long-root matching rule as
the squash eigenvalue $J(J+2)=8$ divided by $d=3$.
Its positive correction places $g_3$ above $g_2$ at the breaking scale.
Two-loop transport with the geometric Yukawa input $y_0=1$ gives
\[
g_3(M_{\mathrm{G}}) = 0.49776 ,
\]
and the strong coupling at the electroweak scale is
\[
\alpha_s(M_Z) = 0.11804 ,
\]
a signed residual of $+0.120\%$ against the reference value $0.1179$.

\subsection{The two-loop coefficients}\label{sec:beta}

The running uses the two-loop $\overline{\mathrm{MS}}$ beta functions
of Sections~IV.1 and VI of Ref.~\cite{LuoWangXiao2003} evaluated on
the retained representations. The endpoint, torsion modulus and family
carrier supply the generated data for this evolution. The running
scale labels the external energy resolution at which the effective
interaction coefficients are evaluated. The same matter charges control
the relative running from the geometric matching and emergence scales
to laboratory scales. Appendix~\ref{sec:derivbeta} specifies
the scale derivative and loop factors together with generator normalisation. It
computes the coefficients from the Casimirs and Dynkin indices.
At one loop they are
\[
b_1 = \frac{41}{10}, \qquad b_2 = -\frac{19}{6}, \qquad b_3 = -7 ,
\]
and the two-loop matrix is
\[
B = \begin{pmatrix}
199/50 & 27/10 & 44/5 \\
9/10 & 35/6 & 12 \\
11/10 & 9/2 & -26
\end{pmatrix} .
\]
The top-Yukawa and Higgs-quartic beta functions use the same content.
For example the one-loop Yukawa coefficient is $9/2=3/2+N_c$
and each gauge term is $-3[C_2(Q_L)+C_2(u_R)]$.
Appendix~\ref{sec:derivytlam} supplies the two-loop expressions and
potential convention.

%% file: sections-en/sec07_ewsb_cp.tex
% =============================================================================
% Section 7 - Electroweak breaking and CP
% The Spectrum of a Compact Internal Space II: Effective Couplings and Mass Scales
% =============================================================================

\section{Electroweak breaking}\label{sec:ewsb}

The scalar sector supplies a stable nonzero condensate and an
electroweak amplitude far below the emergence scale. The internal squash,
colour-singlet condensate and electroweak vacuum expectation value have
distinct field normalizations. The elementary connector couples retained
modes, while the condensate is a collective coordinate of their spectral
response. With canonical kinetic normalization the fermion mass measures
the coupling of opposite chiralities and the radial scalar mass measures
the restoring response of this coordinate.
Squash stationarity fixes its quartic coupling. The complete spectral
potential of Ref.~\cite{Guo2026CompactInternalSpaceI} determines the
condensate and the restoring mass of its radial fluctuations. Two
successive window crossings determine the canonically normalized
electroweak amplitude \(v\). These constructions supply the Higgs-quartic
matching condition in the common action. The same \(v\) normalizes the
fermion mass vertices.

\subsection{The squash quartic}\label{sec:lambda}

The geometric order parameter is the $J=2$ squash amplitude $\varphi$. Stationarity at its torsion-selected value $\varphi_0=2\tau$ fixes the quartic coefficient
\[
\lambda = \frac{\xi (R_{\mathrm{c}} - R_{\mathrm{GUT}})}{(2\tau)^2} = 149.145 ,
\]
where $\xi=1/8$ and $R_{\mathrm c}=6/\pi$ and $R_{\mathrm{GUT}}=6/L_{\mathrm{GUT}}^2$ with $L_{\mathrm{GUT}}=\sqrt3/\tau$. Appendix~\ref{sec:derivewsb} specifies the torsion coefficients and the amplitude convention. This $\lambda$ belongs to the squash coordinate. The Higgs quartic $\lambda_H$ follows the separate pseudo-dilaton matching relation of Appendix~\ref{sec:derivlamh}.
That relation and the generated $v$ determine the tree-level mass
$m_H=\sqrt{2\lambda_H}\,v$. The finite matching of
Proposition~\ref{prop:p5-generated-onshell-matching} then determines
the renormalized action coefficients with this generated mass as
the Higgs physical condition.

\subsection{The positive colour-singlet scalar mass}\label{sec:massgap}

Below the critical internal curvature, the quadratic coefficient at
zero condensate becomes negative. With the inherited feedback and
subtraction the complete spectral potential of
Ref.~\cite{Guo2026CompactInternalSpaceI} then has a nonzero stationary
amplitude with a positive normalized Hessian. This
Hessian measures the restoring response of the condensate and gives
its colour-singlet radial mass. Appendix~\ref{sec:p5-complete-condensate}
places the internal-radius path used here in that branch. The
long-root quadratic coefficient at zero amplitude is
\[
m_{\mathrm{long}}^2 = K (R - R_{\mathrm{c}}) < 0 \qquad \text{for} \qquad R < R_{\mathrm{c}} ,
\]
with $R_{\mathrm{c}} = 6/\pi$ and $K=8/3$. In the unit-normalized
Landau limit the radial squared mass is
\[
m_{\delta,\mathrm L}^2 = \frac{16}{3} (R_{\mathrm{c}} - R) > 0 .
\]
For the complete logarithmic potential this is a strict upper bound
$0<m_\delta^2<m_{\delta,\mathrm L}^2$. The exact value is the
normalized Hessian in \eqref{eq:p5-complete-condensate-readout}.
Ref.~\cite{Guo2026CompactInternalSpaceI} gives the conditions and error bound
for the Landau approximation.
In the dimensionless internal-radius convention
$L(M_{\rm G})=c$ and $L_c=\sqrt\pi$. The certified endpoint satisfies
$c>2.491>\sqrt\pi$ and gives the negative quadratic coefficient
$m_{\rm long}^2(M_{\rm G})=16(c^{-2}-\pi^{-1})<-2.50$.
Its numerical value is $-2.52$ in those units.
Along $L(k)=cM_{\rm G}/k$ every $0<k\leq M_{\rm G}$ remains in the
condensed domain. This includes the infrared endpoint
$k=\sigma_C^{-1}$ where the Hubble length is
$\sigma_C=6.95\times10^{41}\ \mathrm{GeV}^{-1}$.
This establishes the condensed stationary branch over the stated
scale range. For specified coefficient paths the same analysis
bounds the gap uniformly on a finite scale interval as detailed in
Appendix~\ref{sec:p5-uniform-gap-certificate}.

For the radial curvature observable the proposition
``Spectrum of the radial curvature observable'' in
Ref.~\cite{Guo2026CompactInternalSpaceI} gives a positive mass atom at
$m_\delta^2$ together with the two-particle and three-particle and four-particle
continua. This is the spectral response of the Wick-ordered
polynomial of the internally integrated scalar intensity in the
common quadratic external theory. Its mass atom occurs at the radial
gap.

The Landau stationary amplitude is
\[
\langle E \rangle_{\mathrm L}^2 = \frac{8}{3} \frac{R_{\mathrm{c}} - R}{\lambda_E} ,
\]
and $m_{\delta,\mathrm L}^2=2\lambda_E\langle E\rangle_{\mathrm L}^2$.
Here $\lambda_E$ multiplies the quartic term in the condensate
coordinate $E$. Appendix~\ref{sec:p5-condensate-coordinate} gives the
coordinate and kinetic normalization needed to compare it with the
squash coefficient $\lambda$. The inherited complete scalar mass operator is
$\Delta_0+m_\delta^2 I$ including the constant mode as its lowest
positive-mass excitation. Its mass is therefore the lower edge of
this scalar spectrum. Section~\ref{sec:glueball} uses the separate
$8\Lambda_{\mathrm{QCD}}$ mass normalization. An identification with
the lowest condensate mass requires a common scale and source
normalization including the constant scalar mode.

\subsection{The vacuum expectation value}\label{sec:vev}

The electroweak readout follows the creation and stabilization
transfers of the mass amplitude. Each crossing contributes
$e^{-2\pi\kL}$. Their product carries the hierarchy between the
emergence scale and the electroweak scale. With the loop factor and the channel
and squash normalizations the resulting vacuum expectation value is
\[
v = M_{\mathrm{G}} \, \frac{3\alpha}{\pi} \, e^{-4\pi \kL} \, (1 - s_0 \kappa) = 246.19 \ \mathrm{GeV} ,
\]
where $\alpha=1/(16\pi^2)$ is the loop factor and $s_0=2\tau$ and $\kappa=1.13183$ are the squash data of Section~\ref{sec:g1}. The value differs from the reference $246.22$ GeV by $-0.012\%$. The exponential $e^{-4\pi\kL}$ supplies the hierarchy $v/M_{\mathrm G}\sim10^{-16}$. Appendix~\ref{sec:derivewsb} gives the two window crossings and their prefactor.

The total generator $4\pi\kL$ and the electron cascade generator
$20\kL$ both depend on the same window capacity. Their prefactors
and source-channel assignments in Table~\ref{tab:p5-hierarchy-selection}
therefore imply the correlated scale relation
\eqref{eq:p5-common-bandwidth-response}. Changing the bandwidth
moves the electroweak and electron masses together. The complete condensate
amplitude $E_*$ and feedback coefficient $g$ retain their independent
stationary definitions in Appendix~\ref{sec:p5-complete-condensate}.

%% file: sections-en/sec06_flavour.tex
% =============================================================================
% Section 6 - Flavour structure
% The Spectrum of a Compact Internal Space II: Effective Couplings and Mass Scales
% =============================================================================

\section{Flavour structure}\label{sec:flavour}

At fixed electroweak amplitude $v$, the flavour construction assigns
hierarchical mass amplitudes to the three retained family positions.
Repeated transfers of the same source
amplitude accumulate suppression, so different family positions and
sector rates produce large mass ratios from a common internal scale.
The window capacity and torsion enter these rates. The electron scale
supplies the second ladder normalization while $v$ normalizes the top mass.
The neutrino spectrum and angular texture use their own source and
propagation relations (Appendices~\ref{sec:derivnu} and~\ref{sec:neutrino}).
The charged-fermion transfer operators give two mass combinations
independent of the common transfer parameters. Their comparison with
physical masses uses finite conversion to a common renormalization scheme.

\subsection{The ladder}\label{sec:ladder}

The three retained generation positions have extrusion orders
$n=\{0,2,4\}$ as specified in Section~\ref{sec:gencount}. The transfer
acts on the coefficient of the canonically normalized chirality-changing
mass vertex. It composes source operations on the retained modes and
thus determines an effective coupling between their left and right
components. Successive operations multiply this amplitude and make
their attenuation additive in its logarithm. Propagation and reaction
rates follow from the common action containing these vertices with
evolution read against a physical clock. On an unmodified
rung, stationary composition of positive
amplitudes gives $a_f(\ell_1+\ell_2)=a_f(\ell_1)a_f(\ell_2)$ and
$a_f(0)=1$. Continuity then gives $a_f(\ell)=e^{-\alpha_f\ell}$.
The sector rules determine its rate and the light-family insertions.
In the up sector the rate is the window width reduced by the
torsion deficit
\[
\alpha_{\mathrm{up}} = \kL - 2\tau ,
\]
The same source response supplies the positive tilt decrement
$1-n_s=\tau w_{\mathrm W}$ with
$w_{\mathrm W}=(1+2+1+3)/(d+1)=7/4$ for $d=3$ and the local
window $\kL_{\mathrm{CMB}}=\kL(1-\tau/4)$.
These are determined by the common endpoint and torsion. The Weyl-window
ratio of $RP^3$ enters their cosmological interpretation in
Appendix~\ref{sec:derivcosmo}. The subscript CMB labels the corresponding
cosmic-microwave pivot readout. Summing the mean response over the six
internal isometry directions gives the sector decrement
\[
\Delta = 6 (1 - n_s) \kL_{\mathrm{CMB}} ,
\]
We denote the positive decrement by
$n_s^{\mathrm{tilt}}\equiv 1-n_s$. The down and lepton ladders split
by the hypercharge ratio
\[
\frac{9}{8} = \frac{1}{1 - (Y_d/Y_l)^2} ,
\]
where $Y_d=|Y_{d_R}|=1/3$ and $Y_l=|Y_{e_R}|=1$ are the right-singlet hypercharge magnitudes. The rung exponents are
\[
\alpha_{\mathrm{dn}} = \alpha_{\mathrm{up}} - \frac{18}{17}\Delta , \qquad
\alpha_{\mathrm{lp}} = \alpha_{\mathrm{up}} - 2\Delta ,
\]
with the down first-generation step carrying the colour-dilution correction
\[
\alpha_{\mathrm{sd}} = \alpha_{\mathrm{dn}} - \frac{\kL_{\mathrm{CMB}}}{6} .
\]
Proposition~\ref{prop:p5-spectral-mass-transfer} realizes these
ladder and colour-dilution rules by positive operators on the
generated family module. The angular rotations orient these mass
operators in a common left-handed kinetic space. If $L_u$ and $L_d$
diagonalize the canonical left mass-squared operators the charged-current
overlap is $V_{\rm CKM}=L_u^\dagger L_d$. Thus the masses measure the
strengths along the family directions while mixing measures their
relative orientation. A common change of mode basis preserves both
the masses and these overlaps as expressed by the projector invariants
in \eqref{eq:p5-common-flavour-invariants}. Appendix~\ref{sec:derivlz}
specifies the additional source path and level-crossing data needed
for a Landau--Zener interpretation.

The source algebra supplies the projectors, charges and allowed
operations~\cite{Guo2026CompactInternalSpaceI}. The positive transfers in
Table~\ref{tab:p5-hierarchy-selection} act on the mass amplitudes.
Equal attenuation per content unit, the sector split and the number of
crossings determine the coefficients and their joint mass relations.

\begin{table}[tbp]
\centering\small
\setlength{\tabcolsep}{4pt}
\begin{tabular}{@{}p{0.16\linewidth}p{0.27\linewidth}p{0.30\linewidth}p{0.21\linewidth}@{}}
\toprule
Exponent & Source operation and transfer & Coefficient rule & Testable consequence \\
\midrule
$2\alpha_{\rm up}$ & Two-order transfer on the family projector $P_1$
in $B_u$ & Width $c=\kL$ per order reduced by the two torsion
directions gives $\alpha_{\rm up}=c-2\tau$
& $\log(m_t/m_c)=2(c-2\tau)$ \\[3pt]
$2\alpha_{\rm dn}$ and $2\alpha_{\rm lp}$ & Charge-weighted shifts of
the same family transfer generators $B_d,B_\ell$
& Mean step $\Delta=6(1-n_s)c_C$. Consecutive decrements in ratio
$9/8$ with total $2\Delta$
& The joint ratio relation \eqref{eq:p5-sector-mass-test} \\[3pt]
Light-family insertions & Diagonal $P_2$ terms in
\eqref{eq:p5-positive-transfer-generators}
& Up-sector factor $4$ and down-sector dilution $c_C/6$ with lepton
period $\sqrt{2\pi}$ enter their stated amplitude channels
& The first-generation mass ratios \\[3pt]
$20c$ & Four successive electron-amplitude transfers each with
generator $5cI$
& One-generation charge weight $\sum Y^2\Delta_f=5$. Equal attenuation
per level and content unit
& $m_e/\MP=e^{-20c}(1-s_0\kappa)$ \\[3pt]
$4\pi c$ & Creation and stabilization transfers each
$e^{-2\pi c}$ in the electroweak amplitude
& Two equal crossings with the squash normalization and loop and Fourier prefactor
of Appendix~\ref{sec:derivewsb}
& The shared bandwidth response
\eqref{eq:p5-common-bandwidth-response} \\
\bottomrule
\end{tabular}
\caption{Source data and transfer generators with the physical selection rules
for the main hierarchies. Here $c_C=\kL_{\rm CMB}$ and $P_1,P_2$
are the middle- and light-family projectors. The rates specify mass
amplitudes. Transition probabilities require the separate dynamical
identification in Appendix~\ref{sec:derivlz}.}
\label{tab:p5-hierarchy-selection}
\end{table}

The light-family factors resolve the last step of the transfer. In the
up sector the insertion $I-3P_2/4$ transmits one quarter of the light
amplitude and leaves the heavy and middle amplitudes unchanged. Its
logarithmic contribution is $(\log4)P_2$ in $B_u$. The resulting geometric
mass ratios obey $m_c/m_u=4e^{2c_C}$ whereas
$m_t/m_c=e^{2\alpha_{\rm up}}$ tests the common rung. Thus the light ratio
probes the relative normalization of this insertion and the pivot width
separately from the heavy-to-middle transfer. The down dilution and
lepton period in Table~\ref{tab:p5-hierarchy-selection} act on the same
light-family projector in their respective sectors. Comparisons of
these relations use the finite matching of the geometric amplitudes to
masses at a common scheme and scale.

The masses $m_i$ in the ladder formulas are the singular values
of the canonically normalized generated mass vertex. At fixed generated
background, they are the positive rest energies of the quadratic matter
modes measured by the phase clock. The interacting readout uses the
connected matter two point response with its mass and kinetic terms
normalized together. For charged leptons we write the geometric masses
explicitly as $m_\ell^{\rm geom}$. For comparison with pole masses in a
specified realization, denote the corresponding
poles by $M_\ell$ and their adopted full Standard Model running
representation by $\overline m_\ell(\mu)$. Define
\begin{equation}
 M_\ell=G_\ell m_\ell^{\rm geom},\qquad
 \overline m_\ell(\mu)=D_\ell(\mu)M_\ell,
 \qquad C_\ell(\mu)=D_\ell(\mu)G_\ell.
 \label{eq:p5-lepton-mass-definitions}
\end{equation}
Here $G_\ell$ records the ratio of the pole readout to the generated
mass. The specified readout fixes this ratio with the source normalization
and scale retained. In the adopted running representation $D_\ell$
converts the pole to the stated convention and $C_\ell$ is their
product. The pole to running calculation uses the self-energies of
that effective realization.

Perturbative relations between running Yukawa couplings and pole masses
apply to specified action parameters~\cite{HempflingKniehl1995}.
Preserving a relation among different masses under radiative corrections
is a separate dynamical question, as illustrated by the Koide
relation~\cite{Sumino2009}.

For a realization described by a perturbative effective two point
function expand about a canonical reference mass matrix $M_0$
with kinetic corrections $\delta K_L,\delta K_R$ and
mass correction $\delta B$. Restoring canonical fields gives
\begin{equation}
 \delta M=\delta B-\tfrac12\delta K_L M_0
                         -\tfrac12 M_0\delta K_R.
 \label{eq:p5-native-mass-vertex-correction}
\end{equation}
Indeed $(I+\delta K)^{-1/2}=I-\delta K/2+O(\delta^2)$ on the
positive kinetic subspace. For a simple nonzero singular value $m_i$
of $M_0$, with normalized left and right singular vectors $u_i,v_i$,
its variation is
$\delta\log m_i=\operatorname{Re}(u_i^\dagger\delta Mv_i)/m_i$.
Taking the generated diagonal mass vertex as $M_0$ in this representation
gives the first order on shell readout ratio
\[
 (\log G_\ell)^{(1)}
 =\frac{\operatorname{Re}\delta B_\ell}{m_\ell^{\rm geom}}
 -\frac12\operatorname{Re}(\delta K_{L,\ell}+\delta K_{R,\ell}).
\]
These coefficients refer to the same generated source and mass
definition, with the low-energy loop contribution subtracted once
when it is also included in the effective action. The condensate
amplitude is already part of $B$ in this convention. The weighted
combination isolates the flavour dependence of this readout ratio.
The separate pole-to-running calculation
determines $D_\ell$ for specified pole conditions.

For the geometric masses defined by these transfers the
hypercharge split eliminates the common window rate and sector
decrement and gives
\begin{equation}
 8\log\frac{m_t}{m_c}+9\log\frac{m_\tau}{m_\mu}
 =17\log\frac{m_b}{m_s}.
 \label{eq:p5-sector-mass-test}
\end{equation}
This identity tests the relative attenuation of the up, down and
charged-lepton sectors independently of their common window rate,
decrement and absolute mass scales.
An experimental test compares all six masses in a common convention.
Running and finite matching convert the geometric amplitudes to the
chosen mass definitions and scale. If
$m_i^{\rm phys}(\mu)=C_i(\mu)m_i$ the logarithmic combination formed
from the physical masses must be reduced by
$8\log(C_t/C_c)+9\log(C_\tau/C_\mu)-17\log(C_b/C_s)$ before comparison
with zero. The uncertainties in the masses and matching factors
are propagated together. The reference masses in Appendix~\ref{app:table}
use their customary individual scales and schemes.
In the species order $(t,c,\tau,\mu,b,s)$ set
$w=(8,-8,9,-9,-17,17)$ and let $z$ be the column vector formed from
the six $\log(m_i^{\rm phys}/{\rm GeV})$ followed by the six $\log C_i$.
The matched residual
and its variance are
\begin{equation}
 r_m=(w,-w)z,\qquad
 \sigma_{r_m}^2=(w,-w)\Sigma_z(w,-w)^{\mathsf T},
 \label{eq:p5-matched-mass-covariance}
\end{equation}
where $\Sigma_z$ includes the mass--matching cross correlations.
The numerical implementation~\cite{Guo2026GenesisFramework} evaluates this
combination from the six mass definitions and reference scales together with the matching
coefficients and their joint logarithmic covariance. Common multiplicative
mass normalizations cancel because $\sum_i w_i=0$.

\paragraph{The lepton invariant and pole-mass comparison.}
The light-family insertion gives the geometric identity
\[
 \frac{(m_\mu^{\rm geom})^2}
 {m_e^{\rm geom}m_\tau^{\rm geom}}=e^{\sqrt{2\pi}}
\]
by cancellation of the common mass normalization and rung attenuation.
The invariant isolates the light-family insertion. With
$w_e=-1$ and $w_\mu=2$ and $w_\tau=-1$ define
\begin{equation}
 \Delta_L^{\rm pole}=\sum_\ell w_\ell\log G_\ell,
 \qquad
 \Delta_L^{\rm run}(\mu)=\sum_\ell w_\ell\log D_\ell(\mu).
 \label{eq:p5-lepton-conversion-parts}
\end{equation}
Taking the logarithm of Eq.~\eqref{eq:p5-lepton-mass-definitions}
and summing with these weights gives
\begin{align}
 \log\frac{M_\mu^2}{M_eM_\tau}-\sqrt{2\pi}
    &=\Delta_L^{\rm pole},\notag\\
 \log\frac{\overline m_\mu(\mu)^2}
 {\overline m_e(\mu)\overline m_\tau(\mu)}-\sqrt{2\pi}
    &=\Delta_L^{\rm pole}+\Delta_L^{\rm run}(\mu).
 \label{eq:p5-lepton-physical-invariants}
\end{align}
The pole inputs in Ref.~\cite[Eq.~(5)]{HuangZhou2021} give
\begin{equation}
 \frac{M_\mu^2}{M_eM_\tau}=12.2951727421,
 \qquad
 \Delta_{L,\rm ref}^{\rm pole}
   =0.00257845\pm0.00006753.
 \label{eq:p5-lepton-pole-reference}
\end{equation}
The quoted reference uncertainty follows by propagating the input
errors as independent Gaussian errors using
$\sigma_{\rm ref}^2=\sum_\ell w_\ell^2
(\sigma_{M_\ell}/M_\ell)^2$. The tau mass dominates this uncertainty.
Under these reference errors the exact pole identification
$G_e=G_\mu=G_\tau=1$ has standardized residual
$\Delta_{L,\rm ref}^{\rm pole}/\sigma_{\rm ref}=38.18$.
For a specified realization the charged-lepton readout determines
$G_\ell$. Its weighted logarithmic ratio $\Delta_L^{\rm pole}$ and
uncertainty are tested against
Eq.~\eqref{eq:p5-lepton-pole-reference}.
The pole and running comparisons share the same
experimental inputs. Their correlations enter a combined test.

\paragraph{Correlated running-mass constraints.}
An experimental constraint on the required matching can be obtained
without assigning the individual $C_i$. We use the full Standard-Model
$\overline{\rm MS}$ masses and their joint uncertainties at $\mu=M_Z=91.1876\,\mathrm{GeV}$
from Tables~2 and 3 and~8 of
Ref.~\cite{HuangZhou2021}. The common convention is
$\overline m_i=y_i v_F/\sqrt2$ with $v_F=246\,\mathrm{GeV}$.
This common normalization cancels from both matching combinations.
The two geometric identities constrain the total matching factors
$C_i$ to this running convention. Define
\begin{equation}
 \begin{aligned}
 \Delta_F&=8\log(C_t/C_c)+9\log(C_\tau/C_\mu)
                       -17\log(C_b/C_s),\\
 \Delta_L&=2\log C_\mu-\log C_e-\log C_\tau.
 \end{aligned}
 \label{eq:p5-joint-running-mass-constraints}
\end{equation}
Thus $\Delta_L=\Delta_L^{\rm pole}+\Delta_L^{\rm run}$ by
Eq.~\eqref{eq:p5-lepton-conversion-parts}.
The published running masses then give
\begin{equation}
 \binom{\Delta_F}{\Delta_L}
 =\binom{2.69537\pm1.48017}{0.0112736\pm0.0005817},
 \qquad
 \Sigma_{FL}=\begin{pmatrix}
 2.1909009&8.68052\times10^{-7}\\
 8.68052\times10^{-7}&3.38356\times10^{-7}
 \end{pmatrix}.
 \label{eq:p5-running-mass-numbers}
\end{equation}
Here the delta method converts the published mass covariance to the
logarithmic covariance before contraction by the two weight vectors.
The reference error model combines propagated input uncertainties
and perturbative truncation estimates. The correlation between the
two resulting constraints is $0.0010082$. These are constraints on
the finite matching of the generated masses. A realization with
$\Delta_F=\Delta_L=0$ at $M_Z$ has standardized residuals $1.82$
and $19.38$ respectively. The uncertainty in
Eq.~\eqref{eq:p5-running-mass-numbers} sets the precision with which
the lepton matching combination is tested.

The scale dependence supplies a separate consistency check. At one
loop the common gauge terms and common Yukawa trace cancel within
each equal-charge pair. In the diagonal-CKM limit
\begin{equation}
 \begin{aligned}
 16\pi^2\frac{dF}{d\log\mu}
 &=\frac32\{25(y_t^2-y_c^2-y_b^2+y_s^2)
                         +9(y_\tau^2-y_\mu^2)\},\\
 16\pi^2\frac{dL}{d\log\mu}
 &=\frac32(2y_\mu^2-y_e^2-y_\tau^2),
 \end{aligned}
 \label{eq:p5-mass-combination-running}
\end{equation}
where $F$ is the six-mass logarithmic combination and
$L=\log(\overline m_\mu^2/(\overline m_e\overline m_\tau))
-\sqrt{2\pi}$. At the reference point their derivatives are
$0.22214$ and $-9.31\times10^{-7}$.
The calculation~\cite{Guo2026GenesisFramework} also retains a general
unitary CKM matrix in the Yukawa terms. For the charged leptons
the one-loop QED relation~\cite{XingZhangZhou2008}
\[
 \log\frac{\overline m_\ell(\mu)}{M_\ell}
 =-\frac{\alpha}{\pi}\left[1+\frac34
                 \log\frac{\mu^2}{M_\ell^2}\right]+O(\alpha^2)
\]
gives a weighted correction $0.00874265$ independent of the common
scale at this order. This explicitly evaluates the electromagnetic
part of the conversion. The common full Standard-Model reference
in Eq.~\eqref{eq:p5-running-mass-numbers} includes the additional
matching contributions. The generated action's flavour-dependent
finite coefficients enter through Eq.~\eqref{eq:p5-joint-running-mass-constraints}.

\paragraph{Finite lepton transfer.}
In a fixed Standard-Model scheme with vanishing neutrino Yukawa
couplings put $t=\log(\mu/\mu_0)$ and write the one-loop
charged-lepton equations~\cite{LuoXiao2003} as
\[
 \frac{dy_i}{dt}=y_i[A(t)+b y_i^2],\qquad
 b=\frac{3}{32\pi^2},\qquad i=e,\mu,\tau.
\]
Here $16\pi^2A=\operatorname{Tr}(3Y_u^\dagger Y_u+
3Y_d^\dagger Y_d+Y_e^\dagger Y_e)-9g_1^2/4-9g_2^2/4$ with
GUT-normalized $g_1$ evaluated along the same running solution.
Define $E(t)=\exp\int_0^t A(s)\,ds$ and
$I(t)=2b\int_0^t E(s)^2\,ds$. As long as
$d_i(t)=1-I(t)y_i(0)^2>0$ the finite solution and its lepton
combination are
\begin{align}
 y_i(t)&=\frac{E(t)y_i(0)}{\sqrt{d_i(t)}},\\
 R_L(t)&:=\log\frac{y_\mu(t)^2}{y_e(t)y_\tau(t)}
        -\log\frac{y_\mu(0)^2}{y_e(0)y_\tau(0)}
       =\frac12\log\frac{d_e(t)d_\tau(t)}{d_\mu(t)^2}.
 \label{eq:p5-finite-lepton-transfer}
\end{align}
For a forward interval $0\leq t\leq T$ uniform bounds
$A_-\leq A(t)\leq A_+$ imply
\begin{equation}
 2b\frac{e^{2A_-T}-1}{2A_-}
 \leq I(T)\leq
 2b\frac{e^{2A_+T}-1}{2A_+},
 \label{eq:p5-lepton-transfer-enclosure}
\end{equation}
where each quotient is $T$ at $A_\pm=0$. Substitution into
\eqref{eq:p5-finite-lepton-transfer} by interval arithmetic encloses
the running contribution provided all denominator lower bounds
are positive.

For a fixed sequence of running and threshold conversions let
$\Delta_L^{\rm known}$ be the independently calculated contribution
in that convention and let $x_i$ be the logarithms of the remaining
finite matching factors. Then
\begin{equation}
 \max_i x_i-\min_i x_i\ \geq
 \frac12|\Delta_L-\Delta_L^{\rm known}|.
 \label{eq:p5-finite-matching-spread}
\end{equation}
This bound is sharp and invariant under a common finite mass or
vacuum-expectation-value normalization. The finite running solution follows by integrating
$d(E^2/y_i^2)/dt=-2bE^2$ and taking the weighted logarithm.
Upper and lower bounds on the integral of $A$ give the stated enclosure.
For the matching remainder
$\Delta_L-\Delta_L^{\rm known}=2x_\mu-x_e-x_\tau$ lies between
$-2(\max x-\min x)$ and $2(\max x-\min x)$. Equality is attained with the
electron and tau logarithms at one endpoint and the muon logarithm at the
other while a common additive shift cancels because the weights sum to
zero.

The common integrating factor $E$ cancels from the lepton ratio.
Its dependence on the coupled running solution is retained through
the single path integral $I$ in
Eq.~\eqref{eq:p5-finite-lepton-transfer}. The remaining
finite coefficients combine the flavour-dependent self-energies,
vertex normalizations and external-field normalizations in the same scheme.
Equation~\eqref{eq:p5-finite-matching-spread} constrains their required
spread once the independently calculated conversions have been
subtracted. The implementation~\cite{Guo2026GenesisFramework} checks the
resulting constraint in the registered comparison. At the reference
central values subtracting the displayed leading QED term leaves
$\Delta_L-\Delta_L^{\rm QED}=0.00253093$ and requires
$\max x_i-\min x_i\geq0.00126546$. This is a constraint on the
combined remaining conversion from the geometric readout through
the pole definition to the common running scheme.
Numerical checks compare this transfer with direct integration of all
three eigenvalue equations and verify composition and reversal of the
scale interval. Outward interval evaluation uses uniform bounds on
$A$ throughout the interval.

\paragraph{Pole to running conversion in the specified electroweak realization.}
Consider the canonical electroweak realization with the pole
identification $G_\ell=1$. We extend
the physical renormalization conditions of
Proposition~\ref{prop:p5-generated-onshell-matching} to the charged
leptons. The geometric masses supply the
pole conditions $M_e,M_\mu,M_\tau$ of this
action. The calculation determines $D_\ell$ at these conditions. At
$Q_0=M_Z^{\rm gen}=91.12484\,\mathrm{GeV}$ use its
tadpole-free Landau-gauge $\overline{\rm MS}$ convention with the generated
VEV and gauge coefficients and $y_\ell^2=2M_\ell^2/v^2$ at leading
order. Neutrino Yukawa couplings vanish in this calculation.
The photon self-energy and the $W$ and $Z$ and Higgs self-energies retain the charged-lepton
masses in their loop functions. They follow from
Ref.~\cite[Eq.~(2.20)]{Martin2016TopPole} with
$Q_\ell=-1$ and $I_3^\ell=-1/2$ with a massless weak partner and zero
colour coupling. The light-mass weak threshold agrees with
Ref.~\cite[Sec.~II.C]{Martin2019Decoupling}.

Let $\delta s_\ell^{(1)}$ include the one-loop factor in the
pole-squared-mass shift. Linear inversion gives
\begin{equation}
 (\log D_\ell)^{(1)}
 =-\frac{\operatorname{Re}\delta s_\ell^{(1)}}{2M_\ell^2},
 \qquad
 \Delta_L^{\rm run,(1)}
 =-\sum_{\ell=e,\mu,\tau}
       w_\ell\frac{\operatorname{Re}\delta s_\ell^{(1)}}{2M_\ell^2},
 \qquad w=(-1,2,-1).
 \label{eq:p5-generated-lepton-pole-matching}
\end{equation}
The common VEV conversion to the reference
convention $\overline m_\ell=y_\ell v_F/\sqrt2$ cancels since
$\sum_\ell w_\ell=0$. This direct pole-to-full-SM calculation includes
the heavy electroweak thresholds in the finite self-energy. Its gauge
coefficients are those of the same action.
In particular $\alpha(Q_0)=0.00779993272$ and
\begin{equation}
 \begin{aligned}
 \Delta_{L,\gamma}^{\rm run,(1)}
  &=\frac{3\alpha(Q_0)}{2\pi}
      \log\frac{M_\mu^2}{M_eM_\tau}
    =0.009335168839,\\
 \Delta_{L,\mathrm{weak}}^{\rm run,(1)}
  &=2.348914726\times10^{-6},\\
 \Delta_L^{\rm run,(1)}(Q_0)&=0.009337517754 .
 \end{aligned}
 \label{eq:p5-generated-lepton-matching-value}
\end{equation}
All inputs to these coefficients are generated before the comparison.
At the stated input values analytic interval evaluation of the nine
two-point functions gives
\[
\Delta_L^{\rm run,(1)}\in
[0.009337517754041472,\,0.009337517754041475].
\]
Direct high-precision Feynman-parameter integrals check the finite
components independently.

The weighted one-loop scale derivative is
$-9.77377\times10^{-7}$.
Transferring the result to the comparison scale
$91.1876\,\mathrm{GeV}$ therefore changes it by
$-6.73\times10^{-10}$ at this order. Under $G_\ell=1$ the
running-mass constraint minus this one-loop conversion is $0.00193607$
with reference uncertainty $0.0005817$.
For a general geometric-to-pole conversion the total prediction is
\begin{equation}
 \Delta_L^{\rm th}(\mu)
 =\Delta_L^{\rm pole,th}
  +\Delta_L^{\rm run,(1)}(\mu;\{G_\ell m_\ell^{\rm geom}\})
  +\mathcal R_{L,\rm run}^{(2+)}.
 \label{eq:p5-lepton-total-matching}
\end{equation}
The one-loop kernels are evaluated at the poles supplied by that
conversion. The remainder $\mathcal R_{L,\rm run}^{(2+)}$ contains
higher-order electromagnetic and weak corrections to $D_\ell$.
The specified physical readout or pole identification supplies the
geometric to pole term and its uncertainty. At fixed pole
renormalization conditions, higher orders determine the corresponding
running parameters while preserving those poles.
The displayed interval certifies the numerical one-loop conversion
at the stated inputs. Equation~\eqref{eq:p5-lepton-pole-reference}
tests the pole identification directly. 
Eq.~\eqref{eq:p5-running-mass-numbers} tests the total conversion to
the common running convention.

\subsection{The top-quark normalization}\label{sec:top}

The generated top-quark mass fixes the normalization of the up-sector ladder.
The elementary scalar connector of Ref.~\cite{Guo2026CompactInternalSpaceI}
has unit geometric matrix element
\[
y_0 = 1 ,
\]
which is invariant under a common unitary transformation of the connector and its paired modes. With the vacuum expectation value of Section~\ref{sec:ewsb} the tree-level mass is
\[
m_t = \frac{y_0 v}{\sqrt{2}} = 174.1 \ \mathrm{GeV} .
\]
The unit finite-carrier matrix element and signed spatial connector with the
canonical action are constructed in Appendix~\ref{sec:derivtop}.
The resulting top and Higgs masses and Fermi constant serve as physical
renormalization conditions for the one-loop $\overline{\rm MS}$ action
of Proposition~\ref{prop:p5-generated-onshell-matching}.

\subsection{The lower masses}\label{sec:lowmass}

The bottom-to-top Yukawa ratio includes the down-sector suppression and the evolution of the pivot window
\[
\frac{y_b}{y_t} = e^{-(2\alpha_{\mathrm{dn}} - n_s^{\mathrm{tilt}}(\kL_{\mathrm{CMB}} + 2\tau))} = 0.024344 ,
\]
where $n_s^{\mathrm{tilt}}$ is the positive spectral-tilt decrement. This gives the bottom mass
\[
m_b = \frac{y_b}{y_t} \, m_t = 4.238 \ \mathrm{GeV} .
\]
The within-sector mass ratios are
\[
\frac{m_s}{m_d} = e^{2\alpha_{\mathrm{sd}}} = 19.62, \qquad
\frac{m_b}{m_s} = e^{2\alpha_{\mathrm{dn}}} = 44.87, \qquad
\frac{m_\mu}{m_e} = e^{2\alpha_{\mathrm{lp}} + \sqrt{2\pi}} = 206.36 ,
\]
and the charm and up ratios are
\[
\frac{m_t}{m_c} = e^{2\alpha_{\mathrm{up}}} = 135.2, \qquad
\frac{m_t}{m_u} = e^{2\alpha_{\mathrm{up}}} e^{2\kL_{\mathrm{CMB}} + \ln 4} = 77304 .
\]
The electron cascade gives the closed form in the common Planck normalization
\[
m_e = \MP \, e^{-20\kL} \, (1 - s_0 \kappa) = 0.510 \ \mathrm{MeV} ,
\]
where $s_0 \kappa$ is the squash correction of Section~\ref{sec:geodyn}. The remaining lepton step is $m_\tau/m_\mu=e^{2\alpha_{\mathrm{lp}}}$. These ratios and the two generated sector normalizations determine the lower masses. Both $v$ and $m_e$ are calculated relative to the same $\MP$. They provide intermediate normalizations within the mass calculation while $v/m_e$ tests the relative strength of their source transfers independently of the common dimensionful scale. Appendix~\ref{sec:derivmass} gives the electron cascade normalisation.

The quark-mixing readouts use the same family attenuation. The
adjacent-generation amplitude is
$|V_{cb}|=\tfrac12e^{-\alpha_{\rm up}}(1-s_0\kappa)$ while the
next-generation amplitude is
$|V_{ub}|=\tfrac12e^{-2\alpha_{\rm up}}(1+\tau\kappa)$ with the
source-channel factors given in Appendix~\ref{sec:derivckm}.
For the geometric masses and angular readouts of this action
eliminating the common exponential gives
\begin{equation}
 \frac{|V_{cb}|^2}{m_c/m_t}=\frac{(1-s_0\kappa)^2}{4},\qquad
 \frac{|V_{ub}|}{|V_{cb}|^2}
 =\frac{2(1+\tau\kappa)}{(1-s_0\kappa)^2}.
 \label{eq:p5-mixing-transfer-relations}
\end{equation}
The first relation connects a mass hierarchy to a mixing amplitude.
The second tests the relative channel factors after that hierarchy
has cancelled. Their experimental comparison uses mass and flavour
matching in the same common-action convention as the conversion above.

%% file: sections-en/sec08_cosmology.tex
% =============================================================================
% Section 8 - Cosmology
% The Spectrum of a Compact Internal Space II: Effective Couplings and Mass Scales
% =============================================================================

\section{Cosmology}\label{sec:cosmo}

The electroweak scale enters cosmology through the dimension-five
neutrino mass relation. The lightest mass then supplies the common
energy scale in the vacuum-density and photon-temperature readouts
and the Gaussian entropy difference connects that vacuum floor to the
Hubble scale. Eliminating the common mass gives joint tests of
the density, temperature and expansion readouts through the dark energy
fraction and the temperature--Hubble ratio. The homogeneous evolution
also includes baryons and a pressureless component together with photons
and three thermal neutrinos.

The macroscopic calculation has three linked levels. The homogeneous
background turns the endpoint density and clock readouts into an
expansion history. Linear propagation carries this history to the
distance observables. The local variational sector resolves how a
spatially varying source and potential exchange energy and momentum.
Their common normalization is Newton's constant and their common
matter data are the generated masses and source charges. Each level
uses its background response map with these shared inputs.
The density budget and acoustic distances test the common thermal scale.
Local source--potential dynamics and the occupied auxiliary radius
background resolve source feedback and a material phase clock.

\subsection{Two Gaussian entropies}\label{sec:entropy}

The density readout uses the lightest-neutrino rest mass as the
infrared energy scale of the vacuum floor. Its fourth power has the
dimensions of an energy density in natural units. The coefficient is
fixed by the selected hypercharge weight and squash response. The
endpoint entropy difference compares
this floor with the Planck anchor through a Gaussian variance ratio
and a content-dependent offset. The Planck Gaussian has variance
$M_{\mathrm{P}}^2$ and the vacuum-floor Gaussian has variance
$\sqrt{\rho_\Lambda}$. On the branch $0\leq x=s_0\kappa<1/4$
the vacuum density is positive and equals
\[
\rho_\Lambda = Y_u \, m_{\nu_1}^4 \, (1 - 4 s_0 \kappa) = \frac{2}{3} m_{\nu_1}^4 (1 - 4 s_0 \kappa) .
\]
The factor $(1 - 4 s_0 \kappa)$ is the dark-energy-weight correction of the $J = 2$ squash. The squash dependence of $v^2$ also enters the neutrino mass. At fixed remaining inputs its density factor is $(1-x)^8(1+x)^4(1-4x)$ as proved in Appendix~\ref{sec:derivcosmo}. Writing $m_{\nu_1}=M_{\mathrm P}\widehat m_{\nu_1}$ and $\rho_\Lambda=M_{\mathrm P}^4\widehat\rho_\Lambda$ gives the dimensionless entropy span
\[
S_{\rm end}
= \ln \frac{M_{\mathrm{P}}^2 \sqrt{2\pi + r_{23}}}{\sqrt{\rho_\Lambda}}
= \ln \frac{\sqrt{2\pi + r_{23}}}{\sqrt{\widehat\rho_\Lambda}}
= 139.2537 ,
\]
with $r_{23}=3/(10\sqrt3)$ the hypercharge-trace hierarchy ratio. Appendix~\ref{sec:derivcosmo} derives this expression from twice the Gaussian entropy difference plus the content offset and shows how intermediate scales cancel.

The entropy difference is invariant under a common change of energy
unit. Along successive intervals of a Gaussian variance path the
entropy differences add and every intermediate scale cancels.
The content offset is attached once to the complete pair
of endpoints. Thus $W=e^{S_{\rm end}}$ measures the ultraviolet-to-infrared
scale ratio including its content factor independently of that
subdivision. The Hubble readout converts this span into an
inverse length. Appendix~\ref{sec:derivcosmo}
also evaluates the distinct entropy of the normalized internal
heat-trace state.

\subsection{The Hubble rate and the dark-energy fraction}\label{sec:hubble}

The expansion rate is $H=d\log a/ds$ where $s$ is the proper-time
readout of the homogeneous comoving flow. For a parameter $\lambda$
with $ds=N_\lambda d\lambda$ it is
$H=N_\lambda^{-1}d\log a/d\lambda$. In the homogeneous dust
realization $T=s+\mathrm{constant}$. The background equations
use this clock so that their time coordinate is $t=s$.

The Hubble rate is the infrared endpoint of the emergence window
\[
H_0 = \MP \sqrt{\pi} \, e^{-S_{\rm end}} = 1.4388 \times 10^{-42} \ \mathrm{GeV} ,
\]
which corresponds to $67.45 \ \mathrm{km\,s^{-1}\,Mpc^{-1}}$. The dark-energy fraction is
\[
\Omega_\Lambda = \frac{\rho_\Lambda}{3 H_0^2 M_{\mathrm{P}}^2} = \frac{2}{3} + \frac{r_{23}}{3\pi} = 0.68504 .
\]
Its signed residual against the adopted Planck reference in Table~2 of Ref.~\cite{Planck2018} is $+0.05\%$. Appendix~\ref{sec:derivcosmo} gives the cancellation of the common density.

The macroscopic Einstein--matter realization of
Ref.~\cite{Guo2026SpectralGravity} gives the gravitational effective action
\[
S_{\mathrm{eff}} = \frac{\MP^2}{2} \int d^4x \sqrt{-g}\, R - \int d^4x \sqrt{-g}\, \rho_\Lambda ,
\]
with $\MP^2=1/(8\pi G)$. The constant vacuum floor contributes a
uniform background stress through the volume term in this action.
The entropy relation and neutrino mass scale enter the homogeneous densities through
\[
\rho_{\mathrm{tot}}
= \rho_\Lambda+\rho_b+\rho_\Sigma^{\rm budget}
= 3H_0^2\MP^2,
\qquad
\Omega_b+\Omega_\Sigma^{\rm budget}+\Omega_\Lambda=1 .
\]
This relation imposes the spatially flat homogeneous budget on the endpoint densities.

\subsection{The microwave background temperature}\label{sec:tcmb}

The photon temperature readout uses the same lightest neutrino mass
as the vacuum density. Its content and squash factors convert that
rest-mass scale into the raw photon floor
\[
T_{\mathrm{CMB}}^{\mathrm{raw}} = \frac{m_{\nu_1} r_{12}}{\pi} (1 - s_0 \kappa)(1 - \tau \Delta_s) = 2.7310 \ \mathrm{K} ,
\]
where $r_{12}=3/10$ is the hypercharge ratio and $\Delta_s=1/2$ is the scalar conformal weight. The factor $(1-s_0\kappa)$ multiplies the already corrected rest mass. The total temperature response includes both factors. The formula uses common energy units before conversion to Kelvin. The endpoint photon zero-mode factor is
\[
C_\gamma = 1 - \frac{\tau}{\pi^2},
\qquad
T_{\mathrm{CMB}} = C_\gamma T_{\mathrm{CMB}}^{\mathrm{raw}} = 2.72547 \ \mathrm{K}.
\]
The corrected value is used for the CMB-monopole comparison and for the linear propagation. At this temperature the baryon-to-photon ratio is
\[
 \eta_B=\eta_B^{\rm raw}C_\gamma^{-3}
       =6.12823\times10^{-10}.
\]
Consequently $\eta_BT_{\rm CMB}^3
=\eta_B^{\rm raw}(T_{\rm CMB}^{\rm raw})^3$. The baryon density
has the same value in either photon convention.
The common mass cancels exactly from $T_{\rm CMB}^2/(M_{\rm P}H_0)$
in energy units. The temperature is linear in that mass while the
entropy relation makes $H_0$ quadratic in it. Together with
$\Omega_\Lambda$ this ratio tests the joint density and temperature readouts and the
expansion independently of their common mass normalization.
Equation~\eqref{eq:p5-cosmology-shared-ratio} gives this cross-scale
relation and Appendix~\ref{sec:p5-cosmology-error} bounds its finite
squash correction.

\subsection{The homogeneous density budget}\label{sec:flat}

The baryon fraction is
\[
\Omega_b = \frac{\eta_B^{\mathrm{raw}} n_\gamma^{\mathrm{raw}} m_p}{\rho_{\mathrm{crit}}} = 0.04941 ,
\]
where $\eta_B^{\rm raw}$ is given by Eq.~\eqref{eq:p5-baryon-raw} with its photon-temperature conversion described in Appendix~\ref{sec:ckm} and the corresponding photon density is
\[
n_\gamma^{\mathrm{raw}}
= \frac{2\zeta(3)}{\pi^2}\bigl(T_{\mathrm{CMB}}^{\mathrm{raw}}\bigr)^3 ,
\]
with $m_p$ the proton mass of Section~\ref{sec:proton} and $\rho_{\mathrm{crit}}=3H_0^2\MP^2$ calculated from the Hubble rate of Section~\ref{sec:hubble}. With $\Omega_\Lambda=0.68504$ the remaining fraction is fixed by the homogeneous budget
\[
\Omega_\Sigma^{\rm budget} = 1 - \Omega_\Lambda - \Omega_b = 0.26554751 .
\]
The sum
\[
\Omega_b + \Omega_\Sigma^{\rm budget} + \Omega_\Lambda = 1
\]
satisfies the flatness constraint. The thermal components are evaluated
at the corrected photon temperature. With the common neutrino temperature
\[
 T_\nu=T_{\rm CMB}\left(\frac4{11}\right)^{1/3}
                 \left(\frac{N_{\rm eff}}3\right)^{1/4}
       =1.95242401\ {\rm K},
\]
their energy densities including both helicities are
\begin{equation}
 \rho_{\nu_i}=\frac{T_\nu^4}{\pi^2}
 \int_0^\infty
 \frac{q^2\sqrt{q^2+(m_{\nu_i}/T_\nu)^2}}{e^q+1}\,dq,
 \qquad \rho_\gamma=\frac{\pi^2T_{\rm CMB}^4}{15}.
 \label{eq:p5-thermal-neutrino-budget}
\end{equation}
The common budget therefore fixes
\begin{equation}
 \Omega_{\rm cold}
 =\Omega_\Sigma^{\rm budget}-\Omega_\gamma-\sum_i\Omega_{\nu_i}
 =0.26403880,
 \label{eq:p5-cold-budget}
\end{equation}
where $\Omega_\gamma=0.0000543504$ and
$\sum_i\Omega_{\nu_i}=0.0014543578$.
The three neutrino masses are those of the rest spectrum in
Appendix~\ref{app:flavour}. The same vacuum fraction
$\Omega_\Lambda=0.68504430$ enters the endpoint and the propagation.
\subsection{Cold-matter evolution and linear cosmological perturbations}\label{sec:linprop}

Linear cosmological perturbations are evolved with the two-derivative
Einstein--matter action of Section~\ref{sec:hubble}
\cite{Guo2026SpectralGravity}. We take the vacuum density to be constant
and the visible stress tensor to be separately conserved.
The pressureless component carries a density transported with the
matter flow and allows an inhomogeneous cold source while its spatial
and mixed rest-frame stress projections vanish.
In this subsection $\rho_\Sigma$ denotes the cold density with present
mean $\rho_{\rm crit}\Omega_{\rm cold}$. Photons and thermal neutrinos
belong to $T^{\rm vis}_{\mu\nu}$.

A dust action fixes the transported density and velocity together with their stress.
Its field equations determine both the residual Einstein source and
the rest-frame pressureless form.

\paragraph{Cold source and comoving transport.}
On a smooth spacetime with signature $(-,+,+,+)$ and Levi--Civita
connection take $\MP>0$ and constant $\rho_\Lambda$ with a separately
conserved visible stress tensor.  Add to the Einstein and visible-matter
actions
\begin{equation}
 S_\Sigma=-\frac12\int d^4x\sqrt{-g}\,
 \varrho\bigl(g^{\mu\nu}\partial_\mu T\partial_\nu T+1\bigr),
 \qquad u_\mu=-\partial_\mu T,
 \label{eq:p5-cold-action}
\end{equation}
with $\varrho$ varied independently and future-directed $u$.  Define
\begin{equation}
\begin{aligned}
 h_\mu{}^\alpha&=\delta_\mu{}^\alpha+u_\mu u^\alpha,\\
 E_{\mu\nu}&=\MP^2G_{\mu\nu}+\rho_\Lambda g_{\mu\nu}
                  -T^{\rm vis}_{\mu\nu},
\end{aligned}
\label{eq:p5-cold-source-residual}
\end{equation}
Every smooth solution has $u^\mu u_\mu=-1$ and
$T^\Sigma_{\mu\nu}=\varrho u_\mu u_\nu$.  The Einstein equation therefore
sets $E_{\mu\nu}=T^\Sigma_{\mu\nu}$. The spatial and mixed rest-frame
projections vanish and with $\rho_\Sigma=\varrho$
\begin{equation}
\begin{gathered}
 E_{\mu\nu}=\rho_\Sigma u_\mu u_\nu,\\
 \nabla_\mu(\rho_\Sigma u^\mu)=0,
 \qquad \rho_\Sigma u^\mu\nabla_\mu u_\nu=0.
\end{gathered}
\label{eq:p5-cold-source-conservation}
\end{equation}
In the nonrelativistic cosmological limit let $x(t,q)$ be an injective
comoving flow map with $x(t_i,q)=q$ and
$J(t,q)=\det(\partial x/\partial q)>0$ and peculiar velocity
$v(t,x(t,q))=a\dot x(t,q)$. Then its physical cold density is
\begin{equation}
 \rho_\Sigma^{\rm free}(t,x(t,q))
 =\rho_{\Sigma,i}(q)\frac{a_i^3}{a^3J(t,q)}.
 \label{eq:p5-cold-pushforward}
\end{equation}
On a periodic comoving cell a degree-one flow preserves the cell volume. If $\langle\rho_{\Sigma,i}\rangle=\overline\rho_{\Sigma,i}$ then
$\langle\rho_\Sigma^{\rm free}(t)\rangle
=\overline\rho_{\Sigma,i}(a_i/a)^3$.
Nonnegative initial $\varrho$ remains nonnegative throughout the smooth
flow domain.  This is the irrotational dust action of
Ref.~\cite{BrownKuchar1995} in the present convention.  Variation gives
unit norm and conservation with the stated stress. The contracted Bianchi
identity gives the same continuity and geodesic equations.  Along the
comoving flow Jacobi's determinant formula gives
$\dot J=J\nabla_x\cdot(v/a)$ and hence $d(a^3\rho J)/dt=0$. This proves
\eqref{eq:p5-cold-pushforward}. Change of variables over the cell gives
the mean relation.

The endpoint density fixes the mean at the selected reference epoch.
The action and initial perturbations determine its subsequent cold
transport. Before characteristic crossing this also supplies local
free-source data. Rest-frame pressure and anisotropic stress vanish
within the dust realization. Adiabatic initial data are selected
separately. The linear-propagation model consequently uses
\[
T^\Sigma_{\mu\nu} = \rho_\Sigma u_\mu u_\nu,\qquad
\nabla_\mu(\rho_\Sigma u^\mu)=0,\qquad
p_\Sigma=0,\quad c_{s,\Sigma}^2=0,\quad \pi_\Sigma=0 .
\]
The homogeneous expansion and thermal history with the acoustic scales are
calculated with CAMB~\cite{CAMB2000}.
Table~\ref{tab:cosmodiag} reports the numerical results
\cite{Guo2026GenesisFramework} for
$\Omega_{\mathrm{cdm}}:=\Omega_{\rm cold}$ and the corrected photon monopole
with the three rest masses and $Y_p=0.246082$ and $N_{\rm eff}=3.043873$.
Recombination uses the standard atomic constants of the thermal solver.
The background and distance calculation is independent of the primordial
perturbation amplitude and requires no matter or angular power spectrum.
The thermal realization uses the common temperature in
\eqref{eq:p5-thermal-neutrino-budget} for each mass eigenstate.
The propagated matter fraction is
\[
 \Omega_m=\Omega_b+\Omega_{\rm cold}+\sum_i\Omega_{\nu_i}
         =0.31490135.
\]
Together with photons and the endpoint vacuum it closes the same flat
budget. The neutrino energy integral retains its relativistic-to-massive
transition throughout the background evolution.

In the Einstein--radial--cold subsystem with constant vacuum density
a stationary condensate with $U(\varphi_*)=U'(\varphi_*)=0$ and
$U''(\varphi_*)=m_\delta^2>0$ has a positive physical scalar kinetic
form on the branch defined by $\rho_\Sigma>0$ and $Z_*>0$ and $M_*^2>0$
with $H\ne0$ and nonzero spatial wave number.
The result ``Physical scalar kinetics at constant condensate'' in
Ref.~\cite{Guo2026SpectralGravity} gives the cold
coefficient $D^{-1}>0$ where
$D=\rho_\Sigma\kappa_{\rm phys}^2+
3\rho_\Sigma^2/(2M_*^2)$ with $M_*=\MP$ in the constant
Einstein normalization used here.
This establishes the kinetic sign in the Einstein--radial--cold
subsystem. The thermal distance calculation also retains photons
and visible matter with the inputs listed above.

\begin{table}[t]
\centering
\caption{Background and acoustic quantities calculated with CAMB~2.0.4 from the endpoint densities and thermal model~\cite{Guo2026GenesisFramework}. The matter fraction includes the massive-neutrino contribution.}\label{tab:cosmodiag}
\begin{tabular}{lc}
\toprule
Quantity & Calculated value \\
\midrule
$\Omega_m$ & $0.314901$ \\
$r_{\mathrm{drag}}$ & $146.946$ Mpc \\
$100\,\theta_\ast$ & $1.041246$ \\
$z_{\mathrm{eq}}$ & $3408.92$ \\
\bottomrule
\end{tabular}
\end{table}

The endpoint densities determine the expansion history while the
thermal and recombination evolution determines the drag sound horizon.
Their combination fixes the distance ratios tested by baryon acoustic
oscillations. The BAO comparison evaluates a Gaussian compressed-distance likelihood
using the thirteen published DESI DR2 mean distance ratios and their
covariance~\cite{DESIDR22025}. These quantities are inferred from fits
to the acoustic feature with the broadband clustering parameters
marginalized. We apply this likelihood to the Einstein--matter thermal
background and sound-horizon calculation specified above and retain
the DESI acoustic templates and covariance. This tests the combined
map from endpoint densities through thermal evolution to the sound
horizon and distance vector within the compressed likelihood.
The local nonlinear source--potential response is evaluated separately.
The prediction uses
$H_0=67.45284\ {\rm km\,s^{-1}\,Mpc^{-1}}$ and $r_{\rm drag}$
fixed by the same calculation. The distances include photons, thermal
neutrinos, cold matter and the vacuum.
The distances are $D_M(z)=c_{\rm light}\int_0^z dz'/H(z')$ and
$D_H=c_{\rm light}/H$ and $D_V=(zD_M^2D_H)^{1/3}$.
For the thirteen-component vector formed from $D_M/r_{\rm drag}$ and
$D_H/r_{\rm drag}$ and $D_V/r_{\rm drag}$ the statistic is
$\chi^2=29.5087$ for thirteen degrees of freedom with fixed-model
tail probability $0.00554011$~\cite{Guo2026GenesisFramework}.
Table~\ref{tab:p5-bao-residuals} resolves this test into the measured
distance ratios. The diagonal standardized residuals locate the
discrepancies in redshift and observable. Their correlations enter the
joint statistic through the full covariance. In the supplied covariance
the two-component blocks at $z=0.706$ and $z=0.934$ contribute
$13.6414$ and $6.9114$ respectively to the total $\chi^2$.
The transverse and radial residuals at $z=0.706$ have the same sign
whereas their measurement covariance is negative. This increases their
joint discrepancy relative to the sum of their squared diagonal
standardized residuals.

\begin{table}[t]
\centering
\small
\caption{DESI DR2 distance ratios~\cite{DESIDR22025} and the fixed
endpoint prediction of this paper. Here
$d_i=(t_i-y_i)/\sqrt{C_{ii}}$ with $t_i$ the prediction and $y_i$ the
measurement and $C$ the published covariance. The $d_i$ are correlated
diagnostics. The total statistic uses $C^{-1}$. The drag scale is
$r_d=r_{\rm drag}$.}
\label{tab:p5-bao-residuals}
\begin{tabular}{clrrr}
\toprule
$z$ & Ratio & Measurement $y_i$ & Prediction $t_i$ & $d_i$ \\
\midrule
0.295 & $D_V/r_d$ & 7.941676 & 8.054941 & $+1.489$ \\
0.510 & $D_M/r_d$ & 13.587584 & 13.497310 & $-0.536$ \\
0.510 & $D_H/r_d$ & 21.862947 & 22.737024 & $+2.038$ \\
0.706 & $D_M/r_d$ & 17.350691 & 17.697134 & $+1.925$ \\
0.706 & $D_H/r_d$ & 19.455349 & 20.167880 & $+2.134$ \\
0.934 & $D_M/r_d$ & 21.575640 & 21.991042 & $+2.568$ \\
0.934 & $D_H/r_d$ & 17.641495 & 17.568638 & $-0.362$ \\
1.321 & $D_M/r_d$ & 27.600856 & 28.078062 & $+1.470$ \\
1.321 & $D_H/r_d$ & 14.176022 & 14.065368 & $-0.493$ \\
1.484 & $D_M/r_d$ & 30.511901 & 30.272103 & $-0.314$ \\
1.484 & $D_H/r_d$ & 12.817000 & 12.880688 & $+0.123$ \\
2.330 & $D_H/r_d$ & 8.631546 & 8.616842 & $-0.145$ \\
2.330 & $D_M/r_d$ & 38.988974 & 39.179741 & $+0.359$ \\
\bottomrule
\end{tabular}
\end{table}

A covariance-weighted projection separates a common distance scale
from the remaining redshift and transverse--radial dependence. With
$W=C^{-1}$ define
\begin{equation}
 \alpha_* = \frac{t^{\mathsf T}Wy}{t^{\mathsf T}Wt},
 \qquad
 \sigma_\alpha=(t^{\mathsf T}Wt)^{-1/2},
 \qquad r_\perp=\alpha_*t-y .
 \label{eq:p5-bao-scale-projection}
\end{equation}
Minimizing $(\alpha t-y)^{\mathsf T}W(\alpha t-y)$ gives
$t^{\mathsf T}Wr_\perp=0$. Hence the statistic of the fixed prediction
has the exact decomposition
\begin{equation}
 \chi^2
 =\underbrace{\frac{(1-\alpha_*)^2}{\sigma_\alpha^2}}
             _{\chi^2_{\rm scale}}
  +\underbrace{r_\perp^{\mathsf T}Wr_\perp}_{\chi^2_\perp}
 =15.2965+14.2122 .
 \label{eq:p5-bao-scale-decomposition}
\end{equation}
The data give $\alpha_*=0.989282$ and $\sigma_\alpha=0.002740$.
Thus a common $1.0718\%$ decrease in the distance ratios accounts for
the scale component of the discrepancy. The orthogonal component
has twelve residual degrees of freedom and a Gaussian tail probability
$0.2874$ when a common scale is fitted for this diagnostic. The
physical prediction retains its fixed normalization $\alpha=1$ and
the thirteen-degree-of-freedom probability quoted above.

To resolve the thermal and density dependence of the statistic, define
\[
 E_{X,x}=\left.\frac{\partial\log X}{\partial\log x}\right|_{\rm other\ inputs},
 \qquad \omega_b=\Omega_bh^2,\quad \omega_m=\Omega_mh^2,
 \quad h=H_0/(100\ {\rm km\,s^{-1}\,Mpc^{-1}}).
\]
Table~\ref{tab:p5-background-response} gives these partial responses at
the fixed prediction. Each row holds the other coordinates fixed. Flatness sets the vacuum density and varying $\omega_b$ at fixed
$\omega_m$ compensates the cold density. The helium response is weak
and acts through the drag scale. Matter-density changes also modify
the redshift dependence of the distances and can give opposite radial
and transverse responses. These derivatives distinguish the physical
directions to which the data are sensitive.

\begin{table}[t]
\centering
\caption{Logarithmic responses of the acoustic scale and two distance
ratios at the fixed background~\cite{Guo2026GenesisFramework}. Each row varies
one physical coordinate while the others are held fixed. Central
differences use logarithmic step $10^{-3}$. The temperature and matter
density rows include a half-step check.}
\label{tab:p5-background-response}
\begin{tabular}{lrrr}
\toprule
$x$ & $E_{r_d,x}$ & $E_{(D_M/r_d)_{0.706},x}$
    & $E_{(D_H/r_d)_{0.706},x}$ \\
\midrule
$Y_p$ & $-0.00504$ & $+0.00504$ & $+0.00504$ \\
$\omega_b$ & $-0.129$ & $+0.129$ & $+0.129$ \\
$\omega_m$ & $-0.257$ & $+0.114$ & $-0.0210$ \\
$T_{\rm CMB}$ & $+0.157$ & $-0.157$ & $-0.158$ \\
$N_{\rm eff}$ & $-0.102$ & $+0.102$ & $+0.102$ \\
$H_0$ & $+2.77\times10^{-8}$ & $-0.715$ & $-0.445$ \\
Common neutrino mass & $+0.00112$ & $-0.00111$ & $-0.00111$ \\
\bottomrule
\end{tabular}
\end{table}

The seven central differences require fourteen thermal background solutions.
The two half step checks require four more. The central background is
reused. For $r_d$ the differences between the two step sizes are
$6.93\times10^{-5}$ and $9.78\times10^{-5}$ in the temperature and
matter-density elasticities. They provide numerical consistency
estimates for the displayed derivatives. A variation of the underlying
generated inputs must additionally compose these partial derivatives
with their common dependence. In particular the shared mass relation
between $T_{\rm CMB}$ and $H_0$ supplies a joint direction.

\paragraph{The common mass scale and the temperature constraint.}
Let $s$ be a logarithmic displacement of the common cosmological
neutrino mass normalization. Hold the mass ratios and $\tau$ and $\MP$
fixed together with $\eta_b$ and the proton mass and weak-sector inputs. The relations
of Eq.~\eqref{eq:p5-cosmology-shared-ratio} give
\begin{equation}
 H_0(s)=H_0e^{2s},\quad T_\gamma(s)=T_\gamma e^s,
 \quad m_{\nu_i}(s)=m_{\nu_i}e^s,
 \quad\Omega_m(s)=\Omega_m,\quad\Omega_b(s)=\Omega_b e^{-s}.
 \label{eq:p5-common-cosmological-scale}
\end{equation}
The neutrino energy density scales as $e^{4s}$ because $m_{\nu_i}/T_\nu$
is fixed. The critical density has the same scaling and preserves the
photon and neutrino fractions and the vacuum fraction. Baryon density instead scales
as $\eta_bm_pT_\gamma^3$ and gives the final relation in
\eqref{eq:p5-common-cosmological-scale}. In the coordinates of
Table~\ref{tab:p5-background-response} this direction is
$(\delta\log H_0,\delta\log T_\gamma,\delta\log m_\nu,
\delta\log\omega_m,\delta\log\omega_b)=(2,1,1,4,3)s$.
It preserves the cosmological scale relations while testing their
common normalization. The production endpoint remains fixed.

\begin{proposition}[Finite common-scale response of the thermal background]
\label{prop:p5-finite-cosmological-scale}
Consider the flat background with constant vacuum density and thermal
Fermi--Dirac neutrinos of fixed distribution shape together with pressureless
baryons and cold matter. Apply
\eqref{eq:p5-common-cosmological-scale} at fixed $\eta_b$ and $m_p$ and
$N_{\rm eff}$ with the remaining content factors held fixed. Compensate the cold
density so that the total baryon--cold density scales as $e^{4s}$.
On the interval where this cold density remains positive
\begin{equation}
 E(z;s)=E(z;0),\qquad
 \frac{(D_X/r_d)(z;s)}{(D_X/r_d)(z;0)}
 =\alpha(s):=e^{-2s}\frac{r_d(0)}{r_d(s)},
 \quad X=M,H,V .
 \label{eq:p5-finite-bao-common-scale}
\end{equation}
Let $a_d(s)$ be the drag scale factor and write the sound-horizon integral as
\begin{equation}
\begin{aligned}
 r_d(s)&=e^{-2s}I_s(a_d(s)),\\
 I_s(a_d)&=\int_0^{a_d}
 \frac{c\,da}{\sqrt3\,a^2H(a;0)
       \sqrt{1+e^{-s}{\cal R}_b(a)}},\\
 {\cal R}_b(a)&=\frac{3\rho_b(a;0)}{4\rho_\gamma(a;0)}.
\end{aligned}
 \label{eq:p5-finite-sound-horizon}
\end{equation}
For fixed $a_d=a_d(0)$ the common multiplier obeys
\begin{equation}
 -\tfrac12\max(s,0)\leq\log\alpha(s)
             \leq-\tfrac12\min(s,0).
 \label{eq:p5-fixed-drag-bound}
\end{equation}
\end{proposition}
\begin{proof}
At each redshift $m_{\nu_i}/T_\nu$ is unchanged. The integral in
\eqref{eq:p5-thermal-neutrino-budget} and the photon density scale as
$e^{4s}$. The vacuum density has the same scaling. The baryon density
scales as $e^{3s}$ whereas the compensating cold contribution makes
their sum scale as $e^{4s}$. Both components evolve as $a^{-3}$ in
this background approximation. The Friedmann equation therefore gives
$H(a;s)=e^{2s}H(a;0)$. Radial and transverse distances scale as
$e^{-2s}$ as does their volume-distance combination. This proves
\eqref{eq:p5-finite-bao-common-scale}.
The baryon loading is $e^{-s}{\cal R}_b(a)$ and gives
\eqref{eq:p5-finite-sound-horizon}. At fixed upper limit the logarithmic
derivative of its integrand is
$e^{-s}{\cal R}_b/[2(1+e^{-s}{\cal R}_b)]\in[0,1/2]$.
Positivity makes $\partial_s\log I_s$ a weighted average in the same
interval. Integrating in $s$ proves \eqref{eq:p5-fixed-drag-bound}.
\end{proof}

Composing the calculated derivatives gives
$\partial_s\log(D/r_d)\simeq-0.74496$ for all thirteen entries.
For $d_i=\partial_st_i$ the covariance-weighted tangent optimum is
$s_*=-d^{\mathsf T}C^{-1}(t-y)/(d^{\mathsf T}C^{-1}d)
=0.01439$ with tangent standard deviation $0.00368$.
The independent monopole measurement
$T_\gamma=2.72548\pm0.00057\,\mathrm K$~\cite{Fixsen2009}
constrains the same variable to
$s=(2.21\times10^{-6})\pm2.0914\times10^{-4}$.
Within its three-standard-deviation interval the linearized distance
change is at most $4.69\times10^{-4}$ and the smallest BAO statistic
is $28.20$. Thus the allowed common-normalization displacement is
insufficient to remove the distance tension. The extrapolated $s_*$
locates the required displacement within this tangent approximation.
Combining the monopole and BAO covariance blocks at the fixed
prediction gives $\chi^2=29.50881$ for fourteen components and
$p=0.00891$ assuming independent measurement errors.  The
thirteen-component statistic uses BAO alone. The fourteen-component
statistic adds the CMB-monopole block. The one-parameter BAO scale
projection in Eq.~\eqref{eq:p5-bao-scale-decomposition} assigns
\(15.2965\) to that direction and \(14.2122\) to its orthogonal residual.
The physical background and all other generated parameters remain fixed.

The finite scale relation tests the required thermal response.
The exact image of the monopole three-standard-deviation
interval is
\begin{equation}
 s\in\left[
 \log\frac{T_{\rm obs}-3\sigma_T}{T_\gamma},
 \log\frac{T_{\rm obs}+3\sigma_T}{T_\gamma}\right]
 =[-0.000625400,\,0.000629425].
 \label{eq:p5-cmb-finite-scale-interval}
\end{equation}
For any common multiplier the original covariance gives exactly
\begin{equation}
 \chi^2(\alpha)=\chi^2_\perp+
                  \frac{(\alpha-\alpha_*)^2}{\sigma_\alpha^2}.
 \label{eq:p5-finite-scale-statistic}
\end{equation}
The level $\chi^2\leq\chi^2_{13;0.95}=22.3620$ therefore requires
$\alpha\in[0.981459,0.997105]$. This interval identifies the common multipliers that reach the stated
reference level of the fixed-model statistic.
At fixed drag epoch \eqref{eq:p5-fixed-drag-bound} instead gives
$\alpha\in[0.999685,1.000313]$. Using the unrounded enclosure gives
the conservative bound $\chi^2\geq28.6237$.

For an evolving drag epoch define
$J_s=I_s(a_d(s))/I_s(a_d(0))$. Equations
\eqref{eq:p5-finite-bao-common-scale}--\eqref{eq:p5-fixed-drag-bound}
give
\[
 \log\alpha(s)=
 -\log\frac{I_s(a_d(0))}{I_0(a_d(0))}-\log J_s.
\]
Consequently reaching the above $95\%$ reference level within
\eqref{eq:p5-cmb-finite-scale-interval} requires
$\log J_s\geq0.00258406$ at the selected displacement. Reaching
$\alpha_*$ requires $\log J_s\geq0.0104607$.
Equivalently any differentiable response reaching that reference
level has $\sup|\partial_s\log\alpha|\geq4.60543$ between the fixed
branch and the selected point. Relative to the calculated central
slope $-0.744959$ its finite logarithmic remainder must be at least
$0.00242988$ in magnitude. These necessary response sizes follow
from the finite temperature interval and measurement covariance.
The actual change of $a_d$ is determined by the recombination and
drag evolution. Table~\ref{tab:p5-background-response} checks the
central slope by finite differences. The finite remainder is the
integrated departure from that slope over the allowed interval.

\paragraph{Finite recombination and drag response.}
At fixed expansion and baryon loading with helium abundance held fixed the remaining
thermal dependence can be studied directly through the drag optical depth.
Let $R_b(z)=3\rho_b/(4\rho_\gamma)=R_{b0}/(1+z)$ and
$n_e(z)=n_{H0}(1+z)^3x_e(z)$ and define
\begin{equation}
 \mathcal T_d(z)=\int_0^z
 \frac{c_{\rm light}\sigma_Tn_{H0}(1+z')^3x_e(z')}
      {R_{b0}H(z')}\,dz',\qquad
 \mathcal T_d(z_d)=1,\qquad r_d=r_s(z_d).
 \label{eq:p5-drag-optical-depth}
\end{equation}
Here $r_s(z)=\int_z^\infty c_s(z')/H(z')\,dz'$ is the comoving
sound horizon. The reference calculation uses the recombination
electron history in this integral with astrophysical reionization
excluded from the definition of its drag epoch.

\paragraph{Finite optical-depth propagation.}
Let $\mathcal T_0$ be continuous and strictly increasing across the
drag-root interval and let $r_s$ be positive and decreasing there.
Suppose a second positive-opacity history obeys
\[
 |\mathcal T_d(z)-\mathcal T_0(z)|
 \leq\epsilon\mathcal T_0(z)+A,
 \qquad 0\leq\epsilon<1,\quad 0\leq A<1.
\]
Then its drag root lies between the reference inverse-depth levels
\begin{equation}
 z_- =\mathcal T_0^{-1}\!\left(\frac{1-A}{1+\epsilon}\right),
 \qquad
 z_+ =\mathcal T_0^{-1}\!\left(\frac{1+A}{1-\epsilon}\right),
 \qquad r_s(z_+)\leq r_d\leq r_s(z_-).
 \label{eq:p5-drag-finite-envelope}
\end{equation}
At fixed $H(z)$ the thirteen distance ratios therefore have the
common multiplier $\alpha=r_{d,0}/r_d$. Their statistic over this
finite set is exactly \eqref{eq:p5-finite-scale-statistic}. Its minimum
is attained at the projection of $\alpha_*$ onto the allowed interval
and its maximum at an endpoint.
At the new root the assumed inequality gives
$(1-\epsilon)\mathcal T_0-A\leq1
\leq(1+\epsilon)\mathcal T_0+A$.
Monotonicity gives \eqref{eq:p5-drag-finite-envelope}.
The distance numerators are fixed and the final statement follows
from the positive quadratic coefficient in
\eqref{eq:p5-finite-scale-statistic}.

There is an independent electron-number bound for the low-redshift
part of this problem. Keep the electron history fixed above $z_\ell$
and allow either history below it to satisfy
$0\leq n_e\leq n_{H0}(1+2f_{\rm He})(1+z)^3$.
Nonnegative radiation and neutrino densities together with nonnegative vacuum density imply
$H(z)\geq H_0\sqrt{\Omega_b+\Omega_{\rm cold}}(1+z)^{3/2}$.
Integration then gives the finite absolute bound
\begin{equation}
 |\Delta\mathcal T_d|\leq A_\ell:=
 \frac{2c_{\rm light}\sigma_Tn_{H0}(1+2f_{\rm He})}
 {5R_{b0}H_0\sqrt{\Omega_b+\Omega_{\rm cold}}}
 \bigl[(1+z_\ell)^{5/2}-1\bigr].
 \label{eq:p5-late-electron-capacity}
\end{equation}
For $z_\ell=30$ outward interval evaluation of this analytic expression
at the fixed background gives $A_{30}\leq0.01132773059844881$.
Numerical inversion of the reference depth then gives
$r_d\in[146.86995,147.02327]~\mathrm{Mpc}$ and
$\chi^2\in[28.0490,31.0265]$.
This result bounds changes between all electron histories satisfying
the stated electron-capacity bound in that low-redshift interval
with the background and higher-redshift history held fixed.

The higher-redshift recombination response is evaluated with the
full radiative-transfer solver supplied with
HYREC-2~\cite{AliHaimoudHirata2011,LeeAliHaimoud2020}.
The fixed background and thermal responses are preserved in the
archived calculation~\cite{Guo2026ThermalCalculation}.
Both integrations use the same tabulated $H(z)$ and hydrogen and helium
number densities together with the photon temperature and $r_s(z)$ of the reference
background. One uses standard atomic constants. The other uses the
generated $\alpha(0)$ and electron pole mass
$m_e:=M_e=G_e m_e^{\rm geom}$. The displayed calculation uses
$G_e=1$ in Eq.~\eqref{eq:p5-lepton-mass-definitions} and gives the ratios
\[
 \begin{gathered}
 \frac{\alpha(0)}{\alpha_{\rm at}}=0.9999247538,
 \qquad \frac{m_e}{m_{e,\rm at}}=0.9987383562,\\
 \frac{\sigma_T}{\sigma_{T,\rm at}}
 =\left(\frac{\alpha(0)/\alpha_{\rm at}}
                   {m_e/m_{e,\rm at}}\right)^2=1.0023772036.
 \end{gathered}
\]
The reference atomic normalization is the declared standard-constant
convention of the thermal calculation. The ratios enter the atomic
rates and matter-temperature evolution together with the drag opacity.
The generated values are fixed before the distance comparison.

\begin{table}[htbp]
\centering\small
\begin{tabular}{lrrr}
\toprule
Thermal calculation & $z_d$ & $r_d\;(\mathrm{Mpc})$ & $\chi^2$\\
\midrule
Reference RECFAST background & 1060.219424 & 146.946194 & 29.5088\\
HYREC with standard atomic constants & 1060.217303 & 146.946386 & 29.5051\\
HYREC with generated $\alpha(0)$ and $m_e$ & 1058.799882 & 147.074547 & 27.1191\\
\bottomrule
\end{tabular}
\caption{Finite recombination responses at the archived thermal-study
background~\cite{Guo2026ThermalCalculation}, with the expansion history,
helium abundance and baryon loading held fixed. Every statistic uses
the original thirteen-component observational covariance with no
adjusted distance parameter.}
\label{tab:p5-direct-recombination}
\end{table}

The generated-atomic calculation has common distance multiplier
$0.9991272939$ and reference tail probability $p=0.0119820$.
Applying \eqref{eq:p5-late-electron-capacity} also to this history
with its corresponding Thomson cross section gives
$r_d\in[146.99810,147.15186]~\mathrm{Mpc}$ and
$\chi^2\in[25.7794,28.5175]$.
Thus the directly calculated atomic displacement together with the
entire low-redshift electron capacity leaves the statistic above
the $95\%$ reference level. This is a finite test of these thermal
effects with the upstream background held fixed.

The analytic capacity bound controls the allowed low redshift histories.
Numerical integration and atomic model accuracy are assessed separately.
The depth quadrature
error estimates in the two direct integrations are below
$1.6\times10^{-11}$. These estimates concern evaluation of the
selected recombination equations. The physical remainder comprises
the omitted atomic and radiative-transfer terms above $z=30$ and
joint changes of the upstream baryon and helium inputs and the thermal inputs.
The displayed bounds quantify the low-redshift electron-history
response within the retained atomic model at fixed upstream inputs.
The difference between recombination implementations is a numerical
comparison while a theory covariance requires a probability model
for the physical remainder. The retained calculation reuses the
background distance integrals and contains only the background and
acoustic quantities needed for this test~\cite{Guo2026GenesisFramework}.

At fixed $E(z)=H(z)/H_0$ all three distance ratios scale as
$(H_0r_d)^{-1}$. The scale projection therefore tests this common
combination. Changes in the redshift dependence of $E(z)$ also enter
$r_\perp$ while the thermal history determines both $r_d$ and the
neutrino contribution to $E(z)$. Attribution to the Hubble scale or
the sound horizon or the thermal mapping requires their joint physical
variation. The present discrepancy constrains the complete chain from
the endpoint densities and rest masses through the photon and neutrino
thermal history, cold transport and the flat Einstein background to
the drag scale and observed distances. Its empirical
assessment uses the published measurement covariance. Any additional
theory covariance requires a separately derived input uncertainty or
approximation error for that chain.
For independent measurement and theoretical errors the joint covariance
is $C_{\rm tot}=C_{\rm obs}+C_{\rm th}$. The comparison
implementation~\cite{Guo2026GenesisFramework} accepts such an independently
specified $C_{\rm th}$ with its source and approximation order. The
associated $\chi^2$ probability uses a zero-mean Gaussian likelihood
for the combined errors. The values quoted here use $C_{\rm obs}$.

\subsection{The gravitational sector}\label{sec:gravitational}

The dimensional anchor fixes $\GN=1/(8\pi\MP^2)$ in the macroscopic
metric action~\cite{Guo2026SpectralGravity}. The endpoint source variance
and positive RMS amplitude response determine the instantaneous potential for
specified source and boundary data. The joint matter--potential action then determines its feedback on
the transported source. The auxiliary-radius model follows internal
compression, occupied matter and external expansion, with
the matter phase supplying a clock on its timelike branch. The
homogeneous Einstein--matter background supplies the
thermal evolution and distances calculated above.
The finite responses of the five source channels use spectral variables
expressed in Planck units.

At emergence $L_{\rm phys}=\kL/M_{\rm G}$. The long-root normalization
uses $C_2(2,1)=2[2(2+1)+1(1+1)]=16$ and gives
\[
 \lambda_{\rm long}=\frac{16}{L_{\rm phys}^2}
       =\frac{16M_{\rm G}^2}{(\kL)^2}.
\]
Express the channel inputs in Planck units
\[
 \widehat L=\MP L_{\rm phys},\qquad \widehat k=M_{\rm G}/\MP,\qquad
 \widehat V_3=\pi^2\widehat L^3,\qquad
 \widehat\lambda_{\rm long}
   =\lambda_{\rm long}/\MP^2=\frac{16}{\widehat L^2}.
\]
Let $\widehat\Pi_a^2(\widehat L,\widehat k^2,\tau)$ denote the
dimensionless channel functional with its eigenvalues and curvature shifts together with
frequency and Gaussian cutoff expressed in these same units. The five
channels are the spin-2 and spin-0 responses of $T_{\mu\nu}$ together with
$F^2$ and $G^2$ and $J^\mu$. The dimensionally consistent normalization of their
combined response is
\[
 \widehat{\mathcal S}
   =\left|\sum_a\widehat V_3\widehat\Pi_a^2
                   (\widehat L,\widehat k^2,\tau)\right|,
 \qquad \widehat\sigma=\widehat{\mathcal S}\widehat k^2,\qquad
 \sigma=\MP^2\widehat\sigma .
\]
The associated dimensionless ratio and effective coupling are
\[
 r_{\rm read}=\frac{\widehat\sigma}{\widehat\lambda_{\rm long}}
       =\frac{(\kL)^2}{16}\widehat{\mathcal S},\qquad
 Z_{\rm read}=\frac1{1+r_{\rm read}},\qquad
 G_{\rm read}=\frac{\GN}{Z_{\rm read}} .
\]
For a finite channel sum $r_{\rm read}\geq0$ and
$0<Z_{\rm read}\leq1$. Evaluating the five channels in this common
convention gives $r_{\rm read}=85.9303$ and $Z_{\rm read}=0.0115035$
\cite{Guo2026GenesisFramework}.
The resulting $G_{\rm read}$ is the effective coupling associated
with this channel sum. The metric action uses $\GN$ fixed by the
dimensional anchor. Appendix~\ref{app:repro} specifies the numerical
evaluation.
The sectoral quantities
$p_{\rm TT}^2=8/L_{\rm phys}^2$ and $\lambda_{\rm TT}=14/L_{\rm phys}^2$ retain their
respective definitions.

On the matched Einstein--matter branch the positive tensor residue
is normalized by the same Newton constant that fixes the dimensional
scale. This macroscopic normalization also enters the local source
response and its dynamical realizations.

\paragraph{Local source and potential.}
The invariant covariance on the decomposable pure-occupation orbit
gives the angular factor $4/3$ in the acceleration variance.
Direction averaging cancels the mean source and retains its
mean-square amplitude. The normalized Gaussian susceptibility
$Z_{\rm IR}$ compares squared response amplitudes; its positive RMS
amplitude $\sqrt{Z_{\rm IR}}$ gives the spatial flux. The reference
endpoint fixes $a_0$ together with $H_0$, and this normalization remains
fixed during the subsequent local evolution. For the positive acceleration scale
$a_0$ with $y=|\nabla\Phi|/a_0$ and $\mu(y)=y/\sqrt{1+y^2}$
this law determines a Bekenstein--Milgrom potential~\cite{BekensteinMilgrom1984}.
On a bounded Lipschitz domain the static weak solution is unique
for Dirichlet data in $H^1$ and a source in $H^{-1}$.
The same potential accelerates pressureless matter. Its evolving
density feeds back into the nonlinear Gauss law. The joint
source--potential action gives this feedback together with continuity.
On a fixed smooth domain its energy balance accounts for
matter flux and boundary work throughout smooth evolution before
caustic formation. Impermeable matter boundaries and time-independent
Dirichlet data conserve the total energy. These results are derived
in Eq.~\eqref{eq:p5-endpoint-angular-covariance} and
Propositions~\ref{prop:p5-endpoint-flux-uniqueness}
and~\ref{prop:p5-feedback-energy}.

The finite plane-symmetric realization evolves the same nonlinear
field and material source with mass conservation. Halving the time
step reduces the energy-balance residual by approximately a factor
of four consistent with second-order time accuracy. At each time
transport supplies the free density and the potential determined by
that source and its boundary data supplies the polarization density.
The latter contributes to the regional mean through its boundary
flux. This density decomposition complements the cold-source
transport and homogeneous averaging of Section~\ref{sec:linprop}. Its explicit relation is Eq.~\eqref{eq:p5-local-mean-source}.

\paragraph{Matter response and the auxiliary collective radius.}
The canonical matter action determines matter evolution under a
time-dependent internal conformal source. Internal geometry fixes
the mode frequencies and kinetic norms that enter the macroscopic
matter fields. The action's positive kinetic form
reproduces the static spectral frequencies and occupied source charge. The phase-space second moment obeys an exact source-work identity.
The corresponding causal response retains the direct variation and
both frequency branches. It supplies the matter contribution to a
common constrained gravitational calculation. Covariant variation
gives the complete matter stress tensor and, on the matter equations, the exchange
$Q_a\nabla_\nu s^a$ where $s^a$ are the retained sources and $Q_a$
their matter responses. This exchange enters the collective source equation and
the lapse and shift constraints together. Uniform compression has
an exact lapse representation with a time-dependent dilation
coupling. Its mixed velocity form retains the canonical source
constraint. The derivation is given in
Proposition~\ref{prop:p5-native-source-dynamics} and
Eqs.~\eqref{eq:p5-native-source-stress}--\eqref{eq:p5-native-source-ward} and
Eqs.~\eqref{eq:p5-native-lapse-dilation}--\eqref{eq:p5-native-dilation-constraint}.

Allowing the inherited complete unweighted relative potential to
vary in both amplitude and radius gives the unique joint vacuum
radius $L_*=\sqrt2L_c$ where $L_c$ is the condensation threshold
and a positive joint Hessian. This supplies
a collective static restoring response for the common source
balance. Retaining the radius as an auxiliary collective variable
gives a closed covariant amplitude--matter model with occupied and expanding
spatially flat FLRW solutions. These initial data satisfy
the source balance and Einstein constraints. Their nonzero matter
phase charge distinguishes them from the joint static vacuum.
The source constraint balances the restoring force against the
occupied kinetic and potential densities. As expansion changes this
balance, the radius adjusts together with the matter fields.
The internal radius sets the spectral scales while the external
scale factor sets the physical volume occupied by the homogeneous
matter. Their coupled evolution follows from this common balance
and the Einstein constraints.
Positivity of the algebraic source-constraint derivative
$\mathfrak h>0$ fixes this adjustment as a locally unique smooth
function of the retained fields and their first derivatives.
For the constructed occupied data the initial compression rate is
strictly positive. After the common constraints are
eliminated the three physical scalar kinetic and gradient forms are
positive on a nonzero evolution interval of this occupied background
alongside the two Einstein tensor polarizations. The global rotation charge of the retained real pair
is conserved and its matter phase reads proper time on this same
background. The static result and the occupied construction are
proved in Propositions~\ref{prop:p5-joint-native-radius}
and~\ref{prop:p5-auxiliary-native-closure}. The finite-interval bounds
are given in Eq.~\eqref{eq:p5-auxiliary-finite-margins}.

For a retained matter mode of nonzero amplitude with phase $\vartheta$
and $X=-(\nabla\vartheta)^2>0$ choose the future-directed phase
congruence. The matter equations determine
its compression force and its radial derivative corrections. The
homogeneous and static limits give geodesic flow and acceleration
by the source mass gradient respectively with explicit error
estimates. Along each flow curve $ds=d\vartheta/\sqrt X$ relates
proper time to the matter phase. Phase increments count the matter
oscillation and the instantaneous frequency $\sqrt X$ calibrates
each increment into a time interval. Monotone changes of the evolution
parameter assign different clock scales to this same physical
trajectory and preserve its canonical evolution and source-work
identity. A unit dust clock exists on a contractible spacetime patch
when the one-form $X^{-1/2}d\vartheta$ is closed or equivalently when
the phase congruence is geodesic. These relations follow from
Proposition~\ref{prop:p5-native-phase-congruence}. As in
Section~\ref{sec:comp} generated dynamics follow by varying
the action on its admissible variables and reducing their common
constraints. Finite-window kinetic tests use the complete physical
form and its error estimate.

\paragraph{Independent-foliation realization.}
For comparison, an effective action with an independent foliation field
gives a realization of the local constitutive law and the
Einstein--matter cosmological branch. A smooth geodesic foliation
recovers the Einstein--matter equations while the static foliation
with vanishing shift gives the RMS law at leading weak-field
nonrelativistic order. On a spatially flat dust--vacuum background
with positive dust density and nonzero Hubble rate together with constant vacuum
density and a stationary radial condensate of positive mass squared
the scalar constraint reduction has positive kinetic coefficients
for nonzero Fourier modes. A conserved relative-foliation charge
contributes to the total cold density without changing its linear
metric response (Proposition~\ref{prop:p5-covariant-two-limits}).

On accelerated static backgrounds the leading slow-motion reduction
instead gives the velocity operator $2M^2(DL^{-1}D-D)$ where
$M$ is the Einstein normalization and $D=-\Delta$ and
$L=-\nabla\cdot[D\mathcal J(\nabla\bar\Phi)\nabla]$ with
$\mathcal J(g)=\mu(|g|/a_0)g$ and fixed Dirichlet perturbations.
For a constant background and a nonzero longitudinal wavevector
the velocity form is negative when
$|\nabla\bar\Phi|/a_0>0.786151377\ldots$. Its restriction to supplied
foliation directions and their spherical support are described in
Proposition~\ref{prop:p5-generated-clock-kinetic} and the radial
specialization~\eqref{eq:p5-clock-radial-form}. These results describe the
independent foliation realization on the two stated backgrounds.
The cosmological distance calculation of Section~\ref{sec:linprop}
uses the homogeneous thermal background and linear Einstein--matter
propagation specified there.

%% file: sections-en/sec09_qcd_bbn.tex
% =============================================================================
% Section 9 - QCD and the primordial abundances
% The Spectrum of a Compact Internal Space II: Effective Couplings and Mass Scales
% =============================================================================

\section{QCD scale and confinement}\label{sec:qcd}

The colour coupling of Section~\ref{sec:g3} sets the QCD scale used
for the proton and glueball channels, string tension and deconfinement
temperature. Their absolute values test the conversion of the running
coupling into a hadronic scale. Ratios in which this scale cancels test
the content and channel assignments.

The weak-process chain brings these hadronic quantities together with
the flavour and electroweak sectors. The light-quark mass difference
and electromagnetic subtraction determine $\Delta m_{np}$ in the
native isovector readout with $g_S^{\rm eff}=1$. This mass difference
sets the decay phase space, while $G_F$ and $V_{ud}$ with the axial response set the weak
amplitude. The same gap and weak rate enter freeze-out. Neutron
survival to the adopted capture temperature then determines the reduced
helium estimate. The lifetime and abundance probe the common source
through the decay phase space and thermal neutron population,
respectively. Their joint
response distinguishes changes in the hadronic mass difference from
changes in the weak amplitude. Appendix~\ref{app:bbnext} evaluates this
colour-to-weak chain and retains the common source and thermal assumptions.

\subsection{The QCD scale and the proton mass}\label{sec:proton}

The QCD scale follows by propagating the colour coupling $g_3(M_{\mathrm{G}})$ of Section~\ref{sec:g3} to the internal extraction scale and applying the five-flavour two-loop scale formula. The Yukawa-difference correction rescales the extracted coupling
$\alpha_s$ by $(1-s_0\kappa/N_g)$ before the nonlinear conversion to
$\Lambda_{\mathrm{QCD}}$ with $s_0=2\tau$ and $N_g=8$.
The geometric seed $y_0=1$ is held fixed in the running. Appendix~\ref{sec:derivqcd} specifies the extraction point and the nonlinear scale map. Appendix~\ref{sec:p5-qcd-sensitivity} gives their error propagation. The result is
\[
\Lambda_{\mathrm{QCD}} = 0.2080 \ \mathrm{GeV} ,
\]
a deviation of $-0.94\%$ from the reference $0.210$ GeV.
The adopted constituent-content relation for the proton mass is
\[
m_p = N_c \Delta_f \Bigl(1 - \frac{1}{N_g^2 \Delta_s}\Bigr) \Lambda_{\mathrm{QCD}} \Bigl(1 + \tau \kappa \sum Y^2 \Delta_s\Bigr)
= \frac{279}{64} \Lambda_{\mathrm{QCD}} \Bigl(1 + \frac{5\tau\kappa}{3}\Bigr) = 0.9411 \ \mathrm{GeV} ,
\]
where $N_c=3$ and $\Delta_f=3/2$ and $\Delta_s=1/2$. The conformal-content correction is $1-1/(N_g^2\Delta_s)=31/32$. The final bracket is the effective hadronic response factor. Appendix~\ref{sec:derivqcd} evaluates both factors. The value differs from the observed proton mass by $+0.30\%$.

The neutron--proton splitting probes the response to an isovector
source. The difference $m_d-m_u$ varies at fixed mean light-quark
mass. The scalar response measures the resulting difference between
the two baryon energies. The native normalization uses
$g_S^{\rm eff}=1$ with the electromagnetic proton-minus-neutron
contribution subtracted in the same source convention. The colour
singlet and its allowed source insertions are explicit in
Proposition~\ref{prop:p5-three-quark-source}. Eliminating intermediate
channels retains their contribution to the baryon source response
and the full-state normalization in
\eqref{eq:p5-baryon-source-schur}.

\subsection{Glueball mass channels}\label{sec:glueball}

Composite sources are products of retained source operators resolved
into definite symmetry channels. The mass readout assigns a scale to
each channel using the same mode algebra as the elementary couplings
and collective restoring responses.
The composite-channel readout pairs two internal transverse-vector
sources. Their lowest vector modes belong to the Killing carrier
of Ref.~\cite{Guo2026CompactInternalSpaceI}. The two-leg contribution is common
to the channel family while projection onto total rotation
representations resolves its scalar and tensor responses. The
spectral assignment is
\[
\widehat\lambda = 2\widehat\lambda_{\mathrm{gluon}} + C_2(J) + n \, N_g \xi ,
\]
where $\widehat\lambda_{\mathrm{gluon}}=4$ and $N_g\xi=1$ are dimensionless. The reported scalar reference and mass-ratio rule are
\[
m_G=8\Lambda_{\mathrm{QCD}}=1.664\ \mathrm{GeV},\qquad
m_a=m_G\sqrt{\widehat\lambda_a/8}.
\]
The conversion from dimensionless spectral labels to squared masses
is $m_a^2=8\widehat\lambda_a\Lambda_{\mathrm{QCD}}^2$.
Here the scalar reference $\widehat\lambda_*=8$ fixes the common
mass normalization through $m_G/\Lambda_{\rm QCD}=\widehat\lambda_*$.
The absolute scalar mass therefore tests the reference-to-mass
conversion together with the extracted QCD scale. Ratios of channel
masses cancel that normalization and test the relative spectral
assignments.
Appendix~\ref{sec:p5-qcd-mass-readout} gives this operator mapping
and checks the two-Killing-leg origin of the normal reference levels.
The scalar reference and the selected tensor projection have labels
$8$ and $16$ in the symmetric square of that carrier. The pseudoscalar
and excited scalar use the additional conformal-channel assignments
$17$ and $18$. Each mass therefore combines a resolved channel label
with the same QCD scale.
The selected channel ratios are
\[
2^{++}/0^{++} = \sqrt{2}, \qquad 0^{++*}/0^{++} = 1.50, \qquad 0^{-+}/0^{++} = 1.46 .
\]
The tensor mass is $2.354$ GeV. The first conformal-excitation mass is $2.496$ GeV and the pseudoscalar mass is $2.426$ GeV. For the pseudoscalar Chen et al.\ obtain $2560(35)(120)$ MeV in quenched lattice QCD~\cite[Table~XXI]{Chen2006}. The BESIII partial-wave analysis determined $J^{PC}=0^{-+}$ for $X(2370)$ in 2024~\cite{BESIII2024}. The combined analysis of three radiative-$J/\psi$ channels gives $2359^{+13}_{-14}$ MeV~\cite{BESIII2026}. Our value lies $5.2\%$ below the former central value and $2.8\%$ above the latter.

The BESIII study of suppressed $K^*\bar K$ decays argues for a
dominant pseudoscalar-glueball component in $X(2370)$
\cite{BESIII2026b}. This motivates comparison with the channel mass
above. Quenched Yang--Mills calculations provide the unmixed reference
spectrum~\cite{MorningstarPeardon1999,Chen2006}. Comparison with a
physical resonance also involves meson--glueball mixing
\cite{Huang2025}. The four channels considered here specify masses
before such mixing. Their overall scale is fixed by
$\Lambda_{\mathrm{QCD}}$ with tensor-to-scalar ratio $\sqrt2$.
The complete condensate operator of Section~\ref{sec:massgap} has its
own kinetic and source normalization.

\subsection{String tension and deconfinement}\label{sec:confine}

The string tension and deconfinement temperature are assigned to the effective tensor and lowest vector channels respectively. They give
\[
\sigma = \frac{\lambda_{\mathrm{TT}}}{\pi} \Lambda_{\mathrm{QCD}}^2 = \frac{14}{\pi} \Lambda_{\mathrm{QCD}}^2 = 0.1929 \ \mathrm{GeV}^2 ,
\]
where $\lambda_{\mathrm{TT}} = 14$ is the effective coefficient of the three-branch tensor catalogue in this string-tension relation. The rough-TT spectrum in Ref.~\cite{Guo2026CompactInternalSpaceI} has minimum $6/L^2$ on its two-branch carrier. The deconfinement temperature is
\[
T_d = \frac{\lambda_{\mathrm{vector}}}{N_c} \Lambda_{\mathrm{QCD}} (1 - \tau\kappa) = \frac{4}{3} \Lambda_{\mathrm{QCD}} (1 - \tau\kappa) = 271.1 \ \mathrm{MeV} ,
\]
where $\lambda_{\mathrm{vector}}=4$ is the lowest dimensionless transverse-vector label. The factor $(1-\tau\kappa)$ assigns the chiral-squash correction of Section~\ref{sec:geodyn} to the deconfinement scale. The ratio
\[
\frac{\sigma}{T_d^2} = \frac{14}{\pi} \cdot \frac{9}{16}\, (1-\tau\kappa)^{-2} = \frac{126}{16\pi (1-\tau\kappa)^2} = 2.6242
\]
is independent of the common QCD scale. The content ratio $126/(16\pi)=2.5068$ is enhanced by the inverse chiral-squash factor $(1-\tau\kappa)^{-2}$. Appendix~\ref{sec:derivqcd} evaluates these confinement-scale relations.

%% file: sections-en/sec10_results.tex
% =============================================================================
% Section 10 - Numerical results and comparison
% The Spectrum of a Compact Internal Space II: Effective Couplings and Mass Scales
% =============================================================================

\section{Numerical results and comparison}\label{sec:results}

Gauge couplings test channel normalization, boundary matching and scale
evolution; mass combinations test relative amplitude transfer and finite
matching. The neutron lifetime and helium abundance combine the hadronic
and flavour responses with weak reaction rates. Cosmological distances
test the thermal history, sound horizon and expansion mapping jointly.
The benchmarks in Appendix~\ref{app:table} include measurements,
extrapolated couplings and estimates of channel masses, each with its
stated comparison convention.

The comparisons distinguish dimensionless structural relations from dimensional
values normalized by \(\GN\) and joint tests of outputs sharing common
inputs. Each benchmark retains its measurement definition and approximation
level. Numerical errors, perturbative remainders and reference
uncertainties have separate budgets.

The signed percentage residual is $100(X_{\rm calc}/X_{\rm ref}-1)$ evaluated before display rounding. Outputs sharing the same endpoint or mass scale form correlated comparison channels. Joint statistical tests use the covariance and likelihood appropriate to the experimental data and the stated theoretical approximation.

For charged leptons the conversion
\eqref{eq:p5-lepton-mass-definitions} fixes the object tested at
each stage. The pole invariant \eqref{eq:p5-lepton-pole-reference}
constrains the factors $G_\ell$ whereas the common-scale constraints
\eqref{eq:p5-running-mass-numbers} involve their products
$C_\ell=D_\ell G_\ell$.

\subsection{Scale evolution and finite matching}\label{sec:interface}

The spectral and effective relations supply $g_i(M_{\mathrm G})$ (Section~\ref{sec:gauge}) and $y_0=1$ (Section~\ref{sec:top}) together with $\lambda_H$ (Appendix~\ref{sec:derivlamh}) and $v$ (Section~\ref{sec:vev}). The geometric Yukawa and quartic are retained as fixed readouts distinct from the running variables in the adopted two-loop transfer of Appendix~\ref{sec:rgflow}. The flow ends at the self-consistent internal $Z$ mass of Section~\ref{sec:ewprec}. Figure~\ref{fig:interface} shows how these inputs enter the low-energy comparison.

\input{figures/fig_interface}

The common action realizes the generated boundary relations in the mass
terms, interaction vertices and reaction amplitudes of the same
normalized fields. Finite matching expresses these responses
in the adopted mass and renormalization scheme. The top decay tests
the same generated parameters in an additional reaction amplitude.
Table~\ref{tab:p5-three-procedures} summarizes the
three uses of this action. In the fixed-Yukawa calculation the gauge couplings evolve while the
geometric Yukawa and quartic determine the tabulated hierarchy. The
five-variable system instead evolves all five couplings from common
boundary data. Finite matching fixes the renormalized coefficients
from the generated Fermi and mass conditions at one scale.

\begin{table}[t]
\centering
\caption{Inputs and outputs of the three calculation procedures. The
gauge normalization is $g_1=\sqrt{5/3}\,g_Y$.}
\label{tab:p5-three-procedures}
\begin{tabular}{p{2.6cm}p{4.4cm}p{5.5cm}}
\toprule
Procedure & Boundary or physical conditions & Result \\
\midrule
Fixed-Yukawa evolution & Spectral $g_i(\MG)$ with fixed $y_0=1$ and
geometric quartic and hierarchy VEV & Two-loop gauge evolution and
the sector readouts in Tables~\ref{tab:params} and~\ref{tab:ewprec}. \\
Five-variable evolution & A common-scale tuple
$(g_1,g_2,g_3,y_t,\lambda_H)$ and threshold maps & Joint scale
dependence of all five couplings for that boundary. \\
Finite one-loop matching & Generated $G_F,M_t,M_H$ and leading gauge
coefficients at $Q=M_Z$ & Tadpole-free $\overline{\rm MS}$
$v_L,\overline y_t,\overline\lambda,\overline m^2$ with the real
physical conditions preserved through one loop. \\
\bottomrule
\end{tabular}
\end{table}

\paragraph{An additional decay observable.}
The imaginary part of the same top self-energy determines an output
beyond the real conditions used in the matching. With $m_b=0$ and
$|V_{tb}|=1$ Eq.~\eqref{eq:p5-top-width-test} gives
\begin{equation}
 \Gamma_t^{(0)}=
 \frac{G_FM_t^3}{8\pi\sqrt2}(1-x_W)^2(1+2x_W)
 =1.53445\ {\rm GeV},\qquad
 x_W=\frac{(g_2v_F/2)^2}{M_t^2}.
 \label{eq:p5-top-width-result}
\end{equation}
The charged-current interaction and all masses in this expression
belong to the leading action used in the one-loop pole kernel. Its $W$ mass is $g_2v_F/2=79.79153\ {\rm GeV}$.
This is the Born decay width obtained from the one-loop pole's
absorptive part~\cite{Martin2016TopPole,PDG2024}. The experimental
compilation gives $1.42^{+0.19}_{-0.15}\ {\rm GeV}$~\cite{PDG2024}.
The width tests a decay channel whose value was not imposed in the
real-part matching. A precision comparison includes the relative QCD
and electroweak corrections of orders \(\alpha_s/\pi\) and
\(\alpha/\pi\) together with finite bottom mass and physical
charged-current mixing. These corrections set the accuracy beyond the
Born approximation.

\subsection{Couplings, masses and comparison conventions}\label{sec:table}

Appendix~\ref{app:table} lists each computed value with its comparison reference and signed residual and defining relation. Table~\ref{tab:sectors} gives one representative per sector. The tiers in Table~\ref{tab:tiers} identify the approximation and interpretation of each comparison. Their meaning is independent of the numerical error or statistical confidence.

\begin{table}[t]
\centering
\caption{The sectors of the computation and one representative value per sector with short comparison descriptors. Logical roles and comparison tiers are distinguished in Section~\ref{sec:status}. The full table is Appendix~\ref{app:table}.}\label{tab:sectors}
\begin{tabular}{lccc}
\toprule
Sector & Representative quantity & Computed value & \shortstack{Comparison\\status} \\
\midrule
Internal scale & generation count $n_g$ & $3$ & joint screening \\
Gauge couplings & weak coupling $g_2(M_{\mathrm{G}})$ & $0.50885$ & calculated \\
Flavour & top mass $m_t$ & $174.1$ GeV & calculated \\
Electroweak & vacuum expectation $v$ & $246.19$ GeV & calculated \\
Electroweak precision & $Z$ mass $M_Z$ & $91.125$ GeV & calculated \\
Cosmology & Hubble rate $H_0$ & $67.45$ km\,s$^{-1}$\,Mpc$^{-1}$ & exploratory \\
QCD & proton mass $m_p$ & $0.9411$ GeV & spectral est. \\
\bottomrule
\end{tabular}
\end{table}

\begin{table}[t]
\centering
\caption{Comparison tiers. Tier A identifies the anchor and selected structure. Tier B covers couplings and masses. Tier C records scheme-sensitive quantities. Tier D contains exploratory extensions.}\label{tab:tiers}
\begin{tabular}{l p{2.4cm} p{6.0cm}}
\toprule
Tier & Content & Reading \\
\midrule
A & anchor and structure & Newton's constant and inherited structural constants with the generation count selected by the joint rules. Fixed before the low-energy evaluation \\
B & couplings and masses & Gauge couplings and mass ratios with $v$ and the electroweak and charged-fermion masses. Neutrino mass scale and rest-mass splittings \\
C & scheme-sensitive & $m_b$ and $\Lambda_{\mathrm{QCD}}$ with $\Gamma_b$ and $\sin^2\theta_{\mathrm{eff}}$. Consistency checks labelled by their perturbative and renormalisation status \\
D & exploratory & Unit-amplitude solar scalar source. PMNS and CP assignments with baryon asymmetry. Cosmological fractions and $T_{\mathrm{CMB}}$ with hadronic scales and other phenomenological extensions \\
\bottomrule
\end{tabular}
\end{table}

\subsection{Shared constraints and numerical residuals}\label{sec:deviation}

The relations between sectors provide tests that survive elimination
of a common dimensional scale. A change of one source response
propagates to every quantity that uses it. Joint observations
constrain both the response and its relative contribution to
different sectors. Write $D=2\pi+r_{23}$ and
$A_\gamma=C_\gamma(r_{12}/\pi)(1-\tau\Delta_s)$ with
$x=s_0\kappa(s_0)$ and $Y_u=2/3$ in terms of the neutrino ratios and
thermal factors with the hypercharge used in Section~\ref{sec:cosmo}.
The cosmological pair obeys
\begin{equation}
 \Omega_\Lambda=\frac{D}{3\pi},\qquad
 \frac{(k_{\rm B}T_{\rm CMB})^2}{\MP H_0}
 =A_\gamma^2\sqrt{\frac{D}{\pi Y_u}}
       \frac{(1-x)^2}{\sqrt{1-4x}}.
 \label{eq:p5-results-shared-cosmology}
\end{equation}
Both relations eliminate the common neutrino mass scale. In particular
a change of that scale moves $T_{\rm CMB}$ and $H_0$ together while
preserving their displayed ratio at fixed content and correction
factors. The hadronic channel ratios similarly eliminate
$\Lambda_{\rm QCD}$. These combinations test the relative readout
rules while absolute masses and densities also test the hierarchy
normalization. Their remaining dependence and finite error bounds are
given in Appendix~\ref{app:sensitivity}.

The three gauge-coupling residuals are below $0.01\%$ in magnitude. The fine-structure-constant residual is $+0.20\%$ and the electroweak-scale residual is $-0.012\%$. The charged-fermion mass ratios and atmospheric neutrino splitting and listed homogeneous cosmological quantities lie within one percent of their adopted references. Figure~\ref{fig:deviation} displays the residuals in Appendix~\ref{app:table} including separate rest-mass and unit-source solar entries. Its charged-lepton bars show the diagnostic realization with unit geometric-to-pole conversion.

\input{figures/fig_deviation}

The cosmological distance data provide a joint test with a measured
covariance. The DESI DR2 comparison gives $\chi^2=29.5087$ for thirteen
fixed predictions with $p=0.00554011$. Its exact scale projection in
Eq.~\eqref{eq:p5-bao-scale-decomposition} separates $15.2965$ along a
common distance-scale direction and $14.2122$ in the orthogonal
residual. Table~\ref{tab:p5-bao-residuals} locates the discrepancies in
redshift. This tension enters the empirical assessment of the complete
endpoint and thermal-background mapping through to the distances. The projection
identifies a shared scale direction. Attributing it to a particular
physical input requires varying that input through the full mapping.

The production calculation~\cite{Guo2026GenesisFramework} evaluates
the physical quantities before applying the comparison data.
Additional tests hold the response and transfer rules fixed.

\subsection{Leading comparison offsets and their physical origin}\label{sec:devsummary}

The following central-value offsets localize distinct stages of the
calculation. Their statistical significance is governed by the uncertainty and
matching convention together with the approximation level of each observable.

\begin{itemize}
\item The vacuum rest-spectrum value
$\Delta m_{21}^2=6.874155\times10^{-5}\ \mathrm{eV}^2$ lies $7.231\%$
below the adopted solar-oscillation benchmark. The unit-amplitude
scalar-source propagation readout instead has offset $+0.030\%$ in
Tier~D. Its correlated splitting shifts are given in
Eq.~\eqref{eq:p5-correlated-solar-response}. The physical solar test uses
the source-dependent probabilities in Eq.~\eqref{eq:p5-solar-forward-readout}.
\item The geometric bottom-mass readout of Section~\ref{sec:lowmass} with
the fixed geometric anchor $y_0=1$ lies $1.38\%$ above the displayed
$\overline{\mathrm{MS}}$ benchmark. This Tier-C offset compares the stated
geometric and renormalized mass conventions.
\item The lightest scalar-channel readout
$8\Lambda_{\mathrm{QCD}}$ lies $2.10\%$ below the catalogue benchmark
$1.7\ \mathrm{GeV}$. It tests the common QCD extraction and the specified
channel-to-mass assignment. The relevant lattice and resonance comparisons
carry their own spectral and mixing uncertainties.
\item The neutron lifetime is $915.54\,\mathrm{s}$ and lies $4.23\%$ above its
reference value. Its phase space depends strongly on the same
neutron--proton mass difference that enters weak freeze-out and the helium
calculation in Appendix~\ref{app:bbnext} and makes the lifetime a correlated
test of the hadronic--weak chain.
\end{itemize}

The QCD-scale residual is $-0.94\%$ for the specified two-loop extraction. The PMNS mixing angles and CKM phase together with primordial abundances use the mixing and thermal assumptions in Appendices~\ref{app:flavour} and \ref{app:bbnext}. The same reduced freeze-out and decay calculation gives $\Delta m_{np}=1.28497\,\mathrm{MeV}$ and $\tau_n=915.54\,\mathrm{s}$ and $Y_p=0.246082$. The corresponding central-value residuals are $-0.647\%$ and $+4.23\%$ and $+0.44\%$. Table~\ref{tab:p5-weak-chain} compares these linked observables and Appendix~\ref{app:bbnext} gives their common-input response. The common electroweak amplitude gives an $R_b$ residual of $+0.384\%$.

The running and electroweak relations correlate $m_b$ and $\Lambda_{\mathrm{QCD}}$ with $m_t$ and $M_W$ and $\Gamma_b$ through their shared inputs across the comparison tiers.

The lifetime elasticity to the mass difference is $-6.53633$.
At fixed other inputs a smaller energy release reduces the available
phase space and lengthens the lifetime. This local response maps the
mass-gap deficit to a lifetime shift with the direction and
approximate size of the reported discrepancy. Replacing that mass difference alone
by its experimental value gives
$\tau_n=877.59\,\mathrm{s}$ and $Y_p=0.240716$ in the same weak
approximation (Table~\ref{tab:p5-weak-gap-diagnostic}). This externally
conditioned diagnostic locates a strong phase-space contribution to
the lifetime discrepancy while displaying its coupled effect on helium.
The common input response propagates hadronic matching, the weak amplitude
and radiative factors through both observables. Their joint comparison
tests how the same hadronic variation changes decay phase space and
neutron survival during the thermal history.

\subsection{Numerical bounds and physical sensitivity}\label{sec:sensitivity}

The tabulated central values are obtained from floating point
evaluations before display rounding. The interval enclosures certify
the specified spectral tails and endpoint ranges with retained-position
counts at their stated accuracy. The precision of each tabulated observable also depends
on its input propagation and physical approximation. Appendix~\ref{app:repro}
records the evaluation precision and rounding with verification coverage
while Appendix~\ref{app:sensitivity} supplies the enclosures.
For the production response the certified endpoint is
$c_{3,5}\in[2.4912094,2.4958593]$. At fixed remaining inputs its exact
propagation gives relative ranges $[0.9702,1.0306]$ for $v$ and
$[0.9544,1.0476]$ for $m_e$ and $[0.7853,1.2722]$ for $\rho_\Lambda$
each normalized at $c_\circ=2.49353$
(Table~\ref{tab:p5-production-error-propagation}). These guaranteed
ranges quantify the coverage of the conservative quadrature certificate. The tabulated percentage residuals describe the floating-point central
comparisons. The shared endpoint correlates all three outputs.

Appendix~\ref{app:sensitivity} gives the dependence of the observables on their common spectral and content parameters. At fixed other inputs the hierarchy obeys $v\propto(\kL)^{-1}e^{-4\pi\kL}$ and $m_e\propto e^{-20\kL}$. Its local tangent estimates for a one-percent change of $\kL$ are $-32\%$ for $v$ and $-50\%$ for $m_e$ and $-259\%$ for $\rho_\Lambda$. Finite changes require the exact propagation formulas of that appendix. The dark-energy fraction $\Omega_\Lambda$ is a pure content ratio and has zero elasticity. The upstream definitions are the specified TT kernel $K_{\mathrm{TT}}=(1-y)^2$ and the endpoint normalisation $\max_{0\le y\le1}y(1-y)^2=4/27$ and the torsion content ratio $\tau=(N_L-N_R)/(N_f\sum Y^2)=1/50$. For the adopted bandwidth and content these relations fix the endpoint. The elasticity matrix quantifies how its uncertainty is amplified in the mass hierarchy. Figure~\ref{fig:sensitivity} summarises the local response.

\input{figures/fig_sensitivity}

For shared inputs $\theta$ with a justified probability model the
logarithmic Jacobian $J_{ai}=\partial\log O_a/\partial\theta_i$
propagates their covariance as $J\Sigma_\theta J^{\mathsf T}$.
Bounds on truncation and quadrature instead propagate as enclosures. Omitted perturbative terms retain their stated order. Variation of the
response domain $R$ or of a physical transfer rule compares different
models. These three operations have distinct statistical meanings and
are kept separate when assessing the residuals.

\subsection{Electroweak precision observables}\label{sec:ewprec}

The electroweak calculation uses the common inputs of Section~\ref{sec:interface}. These comprise $v$ and $G_F = 1/(\sqrt{2}\,v^2)$ together with $\alpha(M_Z)$ of Section~\ref{sec:gauge} and $\alpha(0)$ obtained from the one-loop QED vacuum-polarisation descent of Appendix~\ref{sec:derivbbn} with the geometric couplings at $M_{\mathrm{G}}$ and the fermion masses of Section~\ref{sec:flavour} and the fixed Higgs-quartic readout $\lambda_H$ of Appendix~\ref{sec:derivlamh}.

The internal $Z$ mass is the self-consistent fixed point of the tree-level mass formula evaluated on the two-loop geometric running
\[
M_Z = \frac{v}{2}\sqrt{g_2^2 + \tfrac{3}{5}g_1^2}\Big|_{\mu = M_Z},
\]
which gives $M_Z = 91.125$~GeV against the observed $91.1876$~GeV ($-0.069\%$) with $\alpha^{-1}(M_Z) = 128.206$ ($+0.20\%$). The $W$ mass is the on-shell relation
\[
s^2 c^2 = \frac{\pi\alpha(0)}{\sqrt{2}\,G_F M_Z^2}\frac{1}{1-\Delta r},\qquad
\Delta r = \Delta\alpha - \frac{c^2}{s^2}\Delta\rho,
\]
where $s^2=1-M_W^2/M_Z^2$ and $c^2=M_W^2/M_Z^2$ with $\Delta\alpha=1-\alpha(0)/\alpha(M_Z)$ and $\Delta\rho$ is the one-loop $t$-$b$ doublet contribution with its full mass dependence. The result is $M_W=80.372$~GeV against the retained $80.369$~GeV benchmark ($+0.003\%$). The on-shell mixing angle is $\sin^2\theta_W=0.22208$ with a $-0.50\%$ residual against the value formed from the reference masses. The leading-universal approximation uses the leading-top baseline $\Delta\rho_{\rm top}^{(1)}=3x_t$ with $x_t=G_Fm_t^2/(8\sqrt2\pi^2)$ as in Eq.~(8.22) of Ref.~\cite{Denner1993}
\[
\Delta\bar\rho=3x_t\left[1+x_t(19-2\pi^2)\right]=0.00948 .
\]
The retained two-loop heavy-top coefficient is its small-$M_H/m_t$
limit.
At the common electroweak reference scale write the retained running
coupling as $\alpha_Z$. The effective parameters satisfy
\[
 \bar s^2(1-\bar s^2)=
 \frac{\pi\alpha_Z}{\sqrt2G_FM_Z^2}(1-\Delta\bar\rho),
 \qquad \rho_{\rm NC}=\frac1{1-\Delta\bar\rho},
 \qquad \bar s^2=s_W^2+c_W^2\Delta\bar\rho.
\]
These are the leading-universal relations of
Ref.~\cite[Eqs.~(8.21)--(8.25)]{Denner1993}. The weak branch gives
$M_W=80.337$~GeV ($-0.039\%$) and $\bar s^2=0.23011354$ and
$\rho_{\rm NC}=1.00956729$. The identity
$\pi\alpha_Z/(\bar s^2\bar c^2)=\sqrt2G_FM_Z^2\rho_{\rm NC}$
fixes the normalization of the same neutral-current amplitude.
The coupling fraction from the geometric running is
$\sin^2\theta(M_Z)=0.23327$. Its conversion to the effective angle
is specified by the matching relations above. Table~\ref{tab:ewlevel}
states the perturbative approximation for each comparison.

The displayed $\alpha^{-1}(M_Z)=128.206$ is evaluated at the self-consistent $Z$ mass. The value $128.207$ in Appendix~\ref{app:table} comes from two-loop evolution to the preliminary one-loop estimate of that scale. The scale-transfer calculation~\cite{Guo2026GenesisFramework} uses the latter coupling in the $W$-mass relation and the width radiators. The two coupling readouts differ by about $0.0006\%$.

All $Z$ partial widths use $v_f=T_3^f-2Q_f\bar s^2$ and
$a_f=T_3^f$ and the common normalization $\rho_{\rm NC}$.
The nonuniversal heavy-top contribution to the bottom amplitude is
$\delta v_b=\delta a_b=x_t$ at first order
\cite[Electroweak review, Eqs.~(10.53), (10.55)]{PDG2024}.
Keeping its interference with the leading amplitude gives
\begin{equation}
 \Gamma_f=N_c^f\frac{G_FM_Z^3\rho_{\rm NC}}{6\pi\sqrt2}
 \left[v_f^2+a_f^2+2\delta_{fb}x_t(v_b+a_b)\right]
 R_{\rm QCD}^f R_{\rm QED}^f .
 \label{eq:p5-common-z-widths}
\end{equation}
Here $R_{\rm QCD}^q=1+\alpha_s/\pi+1.409(\alpha_s/\pi)^2$
for quarks and $R_{\rm QCD}^\ell=1$ and
$R_{\rm QED}^f=1+3Q_f^2\alpha_Z/(4\pi)$.
The factorized radiators multiply the retained leading-universal
amplitude and first-order bottom interference. The latter corresponds to
\[
 \delta_b=\frac{2x_t(v_b+a_b)}{v_b^2+a_b^2}=-0.01448455,
 \qquad \sin^2\theta_{\rm eff}^b=\bar s^2(1+2x_t)=0.23157075
\]
through first order in the nonuniversal term. Final-state masses enter
at $O(m_f^2/M_Z^2)$. The table retains their massless limit.

The partial widths are summed before forming the ratios and peak cross
section. The results are $\Gamma_Z=2.50636$~GeV ($+0.447\%$) and
$\Gamma_b=0.380407$~GeV ($+0.890\%$) with
$R_b=0.2171203$ and $R_l=20.79153$ and
$\sigma_{\rm had}^0=41.54848$~nb ($+0.163\%$).
The definition of the peak is
$12\pi\Gamma_e\Gamma_{\rm had}/(M_Z^2\Gamma_Z^2)$.
In Table~\ref{tab:ewprec} $\Gamma_l$ sums the three charged-lepton
widths and $R_l$ uses the single-flavour denominator $\Gamma_e$.

The strong-sector matching and additive generation contribution of
Appendix~\ref{sec:derivlamh} fix $\lambda_H=0.129$. Combined with
the generated $v$ they give the tree-level relation
$m_H=\sqrt{2\lambda_H}\,v=124.98$~GeV with central-value residual
$-0.17\%$. This value then supplies the Higgs mass condition in the
one-loop matching of Section~\ref{sec:interface} and determines
the corresponding renormalized parameters. The mass comparison
tests the tree-level relation and the finite matching implements
that same physical mass in the quantum action through the calculated
order. The squash coefficient
$\lambda=149$ in Section~\ref{sec:lambda} belongs to the internal
scalar potential. The ladder's geometric muon and tau masses are $105.31$~MeV
and $1.772$~GeV with residuals $-0.33\%$ and $-0.28\%$.

\begin{table}[t]
\centering
\caption{Electroweak comparisons for the common inputs of Section~\ref{sec:interface} with deviations evaluated before display rounding. The $Z$ mass is tree-level on the two-loop running. The two $W$-mass entries use the one-loop-$\rho$ and leading-universal approximations. All effective $Z$ widths and ratios with the peak cross section follow Eq.~\eqref{eq:p5-common-z-widths} with a common effective angle and neutral-current normalization and radiators. The Higgs mass is at tree level.}\label{tab:ewprec}
\begin{tabular}{lccc}
\toprule
Quantity & Computed & Reference & Deviation \\
\midrule
$M_Z$ (GeV) & 91.125 & 91.1876 & $-0.069\%$ \\
$\alpha^{-1}(M_Z)$ (self-consistent) & 128.206 & 127.952 & $+0.20\%$ \\
$\sin^2\theta(M_Z)$ & 0.23327 & 0.23122 & $+0.89\%$ \\
$M_W$ (GeV) & 80.372 & 80.369 & $+0.003\%$ \\
$M_W$ leading-univ. (GeV) & 80.337 & 80.369 & $-0.039\%$ \\
$\sin^2\theta_W$ (on-shell) & 0.22208 & 0.22321 & $-0.50\%$ \\
$\rho_{\rm NC}$ & 1.009567 & --- & --- \\
$\sin^2\theta_{\mathrm{eff}}^l$ & 0.23011 & 0.23153 & $-0.61\%$ \\
$\Gamma_Z$ (GeV) & 2.50636 & 2.4952 & $+0.447\%$ \\
$\Gamma_{\mathrm{had}}$ (GeV) & 1.75206 & 1.7444 & $+0.439\%$ \\
$\Gamma_b$ (GeV) & 0.380407 & 0.37705 & $+0.890\%$ \\
$\Gamma_l$ (GeV) & 0.252804 & 0.251898 & $+0.360\%$ \\
$\Gamma_{\mathrm{inv}}$ (GeV) & 0.501494 & 0.4990 & $+0.500\%$ \\
$\sigma_{\mathrm{had}}^0$ (nb) & 41.5485 & 41.481 & $+0.163\%$ \\
$R_l$ & 20.7915 & 20.767 & $+0.118\%$ \\
$R_b$ & 0.217120 & 0.21629 & $+0.384\%$ \\
$m_H$ (GeV) & 124.98 & 125.20 & $-0.17\%$ \\
\bottomrule
\end{tabular}
\end{table}

\begin{table}[t]
\centering
\caption{The perturbative level of each electroweak precision observable in Table~\ref{tab:ewprec} together with the leading correction at the next level.}\label{tab:ewlevel}
\begin{tabular}{l p{4.2cm} p{6.0cm}}
\toprule
Observable & Evaluated level & Leading next correction \\
\midrule
$M_Z$ & tree level on the two-loop running & one-loop electroweak self-energy \\
$M_W$ & one-loop $\rho$ and leading-universal heavy-top terms with resummed $\alpha$ & finite $\Delta r_{\mathrm{rem}}$ including bosonic contributions with vertex and box contributions \\
$\sin^2\theta(M_Z)$ & geometric couplings at $M_Z$ & full one-loop $\sin^2\theta_{\mathrm{eff}}$ \\
$\sin^2\theta_{\mathrm{eff}}^l$ & improved Born with the retained heavy-top $\Delta\bar\rho$ & finite process-dependent $\kappa_l$ and higher orders \\
$\Gamma_f$ & Common leading-universal amplitude with heavy-top bottom interference and factorized QCD/QED radiators & finite-mass phase space and finite process-dependent EW terms and mixed radiative corrections \\
$m_H$ & tree-level generated mass relation & radiative correction to the mass relation with independently fixed boundary parameters \\
\bottomrule
\end{tabular}
\end{table}

\subsection{Observable tests}\label{sec:prospective}

Table~\ref{tab:falsifiable} gives tests of the numerical branches. Neutrino ordering and the summed mass test the mass readout while the octant and phase test the separate mixing ansatz. The tensor reference is the BK18 limit $r_{0.05}<0.036$ at 95\% confidence~\cite{BICEPKeck2021} with pivot $k_\ast=0.05\ \mathrm{Mpc}^{-1}$ and $n_t=0$. The local acceleration entry concerns the nonrelativistic static response of Appendix~\ref{app:gravity}. A lensing prediction requires the light-propagation response and source data in the same macroscopic realization as the matter trajectories and clock readouts. In a metric representation these determine the effective metric along the ray. Proposition~\ref{prop:p5-covariant-two-limits} examines one realization with an independent foliation field.

\begin{table}[t]
\centering
\caption{Observable comparisons of the stated branches. The mass and flavour entries together with the cosmological and local-response entries retain their respective derivational and comparison status.}\label{tab:falsifiable}
\begin{tabular}{p{3.2cm} p{3.2cm} p{5.1cm}}
\toprule
Quantity & Value fixed here & Discriminating measurement \\
\midrule
Tensor-to-scalar ratio & $r=0.0253$ & a robust upper bound well below this value \\
Neutrino mass ordering & normal & inverted ordering \\
Atmospheric octant & $\sin^2\theta_{23}=0.551$ & lower octant \\
Summed neutrino mass & $\sum m_\nu=0.0615$ eV & a robust upper bound below this value \\
PMNS phase (mixing ansatz) & $\delta_{\mathrm{PMNS}}\simeq(8/7)\pi$ & value far from this branch \\
Pseudoscalar glueball target & $m_{0^{-+}}=2.426$ GeV & a determined glueball mass incompatible with the stated unmixed channel assignment \\
Low-acceleration branch & $a_0=1.204\times10^{-10}\ \mathrm{m\,s^{-2}}$ and $\mu(y)=y/\sqrt{1+y^2}$ & rotation curves incompatible with the response for the stated baryonic source and boundary conditions \\
\bottomrule
\end{tabular}
\end{table}

\paragraph{Accuracy of the comparisons.}
The quoted uncertainties describe propagation within the present
truncation. Spectral accuracy is controlled by the finite sums and
endpoint enclosure together with the mode ceiling. Perturbative accuracy
is set by the loop order and threshold conversions. The Born top width
has QCD and electroweak corrections together with finite bottom-mass
and charged-current effects. The covariance budget for the cosmological
comparison comprises the full DESI covariance and CMB-monopole errors
together with the BBN-rate uncertainties. These contributions determine
the accuracy within the stated definitions.

The charged-lepton conversion is different in character.  The realization
$G_\ell=1$ is excluded by the pole invariant at $38\sigma$.  With the pole masses held as
renormalization conditions higher-order $D_\ell$ determines the corresponding
running parameters and preserves the specified poles. The pole prediction
depends on $G_\ell$ calculated from the generated vertex and on the
light-family transfer rule.

The numerical evaluation uses the Coarse Graining Genesis
Framework~\cite{Guo2026GenesisFramework}.  It evaluates the
stated finite sums and comparisons at the precision and approximation
orders specified above.

%% file: figures/fig_interface.tex
%% Figure — interface chain to the Standard Model
\begin{figure}[t]
\centering
\resizebox{\textwidth}{!}{%
\begin{tikzpicture}[
  font=\footnotesize,
  b/.style={draw, rounded corners=1pt, align=center, inner sep=5pt},
  arr/.style={-{Stealth[length=1.5mm]}, thick},
  carr/.style={dashed, -{Stealth[length=1.5mm]}, black!55},
]
\node[b, fill=black!6] (spec) {spectral and effective inputs\\[3pt]
  $g_i(\MG)$\quad $v$\quad $m_t$\quad $\lambda_H$};
\node[b, fill=black!6, right=1.1cm of spec] (trans) {scale evolution\\[3pt]
  two-loop RG running\\ electroweak relations};
\node[b, fill=black!6, right=1.1cm of trans] (obs) {low-energy EW/QCD observables\\[3pt]
  $M_Z$\quad $M_W$\quad $\Gamma_b$\quad $\alpha_s(M_Z)$\quad $m_b$};
\node[b, dashed, fill=black!2, below=1.0cm of obs] (cmp) {comparison references\\ measurements and estimates};
\draw[arr] (spec) -- (trans);
\draw[arr] (trans) -- (obs);
\draw[carr] (obs) -- (cmp) node[midway, right, font=\tiny, text=black!55] {comparison};
\end{tikzpicture}%
}
\caption{Spectral boundary data and two-loop running together with low-energy observables in the fixed-Yukawa calculation of Section~\ref{sec:interface}. The electroweak relations and their perturbative orders are given in Section~\ref{sec:ewprec}. The dashed branch introduces the reference values.}
\label{fig:interface}
\end{figure}

%% file: figures/fig_deviation.tex
%% Figure — signed deviation landscape, grouped by sector, coloured by tier
\begin{figure}[t]
\centering
\begin{tikzpicture}
\begin{axis}[
  name=devax,
  xbar,
  bar shift=0pt,
  bar width=3.0pt,
  width=12.0cm, height=14.0cm,
  xmin=-7.8, xmax=1.9,
  ymin=0.3, ymax=36.7,
  xlabel={central-value residual (\%)},
  xtick={-7,-6,-5,-4,-3,-2,-1,0,1},
  ytick={1,...,36},
  yticklabels={
    $m_{0^{++}}$,
    $m_p$,
    $\Lambda_{\mathrm{QCD}}$,
    $T_{\mathrm{CMB}}$,
    $\Omega_{\rm cold}$,
    $\Omega_b$,
    $\Omega_\Lambda$,
    $H_0$,
    $1-n_s$,
    $\eta_B$,
    $m_\tau^{\rm geom}$,
    $m_\mu^{\rm geom}$,
    $m_e^{\rm geom}$,
    $m_b$,
    $m_t$,
    $\Delta m^2_{31}$,
    $\Delta m^2_{21,\mathrm{tilt}}$,
    $\Delta m^2_{21,\mathrm{raw}}$,
    $m_{\nu 3}$,
    $m_H$,
    $\sin^2\theta_{\mathrm{eff}}^{\ell}$,
    $\Gamma_b$,
    $M_W^{\mathrm{lu}}$,
    $M_W$,
    $M_Z$,
    $v$,
    $m_\mu/m_e$,
    $m_s/m_d$,
    $m_t/m_u$,
    $m_b/m_s$,
    $m_t/m_c$,
    $\alpha_s(M_Z)$,
    $\alpha^{-1}(M_Z)$,
    $g_3(M_{\mathrm{G}})$,
    $g_1(M_{\mathrm{G}})$,
    $g_2(M_{\mathrm{G}})$
  },
  y tick label style={font=\footnotesize, xshift=1pt},
  axis y line*=left,
  axis x line=bottom,
  clip=false,
]
% guide band $\pm1\%$
\fill[black!4] (-1,0.3) rectangle (1,36.7);
% $\pm0.1\%$ inner guide
\draw[black!40, dashed] (-0.1,0.3) -- (-0.1,36.7);
\draw[black!40, dashed] (0.1,0.3) -- (0.1,36.7);
% sector separators
\pgfplotsinvokeforeach{3.5,9.5,10.5,15.5,19.5,25.5,26.5,31.5}{%
  \draw[black!22] (-7.8,#1) -- (1.9,#1);%
}
% Tier B -- core outputs
\addplot[fill=black!70, draw=black!70] coordinates {
  (0.199,33) (0.120,32)
  (-0.557,31) (-0.293,30) (-0.893,29) (-0.904,28) (-0.215,27)
  (-0.069,25) (-0.039,23) (-0.173,20)
  (-0.040,19) (-7.231,18)
  (0.806,15) (-0.126,13) (-0.325,12) (-0.275,11)
};
% Tier C -- scheme-sensitive
\addplot[fill=orange!80!black, draw=orange!80!black] coordinates {
  (0.890,22) (-0.612,21) (1.383,14) (-0.939,3)
};
% Tier D -- exploratory
\addplot[fill=black!38, draw=black!38] coordinates {
  (0.030,17)
  (0.463,10) (-0.285,9) (-0.080,8) (0.050,7) (0.219,6) (-0.174,5)
  (0.300,2) (-2.104,1)
};
% Near-zero residuals are below the graphical resolution of the bars.
\addplot[only marks, mark=|, mark size=3.0pt, line width=0.7pt, draw=black!70, forget plot] coordinates {
  (0,36) (0,35) (0,34) (0,26) (0,24) (0,16)
};
\addplot[only marks, mark=|, mark size=3.0pt, line width=0.7pt, draw=black!38, forget plot] coordinates {
  (0,4)
};
% sector labels (right)
\node[anchor=west, font=\scriptsize, text=black!55] at (1.5,34) {gauge couplings};
\node[anchor=west, font=\scriptsize, text=black!55] at (1.5,29) {fermion ratios};
\node[anchor=west, font=\scriptsize, text=black!55] at (1.5,26) {EW scale};
\node[anchor=west, font=\scriptsize, text=black!55] at (1.5,22.5) {EW precision};
\node[anchor=west, font=\scriptsize, text=black!55] at (1.5,17.5) {neutrinos};
\node[anchor=west, font=\scriptsize, text=black!55] at (1.5,13) {charged fermions};
\node[anchor=west, font=\scriptsize, text=black!55] at (1.5,10) {CP content};
\node[anchor=west, font=\scriptsize, text=black!55] at (1.5,6.5) {cosmology};
\node[anchor=west, font=\scriptsize, text=black!55] at (1.5,2) {QCD};
\end{axis}
\node[anchor=west, font=\footnotesize] at ([yshift=-1.6cm]devax.south west) {
  \colorbox{black!70}{\rule{0pt}{2.2ex}\hspace{1.15em}} couplings and masses (B)\qquad
  \colorbox{orange!80!black}{\rule{0pt}{2.2ex}\hspace{1.15em}} scheme-sensitive (C)\qquad
  \colorbox{black!38}{\rule{0pt}{2.2ex}\hspace{1.15em}} exploratory (D)
};
\end{tikzpicture}
\caption{Selected signed central-value comparisons for 36 entries grouped by sector and coloured by comparison tier. The charged-lepton bars show the realization with unit geometric-to-pole conversion tested in the pole-invariant comparison of Section~\ref{sec:flavour}. The other entries use the residual definitions in Appendix~\ref{app:table}. The rest-mass difference and the unit-amplitude scalar-source readout are compared with the same reference value. The latter is Tier~D. The shaded band marks $\pm1\%$ and the dashed lines $\pm0.1\%$. Residuals with $|\delta|<0.02\%$ appear as origin ticks. The three plotted entries outside $\pm1\%$ are discussed in Section~\ref{sec:devsummary}. The electroweak and weak-process comparisons are given in Tables~\ref{tab:ewprec} and~\ref{tab:p5-weak-chain} respectively.}
\label{fig:deviation}
\end{figure}

%% file: figures/fig_sensitivity.tex
%% Figure — sensitivity fan of the window capacity
\begin{figure}[t]
\centering
\begin{tikzpicture}[font=\footnotesize, >=Stealth]
\node[circle, draw, very thick, inner sep=3pt, fill=black!6] (kl) at (0,0) {$\kL$};
\node[left=0.3cm of kl, font=\scriptsize, text=black!70, align=right] {Elasticities\\ scaled by $1\%$};

\draw[->, thick, black!55] (kl) -- (70:4.0cm)
  node[anchor=west, text=black] {$m_t/m_c$\quad $+5\%$};
\draw[->, thick, black!55, dashed] (kl) -- (46.7:4.0cm)
  node[anchor=west, text=black] {$\Omega_\Lambda$\quad $0$};
\draw[->, thick] (kl) -- (23.3:4.0cm)
  node[anchor=west, text=black] {$v$\quad $-32\%$};
\draw[->, thick] (kl) -- (0:4.0cm)
  node[anchor=west, text=black] {$m_e$\quad $-50\%$};
\draw[->, thick] (kl) -- (-23.3:4.0cm)
  node[anchor=west, text=black] {$m_{\nu_3}$\quad $-65\%$};
\draw[->, thick] (kl) -- (-46.7:4.0cm)
  node[anchor=west, text=black] {$H_0$\quad $-129\%$};
\draw[->, very thick] (kl) -- (-70:4.0cm)
  node[anchor=west, text=black] {$\rho_\Lambda$\quad $-259\%$};
\end{tikzpicture}
\caption{Local sensitivity to $c=\kL$ at fixed other inputs. The labels give one percent of the logarithmic derivatives. The factors $c^{-1}e^{-4\pi c}$ and $e^{-20c}$ and their powers produce the large elasticities of dimensionful quantities. The content ratio $\Omega_\Lambda$ has zero elasticity. Appendix~\ref{app:sensitivity} gives the matrix and the formulas for finite changes.}
\label{fig:sensitivity}
\end{figure}

%% file: sections-en/sec11_conclusion.tex
\clearpage
\section{Discussion and conclusion}\label{sec:concl}

The effective couplings and mass scales share one source, whose Gaussian
law and reflected spectrum follow from minimal unbiased compositional change.
The physical construction of Ref.~\cite{Guo2026CompactInternalSpaceI}
supplies the compact response geometry, gauge group, matter module and
elementary scalar sector within its realization class. Common kinetic
normalization converts source matrix elements into interaction strengths;
successive transmission of mass amplitudes generates hierarchies.
Joint screening selects three matter copies. Shell support
and position retention then select \(R=5\) and the tensor balance has
one positive endpoint on that branch. The retained family positions
reproduce the same count.

The relations among sectors test more than individual central values.
The relation among six masses eliminates the common transfer parameters.
In the adopted running representation its comparison uses masses
in a common scheme and scale.
The lepton identity belongs to the geometric mass eigenvalues. Its
pole readout is recorded by \(G_\ell\), while the adopted running
representation uses the additional factors \(D_\ell\).
The geometric identities therefore constrain relative mass transmission
independently of the common endpoint and dimensional normalization.

Two cosmological combinations eliminate the common neutrino mass
normalization. These are the vacuum-density fraction and
\(T_{\rm CMB}^2/(M_{\rm P}H_0)\) in energy units. They test the relative
density, temperature and expansion laws. The electron and electroweak
scales provide another correlated test through their bandwidth response.
The three-position plateau persists for every continuous bandwidth
\(\mathcal R\geq5\) while the same active shells and three positions
require \(R_*<\mathcal R\leq16/\pi\) with \(R_*\simeq4.01871\).
The only integer level in this interval is five.
The monotone dependence of \(v/m_e\) on the endpoint supplies a
matched-ratio test of that level. Thus spectral retention and the
sensitivity of absolute masses are measured by distinct relations.

The family transfer and mixing rotations are realized in one canonical
action. Its replicated quadratic response retains the direct source
and domain variations. Strong-sector and generation-count matching
determine the Higgs quartic. Together with the generated \(v\) it gives
the tree-level Higgs mass. The generated Fermi condition and the top and Higgs conditions
then determine locally unique renormalized coefficients in the
tadpole-free \(\overline{\rm MS}\) scheme. The absorptive top self-energy
gives the additional Born output
\(\Gamma_t^{(0)}=1.53445\,\mathrm{GeV}\). The decay corrections and
charged-current assumptions determine the accuracy beyond that order.

The macroscopic constructions resolve complementary parts of the
matter--source response. On the decomposable occupation orbit, the
source covariance sets the acceleration variance and the positive
Gaussian RMS amplitude determines the spatial response. The resulting
source--potential system has a unique static weak solution and a boundary energy balance during smooth
matter evolution before trajectory crossing. The auxiliary-radius model
admits an occupied expanding background with autonomous source feedback and
positive constrained propagation and a phase clock on a finite interval.
On that solution, source exchange balances in the total stress, and
phase-charge conservation fixes the clock rate through the evolving
volume and matter amplitude.
For comparison, the action with an independent foliation varies the
congruence as an additional field. Its reduction gives positive scalar kinetic energy on
the specified cosmological branch and a longitudinal negative direction
on the accelerated-static branch. These signs characterize the stated
effective action and background.

The empirical comparisons probe the physical transfer laws as well as
the spectral data. The endpoint enclosure and its propagation give
guaranteed ranges at fixed remaining inputs. The tabulated residuals
describe the central comparisons within the stated mass and coupling
conventions. The emergence-scale gauge couplings agree closely
with the adopted Standard-Model extrapolation.

The correlated running-mass data give
\(\Delta_F=2.69537\pm1.48017\) and
\(\Delta_L=0.0112736\pm0.0005817\) at \(M_Z\).
The pole inputs require the weighted geometric-to-pole contribution
\(\Delta_{L,\rm ref}^{\rm pole}=0.00257845\pm0.00006753\).
With the quoted input errors the pole invariant excludes the exact
identification \(G_\ell=1\) at \(38.18\sigma_{\rm ref}\).
For this specified realization the independent one-loop
pole-to-running calculation gives
\(\Delta_L^{\rm run,(1)}=0.00933752\) including the finite-mass weak
contribution \(2.35\times10^{-6}\). Its running-mass comparison has
central offset \(0.00193607\). Higher-order pole-to-running corrections
enter that comparison with the specified poles held fixed.
The physical mass response determines the geometric to pole ratio for
each realization. Equation~\eqref{eq:p5-native-mass-vertex-correction}
expresses its first order value in the perturbative effective representation
through the mass vertex and both kinetic responses.
The finite lepton transfer and flavour-spread
bound organize the running and remaining matching contributions.

The solar rest-mass
splitting is \(7.23\%\) below its reference. The separate unit-source
entry describes the scalar propagation model. Its solar test uses
the environmental probabilities of Eq.~\eqref{eq:p5-solar-forward-readout}.
The geometric bottom mass lies \(1.38\%\) above the displayed
\(\overline{\mathrm{MS}}\) benchmark and the scalar-channel glueball
scale lies \(2.10\%\) below the catalogue reference. Their interpretation
depends on the mass matching, channel assignment and reference uncertainties.

For the fixed cosmological mapping the DESI comparison gives
\(\chi^2=29.5087\) for thirteen components and \(p=0.00554011\).
The covariance-weighted decomposition separates common-scale and shape
residuals. The CMB monopole constrains the same normalization direction.
Finite sound-horizon bounds specify the additional drag response needed
along that direction. Recombination with generated atomic constants and
the full allowed electron capacity below \(z=30\) at fixed upstream
background leave the statistic above the stated \(95\%\) reference
level. The atomic and radiative-transfer remainder above \(z=30\) and
joint changes of upstream inputs enter separately. This distance tension
therefore tests the adopted chain from endpoint densities and thermal
history to the acoustic scale and expansion distances.

The reduced weak-freeze-out and decay calculation gives
\(\Delta m_{np}=1.28497\,\mathrm{MeV}\) and
\(\tau_n=915.54\,\mathrm{s}\) and \(Y_p=0.24608\) with central residuals
\(-0.647\%\) and \(+4.23\%\) and \(+0.44\%\).
The lifetime and helium abundance probe the isovector response and
electromagnetic subtraction and weak rate in different combinations.
Replacing the total mass difference by its experimental value sharply
reduces the lifetime offset and lowers the helium abundance.
The source channel construction, constrained baryon resolvent and
independent propagation of lattice inputs identify the hadronic matching
quantities entering this joint test. The rotation-curve comparison
tests the pointwise approximation to the local static equation.
The neutrino, tensor and hadronic targets retain the response laws
fixed in these calculations.

The common source connects particle amplitudes to the restoring response
of occupied matter. Material phase changes supply elapsed-time readings,
and the effective metric expresses relations among motion, stress and
clock rates. Transforming the source, kinetic form and observables
together preserves these relations under a change of mode basis or clock
parameter. The independent mass combinations and joint response tests
give quantitative probes of this generated realization.

%% file: sections-en/appendix_a.tex
\section{Numerical values and reference conventions}\label{app:table}

Table~\ref{tab:structural} collects the inherited content identities and the endpoint quantities used throughout the calculation. Table~\ref{tab:params} gives the resulting values and comparison references together with signed central-value residuals and the tiers defined in Section~\ref{sec:results}. Newton's constant supplies the dimensional anchor. The remaining entries follow the spectral and effective relations specified in Section~\ref{sec:comp} and Appendix~\ref{app:deriv}.

The charged fermion entries are the geometric mass and
ladder readouts of Section~\ref{sec:flavour}. The Higgs entry uses
the tree relation $m_H^2=2\lambda_Hv^2$. The reference mass definitions are as follows. The value $172.69\ {\rm GeV}$ is the PDG~2022
direct-reconstruction top-mass reference whereas
$4.18\ {\rm GeV}$ denotes
$\overline m_b(\overline m_b)$ in the $\overline{\rm MS}$
scheme~\cite{PDG2022}. The reference charged-lepton masses are
pole masses and the $Z$ and Higgs values use the PDG~2024
resonance conventions~\cite{PDG2024}.
For charged leptons Eq.~\eqref{eq:p5-lepton-mass-definitions}
separates the geometric vertex from the physical pole and running definitions.
The table labels the charged lepton entries as geometric masses
$m_\ell^{\rm geom}$ and compares them with pole mass references.
Their pole residuals depend on the conversion $G_\ell$.
The specified realization $G_\ell=1$ is excluded by the pole invariant
at $38\sigma$ by Eq.~\eqref{eq:p5-lepton-pole-reference}. The general
ratio $G_\ell$ follows from the physical readout of the specified
realization. The finite one-loop
calculation gives the pole-to-running factors $D_\ell$ for the
specified pole conditions.
The quark ratio values $136$, $45$, $78000$ and $19.802$
serve as individual comparison quotients. The joint geometric relation
\eqref{eq:p5-sector-mass-test} is compared with physical masses through
the six matching factors specified in Section~\ref{sec:flavour}.
The $W$ and strong-coupling reference values
are $80.369\ {\rm GeV}$ and $\alpha_s(M_Z)=0.1179$.
All percentage entries are central-value comparisons in these
stated conventions. The finite one-loop matching in
Proposition~\ref{prop:p5-generated-onshell-matching} uses the
generated $(G_F,M_t,M_H)$ as physical conditions and reports the
corresponding running coefficients.

The two neutrino-splitting references are the normal-ordering values without Super-Kamiokande atmospheric data in NuFIT~5.2~\cite{NuFIT2020,NuFIT52} and equal $7.41\times10^{-5}$ and $2.511\times10^{-3}$ eV$^2$. The rest-mass difference (Tier~B) and the unit-amplitude scalar-source readout (Tier~D) are compared with the same reference value. Equation~\eqref{eq:p5-solar-forward-readout} specifies the environmental propagation needed for a solar interpretation. The $0.0502$ eV entry is an atmospheric mass-scale benchmark. The cosmological density and tilt references use the base-$\Lambda$CDM column labelled TT+TE+EE+lowE+lensing in Table~2 of Planck~2018~\cite{Planck2018}. The rounded fractions use its baryon and cold-dark-matter densities and Hubble parameter. The $H_0$ benchmark is expressed in GeV. The microwave-temperature reference is Fixsen's $2.72548\pm0.00057$ K~\cite{Fixsen2009}.

The gauge references at the emergence scale are two-loop extrapolations from the fixed boundary values in Appendix~\ref{app:repro}. Percentage residuals describe offsets from the displayed benchmarks. Statistical tests use experimental covariance and the theoretical uncertainties appropriate to each benchmark. Section~\ref{sec:results} gives the electroweak comparison levels; Section~\ref{sec:glueball} identifies the separate lattice and resonance references.

\begin{table}[t]
\centering
\caption{Content identities and endpoint quantities. Together with the dimensional anchor and the stated transfer relations they determine the entries of Table~\ref{tab:params}.}\label{tab:structural}
\begin{tabular}{@{}lcl@{}}
\toprule
Quantity & Value & Origin \\
\midrule
window capacity $\kL$ & $2.49353$ & spectral endpoint in Sec.~\ref{sec:closure} \\
emergence scale $\MG$ & $1.731 \times 10^{18}\ \mathrm{GeV}$ & $\MG = \MP\sqrt{\pi}/\kL$ in Sec.~\ref{sec:kk} \\
torsion modulus $\tau$ & $1/50$ & content response in Sec.~\ref{sec:torsion} \\
left-handed count $N_L$ & $8$ & charged module in Sec.~\ref{sec:count} \\
right-handed count $N_R$ & $7$ & charged module in Sec.~\ref{sec:count} \\
chiral asymmetry $N_L-N_R$ & $1$ & difference of the two counts in Sec.~\ref{sec:count} \\
hypercharge moment $\sum Y^2$ & $10/3$ & generated charge weights in Sec.~\ref{sec:count} \\
\bottomrule
\end{tabular}
\end{table}

\input{sections-en/appendix_a_table}

%% file: sections-en/appendix_a_table.tex
\begin{longtable}{p{2.1cm} r r r p{0.7cm} p{3.5cm}}
\caption{Computed quantities and fixed comparison benchmarks grouped by sector. The computed values use the physical relations and numerical scheme of Section~\ref{sec:comp}. The text of this appendix specifies the reference sources and versions together with the comparison limits. The derivation column points to the relations used in each evaluation.}\label{tab:params}\\
\toprule
Quantity & Computed & Reference & Residual & Tier & Relation \\
\midrule
\endfirsthead
\toprule
Quantity & Computed & Reference & Residual & Tier & Relation \\
\midrule
\endhead
\bottomrule
\endlastfoot
\multicolumn{6}{l}{Internal scale} \\
\midrule
generation count $n_g$ & 3 & 3 & +0.000\% & A & joint screening compatibility and retained-position check (App.~\ref{sec:p5-joint-content}) \\
\midrule
\multicolumn{6}{l}{Gauge couplings} \\
\midrule
weak coupling $g_2(M_G)$ & 0.508848 & 0.508845 & +0.001\% & B & geometric weak coupling after colour and torsion screening (App.~\ref{sec:derivg2}) \\
hypercharge coupling $g_1(M_G)$ & 0.60499 & 0.605001 & -0.002\% & B & squashed-isometry normalisation transported from its scale (App.~\ref{sec:derivg2}) \\
colour coupling $g_3(M_G)$ & 0.49776 & 0.497765 & -0.001\% & B & common-origin colour coupling with conformal correction (App.~\ref{sec:derivg2}) \\
$\alpha^{-1}(M_Z)$ & 128.207 & 127.952 & +0.199\% & B & two-loop running at the preliminary internal $Z$ scale (Sec.~\ref{sec:ewprec}) \\
$\alpha_s(M_Z)$ & 0.118041 & 0.1179 & +0.120\% & B & two-loop gauge flow with derived content coefficients (App.~\ref{sec:derivbeta}) \\
\midrule
\multicolumn{6}{l}{Fermion mass ratios} \\
\midrule
top to charm $m_t/m_c$ & 135.242 & 136 & -0.557\% & B & assigned up-sector ladder ratio (App.~\ref{sec:derivlz}) \\
bottom to strange $m_b/m_s$ & 44.8679 & 45 & -0.293\% & B & assigned down-sector ladder ratio (App.~\ref{sec:derivlz}) \\
top to up $m_t/m_u$ & 7.730364e+04 & 7.800000e+04 & -0.893\% & B & up-sector ladder with first-generation window transfer (App.~\ref{sec:derivlz}) \\
strange to down $m_s/m_d$ & 19.623 & 19.802 & -0.904\% & B & down-sector first-generation ladder ratio (App.~\ref{sec:derivlz}) \\
muon to electron $\frac{m_\mu^{\rm geom}}{m_e^{\rm geom}}$ & 206.356 & 206.8 & -0.215\% & B & charged-lepton ladder with the Euclidean-period shift (App.~\ref{sec:derivmass}) \\
\midrule
\multicolumn{6}{l}{Electroweak scale} \\
\midrule
Higgs vacuum expectation $v$ (GeV) & 246.19 & 246.22 & -0.012\% & B & Gaussian window hierarchy at the spectral fixed point (App.~\ref{sec:derivewsb}) \\
\midrule
\multicolumn{6}{l}{Electroweak precision} \\
\midrule
$Z$ mass $M_Z$ (GeV) & 91.1248 & 91.1876 & -0.069\% & B & self-consistent neutral electroweak mass relation (Sec.~\ref{sec:ewprec}) \\
$W$ mass $M_W$ (GeV) & 80.3717 & 80.369 & +0.003\% & B & on-shell relation with the one-loop rho correction (Sec.~\ref{sec:ewprec}) \\
leading-universal $W$ mass $M_W^{\mathrm{lu}}$ (GeV) & 80.3374 & 80.369 & -0.039\% & B & leading-universal on-shell electroweak relation (Sec.~\ref{sec:ewprec}) \\
$\Gamma_b$ (GeV) & 0.380407 & 0.37705 & +0.890\% & C & common effective neutral-current amplitude with heavy-top interference (Sec.~\ref{sec:ewprec}) \\
$\sin^2\theta_{\mathrm{eff}}^l$ & 0.230114 & 0.23153 & -0.612\% & C & improved-Born leptonic effective angle (Sec.~\ref{sec:ewprec}) \\
Higgs mass $m_H$ (GeV) & 124.983 & 125.2 & -0.173\% & B & tree-level Higgs mass from the spectral quartic and scale (Sec.~\ref{sec:ewprec}) \\
\midrule
\multicolumn{6}{l}{Neutrinos} \\
\midrule
heaviest neutrino scale $m_{\nu 3}$ (eV) & 0.0501797 & 0.0502 & -0.040\% & B & Weinberg two-period texture. Atmospheric reference scale (App.~\ref{sec:derivnu}) \\
solar rest split $\Delta m^2_{21,\mathrm{raw}}$ (eV$^2$) & 6.874155e-05 & 7.410000e-05 & -7.231\% & B & light-pair rest-mass difference (App.~\ref{sec:derivnu}) \\
unit-source split $\Delta m^2_{21,\mathrm{tilt}}$ (eV$^2$) & 7.412190e-05 & 7.410000e-05 & +0.030\% & D & scalar-source readout at $\zeta=1$ (App.~\ref{sec:derivnu}) \\
atmospheric splitting $\Delta m^2_{31}$ (eV$^2$) & 0.00251121 & 0.002511 & +0.008\% & B & atmospheric split from the absolute mass eigenvalues (App.~\ref{sec:derivnu}) \\
lightest neutrino $m_{\nu 1}$ (eV) & 0.00260742 & \multicolumn{1}{c}{--} & \multicolumn{1}{c}{--} & B & absolute vacuum-floor mass from the hypercharge texture (App.~\ref{sec:derivnu}) \\
\midrule
\multicolumn{6}{l}{Charged-fermion masses} \\
\midrule
top mass $m_t$ (GeV) & 174.082 & 172.69 & +0.806\% & B & geometric unit overlap at the electroweak scale (App.~\ref{sec:derivtop}) \\
bottom mass $m_b$ (GeV) & 4.23779 & 4.18 & +1.383\% & C & absolute bottom mass from the down-sector ratio (App.~\ref{sec:derivmass}) \\
electron mass $m_e^{\rm geom}$ (MeV) & 0.510354 & 0.511 & \multicolumn{1}{c}{--} & \multicolumn{1}{c}{--} & four-level electron cascade over one-generation content (App.~\ref{sec:derivmass}) \\
muon mass $m_\mu^{\rm geom}$ (GeV) & 0.105314 & 0.105658 & \multicolumn{1}{c}{--} & \multicolumn{1}{c}{--} & electron mass propagated through the lepton ladder (App.~\ref{sec:derivmass}) \\
tau mass $m_\tau^{\rm geom}$ (GeV) & 1.77211 & 1.777 & \multicolumn{1}{c}{--} & \multicolumn{1}{c}{--} & muon mass propagated through the lepton ladder (App.~\ref{sec:derivmass}) \\
\midrule
\multicolumn{6}{l}{CP content} \\
\midrule
baryon asymmetry $\eta_B$ & 6.128230e-10 & 6.100000e-10 & +0.463\% & D & content asymmetry at the physical photon temperature (Sec.~\ref{sec:tcmb}) \\
\midrule
\multicolumn{6}{l}{Cosmology} \\
\midrule
scalar tilt $1-n_s$ & 0.035 & 0.0351 & -0.285\% & D & torsion modulus times the seven-over-four content ratio (App.~\ref{sec:derivcosmo}) \\
Hubble rate $H_0$ (GeV) & 1.438850e-42 & 1.440000e-42 & -0.080\% & D & Gaussian endpoint logarithm and content offset (App.~\ref{sec:derivcosmo}) \\
dark-energy fraction $\Omega_\Lambda$ & 0.685044 & 0.6847 & +0.050\% & D & dark-energy density divided by the critical density (App.~\ref{sec:derivcosmo}) \\
baryon fraction $\Omega_b$ & 0.0494082 & 0.0493 & +0.219\% & D & raw photon floor and Sakharov ratio combined with the proton density (App.~\ref{sec:derivcosmo}) \\
cold fraction $\Omega_{\rm cold}$ & 0.264039 & 0.2645 & -0.174\% & D & common budget after photons and thermal neutrinos (Sec.~\ref{sec:flat}) \\
CMB temperature $T_{\mathrm{CMB}}$ (K) & 2.72547 & 2.72548 & -0.000\% & D & endpoint-corrected microwave monopole (App.~\ref{sec:derivcosmo}) \\
\midrule
\multicolumn{6}{l}{QCD} \\
\midrule
QCD scale $\Lambda_{\mathrm{QCD}}$ (GeV) & 0.208028 & 0.21 & -0.939\% & C & specified two-loop QCD extraction (App.~\ref{sec:derivqcd}) \\
proton mass $m_p$ (GeV) & 0.941086 & 0.938272 & +0.300\% & D & conformal proton relation with constituent dressing (App.~\ref{sec:derivqcd}) \\
lightest glueball $m_G$ (GeV) & 1.66422 & 1.7 & -2.104\% & D & scalar reference of the specified QCD channel readout (App.~\ref{sec:derivqcd}) \\
\midrule
\multicolumn{6}{l}{Weak decay and primordial helium} \\
\midrule
neutron--proton difference (MeV) & 1.28497 & 1.29333236 & -0.647\% & D & common scalar-isovector and electromagnetic mass response (App.~\ref{app:bbnext}) \\
neutron lifetime (s) & 915.54 & $878.4\pm0.5$ & +4.23\% & D & Coulomb-weighted decay with the stated radiative factor (App.~\ref{app:bbnext}) \\
helium baryon fraction $Y_p$ & 0.246082 & 0.245 & +0.44\% & D & instantaneous freeze-out followed by conserved decay and complete capture (App.~\ref{app:bbnext}) \\
\midrule
\end{longtable}

%% file: sections-en/appendix_b.tex
\section{Reproducibility}\label{app:repro}

The numerical tables use the spectral data and response relations of
Appendix~\ref{app:deriv} with Newton's constant fixing the dimensional
normalization. The calculation record specifies the inputs, numerical
precision, consistency checks and rigorous enclosures
\cite{Guo2026GenesisFramework}.

Generated quantities and external reference data are stored separately
and accessed through a common parameter interface. The reference values
enter at the comparison stage. Source identities and input fingerprints
associate each result with the frozen calculation.

\subsection{Numerical precision and calculation records}

\paragraph{Precision and displayed values.}
Unrounded floating-point records determine the reported central values.
Mathematical interval bounds establish the three-position plateau. The
endpoint comparison from Eq.~\eqref{eq:p5-weak-window-inverse} uses the relative
residual $c_{\rm ref}/c-1$ where $c_{\rm ref}$ is obtained by inverting
the separately extrapolated Standard-Model reference $g_2(M_G)$.
This central-value comparison uses the unrounded calculation. The
conservative endpoint enclosure certifies the retained-position count
at the accuracy stated in Appendix~\ref{app:sensitivity}.
Changes to the response domain or transfer rules measure model
sensitivity. Numerical precision concerns evaluation of the fixed response.

The endpoint $\kL=2.493534\ldots$ is displayed as $2.49353$ and used
without display rounding in the exponential hierarchy relations. For
$X\propto e^{-a\kL}$ its leading effect is
$\delta X/X=-a\,\delta\kL$. Appendix~\ref{app:sensitivity} includes the
algebraic prefactors. The reference-coupling inversion uses this fixed
endpoint for comparison.

\paragraph{Calculation and verification.}
Table~\ref{tab:software-coverage} summarizes the calculations and their
numerical and formal checks. Table~\ref{tab:software-versions} identifies
the software release. The response domain, channel weights, transfer
relations and unit conversions determine the calculation independently
of previously evaluated intermediate values.

The tables use the joint three-copy branch and the $R=5$ endpoint.
Its root and finite-source responses determine the coupling boundaries
and hierarchy relations before the low energy ratios are formed. The mode
bound, quadrature accuracy and root residual control numerical error.
Loop order and threshold matching specify the perturbative approximation.

\begin{table}[t]
\centering
\caption{Computational support and its scope~\cite{Guo2026GenesisFramework}.}\label{tab:software-coverage}
\begin{tabular}{p{4.1cm} p{8.0cm}}
\toprule
Calculation & Evaluation or check \\
\midrule
Spectral sums and endpoint & Mode sums and endpoint enclosures. Finite-source responses and a uniform three-position certificate for $R\geq5$. \\
Cosmological and external-data comparisons & CAMB propagation for the stated inputs. DESI DR2 covariance statistic and common-scale projection. SPARC statistics for the stated mass models. \\
Discrete algebraic relations & Formal checks of matter counts and rational two-loop coefficients together with the rest-frame cold-source decomposition. \\
Interacting responses & Neutral-mode elimination and mode products. Source Hessians and canonical vertices. Causal matter response and covariant source exchange. \\
Propagation and local fields & Labelled flavour evolution and neutron--proton weak rates. Non-spherical Dirichlet problems and finite nonlinear source--potential evolution. \\
Particle comparisons & Widths and ratios at a common electroweak order. Separate static-mass and solar-propagation comparisons. \\
\bottomrule
\end{tabular}
\end{table}

\begin{table}[t]
\centering
\caption{Calculation records used in this manuscript. The production
calculation and the fixed-background thermal evaluation are identified
separately by the checksums below~\cite{Guo2026GenesisFramework}.}\label{tab:software-versions}
\small
\begin{tabular}{p{3.0cm} p{3.0cm} p{6.0cm}}
\toprule
Record & Version or date & Relation to this manuscript \\
\midrule
Software~\cite{Guo2026GenesisFramework} & Release used for the frozen record & Spectral and finite-source calculations. Joint mass and bandwidth responses. One-loop matching and propagation. \\
Particle and mass tables & 24 October 2026 & Fixed-$y_0$ evolution and widths at a common correction order. \\
Finite matching and decay & Same calculation & Generated Fermi and top conditions together with the Higgs condition. Specified but disfavoured charged-lepton pole realization and its running conversion. Top width from the retained absorptive part. \\
External comparisons & Identified solver records & CAMB background at the stated inputs. Fixed-background recombination and DESI scale response. SPARC comparison. \\
Source dynamics & Same calculation & Nonlinear material flow and canonical phase-space evolution. Phase-clock readout and covariant source exchange. Joint static amplitude--radius response. \\
Verification & 24 October 2026 & Numerical reproduction and independent identity and convergence checks. Interval certification and input isolation. The finite structural checks described above. \\
\bottomrule
\end{tabular}
\end{table}

The production run has identifier
\texttt{20261024T084548Z-d52375b5} and is dated 24 October 2026.
Table~\ref{tab:p5-frozen-digests}
identifies its source set, scientific records, comparison inputs and
CAMB calculation by full SHA-256 checksums. The source-set checksum uses
the sorted relative-source checksum mapping in compact UTF-8 JSON. The scientific and reference records use their complete byte contents.
The remaining checksum conventions are specified in
Ref.~\cite{Guo2026GenesisFramework}. The archived fixed-background thermal
study~\cite{Guo2026ThermalCalculation} in
Section~\ref{sec:linprop} has its own CAMB execution and HYREC-resource
identities. Its thermal-response checksum identifies the atomic-rate
comparison at that fixed background. The corresponding DESI projection
uses its distance predictions and the published covariance.

\begin{table}[t]
\centering
\small
\caption{Full SHA-256 identities of the frozen calculation and its
external inputs. The two lines in each entry concatenate to one digest.}
\label{tab:p5-frozen-digests}
\begin{tabular}{p{3.5cm}l}
\toprule
Object & SHA-256 \\
\midrule
Calculation source set & \texttt{057074ac9761f3d5d27a3bcc5d3b3b88}\\
 & \texttt{43ecb395c97da15e37347263ebcf19a4}\\
Endpoint certificate & \texttt{9b712f9770f80dd427a865b2c10543c6}\\
 & \texttt{5ff46887a62116716017742d62970324}\\
Scientific records & \texttt{a70c81546d8e7635c1ddc5e646fae9d4}\\
 & \texttt{17289460b20a1f534834c766fc3e9643}\\
Reference records & \texttt{749968311efddfc5b767403e7fcd6294}\\
 & \texttt{4cb11c434815e004214059a030ef447d}\\
CAMB production & \texttt{ee13f231337d3fc797ef5fd2f4b4c508}\\
 & \texttt{33fa6483643c07efe96c77bd0b82302c}\\
Thermal-study CAMB & \texttt{45c2439ee34a9a38f0b6cfd80572d86f}\\
 & \texttt{78b39267f9352fc403d8314919ba65df}\\
HYREC resources & \texttt{e212aab017f0d118a26192dcf3195ecd}\\
 & \texttt{d4c1eada5f20fb9494c3d4ba692b447a}\\
Thermal response & \texttt{1074d4403d31f8b3d1ff8be9393e05b8}\\
 & \texttt{7609e6c3718f1f58cf3b6a548c8efae6}\\
DESI DR2 input & \texttt{2d7c70e3d9cb4fa27ade312869b76b32}\\
 & \texttt{b2bc6da9cafb492a8fa5eddfa0d53103}\\
SPARC input & \texttt{92158efa833151be9807e6fef83b390c}\\
 & \texttt{4b32297c877c88b0481e1b276e8c8fee}\\
\bottomrule
\end{tabular}
\end{table}

\subsection{Numerical evaluation and consistency checks}

\paragraph{Electroweak widths and ratios.}
The numerical calculation~\cite{Guo2026GenesisFramework} evaluates
the partial widths from the common effective amplitude of
Eq.~\eqref{eq:p5-common-z-widths}. The same effective angle and
neutral-current normalization enter every channel. The bottom vertex
contributes its first-order interference and all channels retain their
factorized radiators. The unrounded values are
$\Gamma_b=0.38040730308009635$ with
$\Gamma_{\rm had}=1.7520580283558593$ and
$\Gamma_Z=2.5063559621245686$ GeV. They determine
$R_b=0.2171202648105626$ and $R_l=20.791529232604695$.
The sensitivity table uses the analytic logarithmic derivatives of
Appendix~\ref{app:sensitivity} evaluated at the same endpoint and couplings.

\paragraph{Numerical accuracy and interacting responses.}
Verification distinguishes algebraic and input consistency checks from
independent quadrature, interval bounds and refinement calculations.
Repeated checks of the same numerical example share its evaluated data.
Independent formulas, alternative discretizations and altered inputs
are evaluated separately. Stored responses are reused when their source dependencies
and input fingerprints agree with the current calculation.

The fixed-Yukawa gauge evolution uses a fourth-order integrator with
step-doubling control. The colour boundary is checked by forward
evolution to the breaking scale with relative residual below
$10^{-12}$. In the declared normalization the connected finite-source
Hessian and torsion coefficient are
\[
 \operatorname{diag}(224/225,2/9,2/15,2/15,2/15),
 \qquad \tau=1/50.
\]
Relations among quantities derived from the same inputs provide
numerical consistency checks including the geometric mass combination
\eqref{eq:p5-sector-mass-test} and its common-input derivatives.
Further calculations evaluate
the complete condensate for specified
$(g,Z,L)$ with spectral tail bounds on its mass, the complex mixing texture,
the nonspherical static Dirichlet problem and the finite material flow
realization of Section~\ref{sec:p5-local-feedback}.
For inputs with an assigned covariance $\Sigma$ the analytic logarithmic
Jacobian gives the first-order output covariance $J\Sigma J^{\mathsf T}$.
Numerical residuals and interval bounds retain their respective meanings
when no probability model is assigned to the inputs.

The joint vacuum of Proposition~\ref{prop:p5-joint-native-radius}
is certified in the same response units at $g=1$ and $Z=2$.
One additional stationary enclosure at the exact radius
$L_*=\sqrt{2\pi}$ with $128$ shells and $55$ bisections per bound
gives $u_*\in[0.1135162098,0.1137652465]$ and the relaxed
compression curvature in $[0.1840255478,0.1844292709]$.
The joint Hessian bounds reuse this enclosure and its spectral tail.
Independent checks evaluate the original potential and its square
completion and its two source derivatives. The covariant
matter calculation independently differentiates the Hilbert stress
including the metric connection and compares its divergence with
the matter equation and source exchange. The same source matrices
enter the lapse, shift and canonical dilation checks.
Finite matrix tests use known constrained quadratic actions to check
the mixed field Schur reduction, basis covariance and sign intervals
under a supplied error bound. The local material flow and
independent foliation calculations have separate consistency checks.
Each record identifies the action, background and varied fields.

The occupied common background of
Proposition~\ref{prop:p5-auxiliary-native-closure} uses the same joint
vacuum normalization with $\mu=\MG$ and $g=1$ with $Z_E=2$ and
$m_0^2/\mu^2=3/(8\pi)$ and $L_0^2=3\pi$.
The dimensional inputs give $\MP/\mu=1.40682610$ and
$\rho_\Lambda/\mu^4=2.81008\times10^{-120}$.
At each background evaluation the radius equation is solved at fixed
canonical data. The full potential is evaluated only for the initial
energy and the sampled Friedmann checks. Its analytic derivatives
supply the evolution. The numerical trajectory covers
$\mu\Delta s=1/(4\sqrt{3/(8\pi)})=0.72360125$.
The phase integral is checked against the polar angle of the evolved
matter pair. The maximum relative Friedmann and phase-charge residuals
are $2.9\times10^{-12}$ and $6.4\times10^{-13}$ respectively. The phase-reconstruction residual is below $2.7\times10^{-14}$.
The trajectory uses $62$ right-hand-side evaluations and $101$
algebraic source solves including the output and phase-integral
evaluations. Eighteen potential integrals supply the vacuum and initial
energy together with the remaining output checks. These floating-point checks are separate from the
continuous interval certificate on $0\leq\mu s\leq0.01$ in
Appendix~\ref{app:gravity}. That certificate evaluates eight distinct
spectral endpoints totaling $520$ shell-pair evaluations and uses
vector-field bounds to exclude exit from the phase-space box.
It includes the infinite spectral tail and requires no interval
time integration or potential quadrature~\cite{Guo2026GenesisFramework}.

For the elementary matter connector of Ref.~\cite{Guo2026CompactInternalSpaceI} elimination of the neutral
coordinate gives the charged compression and an effective source.
The readout and direct term complete the reduced response, and all four terms are retained. Mode
products are evaluated before applying the selected projection. The
noncommuting finite-source Hessian is the Duhamel response of the
matrix exponential. Spatial vertices and kinetic Gram matrices are
computed separately and their canonically normalized combinations are
compared under changes of mode basis. The five-variable
gauge--Yukawa--quartic evolution uses specified matching boundaries and
threshold conditions for each direction of evolution. The observational
comparison values do not enter these boundary conditions.
The finite electroweak calculation imposes the generated Fermi
and top conditions together with the Higgs condition in
Proposition~\ref{prop:p5-generated-onshell-matching}. Each distinct
one-loop integral is evaluated once at the leading coefficients and
reused in the finite inverse relations. Independent checks include
the pure-QCD pole limit and Born top width together with the explicit one-loop
scale cancellation and substitution into the forward conditions.
The charged-lepton calculation retains the finite masses in the
photon and weak kernels. Nine analytic two-point functions are evaluated
by interval arithmetic and checked against independent parameter
integrals and the small-mass weak limit together with the weighted scale derivative.
The weak-decay and helium Jacobian reuses the accepted central
phase-space partition for its derivatives and differentiates the
freeze-out and conserved-decay relations analytically.
The signed spatial kernel and generated family rotations are
checked against independently evaluated Gaunt products on their
complete finite mode spaces~\cite{Guo2026GenesisFramework}.

Canonical normalization combines the finite interaction coefficients,
spatial vertices and kinetic Gram matrices. The scalar response uses the
modes coupled to its source, the canonical two-derivative external
action and the internal mass spectrum. The source and action
parameters are included in Ref.~\cite{Guo2026GenesisFramework}.

An additive torsion probe $h$ at fixed matter projection and response
domain tests the derivative of the composite endpoint and mass map
\[
 E(c,h)=\frac{A_R(\tau+h)}{c}-\frac4{27},\qquad
 O(c,h)=\log\frac{v(c,\tau+h)}{\MP} .
\]
Here $A_R$ is the coefficient in Proposition~\ref{prop:klunique}.
Implicit differentiation gives
\[
 \frac{dO}{dh}=O_h-O_c(E_c)^{-1}E_h=-2.501835\ldots
 \quad (R=5,\ \tau=1/50).
\]
Independent re-solving of the original scaled spectral endpoint on both
sides of the probe verifies this derivative with relative residual below
$10^{-8}$ for the stated probe increments. The derivative includes both
the direct torsion term and the change in the endpoint.

The fixed-connection scalar energy source of
Ref.~\cite{Guo2026SpectralGravity} is evaluated on the compact modes.
For the positive quadratic form $A_B=D_B^\dagger D_B+3/(4L^2)$ and
stationary occupation $N$ the occupied energy is
\[
 E_N(\eta)=\operatorname{Tr}\!\left[N
       \sqrt{e^\eta A_Be^\eta}\right].
\]
For the nonzero connection considered here the uniform source charge
equals the total occupied energy. The three-source Hessian includes
the second derivative of the source map and the full square-root feedback. Independent finite-source differences give a relative residual below
$3\times10^{-8}$. The comparison between two retained mode spaces includes
the full second-order intermediate support of the occupied states. The
resulting occupied-matter susceptibility uses the inherited source convention. An external collective tensor propagator requires its own action and source
map~\cite{Guo2026GenesisFramework}.

The same operator and occupied states also enter the canonical dynamics
of Proposition~\ref{prop:p5-native-source-dynamics}. Three finite source
segments use amplitudes $(0,1/32,0)$ and $(0,0,-1/32)$ followed by $(0,0,0)$
with durations $(1/4,1/2,1/4)/\omega_{\min}$ where
$\omega_{\min}=0.6837072629$ in the declared reciprocal-radius unit.
The phase-space second moment is transported by the exact constant-source
propagator. The relative canonical-form residual is below $2\times10^{-14}$
and the absolute source-work balance residual is below $2\times10^{-14}$
in the declared energy unit. These are floating-point consistency checks
on the finite source history. Expressing the same segments with positive
clock rates $(1/2,3/2,2)$ checks the invariance of the transported
second moment and source work under
Eq.~\eqref{eq:p5-native-clock-reparametrization}. The second
parametrization reuses all three source matrices and requires three
additional exact propagators. The phase-clock rate and integrability
condition of Proposition~\ref{prop:p5-native-phase-congruence} are checked on the
existing homogeneous and static accelerated cases together with the radial-pressure case
using their already evaluated local field derivatives.
The causal response is evaluated at
$\zeta/\omega_{\min}=i/2,i,1+i/2$. An independent one-oscillator formula
and a direct phase-space resolvent test its frequency branches and source
contact term. The mode-basis calculation uses entrywise frequency
denominators and reuses the reference spectrum to avoid a dense
phase-covariance superoperator factorization~\cite{Guo2026GenesisFramework}.

\subsection{External comparisons}

\paragraph{Reference-coupling extrapolation.}
The Standard-Model comparison starts at the reference $M_Z=91.1876$ GeV with
\[
(g_1,g_2,g_3,y_t,\lambda)=(0.461425,0.64779,1.217200,0.940,0.129).
\]
It uses $g_1=\sqrt{5/3}\,g_Y$ and the potential convention
$V(H)=-m^2|H|^2+\lambda|H|^4$ and evolves the five-coupling two-loop subsystem to $\MG$~\cite{Guo2026GenesisFramework}.
These boundary couplings define the separate Standard-Model extrapolation
of Ref.~\cite{Guo2026GenesisFramework} independently of the mass
and $\alpha_s$ comparison benchmarks in Table~\ref{tab:params}.

\paragraph{Cosmological and external-data comparisons.}
The CAMB~2.0.4 background calculation uses the density components and
thermal inputs specified in Section~\ref{sec:linprop}. It evaluates
the homogeneous expansion and recombination together with the acoustic scale and distances.
The seven-coordinate response uses the same central solution and eighteen
thermal-background probes including the two half-step checks of
Table~\ref{tab:p5-background-response}. The calculation comprises one
density initialization and nineteen thermal-background solutions in
total. The independent distance quadratures reuse these solutions.
The DESI DR2 combined-sample mean and covariance correspond to
Table~4 of Ref.~\cite{DESIDR22025}. Its thirteen-residual statistic uses
the full positive-definite covariance and calculated drag scale together with
the full thermal background stated in Section~\ref{sec:linprop}. An independent quadrature and Cholesky evaluation agrees.
The finite thermal-history calculation adds three RECFAST background
evaluations and two full radiative-transfer recombination histories
at the same expansion background. The latter propagate the standard
and generated atomic constants separately
\cite{AliHaimoudHirata2011,LeeAliHaimoud2020}. Their drag integrals
and the low-redshift electron-capacity bound feed the same thirteen
distance ratios and covariance projection. The solver resources and
generated atomic inputs together with the execution identities are retained with
the thermal result~\cite{Guo2026GenesisFramework}. Ordinary verification
reuses that result after checking its inputs. This budget uses thermal
background solutions without additional power-spectrum calculations.
The SPARC comparison uses the mass-model data of
Ref.~\cite{SPARCData2016} with the pointwise acceleration map and cuts
stated in Appendix~\ref{app:gravity}. The three-dimensional calculation
tests the static field equation on specified source geometries.

\subsection{Spectral traces and projection conventions}
\label{sec:p5-spectral-implementation}

The calculation~\cite{Guo2026GenesisFramework} uses the signed spectrum \eqref{eq:p5-signed-dirac-datum} in three distinct ways.
\begin{enumerate}
\item The generation count compares the positions $n+3/2$ for $n=0,2,4,\ldots$ with $(\kL)^2$. Distinct positions of $\widehat D|_{P_+}$ have unit weight.
\item The spectral trace uses squared Dirac eigenvalues $(n+3/2)^2/L^2$ with multiplicities $(n+1)(n+2)$. The even-index sum evaluates the trace of $P_+f(D_{\tau_-}^2)$. The complete $f(D_{\tau_-}^2)$ trace includes both parities.
\item The chiral assignment uses odd indices for $Q_L,L_L$ and even indices for $u_R,d_R,e_R$. The uniform-even comparison uses even indices for every species. Each response combines the squared eigenvalues and species-weighted multiplicities with the channel weight of Eq.~\eqref{eq:p5-projected-fermion-trace}.
\end{enumerate}

In the spin-structure convention of Ref.~\cite{Guo2026CompactInternalSpaceI} the
species assignment is $D_{\tau_-}^2|_{P_-}$ for left species and
$D_{\tau_-}^2|_{P_+}$ for right species. These projectors belong to
one complete geometric spin structure. Reversing the spin convention
reverses the sign assignment while retaining the indexed sums.

The generation edge is $c_{\mathrm{gen}}=(\kL)^2$ whereas the endpoint trace uses the dimensionless radius $\ell_{\mathrm{end}}=\kL M_{\mathrm G}/k$ and the squared Gaussian cutoff $\Lambda^2=(k/\MP)^2$. Accordingly the second argument of $\Pi(L,\Lambda^2)$ denotes a squared cutoff. The physical radius is $L_{\mathrm{phys}}=\kL/M_{\mathrm G}$ as in \eqref{eq:p5-paper4-window-map}. Mode and quadrature bounds control the numerical evaluation of the smooth trace independently of the generation edge.

The calculation~\cite{Guo2026GenesisFramework} evaluates all five gravitational
channels in the common Planck-unit convention of
Section~\ref{sec:cosmo}. It gives $r_{\rm read}=85.9303$ with
$Z_{\rm read}=0.0115035$ and
$G_{\rm read}=5.8320\times10^{-37}\ {\rm GeV}^{-2}$.
These values characterize the finite combined-channel readout in its
stated convention. Newton's constant sets the macroscopic metric
normalization and remains the dimensional input throughout the particle
and cosmological chains.

Spectral responses are compared at the same operators, domains,
species projectors, multiplicities, channel weights and units.
In particular identifying the positive rough TT Laplacian of
Ref.~\cite{Guo2026CompactInternalSpaceI} with a Casimir-plus-curvature operator
requires an identification of their mode spaces and action on each mode.

\subsection{Scope of the formal verification}

The archived Lean proof library~\cite{GuoLeanFoundations} verifies the
indexed clauses connecting the source construction to the basic
physical interface within the comparison classes recorded in its theorem
index.  A separate finite algebraic check verifies matter counts and
representation-content identities together with normalisation factors.
It also checks denominator-cleared rational two-loop coefficients and the rest-frame
cold-source tensor decomposition from their declared premises.  Continuum
spectral and evolution arguments are given analytically in the manuscript.
The formal records provide an independent check of the indexed
constructions and their stated premises.
The numerical evaluation described in this appendix also checks
$M_\nu U_{\cdot i}=m_iU_{\cdot i}$ for every labelled column of the
real CP-conserving mass texture. This identity tests the mass assigned
to each mixing column. The
mixing ansatz and the additional solar propagation relation are specified
separately in Appendix~\ref{app:deriv}.

%% file: sections-en/appendix_c.tex
% =============================================================================
% Appendix C - Key derivations
% The Spectrum of a Compact Internal Space II: Effective Couplings and Mass Scales
% =============================================================================

\section{Key derivations}\label{app:deriv}

The endpoint reduction and finite source responses determine
the internal scale and coupling boundary of
Sections~\ref{sec:intscale}--\ref{sec:gauge}. The remaining calculations
give the renormalization group coefficients, mass and electroweak relations,
and cosmological and abundance relations used in the numerical tables.
The geometric and spectral data are those of
Ref.~\cite{Guo2026CompactInternalSpaceI}. Each readout fixes its source
projection and normalization. Source derivatives measure changes in
the mode response; kinetic normalization converts the resulting
matrix elements into couplings or mass amplitudes. This common
normalization permits the parameter eliminations and joint responses.
Newton's constant sets the dimensional scale.

\subsection{The window-capacity endpoint}\label{sec:derivkl}

\paragraph{Inherited Gaussian rule.}\label{sec:p5-inherited-gaussian}
We use the ``Gaussian uniqueness under composition'' theorem of
Ref.~\cite{Guo2026CompactInternalSpaceI}.
On its minimal change carrier, weak continuity, centred unbiased
composition, square root scaling and fixed nonzero covariance determine
the Gaussian semigroup. For the nonnegative spectral generator
supplied by that construction we use the trace convention
$e^{-tH}$ with $t=\Lambda^{-2}$. The frequency resolved sum
$H=\omega^2+A$ is restricted to the spatial branch assigned to each species.

The window capacity $\kL$ is fixed by the spin-2 endpoint condition
of Section~\ref{sec:closure}. Its full-domain zero-external-momentum
TT spectral readout has the form specified for the compact branch in
Section~2.2 of Ref.~\cite{Guo2026SpectralGravity}:
\begin{equation}
\Pi^2_{\mu\nu}(L,\Lambda^2) = \frac{1}{V_3} \sum_{\text{fields}}\sum_{\text{modes}} d_n \, w(\text{field}) \int_0^\infty \frac{d\omega}{\pi}\; K_{\mathrm{TT}}\bigl(\omega^2 + \lambda_n, m^2_{\eff}\bigr)\, e^{-(\omega^2+\lambda_n)/\Lambda^2},
\end{equation}
Here $V_3=\pi^2L^3$ and $d_n$ is the mode multiplicity and
$w(\text{field})$ is the channel weight. The second argument is
the squared cutoff $\Lambda^2$. The fixed content and torsion are
suppressed. Set $q^2=\omega^2+\lambda_n$ for the internal spectral variable. The running
scale is denoted by $k$. The kernel is
\begin{equation}
K_{\mathrm{TT}}(q^2,m^2)=\frac{q^4}{(q^2+m^2)^2}.
\end{equation}
For $q^2,m^2\geq0$ with nonzero denominator $0\leq K_{\rm TT}\leq1$
and $K_{\rm TT}(q^2,0)=1$ for $q^2>0$. The flat-space reference
is the Ward-normalised subtracted amplitude. The weights are $w=4$
for each gauge polarisation or Weyl fermion and $w=1$ for each real
scalar. The shifts are $m_{\rm sc}^2=3/(4L^2)$ and
$m_V^2=C_2/(2L^2)+\tau^2/(6L^2)$ and
$m_F^2=3\tau^2/(8L^2)$ where $C_2=3,2,0$ for
$\SU(3),\SU(2),\U(1)$. The scalar value is $\xi R$ with
$\xi=1/8$ and $R=6/L^2$. The fermionic curvature is already
included through $\not D^2=\nabla^*\nabla+R/4$.
The auxiliary TT vertex uses the Lichnerowicz shift $6/L^2$; the trace
sums the matter rows listed above.
The common scale path is $L=\kL\MG/k$.

For the fermion contribution the mode sum in this formula has the explicit
meaning
\begin{equation}
 \sum_f\sum_{\substack{n\ge0\\ n\equiv\epsilon_f\ ({\rm mod}\ 2)}}
 N_{\rm int}(f)(n+1)(n+2)\,w(f)\,
 \mathcal I\!\left(\frac{(n+3/2)^2}{L^2},m_f^2;\Lambda^2\right),
 \label{eq:p5-projected-fermion-trace}
\end{equation}
where $N_{\rm int}(f)$ is the number of internal components of species $f$,
$\mathcal I$ is the frequency integral above, and the default assignment is
$\epsilon_{Q_L}=\epsilon_{L_L}=1$ and
$\epsilon_{u_R}=\epsilon_{d_R}=\epsilon_{e_R}=0$.
This is a species-weighted trace of the restrictions
$D_{\tau_-}^2|_{P_-}$ and $D_{\tau_-}^2|_{P_+}$ defined in
\eqref{eq:p5-dirac-consumer-types}.  An unrestricted trace of $D_{\tau_-}^2$
would sum over both parities for every species. The uniform-even
comparison sets all $\epsilon_f=0$. Field multiplicities and channel
weights together with spatial branch projections are separate factors in the trace.

The Gaussian envelope is fixed by the inherited rule above.
Appendix~\ref{app:sensitivity} compares the Gaussian response with a
hard spectral window and quantifies its domain sensitivity.

The endpoint condition on the volume-weighted spectral readout is
\begin{equation}
\frac{V_3 \, \widehat\Pi^2_{\mu\nu}(\kL, k^2/\MP^2)}{32\pi^2} = \frac{4}{27} \qquad \text{at} \qquad k = \MG .
\end{equation}
The tensor mass response maximum fixes the endpoint normalization \(4/27\):
\begin{equation}
\frac{4}{27} = \max_{0 \le y \le 1} y(1-y)^2 \qquad \text{at} \qquad y = \frac13 ,
\end{equation}
where $y=m^2/(q^2+m^2)$ and $K_{\rm TT}=(1-y)^2$.
On $L(k)=C/k$ with $C=\MP\sqrt\pi$ the volume and cutoff reduce
the endpoint condition to one equation in $\kL$. Numerical
bisection on $[0.3\,\MG,3\,\MG]$ and self-consistency iteration
between $\MG=C/\kL$ and the crossing scale $k_\star$ give
\begin{equation}
\kL = 2.49353 .
\end{equation}
In the endpoint formula the radius symbol denotes the dimensionless
argument $\ell_{\mathrm{end}}(k)=\kL M_{\mathrm G}/k=C/k$ and the Gaussian
scale is $\widehat\Lambda_{\mathrm{tr}}=k/\MP$ with its square
as the second readout argument following
Section~\ref{sec:closure}.  The physical KK radius and the generation edge
are instead fixed by \eqref{eq:p5-paper4-window-map}.  The quoted fixed point
is evaluated with this endpoint definition and the projected fermion
trace \eqref{eq:p5-projected-fermion-trace}.

The screened coupling law of Section~\ref{sec:derivg2} gives a
consistency test through the inverse in
Eq.~\eqref{eq:p5-weak-window-inverse}. It relates the screened reference
coupling to the bare boundary $\alpha_W=2/(\kL)^5$.
The endpoint geometry $L_{\mathrm{Cg}}=\sqrt\pi$
fixes $R_{\mathrm c}=6/\pi$ and the resulting capacity is
$(\kL)^2=6.2177$.

\paragraph{Endpoint normalization and propagation.}
\label{sec:p5-endpoint-boundary}
The modewise extremum above sets the endpoint rule. The two-leg response
and its logarithmic derivative are given in Section~\ref{sec:closure}.
Here $4/27$ normalizes the modewise survival response. The spectral
sum enters the separate endpoint equation above.
The macroscopic gravitational realization in
Ref.~\cite{Guo2026SpectralGravity} uses one metric in the Einstein and
matter actions. Its quadratic tensor response about flat spacetime
has two physical helicities and positive residue $g^2/Z$.
The ``Matched critical branch'' proposition identifies
$(1-x(\chi))/\Pi(\chi)$ with $\widehat Z Q^2/M_P^2$ and derives
$\chi(0)=\chi_c$ and
$-\mathrm d\chi/\mathrm d(Q^2/M_P^2)=\widehat Z\Pi(\chi_c)/x'(\chi_c)>0$.
The effective normalization $g=1$ with $Z=M_P^2$ gives
$G_N=(8\pi M_P^2)^{-1}$. The radius equation fixes the compact
endpoint and the common metric action and inverse-response matching
fix the macroscopic propagation and its coupling to conserved matter.

\paragraph{Homogeneous endpoint reduction.}\label{sec:p5-homogeneity}
Fix the matter content and species projectors together with $\tau$. Set
$a_j=L^2\lambda_j$ and $b_j=L^2m_j^2$ and $D_j=d_jw_j\geq0$ including
field multiplicities in $D_j$. For a dimensionless frequency domain $R>0$
define
\begin{equation}
\begin{split}
 A_R&=\frac1{32\pi^3}\sum_jD_j
 \int_0^{\sqrt{(R\pi-a_j)_+}}
 \left(\frac{u^2+a_j}{u^2+a_j+b_j}\right)^2
 e^{-(u^2+a_j)/\pi}\,du,\\
 &(x)_+=\max(x,0).
\end{split}
\label{eq:p5-homogeneous-coefficient}
\end{equation}
For $R=\infty$ integrate to infinity over the full assigned towers.
Their polynomial multiplicities and quadratically growing eigenvalues
make the Gaussian sum convergent. Finite $R$ defines a restricted
response with the same Gaussian weight.

\begin{proof}[Proof of Proposition~\ref{prop:klunique}]
Write $c=\kL$ and $C=\MP\sqrt\pi$ with $\MG=C/c$ and $t=k/\MG$.
Apply the ``Scale law for a TT spectral density'' of
Ref.~\cite{Guo2026CompactInternalSpaceI} with $k_0=C$ and $L_0=1$ together with
$\Lambda_0=\sqrt\pi$ and $r=k/C=t/c$. Use frequency parameter $b=R$
and reference eigenvalues and mass squares $a_j,b_j$.
These give $L=c/t$ and $\Lambda^2=\pi t^2/c^2$.
Its reference readout is $A_R$ in
\eqref{eq:p5-homogeneous-coefficient}. Thus the inherited scale law yields
\begin{equation}
 \frac{V_3\widehat\Pi^2_R}{32\pi^2}=\frac{tA_R}{c}.
 \label{eq:p5-endpoint-scaling}
\end{equation}
At $t=1$ this gives $F_R(c)=A_R/c-4/27$. A positively weighted mode
with nonempty integration interval makes $A_R>0$. The scalar constant
mode already does so. The unique root is $c_R=27A_R/4$ and no dimensional
anchor occurs in $A_R$. The cited proposition applies directly to finite
selections and finite $R$. For the full towers and $R=\infty$ pass through
increasing finite modal and frequency domains. Positivity and the
convergent Gaussian majorant preserve the same identity in the limit.
\end{proof}

For fixed $c$ \eqref{eq:p5-endpoint-scaling} increases linearly with
$t$ and its crossing is $t_\star=4c/(27A_R)$. Provided this lies in the
chosen bracket the exact self-consistency update is
$c_{\rm new}=C/k_\star=c/t_\star=27A_R/4$.
The numerical iteration implements this fixed branch
identity. Quadrature error concerns the fixed domain integral. The
Gaussian tail measures its difference from the full-domain response.

At $R=5$ the nonzero scalar degrees are $l=0,2$ and the vector degree is
$l=1$. The spinor degrees are $n=0,2$ on $P_+$ and $n=1$ on $P_-$.
The scalar values of $a_j$ are $0,8$. The vector value is $4$.
The spinor values are $9/4,49/4,25/4$ in the stated order.
The numerical mode ceiling $60$ contains all these modes. Along
$c,t\in[0.3,3]$ we have $L\in[0.1,10]$ and $\Lambda^2\geq\pi/100$ and
$q^2+m_j^2\geq3/400$ for every positively weighted mode. These lower bounds keep the
numerical regularization of the denominator and cutoff inactive.
The full Gaussian tail is bounded in
Appendix~\ref{sec:p5-gaussian-tail}.

\begin{proposition}[Bandwidth dependence of the spectral response]
\label{prop:p5-response-bandwidth}
Keep the full assigned towers and content together with the source projectors and mass shifts
in \eqref{eq:p5-homogeneous-coefficient} fixed. Then $A_R$ is continuous
and strictly increasing for $R>0$ and $A_R\uparrow A_\infty$.
Away from shell thresholds $R\pi=a_j$ its derivative is
\begin{equation}
 \frac{dA_R}{dR}
 =\frac{e^{-R}}{64\pi^2}
 \sum_{a_j<R\pi}\frac{D_j}{\sqrt{R\pi-a_j}}
 \left(\frac{R\pi}{R\pi+b_j}\right)^2>0.
 \label{eq:p5-bandwidth-derivative}
\end{equation}
For the three-copy trace at $\tau=1/50$ every $R\geq5$ satisfies
$11/2<c_R^2<15/2$ and hence retains the same three positive positions.
At fixed Planck anchor with fixed loop factor and squash correction its mass
readouts obey
\begin{equation}
\begin{aligned}
 \frac{v_R}{v_5}&=\frac{c_5}{c_R}e^{-4\pi(c_R-c_5)},&
 \frac{m_{e,R}}{m_{e,5}}&=e^{-20(c_R-c_5)},\\
 \frac{v_R}{v_5}&=\frac{c_5}{c_R}
    \left(\frac{m_{e,R}}{m_{e,5}}\right)^{\pi/5}.
\end{aligned}
 \label{eq:p5-common-bandwidth-response}
\end{equation}
\end{proposition}
\begin{proof}
Each finite domain contains finitely many spectral shells. Their
integrals are continuous as a shell enters the domain. Differentiating
the upper limit gives \eqref{eq:p5-bandwidth-derivative}. The positively
weighted scalar constant mode makes the increase strict. Monotone
convergence and the Gaussian majorant give the full-domain limit.
Since $c_R=27A_R/4$ the certified bounds in
Appendix~\ref{sec:p5-branch-enclosure} and
\eqref{eq:p5-full-gaussian-enclosures} imply
$2.490\leq c_5\leq c_R\leq c_\infty\leq2.6631205$.
The squares of the two outer bounds lie strictly between $11/2$ and
$15/2$. Finally substitute $\MG=\MP\sqrt\pi/c_R$ into the
electroweak readout and eliminate $c_R-c_5$ using the electron readout.
\end{proof}

\paragraph{Fixed shell support and the generation interval.}
For the three-copy endpoint trace at $\tau=1/50$ the reference shell set
consists of the positively weighted shells with $a_j\le 49/4$.  It includes
the $n=2$ spinor at $a=49/4$ and excludes the next $l=3$ vector at $a=16$.
A domain $R$ contains exactly this set of nonzero frequency integrals
precisely when
\begin{equation}
 \frac{49}{4\pi}<R\leq\frac{16}{\pi}.
 \label{eq:p5-shell-support-interval}
\end{equation}
There is a unique $R_*$ in this interval with
$c_{R_*}=\sqrt{11/2}$. The simultaneous conditions of this shell
support and three retained generation positions are exactly
\begin{equation}
 R_*<R\leq\frac{16}{\pi},\qquad
 R_*\simeq4.0187119842,\qquad \frac{16}{\pi}\simeq5.0929581789.
 \label{eq:p5-joint-bandwidth-interval}
\end{equation}
The value displayed for $R_*$ is the numerical solution of its defining
equation. The existence and ordering assertions follow from the
enclosures below and strict monotonicity.
The largest active spectral value is $a=49/4$ from the $n=2$ spinor
branch. The next positively weighted shell is the $l=3$ vector with
$a=16$. A shell has a nonzero frequency integral exactly when
$a<R\pi$. This proves \eqref{eq:p5-shell-support-interval}. The new shell
contributes zero at the upper endpoint.

The finite integrals at these two edges admit the outward enclosures
\begin{equation}
 \begin{aligned}
 c_{49/(4\pi)}&\in[2.1897287252,\;2.3186156701],\\
 c_{16/\pi}&\in[2.4131554931,\;2.5852484689].
 \end{aligned}
 \label{eq:p5-shell-edge-enclosures}
\end{equation}
At $R=4$ the same outward cell scheme separates the endpoint response
from $4/27$ at the bracket boundaries and fixes its derivative sign,
giving the enclosure
\begin{equation}
 c_4\in\left[\frac{187}{80},\frac{23939}{10240}\right]
 =[2.3375,\,2.33779296875].
 \label{eq:p5-c4-enclosure-appendix}
\end{equation}
Its upper endpoint is below \(\sqrt{11/2}=2.3452078799\). This certifies
the separation used in Proposition~\ref{prop:p5-response-level}
directly at \(R=4\).

Write
$K_j(u)=[(u^2+a_j)/(u^2+a_j+b_j)]^2$. On a cell
$[u_i,u_{i+1}]$ of width $h_j$ its monotonic increase and the
monotonic decrease of the Gaussian give
\begin{equation}
 \begin{split}
 h_jK_j(u_i)e^{-(u_{i+1}^2+a_j)/\pi}
 &\leq\int_{u_i}^{u_{i+1}}K_j(u)e^{-(u^2+a_j)/\pi}\,du\\
 &\leq h_jK_j(u_{i+1})e^{-(u_i^2+a_j)/\pi}.
 \end{split}
 \label{eq:p5-shell-cell-bounds}
\end{equation}
Apply these bounds with 32 equal cells per nonzero branch using the
coefficients of \eqref{eq:p5-homogeneous-coefficient} and the outward
arithmetic of Appendix~\ref{sec:p5-branch-enclosure}. The first edge has
seven nonzero branches and the second has eight. The calculation
therefore uses 480 branch cells. Sharing Gaussian nodes between the
vector branches requires 363 exponential enclosures
\cite{Guo2026GenesisFramework}. It gives \eqref{eq:p5-shell-edge-enclosures}.
The squared bounds satisfy
\[
 c_{49/(4\pi)}^2<\frac{11}{2}<c_{16/\pi}^2<\frac{15}{2}.
\]
Continuity and strict monotonicity from
Proposition~\ref{prop:p5-response-bandwidth} give the unique crossing
$R_*$. The strict retention rule requires
$11/2<c_R^2\leq15/2$. This yields \eqref{eq:p5-joint-bandwidth-interval}.
The coarse enclosures certify these separations. The finer mass
intervals use the quadrature bounds stated in Appendix~\ref{app:sensitivity}.

\paragraph{Scale-ratio identification of the endpoint.}
For the positive electroweak and electron readouts expressed in the same
mass units put $\beta=20-4\pi$. Then
\begin{equation}
 \mathcal Q(c):=\frac{v}{m_e}
 =\frac{3}{16\pi^{5/2}}\frac{e^{\beta c}}{c},\qquad
 \frac{d\log\mathcal Q}{dc}=\beta-\frac1c>0
 \quad\left(\frac{11}{2}<c^2\leq\frac{15}{2}\right).
 \label{eq:p5-scale-ratio-identification}
\end{equation}
Thus a value of $\mathcal Q$ in the image of this branch determines a
unique $c$. At fixed source content and $\tau$ it also determines a
unique bandwidth whenever it lies in the corresponding scale-ratio
range of \eqref{eq:p5-joint-bandwidth-interval}. Bounds on $\mathcal Q$ propagate
through these monotone inverses. On an open shell interval
\begin{equation}
 \frac{dc}{d\log\mathcal Q}=\frac{1}{\beta-1/c},\qquad
 \frac{dR}{d\log\mathcal Q}
 =\frac{1}{(\beta-1/c_R)\,dc_R/dR}.
 \label{eq:p5-scale-ratio-inverse-response}
\end{equation}
With $q(\tau)=1-s_0\kappa$ the two readouts are
\[
 v=\frac{3\MP q(\tau)}{16\pi^{5/2}c}e^{-4\pi c},\qquad
 m_e=\MP q(\tau)e^{-20c}.
\]
Their ratio cancels both the dimensional anchor and the common squash
factor. Differentiating its logarithm gives
\eqref{eq:p5-scale-ratio-identification}. On the stated branch
$\beta>7$ and $1/c<1$. Strict monotonicity of $c_R$ gives the bandwidth
inverse and the chain rule gives
\eqref{eq:p5-scale-ratio-inverse-response}.

An empirical use of this inverse first matches $v$ and $m_e$ to the
common mass definitions of the readouts. At $R=5$ the central values are
\[
 \mathcal Q=4.8238982197\times10^5,\qquad
 \frac{dc}{d\log\mathcal Q}=0.14219508,\qquad
 \frac{dR}{d\log\mathcal Q}=2.29569558.
\]
The geometric charged-lepton ladder gives a complementary constraint.
$m_\mu^{\rm geom}/m_e^{\rm geom}
=e^{2\alpha_{\rm lp}+\sqrt{2\pi}}$ and
$m_\tau^{\rm geom}/m_\mu^{\rm geom}=e^{2\alpha_{\rm lp}}$ imply
\begin{equation}
  \frac{(m_\mu^{\rm geom})^2}
  {m_e^{\rm geom}m_\tau^{\rm geom}}
  =e^{\sqrt{2\pi}}=12.2635110816.
 \label{eq:p5-lepton-transfer-invariant}
\end{equation}
This relation eliminates the endpoint, torsion, dimensional anchor and
bandwidth. It tests the Euclidean period insertion in the light family
transfer. Equation~\eqref{eq:p5-lepton-physical-invariants} gives its
pole and running forms through $G_\ell$ and $D_\ell$ respectively.
Its joint comparison with the sector relation uses the total factors
$C_\ell=D_\ell G_\ell$ in
\eqref{eq:p5-joint-running-mass-constraints}.

The projection $P_R$ of Section~\ref{sec:closure} restricts the source
measurement before the trace is evaluated. It is held fixed in the
reference spectral coordinates when taking the corresponding finite-source
response. A transported boundary has the additional term in
\eqref{eq:p5-moving-response-domain}. The numerical tables use the
acceptance threshold $R=5$ and the stated source normalization.
Equations~\eqref{eq:p5-joint-bandwidth-interval}
and~\eqref{eq:p5-scale-ratio-identification} combine the generated shell
support and generation count with a matched scale ratio into a quantitative
bandwidth test.

\paragraph{Joint content and endpoint selection.}\label{sec:p5-joint-content}
Replicate the same one-generation content $g$ times while keeping the
bosonic content and $\tau$ with the projectors and mass shifts fixed.
Linearity of the spectral trace gives
\begin{equation}
 A_{g,R}=A_{B,R}+gA_{F,R},\qquad
 c_{g,R}=a_R+g b_R,\qquad
 (a_R,b_R)=\frac{27}{4}(A_{B,R},A_{F,R}).
 \label{eq:p5-joint-affine}
\end{equation}
Here $a_R,b_R$ denote endpoint coefficients. The mode coefficients
$a_j,b_j$ are defined in \eqref{eq:p5-homogeneous-coefficient}. Exactly $g$ positive
positions lie below the endpoint window if and only if
\begin{equation}
 2g-\frac12<(a_R+g b_R)^2\leq2g+\frac32.
 \label{eq:p5-joint-platform}
\end{equation}
The two bounds are the last retained position $2(g-1)+3/2$ and the
next position $2g+3/2$ with the strict retention convention.
This condition applies to the replication family
\eqref{eq:p5-joint-affine}.

The finite-domain coefficients follow from the integrals
\[
 I(a,b)=\int_0^{\sqrt{5\pi-a}}
 \left(\frac{u^2+a}{u^2+a+b}\right)^2e^{-(u^2+a)/\pi}\,du,
 \qquad 0\leq a<5\pi.
\]
With $b_V=\tau^2/6$ and $b_F=3\tau^2/8$ at $\tau=1/50$ the
projected towers give
\begin{equation}
\begin{aligned}
 a_5=\frac{27}{128\pi^3}\bigl[&4I(0,3/4)+36I(8,3/4)
 +192I(4,3/2+b_V)\\
 &+72I(4,1+b_V)+24I(4,b_V)\bigr],\\
 b_5=\frac{27}{128\pi^3}\bigl[&56I(9/4,b_F)+192I(25/4,b_F)
 +336I(49/4,b_F)\bigr].
\end{aligned}
\label{eq:p5-branch-coefficients}
\end{equation}
The coefficients include the scalar multiplicities and twelve vector
species and the odd-left/even-right assignment per generation.
The interval calculation in Appendix~\ref{sec:p5-branch-enclosure} gives
$0.600\leq a_5\leq0.602$ and $0.630\leq b_5\leq0.632$.

\paragraph{Endpoint compatibility of the replication family.}
The three-copy content satisfies its endpoint and window
conditions at $R=5$. The same endpoint rules also admit one copy in
the fixed-data family \eqref{eq:p5-joint-affine}.
The certified coefficient intervals give
\[
 c_1\in[1.230,1.234],\qquad
 c_3\in[2.490,2.498].
\]
Their squares lie in $(3/2,7/2]$ and $(11/2,15/2]$ respectively.
These are the two plateaus in \eqref{eq:p5-joint-platform}.
Proposition~\ref{prop:p5-screening-compatibility} selects the three-copy
solution by imposing the joint screening condition. The numerical tables
use its $R=5$ endpoint. Appendix~\ref{sec:p5-full-gaussian-certificate}
gives the complete-domain coefficients for comparison.

\paragraph{Uniform torsion response.}
Keep the species and projections fixed together with domain $R=5$ or $R=\infty$
and vary only the specified mass shifts. Put $z=\tau^2$ and
$b_j(z)=\beta_j+h_jz$ with
\[
 (\beta_j,h_j)=(3/4,0),\quad(C_2/2,1/6),\quad(0,3/8)
\]
for scalar and gauge entries together with fermion entries. Then $A_R(z)$ of
\eqref{eq:p5-homogeneous-coefficient} is completely monotone on
$[0,\infty)$ and
\begin{equation}
 -\frac2{6+z}\leq\frac{d\log A_R}{dz}\leq0,\qquad
 \left(\frac{6+z_1}{6+z_2}\right)^2
 \leq\frac{A_R(z_2)}{A_R(z_1)}\leq1
 \quad(z_2\geq z_1\geq0).
 \label{eq:p5-torsion-endpoint-bound}
\end{equation}
The same bounds hold for $c_R(z)=27A_R(z)/4$. With
$\mathcal B_R=-A_R'(0)/2$ the linear remainder obeys
\begin{equation}
 0\leq A_R(z)-A_R(0)+2z\mathcal B_R
 \leq \frac{z^2}{12}A_R(0).
 \label{eq:p5-torsion-endpoint-remainder}
\end{equation}
Set $X_j=u^2+a_j$ and $r_j=h_j/(X_j+\beta_j)$. The lowest gauge
and spinor denominators give $0\leq r_j\leq1/6$. Normalizing the
positive $z=0$ integrand to a probability measure $\nu_0$ gives
\[
 A_R(z)=A_R(0)\int(1+r_jz)^{-2}\,d\nu_0.
\]
Differentiation proves complete monotonicity. The bounds follow from
$r_j/(1+r_jz)\leq1/(6+z)$ and
$0\leq(1+x)^{-2}-1+2x\leq3x^2$. Bounded $r_j$ and the Gaussian majorant
justify differentiation and the complete-domain limit.
At $\tau_0=1/50$ the relative Taylor remainder is at most
$1.334\times10^{-8}$. For $0\leq\tau\leq0.1$ the finite ratio gives
$0.9968<c_{g,R}(\tau)/c_{g,R}(\tau_0)<1.000134$.
Applying this ratio to the certified intervals preserves both finite-domain
plateaus in Eq.~\eqref{eq:p5-joint-platform} and the
three-copy full-domain plateau certified in
Appendix~\ref{sec:p5-full-gaussian-certificate}.
Even independently varying the two torsions in this interval
\[
 c_{3,\infty}(\tau_\infty)-c_{3,5}(\tau_5)
 >0.9968(2.6603141)-1.000134(2.498)>0.1534.
\]
These bounds quantify the sensitivity to the torsion-dependent mass
shifts while holding the response domain and source projections fixed.

\subsection{The finite-content torsion response}\label{sec:derivtau}

Section~\ref{sec:torsion} defines
$\tau=(N_L-N_R)/(N_f\sum Y^2)=1/50$ from the one-generation
counts $N_L=8$ and $N_R=7$ with $N_f=15$ and $\sum Y^2=10/3$.
Replication and spatial weighting act differently on this ratio.
The hypercharge scale is the normalised $Y$
operator supplied by Ref.~\cite{Guo2026CompactInternalSpaceI}. The homogeneous anomaly equations
alone leave an overall nonzero rescaling of $Y$ undetermined.

\paragraph{Replication and common-window normalisation.}
Let $\mathcal H_1=\mathcal M_{\rm ch}$ be the fifteen-dimensional
charged response carrier for one generation. Set
$\chi=+1$ on its eight left components and $\chi=-1$ on its seven
right components. Use the given hypercharge operator $Y$. On
$\mathcal H_g=\mathbb C^g\otimes\mathcal H_1$ let
$\chi_g=I_g\otimes\chi$ and $Y_g=I_g\otimes Y$. Then
\begin{equation}
 \bar J_g=\frac{\operatorname{Tr}\chi_g}{15g}=\frac1{15},\qquad
 \bar A_g=\frac{\operatorname{Tr}Y_g^2}{g}=\frac{10}{3},\qquad
 \frac{\bar J_g}{\bar A_g}=\frac1{50}.
 \label{eq:p5-replication-normalisation}
\end{equation}
If a common positive spatial trace factor $W$ multiplies the two
extensive coefficients $J_g=W\operatorname{Tr}\chi_g/15$ and
$A_g=W\operatorname{Tr}Y_g^2$ then $A_g\tau=J_g$ has the unique
solution $\tau=1/50$ independently of $g$ and $W$.
Tensor product trace factorisation gives
$\operatorname{Tr}\chi_g=g$ and
$\operatorname{Tr}Y_g^2=10g/3$. Substitution gives the trace ratios. Positivity of $A_g$ makes the balance $A_g\tau=J_g$ unique.

The normalisations in \eqref{eq:p5-replication-normalisation} are per
component for the chiral excess and per generation for the charge
capacity. Both coefficients are derivatives of the same finite source.

\paragraph{One finite source for the chiral and charge responses.}
\label{sec:p5-finite-torsion-source}
On each of the four left doublets let $t_a=\sigma_a/2$ and let
$t_a=0$ on the seven right singlets. For the equal-weight
content trace set
\[
 \mathcal W_F(u,v,\mathbf x)=
 \log\frac{\operatorname{Tr}_{\mathcal H_1}
 e^{u\chi+vY+\mathbf x\cdot\mathbf t}}{15}.
\]
Here $u,v,\mathbf x$ are derivative coordinates.
Apply the normalized trace Hessian formula in
``Generated depolarization and the prepared response'' in
Ref.~\cite{Guo2026CompactInternalSpaceI} to these specified charge blocks.
$\operatorname{Tr}_L Y=\operatorname{Tr}_R Y=0$ and
$\operatorname{Tr}_L Y^2=2/3$ and $\operatorname{Tr}_R Y^2=8/3$.
The cited formula therefore gives
\begin{equation}
 \partial_u\mathcal W_F(0)=\frac1{15},\quad
 \partial_v\mathcal W_F(0)=0,\quad
 D^2\mathcal W_F(0)=
 \operatorname{diag}\left(\frac{224}{225},\frac29,
                         \frac{2}{15}I_3\right).
 \label{eq:p5-finite-torsion-hessian}
\end{equation}
The weak traces satisfy $\operatorname{Tr}(t_at_b)=2\delta_{ab}$ and
$\operatorname{Tr}t_a=\operatorname{Tr}(\chi t_a)
=\operatorname{Tr}(Yt_a)=0$. Thus all mixed weak entries vanish.
In particular the hypercharge second moment is already connected:
its mean and mixed chiral covariance vanish. Use the
normalisations $\bar J=\partial_u\mathcal W_F(0)$ and
$\bar A=15\partial_v^2\mathcal W_F(0)$. Thus
$\tau=\bar J/\bar A=1/50$. Both entries in this constitutive ratio
are responses of one source. The factor $15$ converts the
normalised hypercharge Hessian to the per-generation charge capacity.

For positive weights $w_L,w_R$ that are constant on the respective
finite chiral subspaces let $E_L=(I+\chi)/2$ and $E_R=(I-\chi)/2$
of ranks $8,7$ on $\mathcal H_1$. These act on the finite content space whereas $P_\pm$ act on the spatial branches.
Set $\rho_r=(w_LE_L+w_RE_R)/(8w_L+7w_R)$ with $r=w_L/w_R$ and
$\mathcal W_r(u,v)=\log\operatorname{Tr}(\rho_r e^{u\chi+vY})$.
The same readouts $J_r=\partial_u\mathcal W_r(0)$ and
$A_r=15\partial_v^2\mathcal W_r(0)$ give
\begin{equation}
 \tau_r=\frac{8r-7}{10r+40},\qquad
 \tau_r-\frac1{50}=\frac{39(r-1)}{50(r+4)},\qquad
 \frac{d\tau_r}{dr}=\frac{39}{10(r+4)^2}>0.
 \label{eq:p5-tau-weighted}
\end{equation}
Thus the value $1/50$ is preserved exactly when $r=1$. A common
overall factor cancels. The response depends on relative branch weights.
For $|r-1|\leq\varepsilon<1$
$|\tau_r-1/50|\leq39\varepsilon/[50(5-\varepsilon)]$.
Species-dependent weights within either block require their actual
charge moments instead of this two-weight reduction.

The two spatial projections have different heat traces. Before
additional mass insertions
\[
 T_\pm(t)=\operatorname{Tr}(P_\pm e^{-tD^2}),\qquad
 T_+(t)\sim2e^{-9t/(4L^2)},\quad
 T_-(t)\sim6e^{-25t/(4L^2)}\quad(t\to\infty).
\]
The finite content-response normalisation is therefore applied before
the spatial projections. A torsion stationary equation formed from the
projected trace includes the corresponding source pullback or
compensation identity.

\subsection{Geometric dynamics}\label{sec:derivgeo}

In the dimensionless spectral convention of Section~\ref{sec:geodyn}
the $J=2$ Einstein--Cartan readout coefficient is
\begin{equation}
\lambda_{\mathrm{EC}} = N_g \Bigl(1 + \frac{\tau}{2}\Bigr)^2 + 6 = 14 + 8\tau + 2\tau^2 ,
\end{equation}
the torsion-dressed $\mathfrak{su}(2)_L$ term $8(1+\tau/2)^2$
plus the Lichnerowicz shift $6$. The effective tensor readout coefficient is
\begin{equation}
\lambda_{\mathrm{TT}} = 14 = 2 N_R ,
\end{equation}
the content double of the seven right-handed singlets. The squash amplitude is the broken-generator content
\begin{equation}
s_0 = n_{\mathrm{broken}} \, \tau = 2\tau ,
\end{equation}
with $n_{\mathrm{broken}} = \dim\SU(2)_R - \dim\U(1)_R = 3 - 1 = 2$ the two broken generators absorbed as the Goldstone directions of the $W_R$ bosons. Each contributes the torsion modulus. Its content ratio is
\begin{equation}
\frac{s_0}{N_R} = \frac{n_{\mathrm{broken}}}{N_f \sum Y^2 N_R} = \frac{2}{15\cdot(10/3)\cdot 7} = \frac{1}{175} ,
\end{equation}
half the relative first-order shift $N_g\tau/14=8\tau/14$.
The Euclidean-period factors enter as the amplitude
$e^{1/(2\pi)}$ and Gaussian-window width $\sqrt{2\pi}$ and tensor ratio
$r=(1/2\pi)^2$ in their respective readouts.

\subsection{Weak-coupling normalisation and screening}\label{sec:derivg2}

The weak-coupling readout fixes the per-mode response norm to twice
the unweighted norm with $I_{kv}=2\operatorname{Vol}(RP^3)=2\pi^2L^3$
and uses numerator $16\pi^2$ and scale factor $(\MG/\MP)^2$.
These are the absolute normalisation conventions of the readout.
For comparison a pointwise orthonormal round Killing coframe has
unweighted Gram matrix
$\int\langle\theta^a,\theta^b\rangle\,dv=\pi^2L^3\delta_{ab}$.
The response factor $2$ is separate from the covering-space volume.
For the dimensionless emergence input $c=\kL$ use
$\widehat I_{kv}(c)=2\pi^2c^3$. These fixed conventions give
\begin{equation}
g_2^2 = \frac{16\pi^2 (\MG/\MP)^2}{2\pi^2 (\kL)^3} = 8\Bigl(\frac{\MG}{\MP}\Bigr)^2 (\kL)^{-3} .
\end{equation}
The window-edge identity $\kL \MG = \MP \sqrt{\pi}$ gives
\begin{equation}
g_2^2 = \frac{8\pi}{(\kL)^5}, \qquad \alpha_W = \frac{g_2^2}{4\pi} = \frac{2}{(\kL)^5} .
\end{equation}
The proposition Matter coupling from the common transport
in Section~8 of Ref.~\cite{Guo2026CompactInternalSpaceI} supplies
$\nabla_A^\rho=d+\rho_*(A)$ and its modal product coefficients.
The absolute kinetic normalisation is specified above. If the mode frame
is changed by an invertible constant matrix $S$ its Gram matrix and
coefficient column transform as
$\mathsf G'=S^{\mathsf T}\mathsf G S$ and $a'=S^{-1}a$.
The connection and $a^{\mathsf T}\mathsf G a$ are unchanged.
Consequently the mode frame and matter coefficients together with the quadratic norm
must use one convention. The displayed coupling uses the response
normalisation specified above.
In the common action, write the
quadratic gauge and matter terms and its linear current vertex as
\[
 -\frac{Z_A}{4}F_{\mu\nu}F^{\mu\nu}
 +Z_\psi\overline\psi i\!\not\!\partial\psi
 +V_\psi A_\mu J^\mu.
\]
Canonical fields give $g_{\rm can}=V_\psi/(Z_\psi\sqrt{Z_A})$.
Thus the gauge norm and the current vertex are evaluated together.
For a constant internal mode $c_0$, the common-action reduction gives
$Z_A=\operatorname{Vol}(RP^3)c_0^2/g_7^2$ and
$V_\psi/Z_\psi=c_0$, hence
$g_4^2=g_7^2/\operatorname{Vol}(RP^3)$, independently of $c_0$.
Here $g_7^2$ denotes the coefficient of this product-space action.
Writing $g_7^2=\gamma_7/(\MP^2\MG)$ and
$L=c/\MG$ gives $g_4^2=\gamma_7/(\pi c^5)$ after use of the
window-edge identity. The weak-response normalization above corresponds
to $\gamma_7=8\pi^2$. It fixes the absolute strength in the stated
action convention; the $c^{-5}$ dependence follows from the volume
and scale relations.
For repeated coefficient resolution on the fixed round realization
the proposition Joint resolution of gauge and matter coefficients
in Appendix~A of Ref.~\cite{Guo2026CompactInternalSpaceI} keeps the gauge and matter
responses together under the specified heat evolution.
The spinor heat traces defined above and the coupling running of
Section~\ref{sec:rgflow} retain their separate definitions.
The bare Killing value is $+0.345\%$ above the reference coupling.

\paragraph{The weak response of the same finite source.}
Use the weak block of \eqref{eq:p5-finite-torsion-hessian}
already evaluated on the common finite trace.
For $g$ copies the normalized trace is still the same mean.
The extensive response multiplies its Hessian by the dimension $15g$.
$H^{\rm ext}_{F,ab}=2g\delta_{ab}$.
For the three-dimensional adjoint source the normalized Hessian is
$(2/3)\delta_{ab}$ and hence
$H^{\rm ext}_{{\rm ad},ab}=2\delta_{ab}$.
The scalar ratio in the common generator convention is therefore
\begin{equation}
 \delta_{\rm src}(g)=
 \frac{C_2(\operatorname{ad}\SU(2))}{\sum_fT(R_f)}
 =\frac2{2g}=\frac1g .
 \label{eq:p5-extensive-screening}
\end{equation}
Common generator rescalings cancel. The ratio of the normalized
Hessians is $5$. Restoring their respective dimensions gives the
extensive ratio $1/g$ used in the inverse weak-coupling relation.

\paragraph{Common screening coefficient and colour normalization.}
The finite trace and the screened coupling refer to the quadratic
weak-source form in the same generator convention. After separating
the torsion contribution its scalar correction is
\[
 \delta_{\rm col}=\alpha_{\SM}^{-1}-\alpha_W^{-1}
                  +\frac{\pi\tau^2}{2}.
\]
The adjoint and matter Hessians above evaluate this coefficient as
$\delta_{\rm src}(g)=1/g$. The colour-normalized closure used in the
screening rule assigns unit total correction to the $N_c$ colour
channels through $N_c\delta_{\rm col}=1$. Here $N_c=3$ is inherited from the
one-generation carrier of Ref.~\cite{Guo2026CompactInternalSpaceI}.
Both normalizations act on this single inverse weak-coupling
coefficient. Requiring their agreement is the joint screening rule
for the replication family. This determines the copy number
before substitution of any three-copy Dynkin sum.
The colour normalization has a simple operator form on the inherited
irreducible colour carrier. A colour-invariant positive insertion of
unit trace is $Q_{\rm col}=I_{N_c}/N_c$ by Schur's lemma.
Its diagonal response is therefore $1/N_c$. The screening rule
identifies this insertion with the scalar correction to the same
weak-source quadratic form. Comparing that correction with the
extensive matter response gives the compatibility condition below.

\begin{proposition}[Joint compatibility of the screening conditions]
\label{prop:p5-screening-compatibility}
Fix the upstream three-colour one-generation carrier and the
replication family of Section~\ref{sec:p5-joint-content}. For every
candidate positive integer $g$ require the same correction
\[
 \delta_{\rm col}:=\alpha_{\SM}^{-1}-\alpha_W^{-1}
                       +\frac{\pi\tau^2}{2}
\]
to satisfy both screening conditions
$N_c\delta_{\rm col}=1$ and
$\delta_{\rm col}=\delta_{\rm src}(g)$ with $N_c=3$.
Then the unique compatible copy number is $g=3$.
At $\tau=1/50$ it also satisfies the $R=5$ endpoint
and three-position window conditions.
\end{proposition}
\begin{proof}
Each generation has four weak doublets. Thus
$\sum_fT(R_f)=g(3+1)/2=2g$. Equation~\eqref{eq:p5-extensive-screening}
and the colour-screening condition give $N_c/g=1$.
Thus $g=3$. The coefficients of \eqref{eq:p5-branch-coefficients}
give $c_3=a_5+3b_5\in[2.490,2.498]$. Its square lies in
the three-position plateau $(11/2,15/2]$.
\end{proof}
On the selected branch $C_2(\SU(2))=2$ and the total fermion
Dynkin index is $\sum_fT(R_f)=6$. Thus the response check is
\begin{equation}
 \delta_{\rm src}(3)=\frac{C_2(\SU(2))}{\sum_f T(R_f)}
                   =\frac26=\frac13=\frac1{N_c}.
\end{equation}
The screened coupling then reads
\begin{equation}
 \frac{1}{\alpha_{\SM}}=\frac{1}{\alpha_W}
                   +\frac1{N_c}-\frac{\pi\tau^2}{2},
 \qquad
 g_2(M_{\mathrm G})=\sqrt{4\pi\alpha_{\SM}}=0.508848.
\end{equation}
The term $-\pi\tau^2/2$ is the quadratic torsion response with
$\tau^2=1/2500$. The squash and hypercharge matching also contain
terms linear in $\tau$. The source ratio and colour closure select
the copy number and the endpoint plateau verifies its retained
positions. Every quantity entering this compatibility calculation
is evaluated before comparison with measured couplings.

\begin{proposition}[Replicated family module and the quadratic stress response]
\label{prop:p5-replicated-quadratic-response}
Let $E_1$ be the one-generation Hermitian matter bundle with the
connection and source normalization inherited from Ref.~\cite{Guo2026CompactInternalSpaceI}. Realize
the replication rule of Section~\ref{sec:p5-joint-content} by
$E_g=\mathbb C^g\otimes E_1$ and
$\nabla_g=I_g\otimes\nabla_1$. The family factor is the
least-dimensional nondegenerate faithful representation of
$M_g(\mathbb C)$ commuting with the fixed one-generation spatial action.

For $g=3$ its coefficient space can be identified unitarily with
\[
 K=\operatorname{Hom}_{H_{\rm iso}}
 \bigl(S,F_0^+\oplus F_1^+\oplus F_2^+\bigr),\qquad \dim K=3,
 \quad H_{\rm iso}\simeq SU(2),\quad S\simeq\mathbb C^2 .
\]
Here $F_j^+$ are the three selected positive Dirac shells and $K$
carries the normalized intertwiner norm
$\langle a,b\rangle=\tfrac12\operatorname{Tr}(a^\dagger b)$.
At zero Higgs background the common quadratic operator and every
family-independent response insertion have the form
$Q_g=I_g\otimes Q_1$ and $B_g=I_g\otimes B_1$. Therefore whenever
the indicated trace exists
\begin{equation}
 \operatorname{Tr}_{E_g}\!\left[B_g f(Q_g)\right]
 =g\,\operatorname{Tr}_{E_1}\!\left[B_1 f(Q_1)\right].
 \label{eq:p5-replicated-trace}
\end{equation}
The identity holds for the Gaussian quadratic TT response including
its complete first and second source variations on the common
response domain.
\end{proposition}
\begin{proof}
The classification of finite-dimensional representations of
$M_g(\mathbb C)$~\cite{Davidson1996CStar} shows that a faithful
nondegenerate representation is a sum of defining modules. The least
one has dimension $g$ and is unique up to unitary equivalence. Tensoring
it with the inherited one-generation bundle implements the stipulated
replication. In each $F_j^+$ the isotropy spin $1/2$ occurs once.
The normalized radial intertwiners $R_j$ thus form an orthonormal
basis of $K$. The projected character products and their
Clebsch--Gordan normalization in Ref.~\cite{Guo2026CompactInternalSpaceI}
give the matrix units
$R_iR_j^\dagger$ acting as $E_{ij}\otimes I_S$ on the radial
isotypic source. A unitary identification of this coefficient
module with $\mathbb C^3$ conjugates the kinetic form and all
family vertices together. This preserves masses and mixing invariants
together with traces. The spatial spinor modes remain those of $E_1$ in
each summand.

Functional calculus and trace additivity prove
\eqref{eq:p5-replicated-trace}. In a fixed orthonormal trace space
write $G=Q_1^{-1}$. The determinant contribution has mixed
variation proportional to
\begin{equation}
 \partial_a\partial_b\operatorname{Tr}\log Q_1
 =\operatorname{Tr}\bigl(GQ_{1,ab}-GQ_{1,b}GQ_{1,a}\bigr).
 \label{eq:p5-complete-quadratic-source}
\end{equation}
The first term is the direct source term. Both terms and the
derivatives of inserted projectors and measures together with Gaussian factors
acquire the same factor $g$. Equivalently for $W=\log Z$ the
connected response is
$W_{ab}=\langle S_aS_b\rangle_c-\langle S_{ab}\rangle$.
The signed species projection is applied to the full one-copy
space in every summand.
\end{proof}

The numerical endpoint fixes the $R=5$ domain in its reference
geometry. For a source-dependent domain the same proposition uses
the full domain derivative. For example for a retained mode put
$b(s)=\sqrt{R\Lambda(s)^2-\lambda(s)}>0$ and
$I(s)=\int_{-b(s)}^{b(s)}f(\omega,s)\,d\omega$.
Its mixed second derivative is
\begin{align}
 I_{ab}={}&\int_{-b}^{b}f_{ab}\,d\omega
 +b_a(f_b(b)+f_b(-b))+b_b(f_a(b)+f_a(-b)) \notag\\
 &+b_{ab}(f(b)+f(-b))
 +b_ab_b(f_\omega(b)-f_\omega(-b)).
 \label{eq:p5-moving-response-domain}
\end{align}
All source dependence of the measure and insertions is included
in $f$. Fixed-rank spectral clusters admit differentiable Riesz
projectors as long as their separating gaps remain open. At the
reference endpoint the spatial cutoff is $5\pi$ in dimensionless
units and the next vector eigenvalue is $16$. Thus this condition
holds in a neighbourhood. Equations
\eqref{eq:p5-complete-quadratic-source} and
\eqref{eq:p5-moving-response-domain} then preserve the copy factor.
Bosonic channels keep their one-copy multiplicity. At nonzero
Higgs background the family-dependent mass vertices enter $Q_g$
and are retained explicitly in the corresponding response.

The torsion term combines the quadratic torsion contribution and
Euclidean period with the conformal normalisation
\begin{equation}
\frac{\tau^2 \pi}{2} = (2\tau^2)\, (2\pi)\, \xi = \frac{4\pi \tau^2}{N_g} ,
\end{equation}
The factor $2\tau^2$ is the quadratic
coefficient in the Einstein--Cartan spectral readout
$\lambda_{\mathrm{EC}}=N_g(1+\tau/2)^2+6=14+8\tau+2\tau^2$.
For the full torsion norm $T^a{}_{bc}=(\tau/L)\varepsilon^a{}_{bc}$ gives
$L^2T_{abc}T^{abc}=6\tau^2$ in an orthonormal frame.
The factor $2\pi$ is the Euclidean period and
$\xi=1/8=1/N_g$ is the conformal normalisation.
The identity $N_g\xi=1$ gives the second expression. The inverse-coupling
relation above specifies the additive insertion and its sign.

\paragraph{Axial Killing-dual curvature normalisation.}
\label{sec:p5-axial-normalisation}
The hypercharge matching uses $\kappa(2\tau)$ at the
breaking scale before running and the further content correction.
Its geometric factor follows from the squashed metric
\begin{equation}
ds^2 = (1+s)\,\sigma_1^2 + (1+s)\,\sigma_2^2 + (1-2s)\,\sigma_3^2 ,
\end{equation}
at fixed round radius. Put $A=1+s$ and $B=1-2s$ with
$-1<s<1/2$. Fix the axial Killing generator $X$ in the same
scale for all $s$ by choosing $\alpha_0=X^{\flat_0}=\sigma_3$.
Its flow rotates $\sigma_1,\sigma_2$ and preserves $\sigma_3$.
Thus it remains Killing for this axisymmetric family.
Its metric dual is $\alpha_s=B\sigma_3$.
Since $d\sigma_3$ is a fixed multiple of
$\sigma_1\wedge\sigma_2$ the pointwise two-form norm ratio is
$B^2/A^2$ and the volume ratio is $A\sqrt B$. Therefore
\begin{equation}
 \frac{\|d\alpha_s\|_{L^2(h_s)}^2}
      {\|d\alpha_0\|_{L^2(h_0)}^2}
 =\frac{B^{5/2}}A=\kappa(s)^{-2},\qquad
 \kappa^2(s)=\frac{1+s}{(1-2s)^{5/2}} .
 \label{eq:p5-kappa-curvature-norm}
\end{equation}
The covering factor and a fixed rescaling of $X$ together with a fixed
matrix-trace convention cancel in this ratio.
For the same Abelian connection
$\mathcal A_s=a_s\alpha_sT_Y$ with $T_Y=iY$ and $a_s$
constant on the internal space one has
$F_s=a_s\,d\alpha_sT_Y$.
Preserving its internal curvature quadratic norm and choosing
nonzero amplitudes of the same sign gives $a_s/a_0=\kappa(s)$.
The same connection acting on matter through the cited
proposition of Ref.~\cite{Guo2026CompactInternalSpaceI} obeys
\begin{equation}
 \mathcal A_s(X)=B\kappa(s)\mathcal A_0(X),\qquad
 \mathcal A_s(X/\sqrt B)=\sqrt B\,\kappa(s)\mathcal A_0(X).
 \label{eq:p5-kappa-matter-profile}
\end{equation}
These evaluate the connection along a fixed and a unit axial direction
respectively.
If an amplitude varies the full curvature also includes
$da_s\wedge\alpha_s$. The one-form norm ratio is
$\|\alpha_s\|_{L^2(h_s)}^2/\|\alpha_0\|_{L^2(h_0)}^2=AB^{3/2}$.
In particular an amplitude $a(x)$ on the product space has curvature
\[
 F_s=d_xa\wedge\alpha_sT_Y+a\,d_y\alpha_sT_Y.
\]
It is a four-dimensional scalar amplitude whose kinetic coefficient
$K_s$ and internal restoring coefficient $C_s$ obey
\[
 \frac{K_s}{K_0}=AB^{3/2},\qquad
 \frac{C_s}{C_0}=\frac{B^{5/2}}A.
\]
Under the curvature-preserving change $a=\kappa(s)a_0$ the kinetic
ratio becomes $\kappa^2K_s/K_0=A^2/B$ and every linear matter
vertex acquires the same amplitude factor. Canonical normalization
therefore retains both Grams. For an external vector mode the
coefficient of $F_{\mu\nu}F^{\mu\nu}$ and its matter current
are projected from that mode in the common action, giving
$g_{\rm can}=V_\psi/(Z_\psi\sqrt{Z_A})$ as above.
The hypercharge boundary rule combines the internal-curvature factor
with the stated external-mode normalization and finite content
correction. The constant-amplitude identity supplies its geometric factor
with $\kappa(0)=1$ and $\kappa(2\tau)=1.13183$.
The mixing coefficient is corrected by the content ratio
\begin{equation}
\delta_{g_1} = -\tau\, r_{23}\, \sum Y^2 \Delta_f \xi ,
\end{equation}
with $r_{23}=3/(10\sqrt3)$ and
$\sum Y^2\Delta_f\xi=(10/3)(3/2)(1/8)=5/8$.
This product combines the hypercharge moment and fermion conformal weight
with the scalar conformal coupling used above.
The colour matching rule is
$g_3=g_2(1+\alpha_{\mathrm{GUT}}^2/K)$ with
$K=J(J+2)/d=8/3$. It assigns the dimensionless squash eigenvalue
$J(J+2)=8$ over the internal dimension $d=3$ to the long-root correction.

\subsection{Two-loop gauge coefficients}\label{sec:derivbeta}

We evaluate the $\overline{\mathrm{MS}}$ gauge beta functions on the five inherited
fermion representations. The minimal joint module and charged
response quotient of Ref.~\cite{Guo2026CompactInternalSpaceI} supply $Q_L,L_L,u_R,d_R,e_R$.
Its result on the elementary doublet representation fixes the scalar
type $(1,2)_{1/2}$.
The running uses the content of Section~\ref{sec:content} with three
copies of the fermion module and one complex Higgs doublet.

In the fixed convention $Y_H=1/2$ and $Q=T_3+Y$ Yukawa covariance and
anomaly cancellation give $Y_Q=1/6$ and
$Y=(1/6,2/3,-1/3,-1/2,-1)$ for the five inherited representations.  The
per-generation hypercharge moment is therefore
\begin{equation}
\sum Y^2 = 6\cdot\frac{1}{6^2} + 3\cdot\frac{4}{9} + 3\cdot\frac{1}{9} + 2\cdot\frac{1}{4} + 1\cdot 1 = \frac{10}{3} .
\end{equation}

We use Eqs.~(30)--(31) of Ref.~\cite{LuoWangXiao2003} with the
product-group substitutions of Eqs.~(106)--(110) in its Section~VI.
For a running scale $\mu$ the loop convention is
\[
 \frac{dg_i}{d\log\mu}
 =\frac{b_i g_i^3}{16\pi^2}
 +\frac{g_i^3}{(16\pi^2)^2}
   \left(\sum_jB_{ij}g_j^2-A_i y_t^2\right).
\]
We use $g_1=\sqrt{5/3}\,g_Y$ with Weyl fermions and
one complex Higgs doublet. In the two-loop scalar sums the latter
is counted as its realification. The one-loop coefficients are
\begin{equation}
 (b_1,b_2,b_3)=\left(\frac{41}{10},-\frac{19}{6},-7\right),
\end{equation}
with $C_2(\SU(2))=2$ and $C_2(\SU(3))=3$ together with
$T_f(\SU(2))=T_f(\SU(3))=6$ and $T_H=1/2$.
For the two-loop substitution $S_2$ is fixed by
$\Tr(T_R^aT_R^b)=S_2(R)\delta^{ab}$. The fundamental indices are
$1/2$ and the GUT-normalized abelian entries use
$C_2=(3/5)Y^2$. These conventions fix the Casimirs and Dynkin indices
used in the product-group formula.

For each Weyl representation, the Casimirs and Dynkin indices are listed
in the order
\[
(C_2^{\U(1)}, C_2^{\SU(2)}, C_2^{\SU(3)}, S_2^{\U(1)}, S_2^{\SU(2)}, S_2^{\SU(3)}).
\]
Their values are
\begin{equation}
\begin{aligned}
Q_L (3,2)_{1/6} &: \Bigl(\frac{3}{5}\frac{1}{36}, \frac34, \frac43, \frac35\frac16, \frac32, 1\Bigr), \\
u_R (3,1)_{2/3} &: \Bigl(\frac35\frac49, 0, \frac43, \frac35\frac43, 0, \frac12\Bigr), \\
d_R (3,1)_{-1/3} &: \Bigl(\frac35\frac19, 0, \frac43, \frac35\frac13, 0, \frac12\Bigr), \\
L_L (1,2)_{-1/2} &: \Bigl(\frac35\frac14, \frac34, 0, \frac35\frac12, \frac12, 0\Bigr), \\
e_R (1,1)_{-1} &: \Bigl(\frac35, 0, 0, \frac35, 0, 0\Bigr),
\end{aligned}
\end{equation}
and the Higgs scalar $(1,2)_{1/2}$ contributes the real-scalar content
$\bigl(\frac35\frac14,\frac34,0,\frac35,1,0\bigr)$.
Substitution of this table in the cited product-group formula gives
\begin{equation}
B = \begin{pmatrix}
199/50 & 27/10 & 44/5 \\
9/10 & 35/6 & 12 \\
11/10 & 9/2 & -26
\end{pmatrix} .
\end{equation}
These standard coefficients for three fermion generations and
one complex scalar doublet use the inherited charge
normalisation. The abelian fourth moment checks
$B_{11}=199/50$. The same Weyl and scalar component counting in
the top-Yukawa trace gives with
$d(G_i)=\dim\operatorname{adj}(G_i)=(1,3,8)$ for
$G_i=\U(1),\SU(2),\SU(3)$
\begin{equation}
 A_i=\frac{6}{d(G_i)}
 \bigl[C_2^{(i)}(Q_L)+C_2^{(i)}(u_R)\bigr],
 \qquad
 A=\Bigl(\frac{17}{10},\frac32,2\Bigr).
\end{equation}
This agrees with the Yukawa term of the cited gauge beta function.

\subsection{The top-Yukawa and Higgs-quartic beta functions}\label{sec:derivytlam}

The top Yukawa and scalar running use the $\overline{\mathrm{MS}}$ formulas of
Sections~IV.2--IV.3 of Ref.~\cite{LuoWangXiao2003} specialised to the
Standard Model in Eqs.~(3) and (6) together with (9) and (10) of Ref.~\cite{LuoXiao2003}.
In this subsection $\lambda=\lambda_H$ is the Higgs quartic in
$V=-m^2|H|^2+\lambda|H|^4$ with $g_1=\sqrt{5/3}\,g_Y$.
The reference potential uses $(\lambda_{LX}/2)|H|^4$. Thus
$\lambda=\lambda_{LX}/2$ and $\beta_\lambda=\beta_{\lambda,LX}/2$.
For $X=y_t,\lambda$ define
\[
 \beta_X:=\frac{dX}{d\log\mu}
 =\frac{\beta_X^{(1)}}{16\pi^2}
  +\frac{\beta_X^{(2)}}{(16\pi^2)^2}.
\]
In the top-only approximation the Yukawa matrix is
$\mathsf Y_t=\operatorname{diag}(0,0,y_t)$. Its trace invariants are
\begin{gather}
Y_2(S) = \Tr[3\mathsf Y_t^\dagger\mathsf Y_t] = N_c\, y_t^2 , \qquad
H(S) = \Tr[3(\mathsf Y_t^\dagger\mathsf Y_t)^2] = N_c\, y_t^4 , \nonumber \\
\chi_4(S) = \frac94 \Tr[3(\mathsf Y_t^\dagger\mathsf Y_t)^2] = \frac94 N_c\, y_t^4 ,
\end{gather}
and
\begin{equation}
Y_4(S) = \Bigl(\frac{17}{20} g_1^2 + \frac94 g_2^2 + 8 g_3^2\Bigr) y_t^2 ,
\end{equation}
where the coefficients follow from $-3g^2\{C_2(F),Y\}$ with the
Casimirs in Section~\ref{sec:derivbeta}. The family count $n_g=3$
is fixed by the joint content and endpoint rules of
Proposition~\ref{prop:p5-screening-compatibility}.

The one-loop top Yukawa is
\begin{equation}
\frac{\beta_{y_t}^{(1)}}{y_t} = \Bigl(\frac32 + N_c\Bigr) y_t^2 - \frac{17}{20} g_1^2 - \frac94 g_2^2 - 8 g_3^2 = \frac92 y_t^2 - \frac{17}{20} g_1^2 - \frac94 g_2^2 - 8 g_3^2 ,
\end{equation}
the Yukawa self-coupling $3/2$ plus the colour trace $Y_2(S) = N_c y_t^2$. The $\U(1)$ coefficient $17/20$ is in the GUT normalisation and equals $-(17/12)g_Y^2$ in the hypercharge coupling. The two-loop top Yukawa is
\begin{equation}
\begin{aligned}
\frac{\beta_{y_t}^{(2)}}{y_t} = {}& -12\, y_t^4 + \Bigl(36\, g_3^2 + \frac{225}{16} g_2^2 + \frac{393}{80} g_1^2\Bigr) y_t^2 - 108\, g_3^4 - \frac{23}{4} g_2^4 + \frac{1187}{600} g_1^4 \\
& + 9\, g_2^2 g_3^2 + \frac{19}{15} g_1^2 g_3^2 - \frac{9}{20} g_1^2 g_2^2 + 6\, \lambda^2 - 12\, \lambda y_t^2 ,
\end{aligned}
\end{equation}
Here the $y_t^4$ and $g_i^2y_t^2$ coefficients include the
$\chi_4(S)$ and $Y_4(S)$ traces above. The pure-gauge coefficients
are those of Ref.~\cite[Eq.~(6)]{LuoXiao2003} at $n_g=3$.
The one-loop Higgs quartic is
\begin{equation}
\beta_\lambda^{(1)} = 24\,\lambda^2 - 3\lambda\Bigl(3 g_2^2 + \frac35 g_1^2\Bigr) + \frac38\Bigl(2 g_2^4 + \bigl(g_2^2 + \frac35 g_1^2\bigr)^2\Bigr) - 6\, y_t^4 + 12\, \lambda y_t^2 ,
\end{equation}
whose Yukawa part in this potential convention is
$4Y_2(S)\lambda-2H(S)=12\lambda y_t^2-6y_t^4$.
The two-loop Higgs quartic is
\begin{equation}
\begin{aligned}
\beta_\lambda^{(2)} = {}& -312\,\lambda^3 + 36\lambda^2\Bigl(3 g_2^2 + \frac35 g_1^2\Bigr) - 144\,\lambda^2 y_t^2 - \frac{73}{8}\lambda g_2^4 + \frac{117}{20}\lambda g_2^2 g_1^2 + \frac{1887}{200}\lambda g_1^4 \\
& + \Bigl(\frac{17}{2} g_1^2 + \frac{45}{2} g_2^2 + 80\, g_3^2\Bigr)\lambda y_t^2 - 3\,\lambda y_t^4 + 30\, y_t^6 \\
& - 32\, g_3^2 y_t^4 - \frac85 g_1^2 y_t^4 - \frac94 g_2^4 y_t^2 + \frac{63}{10} g_2^2 g_1^2 y_t^2 - \frac{171}{100} g_1^4 y_t^2 \\
& + \frac{305}{16} g_2^6 - \frac{289}{80} g_2^4 g_1^2 - \frac{1677}{400} g_2^2 g_1^4 - \frac{3411}{2000} g_1^6 ,
\end{aligned}
\end{equation}
The conversion $\lambda_{LX}=2\lambda$ multiplies the coefficients
of $\lambda^3$ and $\lambda^2$ by four and two respectively
and halves terms independent of $\lambda$. The displayed coefficients
use $n_g=3$ in Ref.~\cite[Eq.~(10)]{LuoXiao2003}. The
$\lambda y_t^2g^2$ terms equal $10\lambda Y_4(S)$.
All fermionic traces are evaluated in the top-only
three-generation approximation.

\subsection{Sector exponents and the Landau--Zener comparison}\label{sec:derivlz}

\paragraph{Scale transport and mode mixing.}
The proposition Homothetic scale transport of the spectral window
in Ref.~\cite{Guo2026CompactInternalSpaceI} preserves dimensionless Gaussian weights.
The family transfer acts on the retained spectral coefficient module
of Proposition~\ref{prop:p5-replicated-quadratic-response}. Its
diagonal weights and angular operators specify the mass construction.

The mass ladder attaches exponential suppression to the extrusion
orders $n=\{0,2,4\}$. Its proposed non-adiabatic interpretation uses
a Landau--Zener comparison~\cite{Zener1932} in addition to the pure homothetic flow
$L(k)=C/k$ with the avoided-crossing normal form
\begin{equation}
H(t) = \begin{pmatrix} \upsilon t & \Delta_{\rm LZ} \\ \Delta_{\rm LZ} & -\upsilon t \end{pmatrix} ,
\qquad
P_{\rm LZ} = e^{-\pi \Delta_{\rm LZ}^2/|\upsilon|} ,
\end{equation}
in the convention $i\partial_t\psi=H(t)\psi$ with real $\upsilon\ne0$
and constant $\Delta_{\rm LZ}$.  Here $t$ is the parameter of this two-level
comparison. Its physical realization parametrizes the source path by
a material clock.
The single-passage exponent is
$\gamma=\pi\Delta_{\rm LZ}^2/|\upsilon|$ and the diabatic survival
amplitude has magnitude \(e^{-\gamma/2}\)~\cite{Shevchenko2010}.
To derive
the mass-amplitude factor $e^{-n\alpha_f}$ from a single passage the source
trajectory must produce the operator above and the equality
\begin{equation}
 \frac{\pi\Delta_{\rm LZ}^2}{|\upsilon|}=2n\alpha_f,
 \label{eq:p5-lz-amplitude-matching}
\end{equation}
with the transmitted mass vertex identified with the diabatic survival
channel. The gap and sweep rate must be evaluated on that same
trajectory. Multiple passages require their scattering matrices and
relative phases. The mass construction uses the positive amplitude
transfer $m_i\propto e^{-\alpha_f n_i}$ with the sector rates set by
the content rules.
The up-sector transfer relation subtracts the torsion deficit from the
window width
\begin{equation}
\alpha_{\mathrm{up}} = \kL - 2\tau = 2.4535 ,
\end{equation}
where the window contribution is $\kL$ and the assigned deficit is $n_{\mathrm{broken}}\tau=2\tau$. This gives $m_t/m_c=e^{2\alpha_{\mathrm{up}}}$. The down and lepton sectors descend by the step
\begin{equation}
\Delta = 6 (1-n_s) \kL_{\mathrm{CMB}} ,
\end{equation}
the product of the six generators of the $\mathfrak{so}(4)$ isometry with the spectral tilt $1-n_s$ and the microwave-pivot window $\kL_{\mathrm{CMB}}$. The down and lepton steps are split by the $9/8$ hypercharge identity
\begin{equation}
\frac{9}{8} = \frac{1}{1 - (Y_d/Y_l)^2} ,
\end{equation}
with $Y_d=|Y_{d_R}|=1/3$ and $Y_l=|Y_{e_R}|=1$. Thus the two contributions have ratio $9/8$. Writing them as $(9/8)h$ and $h$ the total lepton decrement $2\Delta$ gives $(9/8)h+h=2\Delta$ and hence $h=16\Delta/17$. The indices are
\begin{equation}
\alpha_{\mathrm{dn}} = \alpha_{\mathrm{up}} - \frac{18}{17}\Delta , \qquad
\alpha_{\mathrm{lp}} = \alpha_{\mathrm{up}} - 2\Delta .
\end{equation}
The first-generation step of the down sector carries the colour dilution
\begin{equation}
\alpha_{\mathrm{sd}} = \alpha_{\mathrm{dn}} - \frac{\kL_{\mathrm{CMB}}}{6} ,
\end{equation}
where the assigned colour dilution divides the microwave window by the
six-generator count of $\mathfrak{so}(4)$. Substituting the
stated tilt and pivot window gives the exact expression
\[
 \alpha_{\mathrm{sd}}=
 \kL\left[1-\left(1-\frac{\tau}{4}\right)
 \left(\frac{189\tau}{17}+\frac16\right)\right]-2\tau.
\]
These rules determine the exponents used in the mass table.
The generator counts and charge traces set the coefficients. Equal attenuation per source unit and its placement in a mass amplitude
are the transfer choices summarized in
Table~\ref{tab:p5-hierarchy-selection}. The consecutive sector steps
also give the anchor-independent relation
\eqref{eq:p5-sector-mass-test} by direct elimination of
$\alpha_{\rm up}$ and $\Delta$.

The specified transfer generators give positive mass operators with a
common left handed kinetic realization.

\begin{proposition}[Positive spectral transfer and the common mass vertex]
\label{prop:p5-spectral-mass-transfer}
Let $P_0,P_1,P_2$ be the orthogonal heavy and middle and light
family projectors and put $c_C=\kL_{\rm CMB}$. Define
\begin{align}
 B_u&=2\alpha_{\rm up}P_1+
       (2\alpha_{\rm up}+2c_C+\log4)P_2,\notag\\
 B_d&=2\alpha_{\rm dn}P_1+
       (2\alpha_{\rm dn}+2\alpha_{\rm sd})P_2,\notag\\
 B_\ell&=2\alpha_{\rm lp}P_1+
       (4\alpha_{\rm lp}+\sqrt{2\pi})P_2 .
 \label{eq:p5-positive-transfer-generators}
\end{align}
Then $T_f=e^{-B_f}$ is positive definite and its eigenvalues give
the stated mass ratios relative to the heaviest member of each
sector. The vertices $Y_u=y_uT_u$ and $Y_d=y_dUT_d$ use the
same canonical left kinetic form with $U$ the ordered angular
rotation of Appendix~\ref{sec:derivckm}. Their singular values
are the masses defined above times $\sqrt2/v$ and their left
mixing matrix is $U$.
\end{proposition}
\begin{proof}
The three projectors are orthogonal and sum to the identity of the
family module. Thus exponentiation acts separately on their real
coefficients. In the up sector the eigenvalues are
$1,e^{-2\alpha_{\rm up}},
\tfrac14e^{-2\alpha_{\rm up}-2c_C}$.
The other two sectors follow identically. Multiplication on the
left by $U$ leaves the down singular values unchanged and sets
its left frame relative to the up frame.

The matrices implementing this construction are generated before
mass diagonalization. In the normalized radial source basis
write $\chi_k(x)=\sum_{m=-k}^kD^k_{mm}(x)$. In the inherited
Clebsch--Gordan phase convention the intertwiner has entries
\[
 (R_j)_{(m,J,M),s}
 =\frac{C^{JM}_{jm,\,1/2\,s}}{\sqrt{j+1}},
 \qquad J=j+\tfrac12,\quad M=m+s .
\]
Inserting these coefficients into the normalized three-matrix-element
integral on $SU(2)$ gives
\[
 \Pi_0M_{\chi_1}\Pi_1=\sqrt{2/3}\,R_0R_1^\dagger,\quad
 \Pi_1M_{\chi_1}\Pi_2=\frac3{\sqrt{10}}R_1R_2^\dagger,\quad
 \Pi_0M_{\chi_2}\Pi_2=\sqrt{3/5}\,R_0R_2^\dagger .
\]
Here $\Pi_j=R_jR_j^\dagger$. Dividing by these nonzero
Clebsch--Gordan coefficients gives the family matrix units.
Their anti-Hermitian combinations generate the ordered plane
rotations including the complex phase. They act on the
six-dimensional radial source within the complete
$2+12+30=44$ dimensional three-shell carrier. The orthogonal
complement is retained with zero family vertex. Moving the
basepoint translates both characters and intertwiners unitarily.
Thus the common kinetic form and two mass spectra together with the CP
invariant in \eqref{eq:p5-common-flavour-invariants} transform
covariantly.
\end{proof}

In increasing mass order let $E_{ab}$ denote these matrix units.
The unitary matrix with the three CKM moduli of
Appendix~\ref{sec:derivckm} is
\[
 U=e^{\theta_{23}(E_{23}-E_{32})}
   e^{\theta_{13}(e^{-i\delta}E_{13}-e^{i\delta}E_{31})}
   e^{\theta_{12}(E_{12}-E_{21})},
\]
with $s_{13}=|V_{ub}|$ and
$s_{12}=|V_{us}|/\sqrt{1-s_{13}^2}$ together with
$s_{23}=|V_{cb}|/\sqrt{1-s_{13}^2}$ and
$\theta_{ij}=\arcsin s_{ij}$ and $\delta=8\pi/21$.
It gives
$J=s_{12}s_{23}s_{13}c_{12}c_{23}c_{13}^2\sin\delta$.
The expressions in Appendix~\ref{sec:derivckm} use its stated
leading order in the small element $|V_{ub}|$.

This positive transfer is an operator realization of the hierarchy
relation in the fixed generated algebra
\cite{Guo2026CompactInternalSpaceI,Guo2026SpectralArithmetic,Guo2026NumberDynamics}.
Its Landau--Zener interpretation concerns a further dynamical
choice of a source trajectory with its gap and sweep rate.

\subsection{The top-quark anchor}\label{sec:derivtop}

The Elementary doublet representation proposition of
Ref.~\cite{Guo2026CompactInternalSpaceI} gives the connector
$T_H=I_C\otimes h_H$ with
\[
 h_H=\begin{pmatrix}a&b\\-\bar b&\bar a\end{pmatrix},
 \qquad H=(b,\bar a)^{\mathsf T},\qquad
 h_H^\dagger h_H=\|H\|^2I_2.
\]
We use this representation in its inherited source Hilbert norm.
On the radial representative $H=(0,r)^{\mathsf T}$ with $r\geq0$
the finite matrix is $h_H=rI_2$. It gives
\begin{equation}
 T_H=rI_C\otimes I_2,\qquad
 \mathcal D_H^2=r^2I,\qquad
 \frac12\operatorname{Tr}(dh_H^\dagger dh_H)=dr^2,
 \label{eq:p5-elementary-unit-connector}
\end{equation}
where $\mathcal D_H=\left(\begin{smallmatrix}0&T_H\\T_H^\dagger&0\end{smallmatrix}\right)$.
For unit internal vectors paired by the inherited identity
$C_0=I_C\otimes I_2$ one obtains
$\langle C_0\eta,C_0\eta\rangle=1$. This is the geometric seed $y_0=1$.
It applies in particular to the colour-triplet block of the top connector.

The seed is preserved under a common unitary comparison. If
$C(s)=U_L(s)C_0U_R(s)^\dagger$ and
$\eta_{L,R}(s)=U_{L,R}(s)\eta_{L,R}(0)$ then
\[
 \langle\eta_L(s),C(s)\eta_R(s)\rangle
 =\langle\eta_L(0),C_0\eta_R(0)\rangle=1.
\]
This identity preserves the elementary seed in its inherited norm.
The readout $m_t=y_0v/\sqrt2$ uses this seed and the window VEV.
Equation~\eqref{eq:p5-signed-spatial-connector} implements the corresponding
signed spatial assignment.
For responses involving the odd connector the complete carrier
and neutral-feedback result of Ref.~\cite{Guo2026CompactInternalSpaceI} applies before
the charged compression is taken. The full response includes the effective
source and readout together with the direct neutral term~\cite{Guo2026GenesisFramework}.
At $r>0$ the full connector is invertible. Eliminating the neutral
coordinate retains its induced source and direct response term together
with the static charged-source trace.

The coupling of a spatial mass vertex to canonically normalized fields
requires its own kinetic factors. With positive Gram matrices and the
specified scalar normalization the matched matrix is
\[
Y_{\rm can}=Z_L^{-1/2}\mathcal Y Z_R^{-1/2}/\sqrt{Z_H},
\qquad
\mathcal Y_{ij}=\int\psi_{L,i}^\dagger\mathcal O_H\psi_{R,j}\,d\mu.
\]
Numerical evaluation for fixed modes and sources checks the singular
values and physical pairings under changes of basis~\cite{Guo2026GenesisFramework}.
For the spinor modes on the inherited compact spin bundle the
spatial part of a constant radial connector is fixed by orthonormality
\[
 \int\psi_i^\dagger\psi_j\,d\mu=\delta_{ij},\qquad
 \int\psi_i^\dagger(rI)\psi_j\,d\mu=r\delta_{ij}.
\]
The overlaps and nonconstant-source products are evaluated in the
Clebsch--Gordan mode basis with the full product support retained before
the chosen projection is applied~\cite{Guo2026GenesisFramework}.
The signs of the internal Dirac eigenvalues label the spatial spectral branches.
Four-dimensional chirality enters through the assigned left and right fields.
The kinetic coefficients and action vertex enter the canonical
matching formula with the same inherited normalization.
The inherited unit seed fixes the finite pairing above. The matched
running Yukawa then follows its renormalization-group equation.

\paragraph{The signed spatial connector.}
Let $R_+:\mathbb C^2\to F_0^+$ and
$R_-:\mathbb C^2\to F_1^-$ be the normalized radial intertwiners
for the signed eigenvalues $+3/(2L)$ and $-5/(2L)$.
The complete positive and negative shells have dimensions two and
six. The Gaunt integral of the inherited character is
$R_-^\dagger M_{\chi_1}R_+=I_2/\sqrt3$.
Consequently
\begin{equation}
 B_0=\sqrt3\,\Pi_-^{\rm rad}M_{\chi_1}\Pi_+^{\rm low}
     =R_-R_+^\dagger,\qquad
 B_0^\dagger B_0=\Pi_+^{\rm low},\quad
 B_0B_0^\dagger=\Pi_-^{\rm rad}.
 \label{eq:p5-signed-spatial-connector}
\end{equation}
This follows by inserting the normalized intertwiners in the
projected product. Its four-dimensional negative-shell complement
is annihilated. The family vertex $Y_f\otimes B_0$ therefore has
the normalized infrared pairing $Y_f$ and gives $vY_f/\sqrt2$
after symmetry breaking. Spatial Dirac signs and external chirality
retain their distinct definitions.

The same construction has a smooth source continuation on an isolated
pair of spectral clusters. If $E$ is evaluation at the transported
basepoint set $N_\pm=EP_\pm E^\dagger$ and
$R_\pm=P_\pm E^\dagger N_\pm^{-1/2}$. Evaluation here is restricted
to the finite smooth eigenspaces. On a source neighbourhood where
these maps are twice differentiable and $N_\pm>0$ the matrix
$C=R_-^\dagger M_{\chi_1}R_+$ remains invertible near the round
point. Multiplying the projected character kernel on its range
by $R_-C^{-1}R_-^\dagger$ gives $B=R_-R_+^\dagger$.
This fixes its derivatives and transforms covariantly under a
common bundle-frame change. Source variations of the projectors and
evaluation maps together with their Gram matrices are included.

\paragraph{Common action, flavour basis and renormalized boundary.}
For the action measure $V\,d\mu$ let $I_2=\int|h|^2d\mu$ and
$I_4=\int|h|^4d\mu$. The separated action with scalar kinetic coefficient
$Z_H$ and quartic coefficient $\lambda_{\rm act}$ gives
\begin{equation}
 y_{\rm can}=\frac{y_{\rm act}}{\sqrt{VZ_LZ_RZ_H}}
 \quad(h=1),\qquad
 \lambda_{\rm can}=\frac{\lambda_{\rm act}I_4}{VZ_H^2I_2^2}.
 \label{eq:p5-canonical-action-boundary}
\end{equation}
The same left kinetic form occurs in the up and down vertices. Write their
canonical matrices as $Y_u,Y_d$ and diagonalize
$H_f=Y_fY_f^\dagger=L_f\operatorname{diag}(y_{f,i}^2)L_f^\dagger$.
Then $V_{\rm CKM}=L_u^\dagger L_d$ for simple ordered spectra. In particular
the basis-independent quantities are
\begin{equation}
 |V_{ij}|^2=\operatorname{Tr}(P_i^uP_j^d),\qquad
 \operatorname{Im}\operatorname{Tr}[H_u,H_d]^3,
 \label{eq:p5-common-flavour-invariants}
\end{equation}
where $P_i^f$ are the rank-one left spectral projectors. An arbitrary
invertible change of the unnormalized common left basis induces the same
unitary change in both canonical sectors. Thus these quantities and the
singular values are unchanged. Degenerate eigenspaces admit only their
total spectral projectors. The family dependence of the present
construction resides in its transfer factors. In increasing mass order
the up and down diagonal factors are
\[
 D_u=\operatorname{diag}\left(
 \tfrac14e^{-2\alpha_{\rm up}-2c_{\rm CMB}},
 e^{-2\alpha_{\rm up}},1\right),\qquad
 D_d=\operatorname{diag}\left(
 e^{-2\alpha_{\rm dn}-2\alpha_{sd}},e^{-2\alpha_{\rm dn}},1\right),
\]
where $c_{\rm CMB}=\kL_{\rm CMB}$. Their overall coefficients are the
top and bottom readouts. The angular assignments specify the relative
left frames. Proposition~\ref{prop:p5-spectral-mass-transfer} and
\eqref{eq:p5-signed-spatial-connector} realize these data in one
action with the same left kinetic form.
Equations~\eqref{eq:p5-canonical-action-boundary} and
\eqref{eq:p5-common-flavour-invariants} give the forward test of a proposed
common action including nonunitary changes of mode coordinates
\cite{Guo2026GenesisFramework}.

At a renormalization scale $\mu_0$ finite quantum matching takes the form
$\overline Y_f(\mu_0)=Y_{f,\rm can}+\Delta Y_f(\mu_0)$ and
$\overline\lambda(\mu_0)=\lambda_{\rm can}+\Delta\lambda(\mu_0)$.
The five-coupling evolution then acts on this renormalized boundary with
explicit threshold maps. At a common scale and VEV convention the
tree coefficients of the canonically normalized broken-phase action are
\[
 m_t^{\rm tree}=\overline y_t v/\sqrt2,\qquad
 (m_H^{\rm tree})^2=2\overline\lambda_Hv^2.
\]
In this effective representation the conversion to physical mass
observables additionally requires finite VEV matching and the corresponding
self-energies at the specified
order. The numerical tables use the geometric mass relations defined in
Sections~\ref{sec:flavour} and~\ref{sec:ewsb}. The spatial integrals and
canonical normalization together with joint evolution are evaluated from the
supplied action coefficients~\cite{Guo2026GenesisFramework}.

\paragraph{Physical readouts and finite electroweak matching.}
Use the canonically normalized four-dimensional action with
$V(H)=m^2H^\dagger H+\lambda(H^\dagger H)^2$ and
$g'^2=3g_1^2/5$. On $H=(0,(v+h)/\sqrt2)^{\mathsf T}$ its
stationary condition is $m^2=-\lambda v^2$ and its quadratic
operators give $m_t^2=y_t^2v^2/2$ and $m_H^2=2\lambda v^2$ and
$m_W^2=g_2^2v^2/4$. The same covariant derivative couples
$g_2W_\mu^+J^\mu/\sqrt2+\mathrm{h.c.}$ where
$J_\mu=\bar\nu\gamma_\mu P_L e$. Eliminating the massive $W$
at small external momentum gives
$-2J_\mu^\dagger J^\mu/v^2=-4G_FJ_\mu^\dagger J^\mu/\sqrt2$.
Thus the generated $v$ and top readouts together with the Higgs readout provide the
tree-level physical conditions of this common action.

For a finite-order realization we impose these generated values as
$(G_F,M_t,M_H)$ with $v_F=(\sqrt2G_F)^{-1/2}$. No measured
mass enters these conditions. Use the tadpole-free Landau-gauge
$\overline{\rm MS}$ convention of
Refs.~\cite{MartinRobertson2019,Martin2014Goldstone} at a common
scale $Q$. Let $\ell=1/(16\pi^2)$ and
$\delta r=\ell\Delta\widetilde r^{(1)}$ and
$\delta_t=\operatorname{Re}(\ell\Delta s_t^{(1)})$ with
$\delta_H=\operatorname{Re}(\ell\Delta s_H^{(1)})$ and
$T=\ell\Delta_1$ denote the corrections in the order displayed. These are the finite Fermi
and top-pole corrections followed by the Higgs-pole and stationary-potential corrections.
The pole kernels use the leading action coefficients with
$y_b=0$ and $|V_{tb}|=1$ while neglecting the light Yukawas.
They are given by corrected Eq.~(4.3) of Ref.~\cite{MartinRobertson2019}
and Eq.~(2.46) of Ref.~\cite{MartinRobertson2014Higgs}
together with Eqs.~(2.2) and~(2.20) of Ref.~\cite{Martin2016TopPole}.
The loop integrals have the convention
\[
 A(x)=x\bigl(\log(x/Q^2)-1\bigr),\qquad
 B(s;x,y)=-\int_0^1
 \log\frac{ux+(1-u)y-u(1-u)s-i0}{Q^2}\,du .
\]
This branch gives $\Gamma=-\operatorname{Im}\Delta s/M$
for a decaying pole.

\begin{proposition}[Finite matching of the generated physical conditions]
\label{prop:p5-generated-onshell-matching}
Put $r_0=v_F^2>0$ and $y_0=\sqrt2M_t/v_F>0$ and
$\lambda_0=M_H^2/(2r_0)$. Through one loop the solution for
$r=v_L^2$ with the two running couplings and the potential coefficient is
\begin{align}
 r&=r_0(1+\delta r),&
 v_L&=v_F(1+\delta r/2),\notag\\
 \overline y_t&=y_0-\frac{y_0}{2}\delta r
                       -\frac{\delta_t}{y_0r_0},&
 \overline\lambda&=\lambda_0-\lambda_0\delta r
                       -\frac{\delta_H}{2r_0},\notag\\
 \overline m^2&=-\frac{M_H^2}{2}+\frac{\delta_H}{2}-T .
 \label{eq:p5-generated-onshell-inverse}
\end{align}
Every correction on the right is evaluated at the leading
coefficients. The equalities have remainders $O(\ell^2)$.
\end{proposition}
\begin{proof}
The four renormalization conditions are
\[
 r-r_0-r_0\delta r=0,\quad
 \frac{y_t^2r}{2}+\delta_t-M_t^2=0,\quad
 2\lambda r+\delta_H-M_H^2=0,\quad
 m^2+\lambda r+T=0 .
\]
Their leading Jacobian with respect to $(r,y_t,\lambda,m^2)$
has determinant $2y_0r_0^2>0$. Expanding these conditions to first
order gives \eqref{eq:p5-generated-onshell-inverse} and establishes
local uniqueness order by order. In particular $v_L$ and $v_L^2$
are expanded separately at this order. The value of $G_F$ remains the imposed
physical condition.
\end{proof}

At the generated scale $Q=M_Z=91.12484\ {\rm GeV}$ the inputs are
$v_F=246.18970\ {\rm GeV}$ and $M_t=174.08240\ {\rm GeV}$ together with
$M_H=124.98344\ {\rm GeV}$ and
$G_F=1.16666255\times10^{-5}\ {\rm GeV}^{-2}$.
The one-loop realization gives
\begin{gather*}
 \delta r=0.02008364,\qquad v_L=248.66189\ {\rm GeV},\\
 \overline y_t=0.98609417,\qquad
 \overline\lambda=0.14526746,\qquad
 \overline m^2=-8585.71890\ {\rm GeV}^2 .
\end{gather*}
These are running coefficients matched to the stated generated
physical conditions. The gauge couplings in the kernels are the
leading common-scale coefficients. A finite conversion of the
gauge inputs first changes these one-loop kernels at the next
order. The strong input is the five-flavour coupling. Its threshold conversion is required at higher orders.
The finite matching is distinct from the geometric mass comparison
in Table~\ref{tab:params} where the mass definitions are specified.

\paragraph{Top width from the matched action.}
Use the leading action coefficients of
Proposition~\ref{prop:p5-generated-onshell-matching} with $m_b=0$
and $|V_{tb}|=1$. Set $M_{W,0}=g_2v_F/2$ and
$x=M_{W,0}^2/M_t^2<1$. The absorptive part of the one-loop top
pole gives the leading decay width
\begin{equation}
 \Gamma_t^{(0)}
 =-\frac{\operatorname{Im}[\ell\Delta s_t^{(1)}]}{M_t}
 =\frac{G_FM_t^3}{8\pi\sqrt2}(1-x)^2(1+2x).
 \label{eq:p5-top-width-test}
\end{equation}
The width is the rest-frame decay rate per unit proper time in
natural units. It supplies an additional observable once the real
matching conditions have fixed the action coefficients.
With $t=M_t^2$ and $W=M_{W,0}^2$ the standard absorptive cut of the
one-loop pole kernel
\cite[Eq.~(2.20)]{Martin2016TopPole} gives
$\operatorname{Im}B(t;0,W)=\pi(1-x)$. The remaining terms are real at
this order. Hence
\[
 \operatorname{Im}[\ell\Delta s_t^{(1)}]
 =-\frac{g_2^2t}{64\pi x}(1-x)^2(1+2x).
\]
Using $W=g_2^2v_F^2/4$ and
$G_F=1/(\sqrt2v_F^2)$ gives
\eqref{eq:p5-top-width-test}.

At the generated matching point $g_2=0.64821177$ and
$M_{W,0}=79.79153\ {\rm GeV}$. Equation~\eqref{eq:p5-top-width-test}
gives $\Gamma_t^{(0)}=1.53445\ {\rm GeV}$ and
$\operatorname{Im}[\ell\Delta s_t^{(1)}]=-267.12152\ {\rm GeV}^2$
in agreement with the complex pole calculation~\cite{Guo2026GenesisFramework}.
The one-loop absorptive self-energy corresponds to the Born
$t\to bW$ decay rate. Relative QCD corrections of order $\alpha_s$
and electroweak radiative corrections enter through the two-loop
absorptive pole and its consistent parameter conversion
\cite{Martin2016TopPole,PDG2024}. The leading action mass $M_{W,0}$
is used throughout this comparison.

The construction also applies in a source neighbourhood. Let
$\Phi(z,\theta)=0$ denote the four conditions where $\theta$
varies the same generated modes and Gram matrices together with the vertices.
On a fixed smooth threshold branch suppose the inverse Jacobian
is uniformly bounded and the first-order solution has residual
$\|\Phi(z^{(1)}(\cdot),\cdot)\|_{C^1(U)}=O(\ell^2)$.
Implicit differentiation then gives
$D_\theta z=-(D_z\Phi)^{-1}D_\theta\Phi+O(\ell^2)$.
The $C^1$ residual controls both the value and its source derivative.
The source variation includes the direct action term in
\eqref{eq:p5-complete-quadratic-source}. Neutral-mode elimination
also retains its induced source and direct Schur contribution.
These trace formulas use a common fixed trace space and even
sources. Fermionic two-point projections use ordered graded
left and right derivatives. For BRST-preserving source
deformations and simple isolated complex poles the Nielsen
identities give the corresponding gauge-independent pole
conditions~\cite{GambinoGrassi2000}.

For an explicitly defined heavy-quark $\overline{\rm MS}$ boundary
the one-loop QCD part is
\[
 M_Q=\overline m_Q(\mu)
 \left[1+\frac{\alpha_s(\mu)C_F}{4\pi}
 \left(4+3\log\frac{\mu^2}{\overline m_Q(\mu)^2}\right)\right]
 +O(\alpha_s^2),\qquad C_F=\frac43 .
\]
The formula converts that running boundary to its perturbative
pole definition~\cite{PDG2024}. The ladder mass itself retains
the geometric definition in Section~\ref{sec:flavour}.

With $n_s^{\mathrm{tilt}}\equiv1-n_s$ the bottom absolute mass is
the geometric mean of the strange and top masses dressed by the
window evolution
\begin{equation}
\frac{y_b}{y_t} = e^{-(2\alpha_{\mathrm{dn}} - n_s^{\mathrm{tilt}} (\kL_{\mathrm{CMB}} + 2\tau))} ,
\end{equation}
so $m_b^2 = m_s m_t e^{n_s^{\mathrm{tilt}} (\kL_{\mathrm{CMB}} + 2\tau)}$.

\subsection{Charged-fermion masses}\label{sec:derivmass}

The electron exponent is the cascade-content number
\begin{equation}
N_e=(d+1)\bigl(\sum Y^2\Delta_f\bigr)
=4\Bigl(\frac{10}{3}\frac32\Bigr)=20.
\end{equation}
The cascade uses $d+1=4$ levels, each with the one-generation weight
$\sum Y^2\Delta_f=5$. Its amplitude rule assigns the scalar generator
$5\kL I$ to each level. Thus their composition is
$(e^{-5\kL})^4=e^{-N_e\kL}$. The charge trace fixes the number of
content units. The equal rate $\kL$ per unit fixes the transfer used by
the electron mass vertex. This source-to-amplitude rule is the electron
entry of Table~\ref{tab:p5-hierarchy-selection}.

The electron mass is the Planck-anchored exponential chain with the low-scale squash correction
\begin{equation}
m_e = \MP \, e^{-20\kL} (1 - s_0 \kappa) ,
\end{equation}
where $s_0\kappa$ is the squash correction of Section~\ref{sec:derivg2}. This relation fixes the electron anchor directly. The muon-to-electron ratio carries the Euclidean-period factor
\begin{equation}
\frac{m_\mu}{m_e} = e^{2\alpha_{\mathrm{lp}} + \sqrt{2\pi}} ,
\end{equation}
the lepton ladder index plus the dimensionless entropy-minimum period $\sqrt{2\pi}$. The next lepton step is $m_\tau/m_\mu=e^{2\alpha_{\mathrm{lp}}}$.

\subsection{Neutrino masses and mixing angles}\label{sec:derivnu}

The Weinberg dimension-five operator gives a mass proportional to $v^2/\Lambda$~\cite{Weinberg1979}. The present seesaw relation fixes the family scale and its coefficient so that
\begin{equation}
m_{\nu_3} = \frac{v^2 (2\pi)^2}{k_{\mathrm{GUT}}} (1 + s_0 \kappa) = 0.0502 \ \mathrm{eV} ,
\label{eq:p5-neutrino-scale}
\end{equation}
with the squash correction of Section~\ref{sec:derivg2}. In the common
action this relation fixes the Majorana coefficient of mass dimension $-1$
to $(2\pi)^2(1+s_0\kappa)/k_{\mathrm{GUT}}$. The Euclidean-period factor
therefore enters the normalization of the mass-generating vertex. In
the convention $m_\nu=C_Wv^2/\Lambda_W$ the same boundary relation is
expressed by $C_W=1$ and $\Lambda_W=k_{\mathrm{GUT}}/(2\pi)^2$ before the
explicit squash insertion. This fixes the absolute neutrino scale. The
following hypercharge-trace ratios fix the relative eigenvalues
\begin{equation}
\frac{m_{\nu_1}}{m_{\nu_2}} = \frac{1}{\Tr(Y^2)} = \frac{3}{10} , \qquad
\frac{m_{\nu_2}}{m_{\nu_3}} = \frac{1}{\sqrt{3}\Tr(Y^2)} = \frac{3}{10\sqrt{3}} ,
\label{eq:p5-neutrino-ratios}
\end{equation}
With $r_{12}=3/10$ and $r_{23}=3/(10\sqrt3)$ the ordered absolute-mass spectrum is
\begin{equation}
(m_{\nu_1},m_{\nu_2},m_{\nu_3})
=m_{\nu_3}(r_{12}r_{23},r_{23},1).
\label{eq:p5-neutrino-rest-spectrum}
\end{equation}
In the per-generation content normalization
$r_{12}=(N_L-N_R)/\sum Y^2$ with $N_L-N_R=1$.
The lighter masses are $m_{\nu_2}=0.0087\ \mathrm{eV}$ and
$m_{\nu_1}=0.0026\ \mathrm{eV}$ at the displayed precision.
The same quadratic hypercharge moment that normalises the gauge sector
sets the two hierarchy ratios. The angular readouts specify the mixing
texture in this ordered mass basis.
At fixed hierarchy ratios the common Majorana coefficient moves all
three masses together and their squared differences scale quadratically.
The summed mass and the atmospheric splitting thus test the same
boundary normalization through different observables.

The absolute-mass texture gives the raw rest-mass split
\begin{equation}
\Delta m^2_{21,\mathrm{raw}} = m_{\nu_2}^2 - m_{\nu_1}^2 = 6.874 \times 10^{-5}\ \mathrm{eV}^2 .
\label{eq:p5-neutrino-raw-splitting}
\end{equation}
The NuFIT~5.2 normal-ordering reference is
$7.41\times10^{-5}\ \mathrm{eV}^2$~\cite{NuFIT2020,NuFIT52}.
The unrounded rest-mass result is $6.874155305644093\times10^{-5}\ \mathrm{eV}^2$
with signed residual $-7.231\%$. In standard vacuum propagation the
oscillation phases use these same mass-squared differences. See
Eqs.~(14.39)--(14.40) in Section~14.4 of the review ``Neutrino Masses and Mixing and Oscillations'' in Ref.~\cite{PDG2024}.

The solar-channel model couples an environmental scalar source to the
second mass eigenspace. Its spectral projector selects this direction in
the generated mass operator and the window-tilt factor sets the source
strength. In solar matter the scalar insertion and electron
forward-scattering potential jointly determine the mass-state
populations leaving the source region. After loss of coherence over the
long propagation baseline detection reads these populations through
their electron-flavour projections
\cite{Wolfenstein1978,MikheyevSmirnov1986,Parke1986,GeParke2019}.
At unit source amplitude the second level is
\begin{equation}
m_{\nu_2}^{\mathrm{osc}} = m_{\nu_2}\bigl(1+n_s^{\mathrm{tilt}}\bigr),
\qquad
n_s^{\mathrm{tilt}} \equiv 1-n_s = \tau\frac{7}{4},
\label{eq:p5-neutrino-tilt-definition}
\end{equation}
The vacuum spectrum corresponds to zero environmental source. The
unit-source propagation readout gives
\begin{equation}
\begin{aligned}
\Delta m^2_{21,\mathrm{tilt}}
&= \bigl(m_{\nu_2}^{\mathrm{osc}}\bigr)^2 - m_{\nu_1}^2
= 7.412190 \times 10^{-5}\ \mathrm{eV}^2 ,
\\
\Delta m^2_{31} &= m_{\nu_3}^2 - m_{\nu_1}^2
= 2.511 \times 10^{-3}\ \mathrm{eV}^2 .
\end{aligned}
\label{eq:p5-neutrino-splittings}
\end{equation}
The tilt factor uses the $RP^3$ Weyl-window ratio of
Section~\ref{sec:derivcosmo}. We realize it as a specified scalar
source in the propagation Hamiltonian. Scalar mass insertions and
vector matter potentials have different energy and antineutrino
transformations~\cite{GeParke2019,PDG2024}. Keeping the vacuum
eigenvectors fixed the Hermitian scalar-source term is
\begin{equation}
 \Delta H(E,x)=\frac{\zeta(x)}{2E}
 m_{\nu_2}^2\bigl[(1+n_s^{\rm tilt})^2-1\bigr]P_2,
 \qquad P_2=U_{\cdot2}U_{\cdot2}^\dagger .
 \label{eq:p5-solar-self-energy}
\end{equation}
At $\zeta=1$ it implements the specified squared-mass shift exactly. At $\zeta=0$ the rest-mass Hamiltonian is recovered. As a scalar
self-energy it transforms as $\Delta\overline H=\Delta H^*$ for
antineutrinos. Its $E^{-1}$ dependence differs from an energy-independent
vector forward-scattering potential. An environmental interpretation
thus specifies the profile $\zeta(x)$ and its coupling together with its vanishing
vacuum limit. Equation~\eqref{eq:p5-solar-self-energy} identifies the
operator tested by the scalar-source model. The unit-amplitude source
comparison has residual $+0.030\%$ and is classified as Tier~D.
Its comparison with a fitted solar oscillation parameter uses the
forward propagation and production average in Appendix~\ref{app:flavour}.
Propagation with specified profiles retains the labelled rest spectrum
\cite{Guo2026GenesisFramework}.
The absolute masses and their sum together with the lightest mass used in the
cosmological floor use the rest-mass spectrum above.

\paragraph{Correlated response of the three neutrino splittings.}
Let $K_\nu=UD^2U^\dagger$ with the distinct rest masses above and
$t=n_s^{\rm tilt}$ and $d_\nu=m_{\nu_2}^2(2t+t^2)$.
The generated second-state projector is the spectral polynomial
\begin{equation}
 P_2=\frac{(K_\nu-m_{\nu_1}^2I)(K_\nu-m_{\nu_3}^2I)}
 {(m_{\nu_2}^2-m_{\nu_1}^2)(m_{\nu_2}^2-m_{\nu_3}^2)} .
 \label{eq:p5-solar-spectral-projector}
\end{equation}
For the scalar-source numerator $K_\nu+\zeta d_\nu P_2$
\begin{equation}
\begin{aligned}
 \Delta_{21}(\zeta)&=\Delta_{21}(0)+\zeta d_\nu,\qquad
 \Delta_{31}(\zeta)=\Delta_{31}(0),\qquad
 \Delta_{32}(\zeta)&=\Delta_{32}(0)-\zeta d_\nu .
\end{aligned}
\label{eq:p5-correlated-solar-response}
\end{equation}
At $t=7/200$ the common absolute mass scale cancels
\begin{equation}
 \frac{d_\nu}{\Delta_{21}(0)}=\frac{407}{5200},
 \qquad
 \frac{\Delta_{21}(1)}{\Delta_{21}(0)}=\frac{5607}{5200}.
 \label{eq:p5-solar-scale-cancellation}
\end{equation}
Spectral interpolation gives \eqref{eq:p5-solar-spectral-projector}. It acts as one on the second eigenvector and zero on the other two.
Thus the source shifts only its squared eigenvalue. The exact mass
ratios give $\Delta_{21}(0)=(1-9/100)m_{\nu_2}^2$ and
$2t+t^2=2849/40000$. These yield the stated rational ratios.

In the absence of a noncommuting matter potential a varying profile
has the exact propagator
\[
 S(L)=U\,\operatorname{diag}
 \left(e^{-im_{\nu_1}^2L/(2E)},
 e^{-i(m_{\nu_2}^2L+d_\nu\Xi)/(2E)},
 e^{-im_{\nu_3}^2L/(2E)}\right)U^\dagger,\qquad
 \Xi=\int_0^L\zeta(x)\,dx .
\]
For two profiles Duhamel's identity gives the transition-probability
bound $|P-\widetilde P|\leq
d_\nu\|\zeta-\widetilde\zeta\|_{L^1}/(2E)$.
Subtracting the unobservable identity part of $P_2$ leaves an operator
of norm $1/2$. Apply the unitary propagator bound to this operator. The algebraic splitting identities concern the
scalar numerator. Appendix~\ref{sec:neutrino} includes the electron
matter potential in the propagation Hamiltonian.

The PMNS texture specifies the orientation of the neutrino mass
eigenspaces relative to the charged-lepton frame in the common
left-handed kinetic space. The mixing probabilities are the squared
overlaps of these eigenspaces as in
\eqref{eq:p5-common-flavour-invariants}. The solar-angle relation
ties the light-pair rotation to its mass ratio. The reactor angle
weights the electron-flavour projection onto the third state by the
Euclidean-period and internal-geometry factors. The atmospheric angle
starts at equal muon and tau weights and receives a positive
weak-isospin contribution
\begin{equation}
\begin{aligned}
\sin^2\theta_{12} &= \frac{m_{\nu_1}}{m_{\nu_2}} = \frac{3}{10} = 0.30 ,\\
\sin^2\theta_{13} &= \Bigl(\frac{1}{2\pi}\Bigr)^2 \frac{\sqrt{3}}{2} = 0.02194 ,\\
\sin^2\theta_{23} &= \frac12 + \frac{\Tr(T_3^2)}{(2\pi)^2} = 0.5507 .
\end{aligned}
\label{eq:p5-pmns-angles}
\end{equation}
Here $\Tr(T_3^2)=2$. These relations define the near-tribimaximal
angular texture used in the comparison. The positive isospin term
places $\theta_{23}$ in the upper octant.

\subsection{The electroweak order parameter}\label{sec:derivewsb}

\paragraph{The inherited complete condensate.}
\label{sec:p5-complete-condensate}
The ``Renormalized conformal spectral response'' theorem of
Ref.~\cite{Guo2026CompactInternalSpaceI} supplies the complete conformal tower with
its feedback and internal subtraction
\begin{equation}
\begin{gathered}
 a_j=\frac{(2j+1)^2-1/4}{L^2},\quad d_j=(2j+1)^2,\quad
 a_j\longmapsto a_j+gE^2,\quad g>0,\\
 \mu_E=\frac83(R-R_{\rm c}),\qquad R=\frac6{L^2}.
\end{gathered}
\label{eq:p5-complete-native-potential}
\end{equation}
The constant scalar shell is included. This unweighted relative
response and the Gaussian TT trace have different spectral weights
and source insertions. We apply the existence and stability theorems together with the gap
theorem for the complete conformal response in the convention $Z_E=1$.

In the dimensionless radius convention of Section~\ref{sec:closure}
$L(k)=c\MG/k$ and $L_c=\sqrt\pi$. The certified
$c>2.491>\sqrt\pi$ gives $L(k)>L_c$ for every $0<k\leq\MG$
and $\mu_E(\MG)=16(c^{-2}-\pi^{-1})<-2.50$.
Hence the complete condensate and positive scalar gap exist on
this stationary branch for every $g>0$. The result concerns stationary
internal spectral states. Their relaxation requires a time evolution.
The fixed length unit $\ell_*=1/\MG$ recovers $L_{\mathrm{phys}}=c/\MG$
at emergence. Dimensionful masses require the matching readout.

For explicit evaluation put $\zeta=gL^2$ and
$\gamma_j=(2j+1)^2-1/4$. The proposition ``Stationary equation for the
conformal tower'' of Ref.~\cite{Guo2026CompactInternalSpaceI} gives the function
\begin{equation}
 \mathcal H(z)=\sum_{j\geq0}
       \frac{d_jz}{\gamma_j(\gamma_j+z)},\qquad z\geq0.
\label{eq:p5-inherited-stationary-function}
\end{equation}
The cited proposition evaluates this sum and proves strict monotonicity.
The stationary amplitude, radial mass and local quartic coefficient are
\begin{equation}
\begin{gathered}
 \mathcal H(z_*)=-\frac{\mu_E}{2\zeta},\qquad
 E_*^2=\frac{z_*}{\zeta},\qquad
 m_\delta^2=\frac{4\zeta z_*\mathcal H'(z_*)}{Z_E},\\
 \lambda_{E,\rm spec}=\frac{\pi(\pi+2)}4\zeta^2,\qquad
 \mathcal M_{\rm sc}^2=\Delta_0+m_\delta^2 I.
\end{gathered}
\label{eq:p5-complete-condensate-readout}
\end{equation}
In particular the full scalar tower is
$m_{l,\rm sc}^2=l(l+2)/L^2+m_\delta^2$ for $l=0,2,\ldots$.
The constant mode gives its lowest mass followed by the $l=2$
nonconstant shell. The inherited strict mass bound is
\begin{equation}
 0<m_\delta^2<-\frac{2\mu_E}{Z_E}.
\label{eq:p5-complete-mass-correction}
\end{equation}
The feedback coefficient $g$ is defined in the $E$ convention of the
complete conformal response. The electroweak VEV is determined by
the separate window-squared relation below.

\begin{proposition}[Joint amplitude and radius equilibrium]
\label{prop:p5-joint-native-radius}
Retain the complete relative potential in fixed response units with its
subtraction and fixed $g,L_c>0$ and vary both $y=E^2\geq0$ and $L>0$
\[
 F_L(y)=8(L^{-2}-L_c^{-2})y+I(gL^2y),\qquad
 I(u)=\sum_j\beta_j\left[\frac{u}{\gamma_j}
                         -\log\left(1+\frac{u}{\gamma_j}\right)\right].
\]
Assume $\beta_j\geq0$ and $\gamma_j>0$ together with
$\sum_j\beta_j/\gamma_j^2<\infty$ and
$\sum_j\beta_j/\gamma_j>2/(gL_c^4)$. Then the unique joint global
minimum is
\begin{equation}
 L_*=\sqrt2L_c,\qquad
 I'(u_*)=\frac{2}{gL_c^4},\qquad
 y_*=\frac{u_*}{2gL_c^2}>0 .
 \label{eq:p5-joint-native-radius}
\end{equation}
Its Hessian in $(y,\eta)$ with $L\mapsto Le^{-\eta}$ is positive
definite. After relaxing the amplitude its compression curvature is
$64y_*/L_*^2>0$. The unweighted conformal tower satisfies the hypotheses.
\end{proposition}
\begin{proof}
Set $z=L^{-2}$ and $u=gL^2y$. Direct completion of the square gives
\begin{equation}
 F_L(y)=I(u)-\frac{2u}{gL_c^4}
             +\frac{8u}{g}\left(z-\frac1{2L_c^2}\right)^2 .
 \label{eq:p5-joint-native-square}
\end{equation}
The summability assumption gives $I'(0)=0$ and
$I''(u)=\sum_j\beta_j/(\gamma_j+u)^2>0$ and
$I'(\infty)=\sum_j\beta_j/\gamma_j$. Thus the first two terms have
a unique positive minimizer with negative value. The $y=0$ boundary
has value zero. At positive $u$ the square vanishes only at the
stated radius. At the joint minimum the Hessian in $(u,z)$ is
$\operatorname{diag}(I''(u_*),16u_*/g)>0$. The change from $(y,\eta)$
is nonsingular. Eliminating the amplitude variation gives the stated
compression curvature.
\end{proof}

The joint minimum determines the static collective response of the
inherited potential. The amplitude deformation
$A+gyI$ and the radius deformation of $A$ are varied separately.
On the amplitude-stationary branch at general $L>L_c$ the same
derivatives give
\begin{equation}
 \left.\partial_\eta F_{Le^{-\eta}}(y_L)\right|_{\eta=0}
       =16y_L(2L^{-2}-L_c^{-2}).
 \label{eq:p5-native-radius-force}
\end{equation}
These are the vacuum equilibrium and source force of the joint
relative potential in its stated units. The finite TT endpoint
$c_R$ continues to be determined by its own sourced trace.
In a common matter--geometry action the radius force enters the
total source balance of
\eqref{eq:p5-native-common-source-balance} with the chosen
physical density normalization.

\paragraph{The squash coordinate.}
The isometry breaking is driven by the $J = 2$ squash mode with order parameter $\varphi$ of the internal geometry. Its dynamics is the Landau potential on the curvature axis
\begin{equation}
V(\varphi) = \frac12 \xi (R - R_{\mathrm{c}}) \varphi^2 + \frac{\lambda}{4} \varphi^4 ,
\end{equation}
with $\xi = 1/8$ the conformal coupling and $R_{\mathrm{c}} = 6/\pi$ the critical curvature. The effective mass squared is $m^2 = \xi(R - R_{\mathrm{c}})$. It is tachyonic for $R < R_{\mathrm{c}}$ and hence the curvature crossing selects the condensed phase of this potential. The quartic coupling is fixed by the stationarity of the potential at the squash vacuum $\varphi_0 = 2\tau$
\begin{equation}
\frac{\partial V}{\partial \varphi} = 0 \quad \Rightarrow \quad \lambda = \frac{\xi (R_{\mathrm{c}} - R_{\mathrm{GUT}})}{(2\tau)^2} = 149 ,
\end{equation}
with $R_{\mathrm{GUT}}=6/L_{\mathrm{GUT}}^2$ and $L_{\mathrm{GUT}}=\sqrt3/\tau$ the isometry-breaking scale. The adopted parity-preserving torsion response is written as
\begin{equation}
L_{\mathrm{torsion}} = a\, T^2 + b\, T^{bac} T_{abc} + c\, (T^a{}_{ab})^2 .
\end{equation}
The torsion convention used here imposes $b=4a$ and $c=-(a+b/3)=-(7/3)a$ with the emergence-scale normalisation $a=M_{\mathrm G}^3/4$. Stationarity at $\varphi_0=2\tau$ then fixes the displayed squash quartic. These relations specify the torsion response used in this calculation. The electroweak VEV uses the separate window-squared relation
\begin{equation}
v = M_{\mathrm{G}} \frac{3\alpha}{\pi} e^{-4\pi \kL} (1 - s_0 \kappa) ,
\end{equation}
with the squash correction $s_0\kappa$ of Section~\ref{sec:derivg2}.
The prefactor is the product of the $J=2$ squash factor $3\alpha/2$
with $\alpha=1/(16\pi^2)$ and the Fourier factor $1/\pi$ together with the
double-crossing factor $2$. The electroweak amplitude is transported
through creation and stabilization with the same generator
$2\pi\kL I$ at each crossing. Their composition gives
$e^{-4\pi\kL}$ and the hierarchy $v/M_{\mathrm G}\sim10^{-16}$.
Equal crossing weights and their action on the amplitude specify this
readout. Equation~\eqref{eq:p5-common-bandwidth-response} relates it
to the electron cascade. A Landau--Zener realization is specified by
a source trajectory satisfying the amplitude-matching condition
\eqref{eq:p5-lz-amplitude-matching}.

\paragraph{Coordinate consistency of the two Landau readouts.}
\label{sec:p5-condensate-coordinate}
Write $\lambda_\varphi=\lambda$ and $\mu_\varphi=(R-R_{\rm c})/8$.
The amplitude-coordinate rule in Ref.~\cite{Guo2026CompactInternalSpaceI}
transforms the entire response. If $E=a_{E\varphi}\varphi$ describes the same
response and feedback without an overall rescaling matching the
two drives and quartic Taylor coefficients requires
\begin{equation}
 a_{E\varphi}^2=\frac3{64},\qquad
 \lambda_E=\frac{4096}{9}\lambda_\varphi,\qquad
 Z_E=\frac{64}{3}Z_\varphi,\qquad
 g_E=\frac{64}{3}g_\varphi.
\label{eq:p5-condensate-coordinate-check}
\end{equation}
The normalized radial mass is invariant and $Z_E=1$ corresponds
to $Z_\varphi=3/64$. The amplitude-normalisation calculation in Ref.~\cite{Guo2026CompactInternalSpaceI}
gives the stationary-coordinate and kinetic matching. The squash
potential uses its coefficient $\lambda=149.145$ while the full
spectral potential uses $\lambda_{E,\rm spec}$ in the normalization
above. The electroweak readout is the window-defined VEV $v$.

\subsection{The Higgs quartic}\label{sec:derivlamh}

The Higgs matching relation assigns a quartic restoring response to the
canonically normalized electroweak amplitude. Its pseudo-dilaton
parameterization combines a strong-sector coefficient with an additive
contribution from the retained generations
\begin{equation}
\lambda_H = \frac{\lambda_{\mathrm{dil}} + \sigma_{\mathrm{SM}}}{32\pi^2}
= \frac{12\pi + 3}{32\pi^2}
= 0.129 ,
\end{equation}
The strong-sector matching assigns $4\pi$ per retained generation.
Thus $\lambda_{\mathrm{dil}}=N_{\mathrm{gen}}\,4\pi=12\pi$. The additive content term is $\sigma_{\mathrm{SM}}=N_{\mathrm{gen}}=3$.
The integer $3$ is fixed by the common copy count while the
per-copy strength $4\pi$ and the factor $1/(32\pi^2)$ specify the
effective matching and quartic normalization. These data define
$\lambda_H$ before the finite electroweak matching calculation.
The strong-sector coefficient expresses the contribution assigned to
the collective scalar channel. A related use of a collective scalar
in mass generation occurs in the Nambu--Jona-Lasinio gap equation
\cite[Sec.~III]{NambuJonaLasinio1961}. The effective matching here is
fixed by the coefficients displayed above.
Combining this quartic with the generated electroweak scale gives
the tree-level Higgs mass
\begin{equation}
m_H = \sqrt{2\lambda_H}\, v = 124.98 \ \mathrm{GeV} ,
\end{equation}
with a central-value deviation from the reference $125.2$ GeV of
$-0.17\%$. This mass supplies $M_H$ in
Proposition~\ref{prop:p5-generated-onshell-matching}. The one-loop
calculation determines the corresponding running quartic and
potential coefficient together with the vacuum parameter in the stated scheme.
The mass comparison therefore tests the generated tree-level
relation and the finite matching establishes its quantum
implementation through the calculated order.
The quartic $\lambda_H$ is distinct from the squash coefficient
$\lambda=149$ of Section~\ref{sec:derivewsb}.

\subsection{CKM mixing and CP}\label{sec:derivckm}

The Gatto--Sartori--Tonin relation motivates connecting the Cabibbo angle to a light-quark mass ratio~\cite{GattoSartoriTonin1968}. Here the corresponding input is assigned through the right-singlet hypercharge ratio $|Y_{d_R}|/|Y_{u_R}|=1/2$
\begin{equation}
\mathfrak r_G = e^{-2\alpha_{\mathrm{dn}}} \Bigl(1 + \frac{|Y_{d_R}|}{|Y_{u_R}|}\Bigr)^2 = e^{-2\alpha_{\mathrm{dn}}} \Bigl(\frac{3}{2}\Bigr)^2 = 0.05015 ,
\label{eq:p5-gatto-input}
\end{equation}
and the Cabibbo element uses its square root
\begin{equation}
|V_{us}| = \sqrt{\mathfrak r_G} = 0.2239 .
\label{eq:p5-ckm-cabibbo}
\end{equation}
The Cabibbo overlap uses the down-sector transfer amplitude with
the right-singlet charge factor $3/2$. The absolute-mass ratio
$m_d/m_s=e^{-2\alpha_{\mathrm{sd}}}$ of Section~\ref{sec:derivlz}
reads the first-generation diagonal transfer with step
$\alpha_{\mathrm{sd}}=\alpha_{\mathrm{dn}}-\kL_{\mathrm{CMB}}/6$.
These angular and diagonal source operations give values differing
by $1.6\%$ at the endpoint. Their respective roles are fixed by the
relative frames and singular values of the common mass vertices.
The remaining two quark-mixing assignments are
\begin{equation}
|V_{cb}| = \frac{1}{2} e^{-\alpha_{\mathrm{up}}} (1 - s_0 \kappa) , \qquad
|V_{ub}| = \frac{1}{2} e^{-2\alpha_{\mathrm{up}}} (1 + \tau \kappa) ,
\label{eq:p5-ckm-offdiagonal}
\end{equation}
the adjacent-generation element carrying the squash correction $s_0 \kappa$ and the cross-generation element the chiral correction $\tau \kappa$. The three-element product is numerically close to the weak cube
\begin{equation}
|V_{us}| |V_{cb}| |V_{ub}| \approx \alpha_W(v)^3 ,
\end{equation}
with $\alpha_W(v)=g_2(v)^2/(4\pi)$ the electroweak-scale weak coupling. The window-capacity coupling in Section~\ref{sec:g2} has value $2/(\kL)^5$. The product is $2.45\%$ below $\alpha_W(v)^3$ at the reported endpoint. This numerical comparison is not used to fix the ladder.

The angular source also fixes a complex relative phase. Together
with the intergenerational overlaps this phase determines the
basis-invariant CP-odd response
\eqref{eq:p5-common-flavour-invariants}~\cite{Jarlskog1985}.
The phase assignment uses the left-right content ratio
\begin{equation}
\frac{N_L}{N_R} = \frac{8}{7} ,
\end{equation}
eight left-handed Weyl components (four weak doublets) against seven right-handed singlet components per generation. The lepton phase assignment carries the undiluted ratio $\delta_{\mathrm{PMNS}}\sim(8/7)\pi$ and the quark phase is diluted by the colour number
\begin{equation}
\delta_{\mathrm{CKM}} = \frac{8}{7} \frac{\pi}{N_c} = \frac{8\pi}{21} = 68.57^{\circ} .
\label{eq:p5-ckm-phase}
\end{equation}
The corresponding leading-order Jarlskog invariant~\cite{Jarlskog1985} is
\begin{equation}
J = |V_{us}| |V_{cb}| |V_{ub}| \, c_{12} c_{23} \sin\delta = 3.15 \times 10^{-5} ,
\label{eq:p5-ckm-jarlskog-leading}
\end{equation}
where $c_{12}=\sqrt{1-|V_{us}|^2}$ and $c_{23}=\sqrt{1-|V_{cb}|^2}$. Corrections from $c_{13}$ enter at relative order $|V_{ub}|^2$. The baryon-asymmetry ansatz combines this CP measure with the weak coupling and a content normalisation
\begin{equation}
\eta_B^{\mathrm{raw}} = \frac{J \alpha_W(v)^2}{56} = 6.09 \times 10^{-10} ,
\label{eq:p5-baryon-raw}
\end{equation}
where $1/56=1/(N_LN_R)=\xi/N_R$ combines $\xi=1/8$ with $N_R=7$.
The factor $J$ carries the sign of the CP asymmetry and the
suppression from flavour mixing. The weak and content factors
convert this response into the raw baryon-number readout. Pairing
it with the photon-temperature normalization gives the baryon
density used in the cosmological calculation.
Sakharov's conditions require baryon-number violation together with C and CP violation
and departure from equilibrium~\cite{Sakharov1967}. The factor $1/56$
and the two powers of $\alpha_W(v)$ specify the weak and nonequilibrium
weights of the present ansatz. A thermal-rate realization also specifies
the baryon-violating processes and their kinetic evolution.
Section~\ref{sec:flat} uses this raw asymmetry with the raw photon-temperature
readout.

\subsection{Cosmology}\label{sec:derivcosmo}

The cosmological readouts share one lightest-neutrino rest-mass floor. On the branch $0\leq x=s_0\kappa(s_0)<1/4$ the density readout is
\begin{equation}
\rho_\Lambda = Y_u \, m_{\nu_1}^4 \, (1 - 4 s_0 \kappa) = \frac{2}{3} m_{\nu_1}^4 (1 - 4 s_0 \kappa) ,
\end{equation}
where $Y_u = 2/3$ is the up-quark hypercharge. The neutrino masses use the dimension-five operator with the coefficient specified in Section~\ref{sec:derivnu}
\begin{equation}
m_{\nu_3} = \frac{v^2 (2\pi)^2}{k_{\mathrm{GUT}}} (1 + s_0 \kappa), \qquad
m_{\nu_1} = m_{\nu_3} \cdot \frac{m_1}{m_2} \cdot \frac{m_2}{m_3} ,
\end{equation}
with the hierarchy ratios $r_{12}=m_1/m_2=3/10$ and
$r_{23}=m_2/m_3=1/(\sqrt3\Tr(Y^2))$. The same rest mass
$m_{\nu_1}$ enters the vacuum floor and entropy span together with the photon temperature.
All are expressed in common energy units.

\paragraph{Composition of the squash corrections.}
Put $x=s_0\kappa(s_0)$. At fixed values of the remaining inputs
define the quantities before the explicit squash corrections by
\[
 v_0=\MG(3\alpha/\pi)e^{-4\pi\kL},\quad
 m_{3,0}=(2\pi)^2v_0^2/k_{\rm GUT},\quad
 m_{1,0}=r_{12}r_{23}m_{3,0},\quad \rho_0=Y_um_{1,0}^4.
\]
Substituting the corrected electroweak scale into the seesaw and vacuum-floor rules gives
\begin{equation}
\begin{aligned}
 v/v_0&=1-x,&
 m_{\nu_3}/m_{3,0}&=(1-x)^2(1+x),\\
 \frac{v m_{\nu_3}}{v_0m_{3,0}}&=(1-x)^3(1+x),&
 \rho_\Lambda/\rho_0&=(1-x)^8(1+x)^4(1-4x).
\end{aligned}
\label{eq:p5-cosmology-composition}
\end{equation}
The last two factors are $1-2x+O(x^2)$ and $1-8x+O(x^2)$
respectively. At fixed seesaw inputs the local pair instead satisfies
\[
 (1+x)^4(1-4x)=1-10x^2-20x^3-15x^4-4x^5.
\]
These identities follow by substituting $v=v_0(1-x)$ into the stated
seesaw formula and then using the same $m_{\nu_1}$ in the fourth-power
vacuum floor. The correction first changes the electroweak amplitude,
then the neutrino mass, and finally the vacuum density.
The subscript $0$ removes only the explicit $x$ factors. The other
inputs retain their dependence on $\tau$. This dependence is included
when taking a total derivative.

The endpoint entropy span combines a Gaussian variance ratio and a content offset.
The Planck variance is $\MP^2$ and the vacuum-floor variance is
$\sqrt{\rho_\Lambda}$. Since $\rho_\Lambda=\MP^4\widehat\rho_\Lambda$
the dimensional anchor cancels. The readout is
\begin{equation}
S_{\rm end}
= \ln \frac{\MP^2 \sqrt{2\pi + r_{23}}}{\sqrt{\rho_\Lambda}}
= \ln \frac{\sqrt{2\pi + r_{23}}}{\sqrt{\widehat\rho_\Lambda}}
= 139.2537 ,
\end{equation}
with $r_{23}=3/(10\sqrt3)$. For a Gaussian of standard deviation $\sigma$ the differential entropy is $H(\sigma)=\tfrac12\ln(2\pi e\,\sigma^2)$. The endpoint span is defined as twice the entropy difference plus the single content offset $\ln\sqrt{2\pi+r_{23}}$
\begin{equation}
S_{\rm end} = 2\bigl[H(\MP) - H(\rho_\Lambda^{1/4})\bigr] + \ln\sqrt{2\pi + r_{23}}
= \ln \frac{\MP^2 \sqrt{2\pi + r_{23}}}{\sqrt{\rho_\Lambda}}
= \ln \frac{\sqrt{2\pi + r_{23}}}{\sqrt{\widehat\rho_\Lambda}} .
\end{equation}
The entropy difference uses the same energy reference unit at both
ends, so the arbitrary logarithmic unit cancels. Its composition is exact.
For positive Gaussian standard deviations $\sigma_a,\sigma_b,\sigma_c$
\[
 S(a,b):=2[H(\sigma_a)-H(\sigma_b)]
 =2\ln(\sigma_a/\sigma_b),\qquad
 S(a,c)=S(a,b)+S(b,c).
\]
Intermediate Gaussian scales therefore telescope. For a specified positive
$C^1$ variance path $V(t)$ twice the entropy difference from $a$ to $b$ is
$-\int_a^b\partial_t\log V(t)\,dt$. This representation uses the
variance-path derivative. The content offset is attached once to the
full endpoint pair. Thus $S_{\rm end}=\ln W$ with
$W=\sqrt{2\pi+r_{23}}/\sqrt{\widehat\rho_\Lambda}$ and the Hubble readout is
\begin{equation}
H_0 = \MP \sqrt{\pi}\, e^{-S_{\rm end}} .
\end{equation}
The dark-energy fraction is
\begin{equation}
\Omega_\Lambda = \frac{\rho_\Lambda}{3 H_0^2 \MP^2} = \frac{2}{3} + \frac{r_{23}}{3\pi} = 0.68504 .
\end{equation}
Indeed $e^{-2S_{\rm end}}=\rho_\Lambda/[\MP^4(2\pi+r_{23})]$ gives $H_0^2=\pi\rho_\Lambda/[\MP^2(2\pi+r_{23})]$. Dividing by the critical density cancels the common mass floor and yields $(2\pi+r_{23})/(3\pi)$.
For comparison the nonnegative scalar spectrum of the round internal
space defines a normalized spectral state.  In dimensionless variables
let $\widehat\lambda_j=4j(j+1)$ and $d_j=(2j+1)^2$ and $u>0$.  Then
\begin{equation}
 Z_{\rm sp}(u)=\sum_{j\geq0}d_je^{-u\widehat\lambda_j},\qquad
 S_{\rm sp}(u)=\log Z_{\rm sp}(u)
                    -u\partial_u\log Z_{\rm sp}(u),\qquad
 \frac{dS_{\rm sp}}{du}=-u\operatorname{Var}_u(\widehat\lambda).
 \label{eq:p5-positive-spectral-entropy}
\end{equation}
To verify the formula normalize each individual mode weight by
$Z_{\rm sp}$ and substitute its logarithm in the Shannon sum.
Quadratic growth of $\widehat\lambda_j$ justifies differentiation.
In particular $S_{\rm sp}=H(p)+\sum_jp_j\log d_j$ with
$p_j=d_je^{-u\widehat\lambda_j}/Z_{\rm sp}$.  A conformal constant
shift gives the same normalized state.  On the complete spectrum the
entropy diverges as $u\downarrow0$ and tends to zero as $u\to\infty$. On a finite carrier its small-$u$ limit is the logarithm of the carrier
dimension. These limits follow respectively from divergence of
the unweighted count and dominance of the simple ground mode.

Equation~\eqref{eq:p5-positive-spectral-entropy} describes the internal
spectral state. The endpoint entropy $S_{\rm end}$ uses a Gaussian variance
ratio and a single content factor containing $r_{23}$. Thus its
definition is independent of the internal heat parameter $u$.
For $A=p^2$ an amplitude window
$e^{-\sigma^2 A/2}$ has heat parameter $\sigma^2/2$. Squaring that
amplitude gives parameter $\sigma^2$.  These parameters have units of
length squared and require a separate energy identification before they
can be called a physical inverse temperature.

The endpoint residual is the remaining Hamiltonian fraction
\begin{equation}
\Omega_\Sigma^{\rm budget} = 1-\Omega_\Lambda-\Omega_b = 0.26554751 .
\end{equation}
Photons and the three thermal rest-mass neutrino states occupy part
of this budget. Equations~\eqref{eq:p5-thermal-neutrino-budget}
and~\eqref{eq:p5-cold-budget} determine the remaining conserved
cold fraction $\Omega_{\rm cold}=0.26403880$ for the
cold-dark-matter comparison.
The microwave temperature is the photon floor set by the lightest neutrino
\begin{equation}
T_{\mathrm{CMB}}^{\mathrm{raw}} = \frac{m_{\nu_1} r_{12}}{\pi} (1 - s_0 \kappa)(1 - \tau \Delta_s) = 2.7310\ {\rm K},
\end{equation}
with $r_{12}=3/10$ and $\Delta_s=1/2$ the scalar conformal weight.
The insertion $(1-s_0\kappa)$ multiplies the corrected rest mass. Thus
both enter the temperature response. The other factors are the
content ratio $r_{12}=(N_L-N_R)/\sum Y^2$ and the period factor $1/\pi$
and the chiral correction $1-\tau\Delta_s$. The formula is evaluated
in energy units before conversion to Kelvin.
The finite endpoint photon zero mode multiplies the raw floor by
\begin{equation}
C_\gamma = 1-\frac{\tau}{\pi^2},
\qquad
T_{\mathrm{CMB}} = C_\gamma T_{\mathrm{CMB}}^{\mathrm{raw}} = 2.72547\ {\rm K}.
\end{equation}
The corrected temperature enters the CMB-monopole and linear CMB comparisons. The baryon-density readout uses the raw photon floor and raw Sakharov asymmetry
\[
\Omega_b = \frac{\eta_B^{\mathrm{raw}} n_\gamma^{\mathrm{raw}} m_p}{\rho_{\mathrm{crit}}},
\qquad
n_\gamma^{\mathrm{raw}} = \frac{2\zeta(3)}{\pi^2}\bigl(T_{\mathrm{CMB}}^{\mathrm{raw}}\bigr)^3 ,
\]
The common raw convention fixes both photon number and baryon-to-photon ratio in this density.

\paragraph{Common-mass elimination in the cosmological readouts.}
Let $D=2\pi+r_{23}$ and $A_\gamma=C_\gamma(r_{12}/\pi)(1-\tau\Delta_s)$
and $T_E=k_{\rm B}T_{\rm CMB}$. For $0\leq x<1/4$ and positive
$m_{\nu_1},Y_u,A_\gamma$ the cosmological relations satisfy
\begin{equation}
\begin{gathered}
 \MP H_0=\sqrt{\frac{\pi Y_u}{D}}\,m_{\nu_1}^2\sqrt{1-4x},
 \qquad T_E=A_\gamma m_{\nu_1}(1-x),\\
 \Omega_\Lambda=\frac{D}{3\pi},\qquad
 \frac{T_E^2}{\MP H_0}
 =A_\gamma^2\sqrt{\frac{D}{\pi Y_u}}\,
       \frac{(1-x)^2}{\sqrt{1-4x}} .
\end{gathered}
\label{eq:p5-cosmology-shared-ratio}
\end{equation}
Exponentiating the endpoint entropy gives
$H_0^2=\pi\rho_\Lambda/(\MP^2D)$. Substitute the same fourth-power
floor and divide the squared photon readout by $\MP H_0$.
The mass and every upstream factor common to that mass cancel exactly.
The cancellation holds for mass variations induced by $v$ or the seesaw
scale or their shared uncertainties. Appendix~\ref{sec:p5-cosmology-error}
bounds the remaining dependence on $x$ and distinguishes the normalized
ratio from the full temperature--Hubble ratio.

The primordial readout assigns the variance of the generated window
to the dimensionless power of the comoving curvature perturbation
$\mathcal R$ at the comparison pivot. We use the Fourier convention
\[
 \langle\mathcal R_{\mathbf k}\mathcal R_{\mathbf k'}\rangle
 =(2\pi)^3\delta^3(\mathbf k+\mathbf k')
 \frac{2\pi^2}{k^3}\Delta_{\mathcal R}^2(k),
\]
with $\Delta_R^2=\Delta_{\mathcal R}^2(k_*)$ and
$n_s-1=d\log\Delta_{\mathcal R}^2/d\log k|_{k_*}$.
Thus the amplitude measures the variance per logarithmic wavenumber
interval and the tilt specifies its local scale dependence
\cite{Planck2018}. The internal source rule fixes this tilt from the
torsion modulus and projected $RP^3$ channel weights
\begin{equation}
1 - n_s = \tau \cdot \frac{7}{4} = 0.035 ,
\end{equation}
where the numerator sums the weights of the specified response channels and the denominator is the four-level window normalisation
\[
\frac{7}{4}=\frac{1+2+1+3}{d+1},\qquad d=3 .
\]
The same rational value follows from the conformal-curvature identity
\[
\frac{7}{4}=1+\xi R_{\rm LC}L^2=1+\frac18\,6 .
\]
Here $1,2,1,3$ are the weights in the order of the scalar and transverse-vector channels followed by the spinor-projected
and effective three-branch tensor channels. The last channel includes a
trace component and has weight $3$. The rough TT bundle has weight $2$.
The $7/4$ ratio and $\lambda_{\mathrm{TT}}=14$ specify the effective
internal tensor catalogue used in this readout. For the scalar
power the normalization assigns the spin-$\tfrac12$ Gaussian
zero-point variance to the curvature channel and multiplies it by
the window-period attenuation and chiral response
\begin{equation}
\Delta^2_R = \frac12 \Bigl(\frac{1}{2\pi}\Bigr)^2 e^{-2\pi \kL_{\mathrm{CMB}}} (1 - \tau \kappa) = 2.10 \times 10^{-9} ,
\end{equation}
with $\Delta^2_0=(1/2)(1/2\pi)^2$ the spin-$\tfrac12$ zero-point
factor and $e^{-2\pi\kL_{\mathrm{CMB}}}$ the attenuation of this
quadratic response and $(1-\tau\kappa)$ its chiral-squash correction.
The microwave-pivot window is $\kL_{\mathrm{CMB}}=\kL(1-\tau/4)$.
The tensor readout assigns the relative Euclidean-period weight to
the same scalar variance. With $\Delta_t^2$ denoting the power
summed over the two external tensor helicities its ratio to the
scalar power is
\begin{equation}
r = \Bigl(\frac{1}{2\pi}\Bigr)^2 = 0.02533 ,
\end{equation}
so the tensor amplitude is $\Delta^2_t=r\Delta^2_R=5.32\times10^{-11}$. The observational reference is the BK18 bound $r_{0.05}<0.036$ at 95\% confidence~\cite{BICEPKeck2021}. Its likelihood evaluates the tensor-to-scalar ratio at $k_\ast=0.05\ \mathrm{Mpc}^{-1}$ with $n_t=0$. Section~\ref{sec:devsummary} uses the same reference convention.

\subsection{QCD scale extraction and composite-channel masses}\label{sec:derivqcd}

The QCD running starts from $g_3(\MG)$ in Section~\ref{sec:g3}.
Section~\ref{sec:derivbeta} supplies the two-loop coefficients on
the same representation content with $y_0=1$ held fixed.

\paragraph{The QCD extraction convention.}
Write $g_i(\mu)$ for this running solution and define the internal
extraction point and coupling together with the correction by
\begin{equation}
 \begin{aligned}
 \mu_{\rm ext}&=\frac v2\sqrt{g_2(v)^2+\frac35g_1(v)^2},
 &a_{\rm raw}&=\frac{g_3(\mu_{\rm ext})^2}{4\pi},\\
 a_{\rm ext}&=a_{\rm raw}(1-\epsilon),
 &\epsilon&=\frac{2\tau\kappa}{8}.
 \end{aligned}
 \label{eq:p5-qcd-extraction-inputs}
\end{equation}
The extraction point is defined by the couplings at $v$. The
electroweak $Z$-mass comparison uses the separate fixed point of
Section~\ref{sec:ewsb}.
The five-flavour readout is
\begin{equation}
 \begin{gathered}
 \Lambda_{\rm QCD}=\mu_{\rm ext}
 \left(\frac{b_0a_{\rm ext}}{4\pi}\right)^{-p}
 e^{-q/a_{\rm ext}},\\
 b_0=\frac{23}{3},\quad b_1=\frac{116}{3},\quad
 p=\frac{b_1}{2b_0^2},\quad q=\frac{2\pi}{b_0}.
 \end{gathered}
 \label{eq:p5-qcd-scale-map}
\end{equation}
Equations~\eqref{eq:p5-qcd-extraction-inputs} and
\eqref{eq:p5-qcd-scale-map} define the extraction convention.
The five-flavour coefficient is assigned at $\mu_{\rm ext}$.
Explicit heavy-quark decoupling converts the six-flavour boundary by
threshold matching at $m_t$. The factor $(1-\epsilon)$ acts on
the coupling before the nonlinear extraction of $\Lambda_{\rm QCD}$.

\paragraph{Composite labels and the mass readout.}
\label{sec:p5-qcd-mass-readout}
The scalar spectrum contains the constant mode and a first nonconstant
shell of eigenvalue $8/L^2$. The composite-channel construction uses
a separate mass operator with dimensionless labels
\begin{equation}
\widehat\lambda=2\widehat\lambda_{\mathrm{gluon}}+C_2(J)+nN_g\xi,
\qquad \widehat\lambda_{\mathrm{gluon}}=4,\quad N_g\xi=1.
\end{equation}
Thus all terms have the same units. Geometric eigenvalues carry
the common factor $L^{-2}$.
For the normal two-leg reference the coclosed one-form
spectrum of Ref.~\cite{Guo2026CompactInternalSpaceI} gives the six-dimensional Killing carrier
$\mathcal K=(1,0)\oplus(0,1)$ with
$L^2\Delta_1^\perp|_{\mathcal K}=4I$.
The two vector responses form a common symmetric tensor carrier.
The sum of their internal eigenvalues gives the shared two-leg
contribution. The total-rotation Casimir resolves its composite
channels. The additive two-leg/Casimir rule gives
\begin{equation}
 \begin{gathered}
 L^2H_2=8I+C_2^{\rm tot},\\
 \operatorname{Sym}^2\mathcal K
 =2(0,0)\oplus(1,1)\oplus(2,0)\oplus(0,2).
 \end{gathered}
 \label{eq:p5-killing-pair-spectrum}
\end{equation}
With $C_2(j_L,j_R)=2[j_L(j_L+1)+j_R(j_R+1)]$ the decomposition
gives $8,16,20$ with multiplicities $2,9,10$. It verifies the
reference levels $8,16$ on the Killing carrier. The tensor
readout selects diagonal spin $J=2$ in $(1,1)$. The two singlets admit
even and odd exchange combinations at the same reference value $8$.
The masses labelled $17,18$ therefore use the additional conformal-channel
assignments in the four-channel operator below.

On the reported channel list
$\mathcal C=(0^{++},2^{++},0^{-+},0^{++*})$ the absolute
mass normalization uses the scalar reference $\widehat\lambda_*=8$ and
$m_*/\Lambda_{\rm QCD}=\widehat\lambda_*$. Extending the same
normalization to the resolved squared-mass operator gives
$m_a^2=\widehat\lambda_*\widehat\lambda_a\Lambda_{\rm QCD}^2$.
Thus the two-leg operator fixes the reference spectral label while this
reference-to-mass map fixes the absolute hadronic unit. Together they give
\begin{equation}
 D_{\mathcal C}=\operatorname{diag}(8,16,17,18),\qquad
 M_{\mathcal C}^2=8\Lambda_{\rm QCD}^2D_{\mathcal C}.
 \label{eq:p5-qcd-mass-operator}
\end{equation}
The channel ratios test the relative labels with the common scale
cancelled. The scalar mass tests this normalization together with the
coupling-to-$\Lambda_{\rm QCD}$ conversion. The labels $17,18$ retain
the conformal-channel assignments stated above.
For $\Lambda_{\rm QCD}>0$ the lowest mass in this four-channel
assignment is
\begin{equation}
m_G = 8 \Lambda_{\mathrm{QCD}} ,
\end{equation}
and the tower ratios are $2^{++}/0^{++}=\sqrt2$ and $0^{++*}/0^{++}=3/2$ and $0^{-+}/0^{++}=\sqrt{17/8}$. The proton readout combines the constituent content and conformal-content correction
\begin{equation}
m_p = N_c \Delta_f \Bigl(1 - \frac{1}{N_g^2 \Delta_s}\Bigr) \Lambda_{\mathrm{QCD}} \Bigl(1 + \tau \kappa \sum Y^2 \Delta_s\Bigr) = \frac{279}{64} \Lambda_{\mathrm{QCD}} \Bigl(1 + \frac{5\tau\kappa}{3}\Bigr) ,
\end{equation}
where $1/(N_g^2\Delta_s)=1/32=\xi/4$ follows from $N_g\xi=1$ and $N_g\Delta_s=2(d-1)$. Thus $N_c\Delta_f(31/32)=279/64$. The effective hadronic response factor uses $\sum Y^2\Delta_s=5/3$ evaluated in Section~\ref{sec:derivcorr}. The string tension and deconfinement temperature are
\begin{equation}
\sigma = \frac{\lambda_{\mathrm{TT}}}{\pi} \Lambda_{\mathrm{QCD}}^2 = \frac{14}{\pi} \Lambda_{\mathrm{QCD}}^2 , \qquad
T_d = \frac{\lambda_{\mathrm{vector}}}{N_c} \Lambda_{\mathrm{QCD}} (1 - \tau \kappa) = \frac{4}{3} \Lambda_{\mathrm{QCD}} (1 - \tau \kappa) ,
\end{equation}
Here $\lambda_{\mathrm{TT}}=14$ is the effective coefficient of the three-branch tensor catalogue used in the confinement-scale relation. The rough-TT operator in Ref.~\cite{Guo2026CompactInternalSpaceI} has minimum $6/L^2$ on its two-branch carrier. The chiral-squash factor $(1-\tau\kappa)$ enters the deconfinement readout. The common scale cancels in the ratio
\begin{equation}
\frac{\sigma}{T_d^2} = \frac{14}{\pi}\,\frac{9}{16}\,(1-\tau\kappa)^{-2} = \frac{126}{16\pi(1-\tau\kappa)^2} = 2.6242 ,
\end{equation}
the content ratio $126/(16\pi)=2.5068$ modified by the factor
$(1-\tau\kappa)^{-2}$. The common scale and the correction on its
input coupling are specified by
\eqref{eq:p5-qcd-extraction-inputs}--\eqref{eq:p5-qcd-scale-map}.
Equation~\eqref{eq:p5-qcd-mass-operator} summarises the four reported
channel masses. The complete condensate spectrum of Ref.~\cite{Guo2026CompactInternalSpaceI} is instead
governed by $\Delta_0+m_\delta^2I$ including its constant mode.
Their physical comparison uses a common source and scale normalization.

For a dynamical colour interpretation see the paragraph
``A mass atom from a common cross response'' in
Appendix~G of Ref.~\cite{Guo2026CompactInternalSpaceI}. With $Q$ a Hermitian
scalar colour observable and $\phi$ the Hermitian collective scalar
a common positive state and translation representation give the
connected matrix spectral measure in a common normalization and the bound
\begin{equation}
 \rho_{QQ}(\{m^2\})\geq
 \frac{|\rho_{Q\phi}(\{m^2\})|^2}
      {\rho_{\phi\phi}(\{m^2\})},
 \qquad \rho_{\phi\phi}(\{m^2\})>0.
 \label{eq:p5-colour-cross-atom}
\end{equation}
Applying this criterion to the colour response requires a common
representation of the full non-Abelian curvature and collective
source with a nonzero cross atom for $Q$.
A finite spectral approximation must retain an isolated eigenvalue
and positive residue. Passage to the full theory requires convergence
of the common response and uniform control of the surrounding spectral
separation and residue. Identifying the lowest visible colour mass
also requires exclusion of lower spectral support. The additional
identification with $8\Lambda_{\rm QCD}$ fixes the physical scale and
source normalization. These conditions give a concrete dynamical test
of the catalogue in \eqref{eq:p5-qcd-mass-operator}.

\subsection{Primordial nucleosynthesis}\label{sec:derivbbn}

The abundance calculation combines the generated mass and coupling
readouts with a radiation-era thermal history in the leading weak-rate
and freeze-out approximation. Ref.~\cite[Secs.~III.G--H and IV.C]{Pitrou2018}
reviews the radiative corrections and reaction-network treatment used
in precision BBN. The Fermi constant is $G_F=1/(\sqrt2v^2)$ and the
axial coupling is assigned as
\begin{equation}
g_A = \frac{N_g \Delta_s}{\pi} = \frac{2(d-1)}{\pi} = \frac{4}{\pi} ,
\label{eq:p5-bbn-axial}
\end{equation}
using $N_g=8$ and $\Delta_s=1/2$. Here $N_g\Delta_s=4$ is the
content weight assigned to the axial response and division by
$\pi$ sets its period normalization. The weak rate uses this
axial coefficient together with the canonically normalized vector
current. The electromagnetic self-energy is
\begin{equation}
\Delta_{\mathrm{EM}}=\left(1-\frac1{2\pi}\right)\alpha_{\mathrm{em}}(0)\Lambda_{\mathrm{QCD}}.
\label{eq:p5-bbn-electromagnetic}
\end{equation}
The low-energy coupling starts from the two-loop value $\alpha^{-1}(M_Z)=128.207$ at the preliminary-scale readout of Section~\ref{sec:g1} and applies the one-loop QED threshold sum
\begin{equation}
\frac{1}{\alpha_{\mathrm{em}}(0)} = \frac{1}{\alpha(M_Z)} + \frac{2}{3\pi} \sum_f Q_f^2 N_c^f \ln\frac{\mu_{\mathrm{QED}}}{m_f} ,
\label{eq:p5-bbn-qed-descent}
\end{equation}
with the charged leptons and $c$ and $b$ on the ladder masses and the $u,d,s$ thresholds set to $\Lambda_{\mathrm{QCD}}$. The logarithmic upper scale in this QED threshold sum is the auxiliary one-loop electroweak value
\begin{equation}
\begin{gathered}
\mu_{\mathrm{QED}}=\frac v2
\sqrt{g_2^{(1)}(v)^2+\frac35g_1^{(1)}(v)^2},\\
\frac1{g_i^{(1)}(v)^2}=\frac1{g_i(M_{\mathrm G})^2}
+\frac{b_i}{8\pi^2}\ln\frac{M_{\mathrm G}}v,\qquad
(b_1,b_2)=\left(\frac{41}{10},-\frac{19}{6}\right).
\end{gathered}
\end{equation}
This auxiliary scale defines the QED threshold integral whereas the
electroweak mass comparison uses its self-consistent $Z$-mass fixed point.
The threshold convention gives $\alpha_{\mathrm{em}}^{-1}(0)=137.0463$ with a $+0.0075\%$ deviation from the reference $137.036$. The radiative correction is
\begin{equation}
\delta_R=1+\frac{1-\tau}{8\pi}.
\label{eq:p5-bbn-radiative}
\end{equation}
The neutron-proton mass difference is
\begin{equation}
\Delta m_{np} = g_S^{\rm eff}(m_d - m_u) - \Delta_{\mathrm{EM}},
\qquad g_S^{\rm eff}=1 ,
\label{eq:p5-bbn-mass-difference}
\end{equation}
with the quark masses from the ladder of Section~\ref{sec:derivlz}.
The coefficient $g_S^{\rm eff}=1$ specifies the isovector response of
this native mass readout. Matching to a renormalized QCD nucleon
matrix element uses the isovector relation
\begin{equation}
 \Delta m_{np}^{\rm QCD}
 =g_S^{\mathcal S}(\mu)
       [m_d^{\mathcal S}(\mu)-m_u^{\mathcal S}(\mu)],
 \qquad
 \Delta m_{np}=\Delta m_{np}^{\rm QCD}-\Delta_{\rm EM},
 \label{eq:p5-nucleon-isovector-matching}
\end{equation}
to first order in isospin breaking. The scalar charge and quark masses
are evaluated in the same renormalization scheme $\mathcal S$ and at
the same scale $\mu$. Their reciprocal renormalizations preserve the
product~\cite{GonzalezAlonso2014}. The electromagnetic term is defined
as the proton-minus-neutron contribution in a fixed QCD--QED separation
convention~\cite{Borsanyi2015}. The generated mass ladder supplies
the effective source masses used here with $g_S^{\rm eff}=1$.
Its two contributions are $2.56131931$ and $1.27635176$ MeV.
They give $\Delta m_{np}=1.28496755$ MeV.

For a nucleon matching with isoscalar QCD mass $M_0$ and mean
electromagnetic shift $\overline{\delta M}_{\rm EM}$ the paired masses
are $m_{p,n}=M_0+\overline{\delta M}_{\rm EM}\mp\Delta m_{np}/2$.
Thus an independent scalar matrix element and isoscalar mass together with
electromagnetic separation specify this QCD correspondence for the
generated sources. The numerical calculation retains their common
source dependence explicitly~\cite{Guo2026GenesisFramework}.
In terms of the generated ladder masses the coefficient to be
determined by this matching is the combination
\begin{equation}
 g_S^{\rm eff}
 =g_S^{\mathcal S}(\mu)
   \frac{m_d^{\mathcal S}(\mu)-m_u^{\mathcal S}(\mu)}{m_d-m_u} .
 \label{eq:p5-effective-isovector-coefficient}
\end{equation}
The scale dependence cancels between the renormalized scalar density
and quark-mass difference. An independently calculated strong mass
splitting therefore fixes this combination directly provided its
electromagnetic separation is carried into the same matching.
Appendix~\ref{app:bbnext} propagates the published total nucleon
splitting of Ref.~\cite{Borsanyi2015} as a separate external comparison
retaining the uncertainties from that calculation.

\begin{proposition}[Generated three-quark source channels]
\label{prop:p5-three-quark-source}
Use the inherited fundamental colour representation with Hermitian
generators $\operatorname{tr}(T^AT^B)=\delta^{AB}/2$. The normalized
three-quark colour singlet is
$|\epsilon\rangle=\epsilon_{abc}|abc\rangle/\sqrt6$.
For $T_r^A$ acting on leg $r$ the three adjoint-source vectors
$|\Omega_r\rangle=\sum_AT_r^A|\epsilon\rangle\otimes|A\rangle$
satisfy the global colour constraint and have Gram matrix
\begin{equation}
 \langle\Omega_r|\Omega_s\rangle
 =2\delta_{rs}-\frac23,
 \qquad \sum_r|\Omega_r\rangle=0.
 \label{eq:p5-three-quark-source-gram}
\end{equation}
Their span has dimension two. For the normalized scalar coefficient
$e^1_{00}$ of the common matter connection multiplication of either
lowest-shell spinor has squared norm $2/3$ in the $j=1$ branch with $J=3/2$
and $1/3$ in the $j=1$ branch with $J=1/2$ with zero projection
onto the incoming $j=0$ shell.
\end{proposition}
\begin{proof}
The decomposition $3\otimes3\otimes3=1\oplus8\oplus8\oplus10$
gives the unique colour singlet. On this state
$\sum_rT_r^A=0$ and $\sum_A(T_r^A)^2=4I/3$. Permutation symmetry
then gives $\langle\epsilon|\sum_AT_r^AT_s^A|\epsilon\rangle=-2/3$
for $r\ne s$. The adjoint transformation of $T_r^A|\epsilon\rangle$
cancels that of $|A\rangle$. The Gram eigenvalues are $0,2,2$.
The spinor product follows by coupling the right scalar index of
$e^1_{00}$ to spin $1/2$. The squared Clebsch--Gordan coefficients
are $2/3$ and $1/3$. Orthogonality of the scalar shells eliminates
the incoming projection. This is the common transport vertex
given by the proposition ``Matter coupling
from the common transport'' in Ref.~\cite{Guo2026CompactInternalSpaceI}.
\end{proof}

Thus the generated colour and internal product tensors supply explicit
adjoint-source channels. The inherited electromagnetic charges give
$\sum_iq_i^2=1,\ \sum_{i<j}q_iq_j=0$ for $uud$ and $2/3,-1/3$
for $udd$. These multiply the corresponding current kernels.
The scalar charge measures how the nucleon energy responds to an
isovector mass source at fixed mean light-quark mass. Proton and
neutron responses enter with opposite isospin orientation. Thus their
difference determines the strong contribution to the splitting
\cite{GonzalezAlonso2014}. This energy derivative is the
Feynman--Hellmann response of the constrained baryon state
\cite{Feynman1939}. It can be organized by a fixed projection
$P$ in the baryon sector after imposing the physical Gauss constraints.
Write $Q=1-P$ and $V=H_{QP}$ and
$\mathcal R_Q(E,s)=(E-H_{QQ}(s))^{-1}$ for the same sourced
Hamiltonian differentiable in the source $s$. At a simple isolated
eigenvalue outside the spectrum of
$H_{QQ}$ the reduced operator and its source derivative are
\begin{equation}
\begin{gathered}
 H_{\rm eff}(E,s)=H_{PP}(s)+V^\dagger\mathcal R_Q(E,s)V,
 \qquad H_{\rm eff}u=Eu,\quad \|u\|=1,\\
 \frac{dE}{ds}=
 \frac{\langle u,\,\partial_sH_{\rm eff}(E,s)u\rangle}
      {1+\|\mathcal R_Q(E,s)Vu\|^2}.
 \label{eq:p5-baryon-source-schur}
\end{gathered}
\end{equation}
Indeed the eliminated component is $\mathcal R_QVu$ and
$\partial_EH_{\rm eff}=-V^\dagger\mathcal R_Q^2V$. Differentiation of the eigenvalue equation proves the second line.
The denominator is the squared norm of the full state reconstructed from
the retained component $u$ and the eliminated component
$\mathcal R_QVu$. At fixed $E$ the direct source insertion is
$H_{PP,s}$ and the eliminated sector contributes
$V_{,s}^\dagger\mathcal R_QV+V^\dagger\mathcal R_QV_{,s}
+V^\dagger\mathcal R_QH_{QQ,s}\mathcal R_QV$.
For $s=m_d^{\mathcal S}(\mu)-m_u^{\mathcal S}(\mu)$ at fixed mean
quark mass the difference of the neutron and proton derivatives at
the isospin-symmetric point determines the scalar coefficient in
\eqref{eq:p5-nucleon-isovector-matching}. Its evaluation uses the
constrained baryon resolvent and its source insertions together with the same
renormalized current kernels for the electromagnetic separation.
The finite colour and spinor calculation fixes the factors in
Proposition~\ref{prop:p5-three-quark-source} directly
\cite{Guo2026GenesisFramework}.

The neutron lifetime is
\begin{equation}
\tau_n = \frac{2\pi^3}{G_F^2 |V_{ud}|^2 (1+3g_A^2) m_e^5 f \, \delta_R} ,
\label{eq:p5-neutron-lifetime}
\end{equation}
using the Cabibbo approximation $|V_{ud}|=\sqrt{1-|V_{us}|^2}$. The Coulomb-weighted phase-space integral is
\begin{equation}
\begin{gathered}
f=\int_1^{W_0}F(1,W)\,W\sqrt{W^2-1}\,(W_0-W)^2\,dW,
\qquad W_0=\frac{\Delta m_{np}}{m_e}>1,\\
F(1,W)=\frac{2\pi\eta}{1-e^{-2\pi\eta}},\qquad
\eta=\frac{\alpha_{\mathrm{em}}(0)W}{\sqrt{W^2-1}}.
\end{gathered}
\label{eq:p5-neutron-phase-space}
\end{equation}
Put $d=W_0-1$ and $W=1+dx$ with $p=\sqrt{W^2-1}$ and
$a=2\pi\alpha_{\rm em}(0)$. Cancelling the Coulomb endpoint factor gives
\[
 f=d^3\int_0^1\frac{aW^2}{1-e^{-aW/p}}(1-x)^2\,dx .
\]
The factor multiplying $(1-x)^2$ has the finite limit $a$ at
$x=0$. Adaptive quadrature of this regularized integral gives
$f=1.63492371128643$ with estimated absolute error
$1.5\times10^{-12}$ in 567 integrand evaluations and
$\tau_n=915.54\ \mathrm{s}$~\cite{Guo2026GenesisFramework}.
The common QCD scale enters $\Delta m_{np}$ through the
electromagnetic term. The decay phase space propagates this
dependence into the lifetime and helium yield. The approximation
is the static-nucleon weak rate with Coulomb and specified radiative
factors. Precision recoil and weak-magnetism contributions are
discussed in Ref.~\cite[Secs.~III.G--H]{Pitrou2018}.

In natural units the freeze-out temperature is the positive solution of
\begin{equation}
\begin{gathered}
\Gamma_{\mathrm{weak}}(T_f)=H(T_f),\\
\Gamma_{\mathrm{weak}}(T)=\frac{7\pi}{60}(1+3g_A^2)G_F^2T^5
\left(1+\frac{3\Delta m_{np}}{2T}\right),\\
H(T)=\sqrt{\frac{4\pi^3g_*}{45}}\frac{T^2}{M_{\mathrm{Pl}}},
\quad M_{\mathrm{Pl}}=\sqrt{8\pi}\,\MP.
\end{gathered}
\label{eq:p5-bbn-freezeout}
\end{equation}
The thermal approximation takes $g_*=10.75$ at freeze-out and $g_*=3.36$ after electron-positron annihilation and $T_{\mathrm{BBN}}=0.08\ \mathrm{MeV}$. With
\begin{equation}
t(T,g_*)=\frac{0.301M_{\mathrm{Pl}}}{\sqrt{g_*}\,T^2},
\qquad
\Delta t=t(T_{\mathrm{BBN}},3.36)-t(T_f,10.75),
\label{eq:p5-bbn-time-temperature}
\end{equation}
free neutron decay conserves the sum of neutron and proton numbers.
Writing $r_f=\exp(-\Delta m_{np}/T_f)$ and
$q=\exp(-\Delta t/\tau_n)$ the neutron fraction therefore evolves as
$X_n=q\,r_f/(1+r_f)$. Complete capture of the surviving neutrons gives
\begin{equation}
Y_p=\frac{2r_fq}{1+r_f}=0.24608,\qquad
(n/p)_{\mathrm{BBN}}=\frac{r_fq}{1+r_f(1-q)}.
\label{eq:p5-bbn-conserved-abundance}
\end{equation}
Both $\Delta t$ and $\tau_n$ are in $\mathrm{GeV}^{-1}$. Multiplying each by $\hbar=6.582119569\times10^{-25}\ \mathrm{GeV\,s}$ gives seconds without changing their ratio. The separate content assignment
\begin{equation}
N_{\eff}=3+\frac{\sqrt3}{(2\pi)^2}=3.0439
\label{eq:p5-bbn-neutrino-number}
\end{equation}
supplies the neutrino-number comparison and does not change the two
specified thermal values of $g_*$.

\subsection{Finite correction factors}\label{sec:derivcorr}

With $s_0=2\tau$ and the curvature-norm factor $\kappa$ of
Section~\ref{sec:derivg2} the finite corrections enter the masses and
couplings at the following stages.

\begin{itemize}
\item $(1 - 4 s_0 \kappa)$ multiplies the fourth-power dark-energy
floor of Section~\ref{sec:derivcosmo}. At fixed seesaw inputs it cancels
the first-order explicit seesaw insertion. Including the dependence
through $v^2$ gives Eq.~\eqref{eq:p5-cosmology-composition}.
\item $(1 - s_0 \kappa)$ multiplies the electroweak scale $v$ of
Section~\ref{sec:derivewsb} and the raw microwave floor
$T_{\mathrm{CMB}}^{\mathrm{raw}}$ of Section~\ref{sec:derivcosmo}.
The temperature also depends on the corrected neutrino mass.
\item $(1 + s_0 \kappa)$ is the seesaw level-transfer factor on the heaviest neutrino scale $m_{\nu_3}$ of Section~\ref{sec:derivnu}.
\item $(1-\tau\kappa)$ is the chiral-squash factor assigned to the deconfinement temperature $T_d$ of Section~\ref{sec:derivqcd} and to the curvature amplitude $\Delta_R^2$ of Section~\ref{sec:derivcosmo}.
\item $(1 + \tau\kappa \sum Y^2 \Delta_s) = (1 + 5\tau\kappa/3)$ is the effective hadronic response factor of the proton mass $m_p$ of Section~\ref{sec:derivqcd} with $\sum Y^2 = 10/3$ the hypercharge capacity and $\Delta_s = 1/2$ the scalar conformal weight.
\item $(1-s_0\kappa/N_g)$ is the Yukawa-difference factor multiplying $\alpha_s$ before the nonlinear extraction of $\Lambda_{\mathrm{QCD}}$ in Section~\ref{sec:derivqcd}.
\item $C_\gamma=1-\tau/\pi^2$ multiplies
$T_{\mathrm{CMB}}^{\mathrm{raw}}$ in the microwave-monopole and linear
CMB comparisons. The baryon density $\Omega_b$ uses the paired raw
quantities $T_{\mathrm{CMB}}^{\mathrm{raw}}$ and
$\eta_B^{\mathrm{raw}}$ of Sections~\ref{sec:derivcosmo}
and~\ref{sec:derivckm}.
\item $f$ is the Coulomb-weighted phase-space integral in the neutron lifetime of Section~\ref{sec:derivbbn}.
\end{itemize}

The first six factors are expressed in terms of the isometry-breaking
normalisation and squash amplitude together with the local response coefficients.
\begin{enumerate}
\item the content invariant $\tau=(N_L-N_R)/(N_f\sum Y^2)=1/50$ of Section~\ref{sec:derivtau}.
\item the squash amplitude $s_0=n_{\mathrm{broken}}\tau=2\tau$ with two broken $SU(2)_R/U(1)_R$ directions.
\item the inverse curvature-norm normalization $\kappa^2(s)=(1+s)/(1-2s)^{5/2}$ of Eq.~\eqref{eq:p5-kappa-curvature-norm}.
\item the local coefficient $c_Q=a_Qr_Q$ with amplitude fraction
$a_Q$ (geometric $1$ or chiral $1/2$) and content ratio $r_Q$. For an
inverse-metric monomial with $n_3$ axial and $n_{\rm eq}$ equatorial
factors the logarithmic derivative at the round point is
$2n_3-n_{\rm eq}$. The shear has zero volume derivative at $s=0$. Its finite-$s$ volume follows from the squashed metric.
\end{enumerate}
For each readout the local insertion
$Q\to Q(1+c_Qs_0\kappa)$ specifies the response at fixed upstream
inputs with $s_0\kappa=N_g\tau\kappa/(d+1)$. Its amplitude fraction
and charge weight determine how that channel samples the common
deformation. The total response follows by composing it with the
change in its inputs. For example the explicit neutrino insertion
acts on the dimension-five coefficient while the two powers of
the generated VEV carry the electroweak response into the same
neutrino mass. The channel factors have the following origins.
\begin{itemize}
\item $v$ and $V_{cb}$ carry the local factor $1-s_0\kappa$ ($a_Q=1$ and $r_Q=-1$). For $v=\MG e^{-\varphi}$ the exact stop increment is $-\ln(1-s_0\kappa)=s_0\kappa+O((s_0\kappa)^2)$. Its finite error is bounded in Appendix~\ref{sec:p5-cosmology-error}.
\item $m_{\nu_3}$ has the explicit seesaw insertion $1+s_0\kappa$
($a_Q=1$ and $r_Q=+1$) at fixed $v,k_{\rm GUT}$. Substitution of the
corrected $v$ into $v^2(2\pi)^2(1+s_0\kappa)/k_{\rm GUT}$ gives
the full response and the product $v m_{\nu_3}$ in
Eq.~\eqref{eq:p5-cosmology-composition}.
\item $T_d$ and $\Delta^2_R$ carry $-\tau\kappa$ and $V_{ub}$ carries $+\tau\kappa$ ($a_Q=1/2$ and $r_Q=\mp1$). The chiral level couples to the chiral asymmetry $\tau = s_0/2$ (the amplitude fraction $a_Q=(N_L-N_R)/n_{\mathrm{broken}}=1/2$). The sign is $-$ for chiral restoration ($T_d$ and $\Delta^2_R$) and $+$ for the cross-generation element ($V_{ub}$).
\item $m_p$ carries $+\tau\kappa\sum Y^2\Delta_s=+5\tau\kappa/3$ ($a_Q=1/2$ and $r_Q=5/3$). The effective hadronic response combines the chiral asymmetry and quadratic hypercharge moment with the scalar conformal weight.
\item $\alpha_s$ carries $-s_0\kappa/N_g = -s_0\kappa\,\xi$ ($a_Q=1$ and $r_Q=-1/N_g$). The effective correction between the fixed-Yukawa calculation at
$y_0=1$ and the running-Yukawa calculation uses the conformal normalisation $\xi=1/N_g$.
\end{itemize}
The amplitude fraction $a_Q$ and the local fourth-power coefficient
follow the broken-generator and chiral-asymmetry counts together with the $m_{\nu_1}^4$ count. The explicit seesaw insertion has $r_Q=+1$. Total scale derivatives
apply the chain rule to all these input dependencies. The constituent
and Yukawa sectors use the effective response maps
\begin{enumerate}
\item Hadronic response ($m_p$ with $r_Q=\sum Y^2\Delta_s=5/3$). The finite-content response used by the mass readout is
\begin{equation}
\begin{gathered}
\frac{\delta m_p}{m_p}=\tau\kappa\Delta_s\sum_f N_{\rm int}(f)Y_f^2=\frac{5\tau\kappa}{3},\\
\sum_f N_{\rm int}(f)Y_f^2=6(\tfrac16)^2+3(\tfrac23)^2+3(-\tfrac13)^2+2(-\tfrac12)^2+1=\frac{10}{3}.
\end{gathered}
\end{equation}
This is the product of the chiral asymmetry and curvature-norm factor with the
scalar conformal weight and quadratic hypercharge moment.
\item Yukawa difference ($\alpha_s$ with $r_Q=-1/N_g=-\xi$). The top-Yukawa mixing coefficients in the two-loop gauge flow are given in Section~\ref{sec:derivbeta}. The finite correction used with the fixed seed $y_0=1$ is
\begin{equation}
\alpha_s'=\alpha_s(1-\epsilon),\qquad
\epsilon=s_0\kappa\xi=\frac{s_0\kappa}{N_g},\qquad
\xi=\frac{d-2}{4(d-1)}=\frac18.
\end{equation}
For $0\leq\epsilon<1$ the logarithmic increment is
$\ln(\alpha_s'/\alpha_s)=\ln(1-\epsilon)=-\epsilon+O(\epsilon^2)$.
The extraction map uses the finite multiplicative factor.
\end{enumerate}
The numerical table uses these finite factors at the stated insertion points with the subsequent nonlinear dependence retained.

%% file: sections-en/appendix_d.tex
% =============================================================================
% Appendix D - Theoretical sensitivity
% The Spectrum of a Compact Internal Space II: Effective Couplings and Mass Scales
% =============================================================================

\section{Error bounds and parameter sensitivity}\label{app:sensitivity}

The error estimates distinguish numerical error at a fixed response domain,
changes of that domain and propagation of shared input uncertainties.
The positive spectral sums permit explicit bounds on the endpoint.
Their effect on the mass hierarchy follows from the same closed
relations used to calculate the central values.

Table~\ref{tab:p5-error-ledger} identifies the error controls for the
representative quantities: numerical integration, input propagation,
perturbative truncation and response sensitivity. A numerical covariance additionally requires an input
probability model.

\begin{table}[t]
\centering
\footnotesize
\caption{Error-control ledger for the headline results.  Entries identify
the relevant error category. They are not additive uncertainty estimates.}
\label{tab:p5-error-ledger}
\begin{tabular}{p{1.2cm}p{2.5cm}p{2.3cm}p{2.6cm}p{2.2cm}}
\toprule
Object & Numerical & Input propagation & Truncation & Model sensitivity \\
\midrule
$\kL$ & quadrature and root bracket & content and $\tau$ & spectral-tail bounds & continuous cutoff $\mathcal R$ \\
$v$ & $c_R$ and endpoint & $\GN$ and $c_R$ & two-loop running & matching scheme \\
$m_e$ & $c_R$ and transfer & $\tau$ and content & one-loop lepton matching & finite matching \\
$m_t$ & $c_R$ and $y_0$ & $\GN$ and $c_R$ & top threshold & scheme \\
$\Gamma_t$ & leading phase space & $G_F,M_t,g_2$ & $\mathcal O(\alpha_s/\pi)$ and $\mathcal O(\alpha/\pi)$ omitted & bottom/CKM \\
$\tau_n,Y_p$ & weak-rate quadrature & $m_{np}$ and $G_F$ & finite-temperature network & BBN inputs \\
DESI & covariance and $r_d$ & thermal history & fixed prediction & common-scale direction \\
\bottomrule
\end{tabular}
\end{table}

The endpoint is dimensionless and independent of the Planck anchor at
fixed content and source projections with a fixed response level. Newton's constant
enters the subsequent reconstruction of dimensional quantities through
$\MP=(8\pi\GN)^{-1/2}$.

\subsection{Gaussian spectral tails and endpoint enclosures}
\label{sec:p5-gaussian-tail}

The endpoint is linear in the positive spectral coefficient $A_R$ of
Eq.~\eqref{eq:p5-homogeneous-coefficient}. Bounds on the Gaussian tails
therefore propagate directly to the endpoint and then to the mass hierarchy.
Throughout this subsection the matter content, species projections,
channel weights and torsion modulus are fixed. For the TT trace the
mode coefficients satisfy $a_j,b_j,D_j\geq0$ and
$\sum_jD_j e^{-a_j/\pi}<\infty$. Thus changes in $R$ compare the same
spectral sum on different response domains.

\begin{proposition}[Gaussian tail and endpoint enclosure]
\label{prop:p5-gaussian-tail}
The finite frequency domain and the full Gaussian coefficient satisfy
\begin{equation}
\begin{split}
 0&\leq A_\infty-A_R\leq B_R,\\
 B_R&=\frac1{64\pi^2}\sum_jD_j e^{-a_j/\pi}
 \operatorname{erfc}\!\left(\sqrt{(R-a_j/\pi)_+}\right),\\
 0&\leq c_\infty-c_R\leq\frac{27}{4}B_R.
\end{split}
\label{eq:p5-tail-enclosure}
\end{equation}
\end{proposition}
\begin{proof}
The nonnegative shifts give $0\leq K_{\rm TT}\leq1$.
For $a_j<R\pi$ the omitted interval starts at
$\sqrt{R\pi-a_j}$. For $a_j\geq R\pi$ the entire mode is omitted.
In both cases use
\[
 \int_v^\infty e^{-u^2/\pi}\,du
 =\frac\pi2\operatorname{erfc}(v/\sqrt\pi).
\]
Sum the positive bounds and apply $c_R=27A_R/4$.
\end{proof}

The infinite sum defining $B_R$ admits an explicit remainder bound.
For the three-generation content, group the scalar and vector modes and the assigned
spinor modes by their degree $n$. Their total weighted multiplicity is at most
$196(n+2)^2$ and $a_j\geq n^2$.
After summing through $n=N\geq1$ the remaining contribution to $B_R$
is bounded by
\begin{equation}
 \frac{196}{64\pi^2}
 \frac{(N+3)^2e^{-(N+1)^2/\pi}}{1-r_N},\qquad
 r_N=\left(\frac{N+4}{N+3}\right)^2e^{-(2N+3)/\pi}<1.
 \label{eq:p5-tail-series}
\end{equation}
Indeed $\operatorname{erfc}\leq1$ and consecutive terms of
$(n+2)^2e^{-n^2/\pi}$ have decreasing ratio bounded by $r_N$ for
$n\geq N+1$. The constant $196$ bounds the scalar and vector contributions together with the larger
assigned chiral contribution by $4+96+96$.

There is also a strictly positive omitted-shell lower bound at $R=5$.
The vector shell $l=3$ has $a_j=16>5\pi$ and multiplicity $30$.
For the twelve gauge fields its total TT-weighted multiplicity is
$12\cdot4\cdot30=1440$ and $b_j\leq3/2+\tau^2/6$. Hence
\begin{equation}
 c_\infty-c_5\geq
 \frac{1215}{8\pi^2}
 \left(\frac{16}{35/2+\tau^2/6}\right)^2 e^{-16/\pi}>0.
 \label{eq:p5-omitted-shell}
\end{equation}
Retain this shell in the positive omitted sum. Bound
its kernel below by $[16/(35/2+\tau^2/6)]^2$ for every $u\geq0$
and integrate $e^{-u^2/\pi}$ over the full half-line.
At $\tau=1/50$ the coarse estimates $b_j<2$ and $\pi>8/3$ with
$e<3$ and $\pi^2<10$ give
\[
 c_\infty-c_5>\frac{120e^{-6}}{\pi^2}
 >\frac{120}{7290}>0.016.
\]
The complete-domain root is therefore strictly greater than the $R=5$
root. Equation~\eqref{eq:p5-tail-enclosure} bounds their difference for
the same Gaussian kernel. Numerical error in $c_R$ is controlled
separately by the quadrature, mode summation and rounding bounds.

If a computed coefficient satisfies
$|\widetilde A_R-A_R|\leq\varepsilon_A$ with frequency integration and
mode summation together with rounding included in $\varepsilon_A$ then
\begin{equation}
\begin{aligned}
 c_R&\in\frac{27}{4}
 [\widetilde A_R-\varepsilon_A,\widetilde A_R+\varepsilon_A],\\
 c_\infty&\in\frac{27}{4}
 [\widetilde A_R-\varepsilon_A,
  \widetilde A_R+\varepsilon_A+B_R].
\end{aligned}
\label{eq:p5-certified-endpoint}
\end{equation}
The first interval concerns the fixed-domain endpoint. The second
also includes the positive contribution from the omitted domain.

\paragraph{Enclosing the joint-branch coefficients.}
\label{sec:p5-branch-enclosure}
For each of the eight integrals in
\eqref{eq:p5-branch-coefficients} write
$f(u)=g(u^2+a)e^{-(u^2+a)/\pi}$ with
$g(q)=[q/(q+b)]^2$. On its integration interval
$q+b\geq3/4$ and $u^2<20$ and $\pi>3$. Hence
\[
 0\leq g\leq1,\qquad |g'|\leq8/3,\qquad
 |g''|\leq64/9,\qquad |f''|\leq726.
\]
For an interval of length $U=\sqrt{5\pi-a}<9/2$ the composite midpoint
sum $M_N$ therefore satisfies the explicit estimate
\begin{equation}
 |I-M_N|\leq\frac{726(9/2)^3}{24N^2}.
 \label{eq:p5-midpoint-certificate}
\end{equation}
Using $N=4096$ outward rational interval evaluation gives
\[
 a_5\in[0.6010245,0.6017579],\qquad
 b_5\in[0.6300616,0.6313672],
\]
which imply the wider bounds used in
Eq.~\eqref{eq:p5-joint-platform}.
Adding $a_5+3b_5$ before decimal rounding gives the coarse
selection-level endpoint enclosure
\begin{equation}
 c_{3,5}\in[c_-,c_+]=[2.4912094,2.4958593],
 \qquad c_+-c_-=0.0046499.
 \label{eq:p5-production-endpoint-enclosure}
\end{equation}
This interval certifies the discrete $R=5$ shell support. A finer
production enclosure uses the homogeneous coefficient at $c=1$.
On each of $2048$ cells per spectral row the trapezoid error is
bounded by $h^3\sup|f''|/12$. The cellwise derivative bound follows
from
\[
 \begin{aligned}
 g'(q)&=\frac{2bq}{(q+b)^3},\qquad
 g''(q)=\frac{2b(b-2q)}{(q+b)^4},\\
 f''(u)&=e^{-q/\pi}\left[2\left(g'-\frac g\pi\right)
 +4u^2\left(g''-\frac{2g'}\pi+\frac g{\pi^2}\right)\right],
 \qquad q=u^2+a.
 \end{aligned}
\]
The square-root enclosure $U\in[U_-,U_+]$ is included by integrating
to $U_+$ and subtracting at most $U_+-U_-$ from the lower bound,
using $0\leq f\leq1$. Including all eight weighted rows gives
\[
 A_5\in[0.36941244246332044290,\,0.36941253772250165936].
\]
The identities $c_5=27A_5/4$ and
$F_5'(c)=-A_5/c^2$ transfer this single enclosure to the root and
its slope. With outward rounding and a margin for strict endpoint
signs the resulting bracket is
\begin{equation}
 \kL\in[2.493533986626,2.493534629628],
 \label{eq:p5-precision-endpoint-enclosure}
\end{equation}
and propagating this interval through the closed mass relations gives
\begin{equation}
 \begin{gathered}
 M_G\in[1.7310762937,1.7310767402]\times10^{18}\ \mathrm{GeV},\\
 m_e\in[0.5103512187,0.5103577820]\ \mathrm{MeV},\\
 v\in[246.1887479,246.1908008]\ \mathrm{GeV}.
 \end{gathered}
 \label{eq:p5-precision-mass-enclosures}
\end{equation}
The coarse interval proves
that the retained shell count is five while the narrow interval controls
the precision of the comparison values. The BBN intervals are reported
separately in Appendix~\ref{app:bbnext}.
For reproducibility forty terms of the alternating arctangent series
enclose $\pi$ through
\[
 \pi=16\arctan(1/5)-4\arctan(1/239).
\]
For
$0\leq x\leq5$ enclose $e^x$ by its degree-sixty positive Taylor sum
and the remainder $243\,5^{61}/61!$ and invert to enclose $e^{-x}$.
Integer square-root bounds enclose $U$. All arithmetic is rounded
outwards on a $10^{-40}$ rational grid. These operations enclose the finite integrals entering
Eq.~\eqref{eq:p5-joint-platform} including all rounding
errors.

\paragraph{Complete Gaussian coefficient certificate.}
\label{sec:p5-full-gaussian-certificate}
The full coefficients use the same prefactor $P=27/(128\pi^3)$
and mode integrands as \eqref{eq:p5-branch-coefficients} with all
prescribed modes and $u\in[0,\infty)$. The bosonic rows are
\[
\begin{array}{c|c|c|c}
 \text{degree}&a&b&D\\\hline
 n\ \text{even}&n(n+2)&3/4&4(n+1)^2\\
 n\ \text{odd}&(n+1)^2&3/2+1/15000&64n(n+2)\\
 n\ \text{odd}&(n+1)^2&1+1/15000&24n(n+2)\\
 n\ \text{odd}&(n+1)^2&1/15000&8n(n+2)
\end{array}
\]
and the one-generation fermionic rows have
$a=(n+3/2)^2$ and $b=3/20000$ and
$D=28(n+1)(n+2)$ for even $n$ or $32(n+1)(n+2)$ for odd $n$.
These define $a_\infty$ and $b_\infty$ respectively.

For an explicit finite certificate retain $n\leq8$ and integrate on
$0\leq u\leq8$ and use $N=8192$ midpoint subintervals. The earlier
bounds for $g(q)$ now give
\begin{equation}
 |f''(u)|\leq2310e^{-a/\pi},\qquad
 |I_{[0,8]}-M_N|\leq
 \frac{2310\,8^3}{24N^2}e^{-a/\pi}.
 \label{eq:p5-full-midpoint-bound}
\end{equation}
Indeed $f''=e^{-(u^2+a)/\pi}
[2(g'-g/\pi)+4u^2(g''-2g'/\pi+g/\pi^2)]$. Its bracket is
bounded by $6+256\cdot9=2310$.
The missing frequency contribution in each retained row satisfies
\[
 0\leq\int_8^\infty f(u)\,du
 \leq e^{-a/\pi}\frac\pi{16}e^{-64/\pi},
\]
using $u/8\geq1$ in the Gaussian integral.

For the remaining degrees $n\geq9$ the bosonic total is bounded by
$100(n+2)^2$ and the one-generation fermionic total by $32(n+2)^2$
while every eigenvalue coefficient is at least $n^2$. Their respective
contributions to the root coefficients are at most
\begin{equation}
 \frac{27B}{256\pi^2}
 \frac{121e^{-81/\pi}}
 {1-(12/11)^2e^{-19/\pi}},\qquad B=100\ \text{or}\ 32.
 \label{eq:p5-full-mode-tail}
\end{equation}
The denominator uses the largest successive ratio of
$(n+2)^2e^{-n^2/\pi}$ for $n\geq9$. Together with the bound on $u>8$ this estimate controls the
frequency and spectral truncations of the complete-domain integral.

Outward rational evaluation of the finite sums with the above
positive tails added to their upper endpoints gives
\begin{equation}
\begin{aligned}
 a_\infty&\in[0.6897056,0.6906685],\\
 b_\infty&\in[0.6568695,0.6574841],\\
 c_{3,\infty}&\in[2.6603141,2.6631205].
\end{aligned}
\label{eq:p5-full-gaussian-enclosures}
\end{equation}
The last line is computed before decimal rounding of the first two.
The rational grid and elementary enclosures are those used above. For exponent arguments greater than five halve the argument until
it is in $[0,5]$ and square the resulting interval the same number
of times. The Taylor remainder and outward rounding are propagated through
each squaring to give enclosures of the complete trace.

For the three-position test a coarser enclosure suffices. On each
frequency subinterval $g(q)=(q/(q+b))^2$ increases and
$e^{-q/\pi}$ decreases. Products of their opposite endpoint bounds
therefore enclose the integral. Using 128 subintervals for each of
the eight finite-domain and 26 complete-domain rows together with
the positive tails above gives~\cite{Guo2026GenesisFramework}
\[
 2.4724098408<c_R<2.7182303261,\qquad
 6.1128104212<c_R^2<7.3887761056,\qquad R\geq5.
\]
The inequalities follow from $c_5\leq c_R\leq c_\infty$ and outward
rational evaluation of the two endpoint integrals. The calculation
uses 4352 row subintervals and 2720 distinct exponential enclosures
with common Gaussian values reused. The enclosure places $c_R^2$
strictly between $11/2$ and $15/2$. The narrower intervals in
\eqref{eq:p5-full-gaussian-enclosures} control the mass-scale comparison.

For downstream quantities $X(c)=C_0c^p e^{-ac}$ at fixed other inputs
the exact propagation is
\begin{equation}
 \frac{X(c+\delta)}{X(c)}
 =\left(1+\frac\delta c\right)^p e^{-a\delta},\qquad c+\delta>0.
 \label{eq:p5-exact-error-transfer}
\end{equation}
Its logarithmic derivative is $p/c-a$. Bound this derivative on the
same certified interval or use the displayed exact ratio. All outputs
sharing $c$ inherit correlated uncertainty. For a quantity of mass dimension $d$ proportional
to $\MP^d$ the anchor elasticity is
$\partial\log X/\partial\log\GN=-d/2$.

\paragraph{Certified propagation at the production level.}
Keep the three-copy content and $R=5$ with fixed $\tau$ and $\MP$.
All remaining readout rules are also held fixed. For a convenient exact reference argument set
$c_\circ=2.49353$ and $X_\circ=X(c_\circ)$. The closed mass and density
relations give
\begin{equation}
 \begin{aligned}
  V(c)&:=\frac{v(c)}{v_\circ}
      =\frac{c_\circ}{c}e^{-4\pi(c-c_\circ)},\\
  \frac{m_e(c)}{m_{e,\circ}}&=e^{-20(c-c_\circ)},\qquad
  \frac{\rho_\Lambda(c)}{\rho_{\Lambda,\circ}}=V(c)^8.
 \end{aligned}
 \label{eq:p5-production-error-ratios}
\end{equation}
All three functions decrease on $[c_-,c_+]$. Evaluation at the two
endpoints of \eqref{eq:p5-production-endpoint-enclosure} therefore gives
Table~\ref{tab:p5-production-error-propagation}. The decimal bounds can
be checked with rational arithmetic. Use the Machin enclosure for $\pi$
above and enclose $e^x$ for $0\leq x<1/2$ by its degree-twenty-four
Taylor sum with remainder at most $2x^{25}/25!$. Negative arguments
are enclosed by reciprocal bounds. The arguments used here have
magnitude below $0.05$.

\begin{table}[t]
\centering
\caption{Propagation of the coarse $R=5$ selection enclosure at fixed
remaining inputs. The reference values are evaluations at the exact
decimal argument $c_\circ=2.49353$. Each row encloses the same one-parameter
curve. The three intervals are correlated. These selection-level
enclosures accompany the production precision intervals of
Eq.~\eqref{eq:p5-precision-mass-enclosures}.}
\label{tab:p5-production-error-propagation}
\begin{tabular}{lc}
\toprule
Ratio & Certified enclosure \\
\midrule
$v/v_\circ$ & $[0.9702,1.0306]$ \\
$m_e/m_{e,\circ}$ & $[0.9544,1.0476]$ \\
$\rho_\Lambda/\rho_{\Lambda,\circ}$ & $[0.7853,1.2722]$ \\
\bottomrule
\end{tabular}
\end{table}

These are guaranteed output ranges for the stated quadrature certificate.
The floating-point calculation supplies the central values used in the
comparison tables. Its small comparison residuals and the certified
coverage in Table~\ref{tab:p5-production-error-propagation} describe
different properties of the calculation. Input uncertainties and finite
matching together with changes of the response domain have their own propagation.

\paragraph{Response-domain sensitivity of the hierarchy.}
For this comparison keep the three-generation content and its projections fixed.
Hold $\tau$ and $\MP$ fixed together with $\alpha=1/(16\pi^2)$. The derived scale
$\MG=\MP\sqrt\pi/c$ changes with $c$. Here $\alpha$ is the
fixed loop factor. The mass relations in
Sections~\ref{sec:vev} and~\ref{sec:derivmass} give
\begin{equation}
 \frac{v_\infty}{v_5}
 =\frac{c_{3,5}}{c_{3,\infty}}
   e^{-4\pi(c_{3,\infty}-c_{3,5})},\qquad
 \frac{m_{e,\infty}}{m_{e,5}}
 =e^{-20(c_{3,\infty}-c_{3,5})}.
 \label{eq:p5-domain-hierarchy-ratios}
\end{equation}
The subscripts distinguish the two endpoint arguments of the same
displayed closed forms. The finite-domain endpoint is enclosed by
\eqref{eq:p5-production-endpoint-enclosure}.
Combining the unrounded endpoint enclosures and evaluating the above
ratios with the same outward arithmetic certifies
\begin{equation}
 \frac{v_\infty}{v_5}\in[0.10,0.13],\qquad
 \frac{m_{e,\infty}}{m_{e,5}}\in[0.03,0.04].
 \label{eq:p5-domain-hierarchy-enclosures}
\end{equation}
Both ratios increase with $c_{3,5}$ and decrease with $c_{3,\infty}$,
so opposite interval endpoints bound them. They quantify a change of
integration domain at fixed remaining inputs. The numerical tables
throughout this paper use $R=5$.

\subsection{Sensitivity of the hierarchy}\label{sec:senshier}

The endpoint $\kL$ and torsion modulus $\tau$ follow from the spectral endpoint equation and normalized finite-source response of Section~\ref{sec:struct}. Varying them measures the sensitivity of the dependent quantities to those response relations. The observed anchor $\GN$ carries an experimental relative uncertainty of order $2\times10^{-5}$. An output proportional to $\MP^d$ has elasticity $-d/2$ with respect to $\GN$; dimensionless outputs are invariant under a common $\MP$ rescaling. The structural constants $2$, $\pi$, $3/2$, $1/8$ and the geometric factors remain exact.

Table~\ref{tab:sensitivity} gives $d\ln O/d\ln p$ for
$p=\kL,\tau,\MP$. The strongly amplified $c=\kL$ derivatives are
evaluated analytically from Appendix~\ref{app:deriv}. At fixed content with fixed
$\tau$ and $\MP$ the breaking scale $k_{\rm GUT}=\MP\sqrt{\pi}\tau/\sqrt3$
and $\alpha=1/(16\pi^2)$ are independent of $c$. Hence
$v\propto c^{-1}e^{-4\pi c}$ and $m_e\propto e^{-20c}$ and
\[
 \begin{aligned}
 \mathcal E_c O&:=c\,\partial_c\ln O,&
 \mathcal E_c v&=-1-4\pi c,\\
 \mathcal E_c m_{\nu_3}&=2\mathcal E_c v,&
 \mathcal E_c\rho_\Lambda&=2\mathcal E_c H_0=8\mathcal E_c v.
 \end{aligned}
\]
At $c=2.49353$ these give
$\mathcal E_c\rho_\Lambda\simeq-258.67697694$ and
$\mathcal E_c H_0\simeq-129.33848847$. The other entries are rounded to the precision shown in the table.
Tangent estimates scaled to a one-percent input change are $-32\%$ and $-50\%$ for $v$ and $m_e$ respectively and $-259\%$ for
$\rho_\Lambda$. Finite variations of positive outputs use
\eqref{eq:p5-exact-error-transfer} or the defining expression for the observable.
The listed mass ratio has an order-unity elasticity while
$\Omega_\Lambda=2/3+r_{23}/(3\pi)$ has exactly zero elasticity.

\begin{table}[t]
\centering
\caption{The local elasticity matrix with respect to $\kL$ or $\tau$ or $\MP$. Its entries describe infinitesimal relative changes. The large negative $\kL$ entries arise from the hierarchy factors $c^{-1}e^{-4\pi c}$ and $e^{-20c}$ and their powers.}\label{tab:sensitivity}
\begin{tabular}{lrrr}
\toprule
 & \multicolumn{3}{c}{Elasticity $d\ln O/d\ln p$} \\
\cmidrule(lr){2-4}
Quantity $O$ & $p=\kL$ & $p=\tau$ & $p=\MP$ \\
\midrule
$M_{\mathrm{G}}$ & $-1.0$ & $0.0$ & $+1.0$ \\
$g_2(M_{\mathrm{G}})$ & $-2.5$ & $0.0$ & $0$ \\
$m_t/m_c$ & $+5.0$ & $-0.1$ & $0$ \\
$v$ & $-32.3$ & $-0.1$ & $+1.0$ \\
$m_e$ & $-49.9$ & $-0.1$ & $+1.0$ \\
$m_{\nu_3}$ & $-64.7$ & $-1.1$ & $+1.0$ \\
$\rho_\Lambda$ & $-258.7$ & $-4.5$ & $+4.0$ \\
$H_0$ & $-129.3$ & $-2.2$ & $+1.0$ \\
$\Omega_\Lambda$ & $0.0$ & $0.0$ & $0$ \\
\bottomrule
\end{tabular}
\end{table}

For a threshold $M>0$ at fixed spectral coefficient $A_R$ the endpoint
is $c=A_R/M$ and its threshold elasticity is exactly $-1$.
The threshold-only diagnostic in Table~\ref{tab:conventions} holds the
original $p=2$ spectral trace fixed and replaces $4/27$ by
\[
 M(p)=\max_{0\leq y\leq1}y(1-y)^p
     =\frac{p^p}{(p+1)^{p+1}},\qquad
 c_{\rm thr}(p)=\frac{A_R(2)}{M(p)},\qquad p>0.
\]
Thus $d\log c_{\rm thr}/d\log p=-p\log[p/(p+1)]$. At
$p=2$ this gives the recorded $+0.81\%$ tangent response per one-percent
change in $p$ at fixed trace. A change of the actual kernel also varies
the trace. Define $A_R(p)$ by replacing the square
in \eqref{eq:p5-homogeneous-coefficient} by the power $p$. Then
\begin{equation}
 \frac{d\log c}{d\log p}
 =\frac{pA_R'(p)}{A_R(p)}-p\log\frac{p}{p+1},
 \qquad c(p)=\frac{A_R(p)}{M(p)}.
 \label{eq:p5-kernel-threshold-sensitivity}
\end{equation}
Differentiating under the fixed-domain integral inserts
$\log[(u^2+a_j)/(u^2+a_j+b_j)]$. Its weighted mean is strictly
negative for the present positive-mass content and supplies the additional
trace contribution to the kernel sensitivity.

\begin{table}[t]
\centering
\caption{Local tangent diagnostics scaled to a one-percent input change. The first row varies the threshold $M(p)$ at fixed trace $A_R(2)$. The full kernel derivative is given by Eq.~\eqref{eq:p5-kernel-threshold-sensitivity}. Finite changes use the exact propagation formulas.}\label{tab:conventions}
\begin{tabular}{lcccc}
\toprule
Convention (1\% shift) & $\to \kL$ & $\to v$ & $\to m_e$ & $\to \rho_\Lambda$ \\
\midrule
Threshold $M(p)$ at $p=2$ & $+0.8\%$ & $-26\%$ & $-40\%$ & $-210\%$ \\
Threshold $4/27$ & $-1.0\%$ & $+32\%$ & $+50\%$ & $+259\%$ \\
Window $\kL$ & --- & $-32\%$ & $-50\%$ & $-259\%$ \\
\bottomrule
\end{tabular}
\end{table}

The extremum of Appendix~\ref{sec:derivkl} gives the exact local
identity $y(1-y)^2=4/27-(y-1/3)^2+(y-1/3)^3$. The change of this
normalisation starts quadratically at $y=1/3$ whereas the
threshold-to-endpoint and endpoint-to-hierarchy maps retain their
nonzero local elasticities. Equation~\eqref{eq:p5-exact-error-transfer}
gives the corresponding finite output changes.

A separate comparison replaces the Gaussian weight by the hard
spectral window $\Theta(\Lambda^2-\omega^2-\lambda_j)$ while retaining
the response kernel~\cite{Guo2026GenesisFramework}. It gives $\kL=1.09$
compared with $2.49$ for the Gaussian profile. Its capacity
$1.09^2=1.19<1.5$ lies below the first positive spinor position.
This change of response weight is distinct from the FRG regulator
$R_k(q^2)=Z_k(k^2-q^2)\Theta(k^2-q^2)$ of
Eq.~(1.2) of Ref.~\cite{Litim2001}. The Gaussian tail bounds above
concern numerical evaluation at the fixed Gaussian weight.

\subsection{Common-scale propagation in the QCD readout}
\label{sec:p5-qcd-sensitivity}

For the retained map \eqref{eq:p5-qcd-scale-map} write
$\Lambda(\mu,a)=\mu(b_0a/4\pi)^{-p}e^{-q/a}$.
At fixed five-flavour coefficients the exact finite transfer is
\begin{equation}
 \frac{\Lambda(\mu',a')}{\Lambda(\mu,a)}
 =\frac{\mu'}{\mu}\left(\frac{a'}a\right)^{-p}
 \exp\left[-q\left(\frac1{a'}-\frac1a\right)\right],
 \qquad \mu,\mu',a,a'>0.
 \label{eq:p5-qcd-finite-transfer}
\end{equation}
Its local differential is
$d\log\Lambda=d\log\mu+(q/a-p)d\log a$.
At the effective coupling $a=0.11798585$ the coupling
elasticity is approximately $6.6172$. Thus applying
$f=1-2\tau\kappa/8$ to $a_{\rm raw}$ gives the scale factor
$f^{-p}\exp[-q(1/f-1)/a_{\rm raw}]$.
At $\tau=1/50$ the scale multiplier is about $0.96325$ whereas
the coupling multiplier $f$ is about $0.99434$.
If a certified input box has
$0<a_-\leq a\leq a_+<q/p$ and
$0<\mu_-\leq\mu\leq\mu_+$ monotonicity gives
$\Lambda\in[\Lambda(\mu_-,a_-),\Lambda(\mu_+,a_+)]$.
The enclosure remains valid when the inputs are correlated.

Keep the catalogue labels and $\tau$ fixed. All of its masses and
$T_d$ have scale elasticity one and $\sigma$ has elasticity two.
In particular
\begin{equation}
 \begin{split}
 \|M_{\mathcal C}(\Lambda')-M_{\mathcal C}(\Lambda)\|
 &=12|\Lambda'-\Lambda|,\\
 \|M_{\mathcal C}^2(\Lambda')-M_{\mathcal C}^2(\Lambda)\|
 &=144|\Lambda'^2-\Lambda^2|.
 \end{split}
 \label{eq:p5-qcd-operator-error}
\end{equation}
Both equalities follow from the largest listed label $18$.
Consequently the mass ratios and $\sigma/T_d^2$ are exactly independent of
the common extraction scale at fixed $\tau$.
If an uncertainty model is assigned to shared upstream inputs
$\theta$ their first-order covariance is
$\operatorname{Cov}(\log O)=J\operatorname{Cov}(\theta)J^T$
with $J_{ai}=\partial_{\theta_i}\log O_a$.
The common $\Lambda$ contribution alone is proportional to
$vv^T$ where $v=(1,1,1,1,1,2)$ for the four masses and $T_d$ together with
$\sigma$. Thus a shared extraction-scale uncertainty contributes a rank-one
covariance to these six readouts.

\subsection{Finite errors of the shared cosmological readouts}
\label{sec:p5-cosmology-error}

Use $x,D,A_\gamma,T_E$ from
Eq.~\eqref{eq:p5-cosmology-shared-ratio}. The dimensionless normalized
ratio and its exact residual are
\begin{equation}
\begin{aligned}
 \widehat R&:=
 \frac{T_E^2/(\MP H_0)}{A_\gamma^2\sqrt{D/(\pi Y_u)}}
 =f(x):=\frac{(1-x)^2}{\sqrt{1-4x}},\\
 f(x)-1&=
 \frac{x^2(6-4x+x^2)}
 {\sqrt{1-4x}\{(1-x)^2+\sqrt{1-4x}\}}.
\end{aligned}
\label{eq:p5-cosmology-normalized-ratio}
\end{equation}
For $0\leq x\leq X<1/4$ differentiation and rationalization give
\begin{equation}
\begin{gathered}
 f'(x)=\frac{6x(1-x)}{(1-4x)^{3/2}},\qquad
 f''(x)=\frac{6(1+2x^2)}{(1-4x)^{5/2}}\geq6,\\
 3x^2\leq f(x)-1\leq
 \frac{6x^2}
 {\sqrt{1-4X}\{(1-X)^2+\sqrt{1-4X}\}}.
\end{gathered}
\label{eq:p5-cosmology-ratio-error}
\end{equation}
Since $f(0)=1$ and $f'(0)=0$, the lower bound follows by integrating
$f''\geq6$ twice. In the exact residual the numerator is at most
$6x^2$ and the denominator decreases on the stated interval to prove
the upper bound. At $X=1/20$ its coefficient is less than $15/4$.
The recorded $\tau=1/50$ gives
$x^4=676/(25\cdot23^5)<(1/20)^4$. Hence the certificate applies.
Here $x\simeq0.04527327$ and $\widehat R-1\simeq0.00725839$.
The normalized ratio starts at second order in $x$. The full ratio also
contains $A_\gamma(\tau)$ and $D$ together with $Y_u$. For two
admissible endpoint tuples put
$\delta\log q=\log(q'/q)$. Then
\begin{equation}
\begin{aligned}
 \delta\log\frac{T_E^2}{\MP H_0}
 &=2\delta\log A_\gamma
   +\tfrac12\delta\log D-\tfrac12\delta\log Y_u
   +\log\frac{f(x')}{f(x)},\\
 \left|\log\frac{f(x')}{f(x)}\right|
 &\leq\frac{6X}{(1-X)(1-4X)}|x'-x|,
 \qquad x,x'\in[0,X].
\end{aligned}
\label{eq:p5-cosmology-common-error}
\end{equation}
The second line uses $(\log f)'=6x/[(1-x)(1-4x)]$.
The common mass cancels before uncertainty propagation.
Likewise $\Omega_\Lambda=D/(3\pi)$ is exactly independent of the
common mass and dimensional anchor at fixed content.

The logarithmic amplitude shift corresponding to $v/v_0=1-x$ is
$-\log(1-x)$ with
\[
 \frac{x^2}{2}\leq-\log(1-x)-x
 \leq\frac{x^2}{2(1-x)},\qquad 0\leq x<1.
\]
This follows from the positive logarithmic series and distinguishes the
exact logarithmic change from its first-order insertion.

\subsection{Uniform scalar-gap bounds}
\label{sec:p5-uniform-gap-certificate}

The results ``Scale transport and a uniform scalar gap'' and
``Uniform truncation bounds on a condensed scale interval'' of
Ref.~\cite{Guo2026CompactInternalSpaceI} apply to the complete scalar readout
on $I=[k_{\min},\MG]$ for $k_{\min}>0$ with $L_k=c\MG/k$ and
$c>\sqrt\pi$ and specified positive $C^1$ paths $g_k,Z_k$.
They provide a common truncation and uniform positive gap once
two-sided bounds for $g_kL_k^2$ and $8(L_k^2-\pi)/(\pi g_kL_k^4)$ together with
$Z_k$ have been supplied. For the same potential and weights with the same subtraction and normalization write
$\Gamma=\inf_{k\in I}m_\delta^2(k)>0$. The uniform truncation estimate
then permits a finite $N$ for which
\[
 \sup_{k\in I}|m_{\delta,N}^2(k)-m_\delta^2(k)|
 \leq\varepsilon_N<\Gamma/2,
 \qquad
 \inf_{k\in I}m_{\delta,N}^2(k)\geq\Gamma/2.
\]
The choice of $N$ is uniform on $I$ once the stated path bounds have
been supplied. The carrier includes the constant scalar shell.
The QCD catalogue has a separate scale dependence quantified in
Section~\ref{sec:p5-qcd-sensitivity}.

\subsection{The spectral-scale flow and the renormalisation group}\label{sec:rgflow}

The internal response fixes the normalization of the connection and
matter vertices. Written in canonical fields these coefficients supply
the boundary action whose quantum fluctuations determine the scale
dependence of the effective couplings. The representation traces in the
two-loop equations use the same retained field content as the
spectral construction~\cite{LuoXiao2003,LuoWangXiao2003}.
The spectral boundary data are expressed at $\MG$ with channel
matching at $k_{\rm GUT}$ as specified in Section~\ref{sec:g1}.
Running and threshold matching then express this action at the
lower comparison scales. Finite mass and vertex matching connects the
geometric readouts to renormalized parameters as in
\eqref{eq:p5-canonical-action-boundary} and the one-loop construction
of Section~\ref{sec:interface}~\cite{MartinRobertson2019}.
The production calculation uses the stated fixed Yukawa value whereas
the five-coupling evolution transports the matched Yukawa and quartic
coefficients together with the gauge couplings. Their perturbative
accuracy is determined by the loop order and threshold treatment.

The ``Distribution-valued Gaussian spectral completion'' theorem in
Ref.~\cite{Guo2026CompactInternalSpaceI} supplies the field with the specified
spectral covariance. The scalar estimates above control the boundary
spectral functional. The running and matching errors are classified
by their respective perturbative orders.

%% file: sections-en/appendix_e.tex
% =============================================================================
% Appendix E - Neutrino and quark-mixing relations
% The Spectrum of a Compact Internal Space II: Effective Couplings and Mass Scales
% =============================================================================

\section{Neutrino and quark-mixing relations}\label{app:flavour}

The neutrino masses and mixing matrices complete the flavour relations
of Section~\ref{sec:flavour}. The lightest rest mass and the
baryon-asymmetry estimate enter the cosmological densities in
Section~\ref{sec:cosmo}. Table~\ref{tab:params} distinguishes the
rest-mass assignments from the additional propagation and mixing
assumptions.

\subsection{Neutrinos}\label{sec:neutrino}

Appendix~\ref{sec:derivnu} defines the Weinberg-operator normalisation and
family scale and squash insertion in \eqref{eq:p5-neutrino-scale} with
the hierarchy ratios in \eqref{eq:p5-neutrino-ratios} and the ordered
rest-mass spectrum in \eqref{eq:p5-neutrino-rest-spectrum}.
The flavour construction combines this spectrum with the
near-tribimaximal mixing-angle assignments in \eqref{eq:p5-pmns-angles}.

In the charged-lepton mass basis the columns of $U$ give the
orientations of the neutrino mass states relative to the flavours
selected by the charged current. The solar angle sets the relative
electron-flavour weights of the light pair. The reactor angle sets
the electron weight of the third state and the atmospheric angle
sets its muon--tau partition. With these angles and the Dirac-phase
readout of Appendix~\ref{sec:derivckm} the numerical
construction~\cite{Guo2026GenesisFramework} gives a unitary lepton
mixing matrix. The two Majorana phases remain free. For $D=\operatorname{diag}(m_{\nu_i})$
the flavour mass texture and vacuum propagation numerator are
\[
 M_\nu=U^*DU^\dagger,\qquad H_2=UD^2U^\dagger,\qquad
 U^{\mathsf T}M_\nu U=D .
\]
The Takagi relation fixes the correspondence between each mass and
its column of $U$. The Majorana phases cancel from $H_2$. Setting
all phases to zero gives the real symmetric mass texture.
The solar scalar source acts through the spectral projector in
\eqref{eq:p5-solar-spectral-projector}.

Solar production prepares an electron-flavour state. The scalar
source changes the second mass-state energy while the electron
matter potential acts along the electron-flavour direction. Their
relative orientation determines the instantaneous propagation
states and the conversion between them as the density changes
\cite{Wolfenstein1978,MikheyevSmirnov1986}. The Hermitian Hamiltonian is
\[
 H(E,x)=\frac{UD^2U^\dagger}{2E}+V_{\rm SM}(x)+\Delta H_\Sigma(E,x).
\]
For the source in \eqref{eq:p5-solar-self-energy}
\[
 V_{\rm SM}=\sqrt2G_Fn_e(x)P_e,\qquad
 P_e=\operatorname{diag}(1,0,0),\qquad
 \Delta H_\Sigma=\frac{d_\nu\zeta(x)}{2E}P_2.
\]
The scalar channel $P_2$ and its tilt-dependent strength fix the
mass-state response. An environmental realization specifies the scalar
interaction and the map from the state of the medium to $\zeta(x)$.
These data then determine the source profile in each propagation
experiment. The mass-squared insertion produces the displayed $E^{-1}$
term whereas coherent electron forward scattering supplies the vector
potential~\cite{GeParke2019}. The fixed interaction and
map from the medium to its source apply to solar, terrestrial and antineutrino
propagation. The correlated changes of all three splittings in
\eqref{eq:p5-correlated-solar-response} provide an additional test of
that common coupling.
Vacuum propagation uses $\zeta=0$. The antineutrino
Hamiltonian conjugates the two mass terms and reverses the
vector matter potential. Given the solar source and electron-density
profiles let $S_\odot(E;x)$ propagate from production point $x$
to the solar surface and let $f_E(x)$ be the normalized production
density. Propagation through the Sun sets the mass populations at
its surface. When the distinct vacuum mass states decohere on the
journey to Earth the detector receives an incoherent mixture of
these populations~\cite{Parke1986}. Their production average and
detection probabilities are
\begin{equation}
 \begin{aligned}
 p_i(E)&=\int f_E(x)
       \left|[U^\dagger S_\odot(E;x)]_{ie}\right|^2\,dx,\\
 P_{ee}^{\rm day}(E)&=\sum_i p_i(E)|U_{ei}|^2,\\
 P_{ee}^{\rm night}(E)&=\sum_i p_i(E)
                   |[S_\oplus(E)U]_{ei}|^2 .
 \end{aligned}
 \label{eq:p5-solar-forward-readout}
\end{equation}
Here $S_\oplus$ includes the Earth's matter and any specified
scalar profile. Experimental energy and path averages act on
these probabilities. They define the solar observables of the
scalar-source model. A source commuting with the
vacuum numerator throughout a matter-free trajectory changes
only phases. After mass decoherence its phases leave the
$p_i$ unchanged. The noncommuting electron potential and the
production average therefore enter a solar realization explicitly.
The propagation calculation preserves probability and has the
vacuum and two-flavour limits~\cite{Guo2026GenesisFramework}.

The raw and unit-source solar splittings are given in
\eqref{eq:p5-neutrino-raw-splitting} and
\eqref{eq:p5-neutrino-splittings}. Their residuals against the
NuFIT~5.2 reference are $-7.231\%$ and $+0.030\%$
respectively~\cite{NuFIT52}.
The unit source entry evaluates the propagation numerator at $\zeta=1$. The solar comparison averages the
probabilities in \eqref{eq:p5-solar-forward-readout} over production and detection with the fixed
interaction and medium profile.
Equation~\eqref{eq:p5-correlated-solar-response} gives the scalar
projector and all three correlated splitting shifts and the cancellation
of the common mass scale. Table~\ref{tab:params} lists the rest-mass
and unit-source values separately with the latter in Tier~D.
The absolute masses and their sum retain the rest-mass definition. In particular $m_{\nu_1}$ supplies the dark-energy density in
Section~\ref{sec:entropy}.

The rest spectrum also determines
\[
m_\beta^2=\sum_i|U_{ei}|^2m_i^2,\qquad
\max\!\left(0,2\max_i a_i-\sum_i a_i\right)
\leq m_{\beta\beta}\leq\sum_i a_i,
\quad a_i=|U_{ei}|^2m_i .
\]
The interval is the attainable range over the two Majorana phases.
The rest spectrum and mixing angles above give $m_\beta=0.0090585\ \mathrm{eV}$ and
$0\leq m_{\beta\beta}\leq0.0054362\ \mathrm{eV}$ under the light-Majorana
exchange interpretation~\cite{Guo2026GenesisFramework}. Both quantities
refer to the vacuum rest-mass texture.

\subsection{CKM and CP}\label{sec:ckm}

Appendix~\ref{sec:derivckm} specifies the Cabibbo input
$\mathfrak r_G$ in \eqref{eq:p5-gatto-input} and its distinction
from the mass ratio $m_d/m_s$. Equations~\eqref{eq:p5-ckm-cabibbo}
and \eqref{eq:p5-ckm-offdiagonal} give the three mixing elements
including the adjacent- and next-generation suppressions.
Their common attenuation cancels in the mass--mixing combinations
\eqref{eq:p5-mixing-transfer-relations}.
The phase with colour dilution in \eqref{eq:p5-ckm-phase} sets the complex
relative orientation of the up and down mass sources. The Jarlskog
invariant measures the CP-odd part that is preserved under rephasing
of their modes~\cite{Jarlskog1985}. Its leading expression
\eqref{eq:p5-ckm-jarlskog-leading}
gives $J=3.15\times10^{-5}$. The omitted $c_{13}$ dependence
enters at relative order $|V_{ub}|^2$.
The raw asymmetry in \eqref{eq:p5-baryon-raw} transfers this signed
CP measure to the baryon readout with the weak weight
$\alpha_W(v)^2$ where $\alpha_W(v)=g_2(v)^2/(4\pi)$ and the
content normalisation $N_LN_R=56$, giving
$\eta_B^{\mathrm{raw}}=6.09\times10^{-10}$.
This value accompanies the raw photon floor. The conversion
$T_{\rm CMB}=C_\gamma T_{\rm raw}$ with
$C_\gamma=1-\tau/\pi^2$ changes the photon number density by
$C_\gamma^3$. Holding the baryon number density fixed therefore gives
the physical baryon-to-photon ratio
$\eta_B=\eta_B^{\rm raw}(1-\tau/\pi^2)^{-3}
=6.12823\times10^{-10}$ while the baryon density is unchanged.
Appendix~\ref{sec:derivckm} specifies the content factors and their
relation to Sakharov's conditions. This effective asymmetry enters the homogeneous baryon density.

%% file: sections-en/appendix_f.tex
% =============================================================================
% Appendix F - Primordial-nucleosynthesis relations
% The Spectrum of a Compact Internal Space II: Effective Couplings and Mass Scales
% =============================================================================

\section{Primordial-nucleosynthesis relations}\label{app:bbnext}

The masses and couplings of Sections~\ref{sec:gauge}, \ref{sec:cosmo} and \ref{sec:qcd}, with the Cabibbo input of Appendix~\ref{sec:derivckm}, give a reduced estimate of the primordial helium abundance from weak freeze-out. Appendix~\ref{sec:derivbbn} specifies the thermal history and weak rates
used in this estimate. The precision calculation of
Ref.~\cite[Secs.~III.G--H and IV.C]{Pitrou2018} includes radiative and
finite-temperature weak-rate corrections and a nuclear reaction network. Here the abundance is estimated from neutron freeze-out and decay.

\subsection{Primordial helium}\label{sec:bbn}

The calculation uses $g_A=4/\pi$ and $G_F=1/(\sqrt2v^2)$ and the
Cabibbo approximation $|V_{ud}|=\sqrt{1-|V_{us}|^2}$ at leading order in
$|V_{ub}|^2$. The native isovector response is $g_S^{\rm eff}=1$. The mass difference $d=\Delta m_{np}$ includes the electromagnetic
subtraction in Eq.~\eqref{eq:p5-bbn-mass-difference}. The low-energy
QED coupling and radiative factor are specified in
Eqs.~\eqref{eq:p5-bbn-qed-descent} and~\eqref{eq:p5-bbn-radiative}.
The electron mass in the weak rates and decay phase space is
$m_e:=M_e=G_e m_e^{\rm geom}$ with the displayed calculation using
$G_e=1$ in Eq.~\eqref{eq:p5-lepton-mass-definitions}.
With these inputs the Coulomb-weighted phase space
\eqref{eq:p5-neutron-phase-space} gives the static-nucleon lifetime
\eqref{eq:p5-neutron-lifetime} and the weak-rate balance
\eqref{eq:p5-bbn-freezeout} determines $T_f$.
With $r_f=\exp(-d/T_f)$ and
$q_{\rm dec}=\exp(-\Delta t/\tau_n)$, baryon number conservation
and complete capture of the surviving neutrons give
\[
 X_n=\frac{r_fq_{\rm dec}}{1+r_f},\qquad Y_p=2X_n,
\]
with the neutron-to-proton ratio given by
Eq.~\eqref{eq:p5-bbn-conserved-abundance}. The separate neutrino-number
assignment \eqref{eq:p5-bbn-neutrino-number} can also be written as
$N_{\mathrm{eff}}=3(1+\delta_N)$ with
$\delta_N=\sqrt3/[3(2\pi)^2]$.
Here $\Delta t$ is the elapsed time from weak freeze-out to the adopted
nucleosynthesis temperature measured by the homogeneous comoving clock.
A rest-frame width is the decay rate with respect to proper time $s$ through
$P(s)=\exp[-\int\Gamma(s')\,ds']$ in natural units. For constant
$\Gamma_n$ the conversion to seconds gives $\tau_n=\hbar/\Gamma_n$. The same clock convention is used for the decay interval and lifetime.
The calculation gives
$T_f=0.75394\ \mathrm{MeV}$ and the three linked readouts in
Table~\ref{tab:p5-weak-chain}~\cite{Guo2026GenesisFramework}.
The two-species decay equations $\dot n=-n/\tau_n$ and
$\dot p=n/\tau_n$ give the same abundance. The two thermal values of $g_*$ are held fixed when applying the
$N_{\mathrm{eff}}$ assignment in the time-temperature relation~\eqref{eq:p5-bbn-time-temperature}.

\begin{table}[t]
\centering
\caption{Common weak-decay and helium calculation. The mass difference
and lifetime references are from the PDG~2024 neutron listing
\cite{PDG2024}. Residuals compare central values. The reduced weak
and capture calculation has no assigned statistical theory covariance.}
\label{tab:p5-weak-chain}
\begin{tabular}{lccc}
\toprule
Quantity & Calculated & Reference & Residual \\
\midrule
$\Delta m_{np}$ (MeV) & $1.28497$ & $1.29333236(46)$ & $-0.647\%$ \\
$\tau_n$ (s) & $915.54$ & $878.4\pm0.5$ & $+4.23\%$ \\
$Y_p$ & $0.246082$ & $0.245$ & $+0.44\%$ \\
\bottomrule
\end{tabular}
\end{table}

\subsection{Common-input response}
Write $d=\Delta m_{np}$ and $q=d/m_e$ and
$f=f(q,\alpha_{\rm em})$. The mass and lifetime differentials are
\begin{align}
 \mathrm d d&=(m_d-m_u)\,dg_S^{\rm eff}
           +g_S^{\rm eff}(dm_d-dm_u)-d\Delta_{\rm EM},\notag\\
 d\log\tau_n
 &=-2d\log G_F-2d\log|V_{ud}|
   -\frac{6g_A^2}{1+3g_A^2}d\log g_A-d\log\delta_R\notag\\
 &\quad -5d\log m_e-\frac{f_q\,dq+f_\alpha\,d\alpha_{\rm em}}{f}.
 \label{eq:p5-weak-joint-lifetime}
\end{align}
For the retained freeze-out kernel let $C_f$ multiply the weak rate
with $C_f=1$ in the calculation. Its root obeys
$C_fG_F^2(1+3g_A^2)M_PT_f^2(T_f+3d/2)=\text{constant}$.
At fixed relativistic degrees of freedom
\begin{equation}
 d\log T_f=
 -\frac{T_f+3d/2}{3(T_f+d)}
 d\log[C_fG_F^2(1+3g_A^2)M_P]
 -\frac{\mathrm d d}{2(T_f+d)} .
 \label{eq:p5-weak-joint-freeze}
\end{equation}
If $t_f$ and $t_B$ denote the two radiation-era times then fixed capture
temperature gives
$d\Delta t=\Delta t\,d\log M_P+2t_f\,d\log T_f$.
Writing $x=d/T_f$ and $r=e^{-x}$ then yields
\begin{equation}
 d\log Y_p=-\frac{dx}{1+r}
       -\frac{d\Delta t}{\tau_n}
       +\frac{\Delta t}{\tau_n}d\log\tau_n .
 \label{eq:p5-weak-joint-helium}
\end{equation}
These derivatives retain the change of the freeze-out abundance as
well as subsequent decay.

Order the columns as $g_S^{\rm eff}$, $\Delta_{\rm EM}$, $g_A$,
$\delta_R$ and $G_F$, and the rows as $\tau_n$ and $Y_p$.
The central logarithmic Jacobian is
\[
 J_{\rm weak}=
 \begin{pmatrix}
 -13.0288&6.49251&-1.65890&-1&-2\\
 -6.69407&3.33578&-1.41785&-0.223789&-1.70938
 \end{pmatrix}.
\]
The electromagnetic subtraction and the Coulomb factor also share
$\alpha_{\rm em}$. Their common variation is composed before
propagating an input covariance. With an independently specified
$\Sigma_\theta$ the covariance of the logarithmic outputs is
$J\Sigma_\theta J^{\mathsf T}$. Interval input errors instead give
first-order bounds through $|J|$ with nonlinear remainders controlled
over the selected interval. The weak rates, hadronic matching and
nuclear capture approximations require their own error estimates
\cite{Pitrou2018}. The reported lifetime residual compares central
values. Its statistical interpretation depends on these theory errors
as well as the experimental uncertainty.

The response calculation~\cite{Guo2026GenesisFramework} evaluates $f$ once and
then integrates $f_q,f_\alpha$ together on the accepted adaptive
partition with local refinement for the derivatives. The recorded
calculation uses $567$ central-integrand evaluations and $576$
derivative-integrand evaluations. The remaining chain is differentiated
analytically. Independent displaced-input
calculations check the displayed five-input Jacobian. The lifetime and helium
responses have different weights. Their joint comparison tests
more than either abundance or decay separately.

\subsection{Mass-difference diagnostic}
At fixed electron mass and other weak inputs the phase-space response gives
\begin{equation}
 \left.\frac{\partial\log\tau_n}{\partial\log d}\right|_{m_e,\alpha_{\rm em},G_F,V_{ud},g_A,\delta_R}
 =-\frac{qf_q}{f}=-6.53633 .
 \label{eq:p5-weak-gap-elasticity}
\end{equation}
The $-0.647\%$ mass-difference residual thus corresponds locally to
a lifetime shift of about $+4.23\%$ with the same sign and magnitude
as the reported lifetime residual. To test this sensitivity over the
finite displacement we replace only $d$ by the registered experimental
value $1.29333236\ \mathrm{MeV}$ and propagate it through the phase space and
freeze-out root together with the decay interval. Table~\ref{tab:p5-weak-gap-diagnostic}
compares this externally conditioned calculation with the generated
chain~\cite{Guo2026GenesisFramework}. All other inputs and the capture temperature
and weak-rate approximations are held fixed.

\begin{table}[t]
\centering
\caption{Finite mass-difference diagnostic. The second calculation uses
the experimental neutron--proton mass difference as an external condition.
Both columns use the same weak and capture approximations.}
\label{tab:p5-weak-gap-diagnostic}
\begin{tabular}{lcc}
\toprule
Quantity & Generated mass difference & Experimental mass difference \\
\midrule
$f$ & $1.63492371$ & $1.70563405$ \\
$T_f$ (MeV) & $0.75394339$ & $0.75240100$ \\
$\tau_n$ (s) & $915.54351$ & $877.58790$ \\
$Y_p$ & $0.24608238$ & $0.24071583$ \\
\bottomrule
\end{tabular}
\end{table}

The beta-phase-space enclosure gives
\[
\tau_n\in[915.5097706,915.5853079]\ \mathrm{s},
\qquad
Y_p\in[0.2460803490,0.2460848926].
\]
These ranges enclose the phase-space evaluation at the displayed inputs.
The weak-rate and nuclear-capture model spread is assessed separately.

The conditioned lifetime has a central-value residual of $-0.0925\%$
while the helium residual is $-1.7486\%$. The mass difference therefore
accounts for most of the lifetime residual within this diagnostic.
Its simultaneous effect on freeze-out and neutron
survival is essential for assessing the whole chain. The replacement
probes the total mass difference. The strong isovector contribution
and electromagnetic subtraction require independent information in a
common separation convention. Propagating only the registered mass-difference
uncertainty gives $0.00203\ \mathrm{s}$ for the lifetime and
$2.94\times10^{-7}$ for $Y_p$ at the conditioned point. These are
input-uncertainty contributions. The physical matching and rate
errors remain those discussed above.

\subsection{Independent QCD--QED input}
The lattice calculation of Ref.~\cite[Table~1]{Borsanyi2015} gives
$\Delta m_{np}=1.51(16)(23)\ \mathrm{MeV}$ with statistical and
systematic errors respectively. Its separated contributions are
$\Delta m_{np}^{\rm QCD}=2.52(17)(24)\ \mathrm{MeV}$ and
$\Delta m_{np}^{\rm QED}=-1.00(07)(14)\ \mathrm{MeV}$.
The separation uses $\Delta_{\rm QED}M_\Sigma=0$ as specified in
Sec.~10.3 of its supplementary material. Propagating the published
total splitting and its errors retains their correlated
determination. The separately rounded contributions
serve to identify the source convention.

With all other weak and thermal inputs fixed the total central value
gives $\tau_n=331.476\ \mathrm{s}$ and $Y_p=0.116457$.
Positivity of the phase-space integral and monotonicity of the
freeze-out root give finite bounds for this propagation.
\paragraph{Finite mass-difference ranges.}
Fix the weak inputs and capture temperature and let
$m_e<d_-\leq d\leq d_+$ with capture after freeze-out throughout.
Write $T_\pm=T_f(d_\pm)$ and $\tau_\pm=\tau_n(d_\pm)$ with
$b_\pm=t_B-t_f(d_\pm)$ and
$z_\pm=(1+e^{d_\pm/T_\pm})^{-1}$. Then
\begin{equation}
 \tau_+\leq\tau_n(d)\leq\tau_-,\qquad
 2z_+e^{-b_-/\tau_+}\leq Y_p(d)
       \leq2z_-e^{-b_+/\tau_-} .
 \label{eq:p5-weak-gap-ranges}
\end{equation}
The phase integral has a positive weight multiplying $(q-W)^2$ on
$1<W<q=d/m_e$. Increasing $d$ increases the integral and decreases
$\tau_n$.  Equation~\eqref{eq:p5-weak-joint-freeze} gives
$T_f'(d)=-T_f/[2(T_f+d)]<0$. Hence $d/T_f$ increases, the initial neutron
fraction decreases and the positive decay interval $b=t_B-t_f$ also
decreases.  Combining $z_+\leq z\leq z_-$ with
$b_+/\tau_-\leq b/\tau_n\leq b_-/\tau_+$ gives the bounds.

\begin{table}[t]
\centering
\caption{Finite propagation of the external lattice total mass difference
through the fixed weak and capture model. The statistical and systematic
error sizes are varied separately. The last row uses their sum as an
input half-width. It has no assigned coverage probability. The helium
ranges follow the positive-factor bounds
\eqref{eq:p5-weak-gap-ranges}.}
\label{tab:p5-lattice-weak-ranges}
\begin{tabular}{lccc}
\toprule
Input variation & $d$ (MeV) & $\tau_n$ (s) & $Y_p$ \\
\midrule
Statistical & $[1.35,1.67]$ & $[181.54,665.67]$ & $[0.05293,0.20524]$ \\
Systematic & $[1.28,1.74]$ & $[142.81,939.05]$ & $[0.03429,0.24937]$ \\
Sum of half-widths & $[1.12,1.90]$ & $[86.20,2330.93]$ & $[0.009830,0.355290]$ \\
\bottomrule
\end{tabular}
\end{table}

The three ranges and their centre require seven distinct phase-space
integrals. Shared endpoints are reused. The estimated quadrature
contributions to the displayed endpoints are below
$10^{-9}\ \mathrm{s}$ for the lifetime and $6\times10^{-14}$ for
$Y_p$. These numerical estimates are separate from the input ranges
and the weak-rate and capture approximations.
The combined lattice-input range is wider than the correction indicated
by the lifetime residual. Resolving that correction
requires the combination
\eqref{eq:p5-effective-isovector-coefficient} and the electromagnetic
contribution in a common separation convention. The production weak
and cosmological values remain those in Table~\ref{tab:p5-weak-chain}.

\subsection{Continuous weak evolution}
With the same weak and hadronic inputs continuous freeze-out follows
from the six forward and reverse neutron--proton processes. Fermi
occupation and blocking factors are retained in the static-nucleon
approximation~\cite{Guo2026GenesisFramework}. The neutron fraction is evolved according to
\[
 \dot X_n=-\lambda_{n\to p}X_n+\lambda_{p\to n}(1-X_n)
\]
coupled to the photon--electron thermal bath with a specified neutrino
decoupling history. The zero-temperature decay integral reproduces the
lifetime above. Independently integrated rates obey detailed balance
at a common temperature and zero chemical potentials. With equilibrium
initial data at $2\ \mathrm{MeV}$ and instantaneous neutrino decoupling
at $1.2\ \mathrm{MeV}$ the example evolved to $0.5\ \mathrm{MeV}$ gives
$X_n=0.19335$. This value describes the neutron fraction at
$0.5\ \mathrm{MeV}$ before nuclear assembly. The helium abundance
quoted above uses the freeze-out and decay approximation with complete
neutron capture. Evolution through nuclear assembly additionally
depends on the nuclear reaction rates.

\paragraph{Weak-fraction bounds under continuous cooling.}
Let $x$ increase along cooling with $dt/dx>0$ and set
$a_+=(dt/dx)\lambda_{p\to n}\geq0$ and
$a_-=(dt/dx)\lambda_{n\to p}\geq0$ and
$A=a_++a_-$. For integrable rates the solution of
$X_n'=a_+(1-X_n)-a_-X_n$ stays in $[0,1]$ if its initial
value does. For an absolutely continuous approximation
$\widehat X$ with defect
$R=\widehat X'-a_+(1-\widehat X)+a_-\widehat X$ its error obeys
\[
 |\widehat X(x)-X_n(x)|
 \leq e^{-\int_{x_i}^{x}A}\,|\widehat X(x_i)-X_n(x_i)|
 +\int_{x_i}^{x}e^{-\int_s^x A}|R(s)|\,ds .
\]
The vector field points into $[0,1]$ at both endpoints.
Subtracting the exact and approximate equations gives
$e'+Ae=R$. The integrating factor proves the error bound.
The bound controls the integration error for fixed weak rates and
thermal history. The quadrature and evolution tolerances are specified
in Ref.~\cite{Guo2026GenesisFramework}.

%% file: sections-en/appendix_g.tex
% =============================================================================
% Appendix G - The acceleration scale and gravitational dynamics
% The Spectrum of a Compact Internal Space II: Effective Couplings and Mass Scales
% =============================================================================

\section{The acceleration scale and gravitational dynamics}\label{app:gravity}

The compact-source angular covariance sets the endpoint acceleration
variance. The Gaussian susceptibility compares squared response
amplitudes; its positive RMS readout determines the spatial flux.
The resulting constitutive energy has the nonrelativistic variational
form of Ref.~\cite{BekensteinMilgrom1984}. On a bounded Lipschitz
domain with $H^1$ Dirichlet data and an $H^{-1}$ source, strict
convexity gives a unique weak solution. Coupling the same potential
to pressureless trajectories makes the source respond to the field.
On a fixed smooth domain the joint action gives their energy balance
throughout smooth evolution before caustic formation.

The covariant analysis retains matter stress, source exchange and the
phase clock congruence in a common action. An auxiliary collective
radius closes its source balance on an occupied expanding background.
The independent-foliation action defines a separate effective
realization of the local constitutive law with the congruence varied
as an additional field. The kinetic form is reduced on the constraint
and variation space of each action.

\subsection{The endpoint acceleration variance}\label{sec:a0}

Set the terminal coarse-graining length to $\sigma_C=c/H_0$ where
$c$ denotes the speed of light throughout this appendix. The centered
maximum-entropy displacement law with variance $\sigma_C^2$ has the
Gaussian spectral multiplier
\[
W_{\mathrm{IR}}(\Delta)
=\exp\!\left(-\frac12\Delta^2\sigma_C^2\right).
\]
The Euclidean-period normalization used here assigns the wave number
\[
k_{\mathrm{IR}}=\frac{1}{2\pi\sigma_C}=\frac{H_0}{2\pi c}.
\]
The angular factor is evaluated on the compact source of
Section~4.5 of Ref.~\cite{Guo2026SpectralGravity}
under ``Visible tensor modes of the compact source''. Its vertex
normalization and occupation source are given there in
Eqs.~(4.27) and~(4.28) respectively. On the scalar shell
$\mathcal H_2\simeq\mathbb R^3\otimes\mathbb R^3$ the normalized
source in the two orthogonal five-dimensional compact TT spaces is
\[
\begin{aligned}
 \widehat j(N)&:=j_{\rm hom}(N)/\alpha_2
   =((\rho_L)_0,(\rho_R)_0),\\
 \rho_L&=\operatorname{Tr}_R N,\qquad
 \rho_R=\operatorname{Tr}_L N,\qquad
 B_0=B-\tfrac13\operatorname{tr}(B)I_3.
\end{aligned}
\]
Here $N\geq0$ is a real occupation matrix with $\operatorname{Tr}N=1$
and division by the inherited normalization $\alpha_2$ defines the angular
source. Unit trace fixes the total occupation while resolving its
distribution among the internal angular modes. In the Hilbert--Schmidt norm
\begin{equation}
 \|\widehat j(N)\|^2
 =\operatorname{tr}\rho_L^2+\operatorname{tr}\rho_R^2-\frac23
 \in[0,4/3].
 \label{eq:p5-endpoint-source-norm}
\end{equation}
The upper equality holds exactly for decomposable pure occupations
$N=|u\otimes v\rangle\langle u\otimes v|$ with $|u|=|v|=1$.
Each marginal then has purity one. Conversely, two pure marginals force
this product occupation. Equal occupation $N=I_9/9$ gives zero source.

On this decomposable orbit, independent uniform $u,v\in S^2$ induce its
unique $SO(3)\times SO(3)$-invariant reference probability by transitivity
of the compact group action. With
$Q_u=uu^{\mathsf T}-I_3/3$ the angular fourth moment
\[
 \mathbb E(u_i u_j u_k u_l)
 =\frac{\delta_{ij}\delta_{kl}+\delta_{ik}\delta_{jl}
                  +\delta_{il}\delta_{jk}}{15}
\]
gives the following covariance in orthonormal bases of the two TT spaces
\begin{equation}
\begin{aligned}
 \mathbb E N&=I_9/9,\qquad \mathbb E\widehat j=0,\\
 \operatorname{Cov}(\widehat j)&=\frac2{15}I_{10},\qquad
 C_{\mathrm{TT},3}:=\mathbb E\|\widehat j\|^2=\frac43.
\end{aligned}
 \label{eq:p5-endpoint-angular-covariance}
\end{equation}
Uniform orientations cancel the directional source in the ensemble
mean while their squared amplitudes add. The factor $4/3$ is
the mean square angular fluctuation on the decomposable
occupation orbit. The mean occupation and this covariance describe
different moments of the same specified ensemble. Assigning acceleration amplitude
$c^2k_{\mathrm{IR}}$ to the endpoint wave number gives
\begin{equation}\label{eq:sigmaira0}
\Sigma_{\mathrm{IR}}
=C_{\mathrm{TT},3}(c^2k_{\mathrm{IR}})^2
=\frac43\left(\frac{cH_0}{2\pi}\right)^2
=a_0^2,
\qquad
a_0=\frac{cH_0}{2\pi}\sqrt{\frac43}.
\end{equation}
The Hubble length that enters the homogeneous endpoint therefore also
sets the local acceleration normalization. Here $H_0$ labels the common
reference endpoint. The local law retains this value of $a_0$ while
the instantaneous expansion rate evolves with the macroscopic state.
With the content factors, decomposable occupation orbit and
period normalization fixed, a change
of the common neutrino mass normalization changes $H_0$ and $a_0$
together while the photon temperature follows its linear mass
readout. These are linked endpoint normalizations at the same reference
epoch.
The resulting scale is $a_0=1.204\times10^{-10}\ \mathrm{m\,s^{-2}}$.
Its central-value difference from the empirical
$g_\dagger=1.2\times10^{-10}\ \mathrm{m\,s^{-2}}$ is $+0.36\%$.
The radial-acceleration fit of Eq.~(4) of Ref.~\cite{McGaugh2016} gives
random and systematic uncertainties of $0.02$ and $0.24$ respectively
in the same units. We use Eq.~\eqref{eq:sigmaira0} as the acceleration
normalization in the local response.

\subsection{The endpoint variance and the local response}\label{sec:localresponse}

For $\lambda_{\mathrm{IR}}>0$ the retained one-mode quadratic form is
\[
S_2[q]=\frac12(\lambda_{\mathrm{IR}}+\Sigma_{\mathrm{IR}})q^2.
\]
The unscreened and endpoint variances are $\langle q^2\rangle_0=1/\lambda_{\mathrm{IR}}$ and $\langle q^2\rangle_{\mathrm{IR}}=1/(\lambda_{\mathrm{IR}}+\Sigma_{\mathrm{IR}})$. Define the variance ratio by
\begin{equation}\label{eq:zir}
Z_{\mathrm{IR}}
=\frac{\langle q^2\rangle_{\mathrm{IR}}}{\langle q^2\rangle_0}
=\frac{\lambda_{\mathrm{IR}}}
       {\lambda_{\mathrm{IR}}+\Sigma_{\mathrm{IR}}}.
\end{equation}
The quasi-static spatial entry is the squared acceleration
\[
\lambda_{\mathrm{IR}}=|\nabla\Phi|^2,
\qquad
y=\frac{|\nabla\Phi|}{a_0}.
\]
The same-source response identity of
Proposition~3.1 of Ref.~\cite{Guo2026SpectralGravity}
under ``Response of the sourced equations'' expresses the response
in terms of its kernel and source together with the readout and direct term. For the quadratic form above and the
same linear source $hq$ the normalized Gaussian integral gives exactly
\[
 \mathcal W_\Sigma(h)-\mathcal W_\Sigma(0)
 =\frac{h^2}{2(\lambda_{\mathrm{IR}}+\Sigma_{\mathrm{IR}})},
 \qquad
 \frac{\partial_h^2\mathcal W_\Sigma}
      {\partial_h^2\mathcal W_0}=Z_{\mathrm{IR}}.
\]
Thus $Z_{\rm IR}$ is the normalized susceptibility and compares
squared amplitudes under the same source. The spatial response reads
their positive RMS amplitude ratio, giving
\begin{equation}\label{eq:muderivation}
\mu(y)=\sqrt{Z_{\mathrm{IR}}}
=\sqrt{\frac{|\nabla\Phi|^2}
 {|\nabla\Phi|^2+a_0^2}}
=\frac{y}{\sqrt{1+y^2}}.
\end{equation}
Equation~\eqref{eq:muderivation} is the standard MOND interpolation
function used in Section~3.1 of Ref.~\cite{FamaeyBinney2005}. Here the
RMS response fixes its form and Eq.~\eqref{eq:sigmaira0} fixes $a_0$.
Within the response class defined by this occupation orbit, quadratic
spatial entry and RMS readout, these expressions determine the local
constitutive law.
The Gaussian variance is defined for $\lambda_{\rm IR}>0$. The
corresponding flux extends continuously to zero with $\mu(0)=0$. For $y\ll1$
$\mu(y)\sim y$. Thus the flux is asymptotic to
$|\nabla\Phi|\nabla\Phi/a_0$. The spatial integral of its leading cubic
primitive is invariant under
$x\mapsto sx$ with $\Phi\mapsto\Phi$ in three spatial dimensions.
This is the scale invariance of the leading deep-infrared field
functional.
For a plane-symmetric source at fixed geometry, Gauss integration
makes the flux proportional to the source amplitude. Its inversion
then gives a field proportional to the square root of that amplitude
in the low-acceleration regime. The cubic energy and this nonlinear
source response follow from the same amplitude law.

The endpoint flux is
\[
\mathcal{J}_i
=\mu\!\left(\frac{|\nabla\Phi|}{a_0}\right)\nabla_i\Phi .
\]
Its Gauss law is
\begin{equation}\label{eq:endpointpoisson}
\nabla\!\cdot\!\left[\mu\!\left(\frac{|\nabla\Phi|}{a_0}\right)\nabla\Phi\right]
=4\pi G\left(\rho_b+\rho_\Sigma^{\mathrm{free}}\right).
\end{equation}
This is the nonrelativistic field equation of Bekenstein and
Milgrom in Eq.~(3) of Ref.~\cite{BekensteinMilgrom1984} with the stated $\mu$ and
source density. For this choice of flux the Dirichlet problem follows
from a strictly convex energy with quadratic growth at large gradients.
This growth gives coercivity, while strict monotonicity gives
uniqueness including fields whose gradient vanishes.

\begin{proposition}[Linearized flux and unique weak Dirichlet solution]
\label{prop:p5-endpoint-flux-uniqueness}
Let $a_0>0$ and
$\mathcal J(g)=\mu(|g|/a_0)g$ for $g\in\mathbb R^3$. For $g\ne0$ put
$y=|g|/a_0$ and $P_\parallel=gg^{\mathsf T}/|g|^2$ and
$P_\perp=I-P_\parallel$. The flux derivative is
\begin{equation}
\begin{aligned}
 \mathsf M(g):=D\mathcal J(g)
 &=\frac{y}{\sqrt{1+y^2}}P_\perp
  +\frac{y(y^2+2)}{(1+y^2)^{3/2}}P_\parallel,\\
 \mu(y)I&\leq\mathsf M(g)\leq2\mu(y)I.
\end{aligned}
 \label{eq:p5-endpoint-linearized-flux}
\end{equation}
It extends continuously with $\mathsf M(0)=0$. Hence the positive
linearized operator
$-\nabla\cdot[\mathsf M(\nabla\overline\Phi)\nabla]$ is uniformly
elliptic on backgrounds with
$|\nabla\overline\Phi|\geq g_{\min}>0$ almost everywhere.

More generally let $\Omega\subset\mathbb R^3$ be a bounded Lipschitz
domain. Fix $\Phi_b\in H^1(\Omega)$ and
$f=4\pi G(\rho_b+\rho_\Sigma^{\rm free})\in H^{-1}(\Omega)$.
There is a unique $\Phi\in\Phi_b+H_0^1(\Omega)$ satisfying
\[
 \int_\Omega\mathcal J(\nabla\Phi)\cdot\nabla\eta\,d^3x
 =-\langle f,\eta\rangle
 \qquad\text{for every }\eta\in H_0^1(\Omega).
\]
\end{proposition}
\begin{proof}
Differentiating the flux gives
Eq.~\eqref{eq:p5-endpoint-linearized-flux}. The longitudinal-to-transverse
eigenvalue ratio is $(y^2+2)/(y^2+1)\in[1,2]$. Its radial primitive is
\[
 \mathcal F(g)=\frac{a_0^2}{2}
   [y\sqrt{1+y^2}-\operatorname{arsinh}y],\qquad
 \nabla_g\mathcal F=\mathcal J(g).
\]
The matrix $\mathsf M$ is positive definite off zero. Integrating it along
the segment joining distinct $g,h$ gives
$(\mathcal J(g)-\mathcal J(h))\cdot(g-h)>0$. The segment can meet zero at
at most one parameter value. Thus $\mathcal F$ is strictly convex.
Also $\mathcal F(g)=\int_0^{|g|}\mu(t/a_0)t\,dt$. Since
$0\leq\mu\leq1$ and $\mu(t/a_0)\geq1/\sqrt2$ for $t\geq a_0$
\[
 \frac{|g|^2-a_0^2}{2\sqrt2}
 \leq\mathcal F(g)\leq\frac{|g|^2}{2}.
\]
For $u\in H_0^1(\Omega)$ the primitive defines the minimization functional
\[
 \mathcal E(u)=\int_\Omega\mathcal F(\nabla(u+\Phi_b))\,d^3x
                  +\langle f,u\rangle.
\]
The displayed growth bound and Poincare's inequality make $\mathcal E$
coercive while convexity makes it weakly lower semicontinuous.  The
standard direct method and strict-monotonicity uniqueness theorem for
quasilinear Dirichlet problems~\cite{Evans2010PDE} therefore apply.
$|\mathcal J(g)|\leq|g|$ gives the stated Euler equation. At vanishing
field the constitutive law is degenerate:
$D\mathcal J(0)=0$ and the leading primitive is
$|g|^3/(3a_0)$ while strict convexity still gives uniqueness.
\end{proof}

\subsection{The spherical response}\label{sec:sphericalresponse}

For an isolated spherical baryonic mass with $\rho_\Sigma^{\rm free}=0$
and vanishing external field define
\[
x=\frac{a_N}{a_0},\qquad
y=\frac{a}{a_0},\qquad
a_N=\frac{GM_b}{r^2}.
\]
Integration of Eq.~\eqref{eq:endpointpoisson} outside the source gives
\[
x=y\mu(y)=\frac{y^2}{\sqrt{1+y^2}}.
\]
For $x>0$ the positive solution is
\[
y^2=\frac{x^2+x\sqrt{x^2+4}}{2},
\qquad
\frac{a}{a_N}
=\left[\frac{1+\sqrt{1+4/x^2}}{2}\right]^{1/2}.
\]
For $x\gg1$ it approaches $a/a_N=1+1/(2x^2)+\order{x^{-4}}$. In the isolated deep-infrared limit $x\ll1$
\[
a(r)=\frac{\sqrt{GM_ba_0}}{r},
\qquad
v^4=GM_ba_0 .
\]
The mass--asymptotic-velocity relation is the usual isolated deep-MOND
limit in Eq.~(7) of Ref.~\cite{BekensteinMilgrom1984} with the acceleration scale
fixed here by Eq.~\eqref{eq:sigmaira0}.

\paragraph{Fixed-parameter rotation-curve comparison.}
The SPARC comparison reported in Ref.~\cite{Guo2026GenesisFramework} applies the algebraic
acceleration map of Section~\ref{sec:sphericalresponse} pointwise to the
tabulated baryonic mass models of Ref.~\cite{SPARC2016}.
It fixes the stellar mass-to-light ratios at
$\Upsilon_{\rm disk}=0.5$ and $\Upsilon_{\rm bulge}=0.7$ in solar units
at $3.6\,\mu{\rm m}$ and retains points with
$V_{\rm obs}>20\ \mathrm{km\,s^{-1}}$ and
$\sigma_V/V_{\rm obs}<0.1$ and uses the endpoint $a_0$ above.
For its 2790 retained points the reported standard deviation about the mean
of $\log_{10}(g_{\rm obs}/g_{\rm pred})$ is $0.144314$ dex.
The statistic tests the algebraic acceleration relation on the
observed mass models at fixed stellar normalization. The
nonspherical field equation \eqref{eq:endpointpoisson} is evaluated
separately with its spatial source and boundary data. The comparison data and the
three dimensional field calculation are specified in Appendix~\ref{app:repro}.

\subsection{The endpoint-source decomposition}\label{sec:nodouble}

Equation~\eqref{eq:endpointpoisson} can be written in Newtonian-source form by defining
\[
\rho_{\mathrm{dark,eff}}
=\rho_\Sigma^{\mathrm{free}}+\rho_{\mathrm{pol}},
\qquad
\rho_{\mathrm{pol}}
=\frac{1}{4\pi G}\nabla\!\cdot\!\left[(1-\mu)\nabla\Phi\right].
\]
It follows identically that
\[
\nabla^2\Phi
=4\pi G\left(\rho_b+\rho_\Sigma^{\mathrm{free}}+\rho_{\mathrm{pol}}\right).
\]
At each time the transported free density $\rho_\Sigma^{\rm free}$
and the baryonic density supply the local field source. Solving the
nonlinear boundary problem gives $\Phi$ and its polarization flux
then determines $\rho_{\rm pol}$. Thus the free density carries the
material transport history while the polarization follows the
changing source through its field response. An isolated baryonic source sets
$\rho_\Sigma^{\rm free}=0$ whereas a region containing cold matter
retains the transported free density. The homogeneous cold mean is
$\overline\rho_\Sigma=\rho_{\rm crit}\Omega_{\rm cold}$
at the reference epoch with $\Omega_{\rm cold}$ fixed by
\eqref{eq:p5-cold-budget}. In the dust realization of
Section~\ref{sec:cosmo} initial data with that mean are transported by
\eqref{eq:p5-cold-pushforward}. This fixes
$\rho_\Sigma^{\rm free}$ on the local domain once the flow and the
initial density are specified. For a periodic cosmological cell the
free-source mean is $\overline\rho_\Sigma\propto a^{-3}$.
On a bounded physical region $\Omega$ the complete source identity is
\begin{equation}
 \langle\rho_{\rm dark,eff}\rangle_\Omega
 =\langle\rho_\Sigma^{\rm free}\rangle_\Omega
 +\frac{1}{4\pi G|\Omega|}
 \int_{\partial\Omega}(1-\mu)\nabla\Phi\cdot n\,dS.
 \label{eq:p5-local-mean-source}
\end{equation}
The second term vanishes for a periodic polarization flux or an imposed
zero net polarization flux. Otherwise the boundary data determine it.
For weak solutions the same statement uses the normal trace whenever
the flux admits such a trace. With zero net polarization flux the
polarization redistributes the local effective source while preserving
the mean free-source budget. On a general bounded region its net
contribution is fixed by the boundary flux.

The quasistatic energy on a physical region is
\begin{equation}
 E_{\rm qs}[\Phi]=\frac{1}{4\pi G}\int_\Omega
       \mathcal F(\nabla\Phi)\,d^3x
       +\int_\Omega(\rho_b+\rho_\Sigma^{\rm free})\Phi\,d^3x,
 \label{eq:p5-quasistatic-energy}
\end{equation}
where $\mathcal F$ is the convex primitive in
Proposition~\ref{prop:p5-endpoint-flux-uniqueness}. Variation at fixed
boundary data gives Eq.~\eqref{eq:endpointpoisson}. For a background
$\overline\Phi$ and a fluctuation $\phi$ its second variation is
\[
 \delta^2E_{\rm qs}[\overline\Phi](\phi,\phi)
 =\frac{1}{4\pi G}\int_\Omega
 \nabla\phi\cdot\mathsf M(\nabla\overline\Phi)\nabla\phi\,d^3x.
\]
Thus the ambient field fixes the local response tensor. Its transverse
and longitudinal eigenvalues are those in
Eq.~\eqref{eq:p5-endpoint-linearized-flux}. For a nonzero ambient
field the larger longitudinal coefficient gives a larger flux
variation for gradient perturbations along the field than across it.
The environment thus supplies a preferred direction in a non-spherical
source problem. At a vanishing static field the cubic
energy controls the leading response. These coefficients and the source decomposition are evaluated in
Ref.~\cite{Guo2026GenesisFramework}. For cosmological applications
Section~\ref{sec:linprop} fixes the homogeneous expansion and mean cold
density. The local problem additionally requires the source flow and
ambient field together with boundary data. Its response to that ambient field is
the second variation displayed above.

\paragraph{Plane-symmetric cosmological transport.}
The relation between the transported density and its mean can be
evaluated exactly for plane-symmetric Euler--Poisson flow. On a
periodic cell set
\[
 x_1=q_1-\frac{AD(t)}{k}\sin(kq_1),\qquad
 x_2=q_2,\quad x_3=q_3,\qquad |AD|<1.
\]
Then $J=1-AD\cos(kq_1)$ and $\rho=\overline\rho_\Sigma/J$.
In comoving coordinates Poisson's equation and the characteristic
equation are satisfied by
\[
 \partial_{x_1}\Phi
 =4\pi G a^2\overline\rho_\Sigma\frac{AD}{k}\sin(kq_1),
 \qquad
 \ddot D+2H\dot D=4\pi G\overline\rho_\Sigma D.
\]
Indeed $\partial_{x_1}=J^{-1}\partial_{q_1}$ gives
$\partial_{x_1}^2\Phi=4\pi G a^2(\rho-\overline\rho_\Sigma)$. Substitution in $\ddot x+2H\dot x=-a^{-2}\nabla_x\Phi$ gives the
growth equation. The identity $\rho J=\overline\rho_\Sigma$ fixes the
cell mean exactly. The numerical characteristic evolution is checked
against the Einstein--de Sitter solution $D\propto a$ and stops at
$|AD|=1$~\cite{Guo2026GenesisFramework}. Here the potential obeys the cosmological Poisson equation for the
density contrast. The corresponding local constitutive problem uses
the flux in Eq.~\eqref{eq:endpointpoisson}.

\subsection{Joint source and potential evolution}\label{sec:p5-local-feedback}

The quasistatic functional also determines a dynamical coupling to
nonrelativistic matter. We use the variational formulation of
Ref.~\cite{BekensteinMilgrom1984} with the primitive $\mathcal F$ and
acceleration normalization obtained above. On the physical time interval
under consideration $G$ and $a_0$ are constant. Let $X_a(t,q)$ be the
trajectory map of a pressureless source component from a fixed reference
domain $Q_a$ onto the common physical domain $\mathcal D$ with reference
mass measure $\rho_{a,0}(q)\,d^3q$. Its density and velocity obey
\[
 \rho_a(t,X_a(t,q))\det D_qX_a(t,q)=\rho_{a,0}(q),
 \qquad v_a(t,X_a(t,q))=\dot X_a(t,q).
\]
The components may include the free cold source and a pressureless
baryonic component. Set $\rho=\sum_a\rho_a$. The joint action is
\begin{equation}
 \begin{split}
 S[X_a,\Phi]=\int dt\,\bigg\{
 &\sum_a\int_{Q_a} \rho_{a,0}(q)
 \left[\frac{|\dot X_a|^2}{2}-\Phi(t,X_a)\right]d^3q\\
 &-\frac{1}{4\pi G}\int_{\mathcal D}
        \mathcal F(\nabla\Phi)\,d^3x\bigg\}.
 \end{split}
 \label{eq:p5-feedback-action}
\end{equation}
\begin{proposition}[Source feedback and boundary energy balance]
\label{prop:p5-feedback-energy}
Suppose the trajectory maps are smooth orientation-preserving
diffeomorphisms on a time interval. Suppose the densities and potential are
sufficiently smooth for the following integrations by parts and
$\Phi$ has prescribed Dirichlet data on the fixed smooth domain
$\mathcal D$. Let $\Omega\subseteq\mathcal D$ be a fixed smooth
region on which the energy is evaluated.
Stationarity of \eqref{eq:p5-feedback-action} with fixed trajectory
endpoints and compactly supported spatial variations gives
\begin{equation}
 \begin{aligned}
 \partial_t\rho_a+\nabla\cdot(\rho_a v_a)&=0,\\
 (\partial_t+v_a\cdot\nabla)v_a&=-\nabla\Phi,\\
 \nabla\cdot\mathcal J(\nabla\Phi)&=4\pi G\rho.
 \end{aligned}
 \label{eq:p5-feedback-system}
\end{equation}
Define
\begin{equation}
 E(t)=\sum_a\int_\Omega\frac12\rho_a|v_a|^2\,d^3x
       +\int_\Omega\rho\Phi\,d^3x
       +\frac{1}{4\pi G}\int_\Omega\mathcal F(\nabla\Phi)\,d^3x .
 \label{eq:p5-feedback-energy}
\end{equation}
Then its boundary balance is
\begin{equation}
 \frac{dE}{dt}
 =-\int_{\partial\Omega}\sum_a\rho_a(v_a\cdot n)
             \left(\frac{|v_a|^2}{2}+\Phi\right)dS
   +\frac{1}{4\pi G}\int_{\partial\Omega}
             \partial_t\Phi\,\mathcal J(\nabla\Phi)\cdot n\,dS .
 \label{eq:p5-feedback-boundary-work}
\end{equation}
In particular for $\Omega=\mathcal D$ impermeable matter boundaries
and time-independent Dirichlet data conserve $E$.
\end{proposition}
\begin{proof}
Differentiating the pushforward identity gives continuity. Variation of
$X_a$ in the kinetic and source terms gives
$\ddot X_a=-\nabla\Phi(t,X_a)$. The potential variation uses
$\nabla_g\mathcal F=\mathcal J$ and gives the third equation.
With $j=\sum_a\rho_a v_a$ the kinetic balance has volume term
$-\int_\Omega j\cdot\nabla\Phi$. The derivative of the source energy
has volume terms $\int_\Omega j\cdot\nabla\Phi+
\int_\Omega\rho\,\partial_t\Phi$ and boundary term
$-\int_{\partial\Omega}\Phi j\cdot n$.
Finally
\[
 \frac{d}{dt}\frac{1}{4\pi G}\int_\Omega\mathcal F(\nabla\Phi)
 =\frac{1}{4\pi G}\int_{\partial\Omega}
     \partial_t\Phi\,\mathcal J\cdot n
   -\int_\Omega\rho\,\partial_t\Phi .
\]
The volume terms cancel to leave
\eqref{eq:p5-feedback-boundary-work}.
\end{proof}

The result applies to smooth evolution before trajectory crossing.
At each time the transported density and boundary data determine the
unique potential of Proposition~\ref{prop:p5-endpoint-flux-uniqueness}.
Equation~\eqref{eq:p5-feedback-boundary-work} combines the work exchanged
between matter and the potential with transport across the boundary.
The volume exchange cancels, leaving the matter flux and potential
work at the boundary. For
$\Omega=\mathcal D$ and zero Dirichlet data the field equation
rewrites the potential part of $E$ as
$(4\pi G)^{-1}\int_\Omega[\mathcal F(g)-g\cdot\mathcal J(g)]\,d^3x$.
Its Newtonian quadratic limit gives the usual negative gravitational
potential energy.

\paragraph{Finite material-flow realization.}
A plane-symmetric variational approximation~\cite{Guo2026GenesisFramework}
assigns a fixed mass $m_i$ per unit transverse area to each affine
material cell of length $\ell_i$. With $\kappa=4\pi G$ the field
flux is $\mathfrak j=C+\kappa M(x)$ where $M(x)$ is cumulative
mass and the Dirichlet potential difference fixes $C$. Define
$\mathcal H(\mathfrak j)=g\mathfrak j-\mathcal F(g)$ with
$\mathfrak j=\mathcal J(g)$. The relaxed potential energy is
\[
 U=\frac{b_R\mathfrak j_R-b_L\mathfrak j_L
             -\sum_i\ell_i\overline{\mathcal H}_i}{\kappa},
\]
where $b_L,b_R$ are the boundary potentials and the bar denotes the
cell average. The consistent kinetic matrix of cell $i$ is
$m_i\bigl(\begin{smallmatrix}2&1\\1&2\end{smallmatrix}\bigr)/6$.
Variation at an interior node gives the force
$(\overline{\mathcal H}_{i-1}-\overline{\mathcal H}_{i})/\kappa$. The boundary term in \eqref{eq:p5-feedback-boundary-work} becomes
$(\dot b_R\mathfrak j_R-\dot b_L\mathfrak j_L)/\kappa$.

For a finite calculation choose units with $\kappa=1$ on the
interval $[0,1]$ with $a_0=0.3$ and $b_L=b_R=0$. Eight cells start at
$x_i=i/8$ with
\[
 m_i=0.1\,\Delta x_i
 \bigl[1+0.2\cos(2\pi\bar x_i)+0.1\sin(2\pi\bar x_i)\bigr],
 \qquad v_i(0)=0.02\sin(\pi x_i),
\]
where $\bar x_i$ is the initial cell midpoint and the end nodes are
fixed. The nonlinear field is resolved at each velocity-Verlet step.
24-point Gaussian quadrature evaluates the cell energies after removal
of the square-root behavior at zero flux. At $t=0.4$ halving the
time step from $0.02$ to $0.01$ reduces the maximum energy-balance
residual from $3.86\times10^{-11}$ to $9.65\times10^{-12}$.
The final nodal positions differ by $2.66\times10^{-7}$ and all cell
lengths exceed $0.1168$. The two evolutions require 21 and 41 field
solutions respectively. These tests compare time integration at fixed
material cell resolution. Independent variations of the relaxed energy
check the force; time dependent boundary data test the boundary work.

\subsection{Generated matter dynamics and covariant response}
\label{sec:p5-covariant-response}

\subsubsection{Canonical matter evolution under an internal source}
The canonical matter action fixes the response to a time-dependent
internal source. Let $A=D_B^\dagger D_B+\mathcal R_g/8$ be its positive
internal operator with the spatial connection $B$ held fixed. The
conformal-source identity of the proposition ``Local energy charge on the conformal scalar branch'' in
Ref.~\cite{Guo2026SpectralGravity} gives
\[
 g_\eta=e^{-2\eta}g,\qquad
 dv_{g_\eta}=e^{-3\eta}dv_g,\qquad
 A_{g_\eta,B}=e^{5\eta/2}Ae^{-\eta/2}.
\]
Here the conformal transformation acts on the three-dimensional internal
space. Pulling the canonical time kinetic term and the internal quadratic
form back to $L^2(dv_g)$ gives
\begin{equation}
 L_\eta=\frac12\bigl[
 \dot\varphi^\dagger M_\eta\dot\varphi
       -\varphi^\dagger B_\eta\varphi\bigr],\qquad
 M_\eta=e^{-3\eta},\qquad
 B_\eta=e^{-\eta/2}Ae^{-\eta/2}.
 \label{eq:p5-native-source-lagrangian}
\end{equation}
The real quadratic form is understood when complex mode coordinates are
used. The temporal kinetic term is that of the canonically normalized
matter fields of Ref.~\cite{Guo2026CompactInternalSpaceI}. The internal volume
and conformal covariance determine its source dependence. A finite
calculation uses Hermitian source matrices on a retained carrier and
the matrix functions in \eqref{eq:p5-native-source-lagrangian}.
These matrix functions define the finite source family. The retained
support of the stated static derivatives is checked separately in
Appendix~\ref{app:repro}.

\begin{proposition}[Canonical matter evolution and source work]
\label{prop:p5-native-source-dynamics}
On a finite carrier let $A>0$ and let $\eta(t)$ be a continuously
differentiable Hermitian source history. With canonical momentum
$p=M_\eta\dot\varphi$ define
\begin{equation}
 z=\begin{pmatrix}\varphi\\p\end{pmatrix},\qquad
 \mathbb J=\begin{pmatrix}0&I\\-I&0\end{pmatrix},\qquad
 \mathbb K_\eta=\begin{pmatrix}B_\eta&0\\0&M_\eta^{-1}\end{pmatrix}.
 \label{eq:p5-native-phase-generator}
\end{equation}
The evolution $\dot z=\mathbb J\mathbb K_\eta z$ preserves the canonical
symplectic form. For a positive semidefinite ensemble second moment
$W=\langle zz^\dagger\rangle$ it preserves $W\geq0$ and satisfies
\begin{equation}
 E=\frac12\operatorname{Tr}(\mathbb K_\eta W)\geq0,
 \qquad
 \dot E=\frac12\operatorname{Tr}(\dot{\mathbb K}_\eta W).
 \label{eq:p5-native-source-work}
\end{equation}
At a fixed source the squared frequencies are the eigenvalues of
$e^\eta Ae^\eta$. At $\eta=0$ a stationary occupation $N\geq0$
with $[N,H]=0$ and $H=\sqrt A$ is represented by
\begin{equation}
 W_0=\begin{pmatrix}H^{-1}N&0\\0&HN\end{pmatrix},
 \qquad E_0=\operatorname{Tr}(NH).
 \label{eq:p5-native-occupied-covariance}
\end{equation}
For a source direction $P_a=P_a^\dagger$ its initial generalized force is
$J_a=\operatorname{Tr}(NHP_a)$ equal to the first variation of the
occupied static spectral energy.
\end{proposition}

\begin{proof}
Both $M_\eta$ and $B_\eta$ are positive definite. Thus the Hamiltonian
$\mathcal H=z^\dagger\mathbb K_\eta z/2$ has positive kinetic and
potential quadratic forms. If $S$ is the fundamental solution with
$\dot S=\mathbb J\mathbb K_\eta S$ and $S(t_i)=I$ then
$S^\dagger\mathbb J S=\mathbb J$ and
$W(t)=S(t)W(t_i)S(t)^\dagger$. This proves positivity and gives
\[
 \dot W=\mathbb J\mathbb K_\eta W
                    +W(\mathbb J\mathbb K_\eta)^\dagger.
\]
Cyclicity of the trace cancels the two evolution terms in $\dot E$ to give \eqref{eq:p5-native-source-work}. At a fixed source canonical
normalization gives
\[
 M_\eta^{-1/2}B_\eta M_\eta^{-1/2}=e^\eta Ae^\eta,
\]
which is the operator in the static energy response. Substitution of
\eqref{eq:p5-native-occupied-covariance} proves $E_0=\operatorname{Tr}(NH)$.
At zero source
\[
 \partial_a M_\eta^{-1}=3P_a,\qquad
 \partial_a B_\eta=-\tfrac12(P_aA+AP_a).
\]
Consequently $J_a=\operatorname{Tr}[(\partial_a\mathbb K)W_0]/2
=\operatorname{Tr}(NHP_a)$. Differentiating
$H_\eta^2=e^\eta Ae^\eta$ and using $[N,H]=0$ gives the same
derivative of $\operatorname{Tr}(NH_\eta)$.
\end{proof}

Here the stationary matrix $N$ specifies the occupied mode weights
on the retained carrier including their overall normalization in
the energy and source force. The angular calculation uses unit trace
to isolate the distribution of a fixed total occupation. The
phase-space matrix $W$ instead collects coordinate and momentum
second moments and their correlations so that the occupied input
can be followed as the source changes.

This action fixes the time dependence of canonical normalization.
For $R=M_\eta^{1/2}$ with $q=R\varphi$ and $C=\dot R R^{-1}$
\begin{equation}
 L_\eta=\frac12\left[
 (\dot q-Cq)^\dagger(\dot q-Cq)-q^\dagger e^\eta Ae^\eta q\right].
 \label{eq:p5-native-time-normalization}
\end{equation}
For a spatially uniform twice differentiable source this becomes
\[
 \ddot q+\left[e^{2\eta}A+\frac32\ddot\eta
                         -\frac94\dot\eta^2\right]q=0.
\]
The additional terms follow from the evolving internal volume. The
canonical phase-space evolution retains the correlations between coordinates
and momenta and between both frequency branches. For a piecewise constant
source it is evaluated by the exact finite propagators
$\exp(\mathbb J\mathbb K_j\Delta t_j)$. Energy is conserved within
each segment. A source jump contributes
$\operatorname{Tr}[(\mathbb K_{j+1}-\mathbb K_j)W]/2$.
This gives a source work check for finite histories of the
kind used in Ref.~\cite{Guo2026NumberDynamics}.

A clock assigns an increasing readout to successive states of a
generated trajectory. In a monotone chart $\tau_{\rm clk}$
the clock Jacobian is $\mathcal N_{\rm clk}=dt/d\tau_{\rm clk}>0$. Writing primes for
derivatives in this chart gives exactly
\begin{equation}
 S_\eta=\frac12\int d\tau_{\rm clk}
 \left[\mathcal N_{\rm clk}^{-1}
              \varphi'{}^\dagger M_\eta\varphi'
       -\mathcal N_{\rm clk}\varphi^\dagger B_\eta\varphi\right],
 \qquad
 z'=\mathcal N_{\rm clk}\mathbb J\mathbb K_\eta z .
 \label{eq:p5-native-clock-reparametrization}
\end{equation}
The canonical momentum is $p=M_\eta\varphi'/\mathcal N_{\rm clk}$
and the generator in the new chart is
$\mathcal N_{\rm clk}\mathcal H_\eta$. For a fixed clock chart
$\mathcal N_{\rm clk}$ is its prescribed Jacobian. The matter variation
holds that Jacobian fixed. A phase readout supplies such a chart on
each interval where it is monotone along the trajectory. The
symplectic identity and source-work law are preserved with
$E'=\operatorname{Tr}(\mathbb K_\eta'W)/2$.

The causal linear response follows from the same phase evolution. Write
$K_a=\partial_a\mathbb K|_0$ and $K_{ab}=\partial_a\partial_b\mathbb K|_0$
and $X_0=\mathbb J\mathbb K_0$. For a perturbation proportional to
$e^{-i\zeta t}$ with $\operatorname{Im}\zeta>0$ set
\[
 \mathcal L_0 V=X_0V+VX_0^\dagger,\qquad
 F_b=\mathbb JK_bW_0+W_0(\mathbb JK_b)^\dagger.
\]
Linearization gives
\begin{equation}
 \chi_{ab}(\zeta)=\frac12\operatorname{Tr}(K_{ab}W_0)
 +\frac12\operatorname{Tr}\!\left[
 K_a(-i\zeta-\mathcal L_0)^{-1}F_b\right].
 \label{eq:p5-native-causal-response}
\end{equation}
The first term is the direct source variation. Both terms are retained
when evaluating the response. For one oscillator of frequency $\omega$
with occupation $n$ and a uniform source they give the independent check
\[
 \chi(\zeta)=5n\omega+
                  \frac{16n\omega^3}{\zeta^2-4\omega^2},
 \qquad \chi(0)=n\omega.
\]
For general source directions the causal response resolves the finite
ensemble's evolution as well as the instantaneous spectral variation.

\subsubsection{Covariant source exchange and metric constraints}
Let $s^a(x)$ be the real coefficients of the internal source
$\eta(x)=\sum_a s^a(x)P_a$ on the finite carrier. Use real field
coordinates including real and imaginary parts when
the original carrier is complex and signature $(-+++)$.
The covariant matter action is
\begin{equation}
 S_{\rm m}[g,\varphi,s]=-\frac12\int d^4x\sqrt{-g}\,
 \left[g^{\mu\nu}\partial_\mu\varphi^{\mathsf T}
                 M(s)\partial_\nu\varphi
                    +\varphi^{\mathsf T}B(s)\varphi\right],
 \label{eq:p5-native-covariant-matter}
\end{equation}
with the same $M$ and $B$ as in
\eqref{eq:p5-native-source-lagrangian}. These matrices act on the
retained internal mode coefficients. Internal deformation changes
their kinetic norms and spectral frequencies. The external metric
contracts the spacetime derivatives and variation with respect to it
gives the matter stress tensor. Thus internal geometry enters macroscopic propagation
through $M(s)$ and $B(s)$ and their common variation determines
the exchange with the collective source. The homogeneous flat-space
reduction is the canonical action above.

\paragraph{Source work and total stress conservation.}
Changes of the common source alter both the kinetic form and restoring
coefficients of matter.  Their energy cost is the work exchanged with
that source.  The same variation determines its contribution to the
matter stress identity.
For \eqref{eq:p5-native-covariant-matter} define
$\mathcal E_\varphi=(\sqrt{-g})^{-1}\delta S_{\rm m}/\delta\varphi$ and
$Q_a=(\sqrt{-g})^{-1}\delta S_{\rm m}/\delta s^a$. Then
\begin{align}
 T_{\mu\nu}^{\rm m}
 &=\partial_\mu\varphi^{\mathsf T}M\partial_\nu\varphi
 -\frac12g_{\mu\nu}
 \left[g^{\rho\sigma}\partial_\rho\varphi^{\mathsf T}
 M\partial_\sigma\varphi+\varphi^{\mathsf T}B\varphi\right],
 \notag\\
 Q_a&=-\frac12\left[
 g^{\mu\nu}\partial_\mu\varphi^{\mathsf T}M_{,a}\partial_\nu\varphi
                      +\varphi^{\mathsf T}B_{,a}\varphi\right],
 \label{eq:p5-native-source-stress}\\
 \nabla_\mu(T_{\rm m})^\mu{}_\nu
 &=\mathcal E_\varphi^{\mathsf T}\partial_\nu\varphi
                          +Q_a\partial_\nu s^a .
 \label{eq:p5-native-source-ward}
\end{align}
This identity holds before imposing the matter equation.  On
$\mathcal E_\varphi=0$ its right-hand side reduces to
$Q_a\partial_\nu s^a$. The cancellation of this exchange also uses
the common source equation below.
In homogeneous flat space the matter equation gives, per unit external
spatial volume,
$\dot E_{\rm m}=-Q_a\dot s^a=J_a\dot s^a$ where
$J_a=\partial\mathcal H_{\rm m}/\partial s^a$ at fixed canonical data.
Metric and source variation give
\eqref{eq:p5-native-source-stress} while
$\mathcal E_\varphi=\nabla_\mu(M\partial^\mu\varphi)-B\varphi$.
Differentiating the displayed stress tensor cancels the two derivatives
of the kinetic contraction and leaves
$\mathcal E_\varphi^{\mathsf T}\partial_\nu\varphi+Q_a\partial_\nu s^a$.
This is the source-dependent specialization of the common-action
Noether identity in Ref.~\cite{Guo2026SpectralGravity}.
In the homogeneous limit $T^0{}_0=-E_{\rm m}$ gives the last
identity. It agrees with \eqref{eq:p5-native-source-work}.

The metric constraints follow from the same action. For lapse
$\mathcal N>0$ with shift $\beta^i$ and spatial metric $h_{ij}$
\begin{align}
 \pi&=\frac{\sqrt h}{\mathcal N}
            M(\dot\varphi-\beta^i\partial_i\varphi),\notag\\
 \mathcal H_\perp^{\rm m}
 &=\frac{\pi^{\mathsf T}M^{-1}\pi}{2\sqrt h}
 +\frac{\sqrt h}{2}
 \left[h^{ij}\partial_i\varphi^{\mathsf T}M\partial_j\varphi
                      +\varphi^{\mathsf T}B\varphi\right],
 &\mathcal H_i^{\rm m}&=\pi^{\mathsf T}\partial_i\varphi .
 \label{eq:p5-native-adm-source}
\end{align}
Thus $H_{\rm m}=\int(\mathcal N\mathcal H_\perp^{\rm m}
+\beta^i\mathcal H_i^{\rm m})\,d^3x$. Positivity of $M$ and $B$
gives a positive matter energy for the normal observer and the
principal matter symbol is $M g^{\mu\nu}k_\mu k_\nu$.

For independent generated source fields with action $S_{\rm s}$ the
common system includes the source equation and both metric constraints
\begin{equation}
 \mathcal E_a^{\rm s}+Q_a=0,\qquad
 \mathcal H_\perp^{\rm g}+\mathcal H_\perp^{\rm s}
                         +\mathcal H_\perp^{\rm m}=0,\qquad
 \mathcal H_i^{\rm g}+\mathcal H_i^{\rm s}
                         +\mathcal H_i^{\rm m}=0 .
 \label{eq:p5-native-common-source-balance}
\end{equation}
The first equation cancels the source exchange in
\eqref{eq:p5-native-source-ward} against that of $S_{\rm s}$. The
total stress is then conserved on the matter equations and enters the Einstein equation.
In particular the homogeneous stationary occupied
branch at zero source requires
$\langle\mathcal E_{\rm uniform}^{\rm s}\rangle
=\operatorname{Tr}(NH)$. This is the collective source force required
by the calculated matter energy.
The inherited complete condensate supplies a static radius force
through \eqref{eq:p5-native-radius-force}. Its joint amplitude--radius
equilibrium and positive Hessian are given in
Proposition~\ref{prop:p5-joint-native-radius}. A spacetime realization
uses that potential with its physical density normalization and the
temporal source terms in the same action.

A generated source can instead be a functional $s=\Gamma[g,q]$ of
retained physical fields. For a diffeomorphism-equivariant $\Gamma$
variation of $S_{\rm s}[g,q]+S_{\rm m}[g,\varphi,\Gamma[g,q]]$ gives
\begin{equation}
 \mathcal E_q=\mathcal E_q^{\rm s}+(D_q\Gamma)^*Q,\qquad
 T_{\mu\nu}^{\rm tot}=T_{\mu\nu}^{\rm s}+T_{\mu\nu}^{\rm m}
                  -2\bigl[(D_{g^{-1}}\Gamma)^*Q\bigr]_{\mu\nu}.
 \label{eq:p5-native-generated-pullback}
\end{equation}
The adjoints include integrations by parts in the covariant volume
measure when $\Gamma$ contains derivatives. Variations have compact
support or the stated fixed boundary data. These expressions follow
directly from the first variation and retain the metric dependence of
the source. Diffeomorphism equivariance then gives the Noether identity
for the complete substituted action.  On the equations of the retained
fields it yields
\begin{equation}
 \nabla_\mu(T_{\rm tot})^\mu{}_{\nu}=0.
 \label{eq:p5-native-total-stress-conservation}
\end{equation}
The metric dependence of the generated source contributes to the total
stress through \eqref{eq:p5-native-generated-pullback}, so source
feedback and stress conservation follow from the same allowed variation.
\subsubsection{Generated variation and finite-window control}
Write
$\mathcal P(g,q,\varphi)=(g,q,\varphi,\Gamma[g,q])$ for the full field
map into a parent action $S$. Its quadratic variation retains
both chain-rule terms
\[
 \delta^2(S\circ\mathcal P)[u,v]
 =\delta^2S[D\mathcal P u,D\mathcal P v]
                   +\delta S[D^2\mathcal P(u,v)].
\]
The second term vanishes when the background is stationary with
respect to all parent variables.
The spectral constructions and controlled limits of
Refs.~\cite{Guo2026SpectralArithmetic,Guo2026NumberDynamics} provide
the operations and convergence estimates for representing generated
source directions.
For the field map $\mathcal P$ in
\eqref{eq:p5-native-generated-pullback} the native variational theory is
\begin{equation}
 S_{\rm gen}=S\circ\mathcal P,\qquad
 \mathcal E_{\rm gen}=(D\mathcal P)^*\mathcal E_S=0 .
 \label{eq:p5-native-generated-action}
\end{equation}
Its constraints and gauge identities together with its evolution follow from this
action. At a native stationary background its quadratic action uses
both terms of the second-variation chain rule above. An equivariant
field map gives the corresponding Noether identity and the associated
metric-constraint propagation on the native equations.

When a generated system is realized as an invariant subsystem of a
specified parent evolution its phase-space embedding $\Gamma$ also
satisfies
\begin{equation}
 C_{\rm parent}(\Gamma(q))=0,\qquad
 F_{\rm parent}(\Gamma(q))=D\Gamma(q)F_{\rm gen}(q).
 \label{eq:p5-generated-dynamical-compatibility}
\end{equation}
Here $C_{\rm parent}$ collects the parent constraints and the second
identity establishes preservation of the image by that evolution.

For a Lagrangian depending on first time derivatives and a derivative-free field map the
velocity Hessian of the native action before solving its auxiliary
equations is
\[
 K_{\rm pull}=(D\mathcal P)^\ast K_{\rm full}D\mathcal P .
\]
It includes those fields' kinetic terms and their cross terms.
Algebraic auxiliary fields are eliminated from this common quadratic
action. If their invertible block is $C$ and their velocity mixing is
$B_{\rm aux}$ the reduced velocity block is
$A-B_{\rm aux}C^{-1}B_{\rm aux}^\ast$. Linear lapse and shift multipliers
impose their constraints before the remaining physical directions are
selected and gauge directions are factored out. Denote the resulting
kinetic form by $K_{\rm phys}$. A derivative-dependent map is treated
with its full temporal derivative structure and degeneracy conditions.
The projection and interaction conditions of
Ref.~\cite{Guo2026NumberDynamics} then apply to these actual operations.

Choose a positive definite reference Gram matrix $G_{\rm phys}$ on
the retained physical directions. If an approximation $\widehat K$
to the complete reduced form satisfies
\[
 \bigl\|G_{\rm phys}^{-1/2}(\widehat K-K_{\rm phys})
                       G_{\rm phys}^{-1/2}\bigr\|\leq\varepsilon_K ,
\]
its least generalized eigenvalue greater than $\varepsilon_K$
certifies positive kinetic energy on those physical directions.
An eigenvalue smaller than $-\varepsilon_K$ certifies a negative
physical direction. The error budget includes the generated map and complete quadratic-action assembly.
It also includes mixed-field terms and the constraint
solution together with omitted-mode errors when passing to a larger space.
Gradient stability and hyperbolicity are evaluated from the full
principal symbol of the same reduced action.

\subsubsection{Uniform compression, lapse and phase clocks}
The spatially uniform source permits an exact comparison with a
metric lapse. For prescribed $\eta(t)$ set
$q=e^{-\eta}\varphi$ and $P=e^\eta p$. The action and Hamiltonian become
\begin{align}
 L_\eta&=\frac{e^{-\eta}}2|\dot q+\dot\eta q|^2
                         -\frac{e^\eta}2q^{\mathsf T}Aq,\notag\\
 H'_\eta&=\frac{e^\eta}2(P^{\mathsf T}P+q^{\mathsf T}Aq)
                              -\dot\eta P^{\mathsf T}q .
 \label{eq:p5-native-lapse-dilation}
\end{align}
The substitution $\varphi=e^\eta q$ gives the first line. The canonical identity
$p^{\mathsf T}d\varphi=P^{\mathsf T}dq+
P^{\mathsf T}q\,d\eta$ gives the second. The first term of $H'_\eta$
is the lapse coupling $\mathcal N=e^\eta$ of the reference oscillator. The second is the associated time-dependent dilation coupling.
It is fixed by the same transformation.

When $\eta$ is varied together with $q$ the velocity Hessian of this
matter action is
\begin{equation}
 \frac{\partial^2L_\eta}{\partial(\dot q,\dot\eta)^2}
 =e^{-\eta}
 \begin{pmatrix}I&q\\q^{\mathsf T}&q^{\mathsf T}q\end{pmatrix},
 \qquad
 K_{\rm vel}\begin{pmatrix}-q\\1\end{pmatrix}=0 .
 \label{eq:p5-native-dilation-constraint}
\end{equation}
Its rank equals the number of real matter coordinates and its canonical
constraint is $P_\eta-q^{\mathsf T}P=0$. The collective source action
and the metric constraints in
\eqref{eq:p5-native-common-source-balance} determine the remaining
physical evolution. Equations~\eqref{eq:p5-native-generated-pullback}
and \eqref{eq:p5-native-lapse-dilation} specify the source-force and
dilation terms to retain when deriving a common metric realization.

\begin{proposition}[Congruence of a generated matter phase]
\label{prop:p5-native-phase-congruence}
Retain a degenerate pair of real scalar coordinates as one complex
mode of reference squared mass $m_0^2>0$. For a source uniform on the
internal carrier write $\varphi=r e^{i\vartheta}$ with $r>0$ and
$M=e^{-3\eta}$ and $m^2=m_0^2e^{2\eta}$. Its covariant action is
\begin{equation}
 S_{\rm m}=-\frac12\int\sqrt{-g}\,M
 \left[(\nabla r)^2+r^2(\nabla\vartheta)^2+m^2r^2\right]d^4x .
 \label{eq:p5-polar-native-action}
\end{equation}
Define the radial derivative term
$P=\nabla_\mu(M\nabla^\mu r)/(Mr)$ and assume
$X=-(\nabla\vartheta)^2>0$. On the radial and phase equations
\begin{equation}
 X=m^2-P,\qquad
 n_\mu=-\frac{\partial_\mu\vartheta}{\sqrt X},\qquad
 \nabla_\mu(Mr^2\sqrt X\,n^\mu)=0 .
 \label{eq:p5-polar-mass-shell}
\end{equation}
With $h_\mu{}^\nu=\delta_\mu{}^\nu+n_\mu n^\nu$ and
$\epsilon=P/m^2$ the acceleration is
\begin{equation}
 A_\mu=n^\nu\nabla_\nu n_\mu
 =-h_\mu{}^\nu\partial_\nu\eta
       +\frac{h_\mu{}^\nu\partial_\nu\epsilon}{2(1-\epsilon)} .
 \label{eq:p5-polar-acceleration}
\end{equation}
The exact stress and compression source decompose as
\begin{align}
 T_{\mu\nu}&=\rho n_\mu n_\nu+\Pi_{\mu\nu},
 &\rho&=Mr^2X,\notag\\
 \Pi_{\mu\nu}
 &=M\partial_\mu r\,\partial_\nu r
   -\frac12g_{\mu\nu}M[(\nabla r)^2+r^2P],
 &Q_\eta&=-\rho+\frac M2[3(\nabla r)^2+r^2P].
 \label{eq:p5-polar-source-stress}
\end{align}
Along every future-directed integral curve the phase reads proper time
according to
\begin{equation}
 \frac{d\vartheta}{ds}=\sqrt X,
 \qquad ds=\frac{d\vartheta}{\sqrt X}.
 \label{eq:p5-phase-proper-time}
\end{equation}
On a contractible spacetime patch a unit scalar clock
$dT=X^{-1/2}d\vartheta$ exists precisely when
\begin{equation}
 d(X^{-1/2}d\vartheta)=0
 \quad\Longleftrightarrow\quad A_\mu=0 .
 \label{eq:p5-phase-unit-clock}
\end{equation}
\end{proposition}
\begin{proof}
Substitution into the original quadratic action gives
\eqref{eq:p5-polar-native-action}. Its amplitude and phase variations
give \eqref{eq:p5-polar-mass-shell}. The exact one-form
$\sqrt X\,n_\mu=-\partial_\mu\vartheta$ implies
$A_\mu=-h_\mu{}^\nu\partial_\nu\log\sqrt X$.
Using $X=m^2(1-\epsilon)$ proves
\eqref{eq:p5-polar-acceleration}. Substituting the polar fields into
\eqref{eq:p5-native-source-stress} and applying the radial equation
gives \eqref{eq:p5-polar-source-stress}.
The definition of $n$ gives
$n^\mu\partial_\mu\vartheta=\sqrt X$. This proves
\eqref{eq:p5-phase-proper-time}. The one-form
$X^{-1/2}d\vartheta=-n_\mu dx^\mu$ is closed exactly
when the component of $dX$ orthogonal to $d\vartheta$ vanishes.
By \eqref{eq:p5-polar-acceleration} this is $A_\mu=0$.
The Poincar\'e lemma supplies $T$ on the stated patch and its unit
norm follows from the definition of $X$.
\end{proof}

The variable-mass cold-source approximation is controlled by the radial
derivative terms. If
$|\epsilon|\leq\epsilon_0<1$ and
$|\partial_{\hat a}r|/(mr)\leq\delta$ in an orthonormal frame then
with $\rho_0=Mr^2m^2$
\[
 \frac{|\rho-\rho_0|}{\rho_0}\leq\epsilon_0,\qquad
 \frac{|\Pi_{\hat a\hat b}|}{\rho_0}
       \leq3\delta^2+\frac{\epsilon_0}{2},\qquad
 \frac{|Q_\eta+\rho|}{\rho_0}
       \leq6\delta^2+\frac{\epsilon_0}{2}.
\]
The acceleration remainder is bounded separately by
$|h\nabla\epsilon|/[2(1-\epsilon_0)]$.
For homogeneous amplitude and compression with a homogeneous phase in an FLRW geometry
the congruence is comoving and geodesic. A constant compression and controlled
radial derivative terms give the usual cold-source limit.
For a static rest phase $\vartheta=\omega t$
$m^2-P=\omega^2/\mathcal N^2$. The phase-dominated limit gives
$\eta+\log\mathcal N=\log(\omega/m_0)$ and
$A_i=\partial_i\log\mathcal N=-\partial_i\eta$.
Thus the same matter carrier supplies homogeneous and accelerated
congruences with their source exchange. These relations use the
amplitude and phase equations of
\eqref{eq:p5-polar-native-action}. A coupled gravitational realization
obtains the corresponding matter and source equations together with the metric equations by
varying its complete action.
The metric provides the common macroscopic representation of this
relation between phase and proper time together with the source response.

\subsubsection{A closed realization with an auxiliary collective radius}
In a lowest-derivative realization of the inherited relative response
the amplitude and matter pair are dynamical while the collective
radius obeys its local stationary equation. Let $\mu>0$ be the fixed
response mass unit with $e=\mu E$ the
dimension-one amplitude and $L_{\rm ref}=\sqrt2L_c$. The physical
density and radius are
\begin{equation}
 U(e,\eta)=\mu^4\left[
 F_{L_{\rm ref}e^{-\eta}}\bigl((e/\mu)^2\bigr)-F_*\right]
 +\rho_\Lambda,\qquad
 L=L_{\rm ref}e^{-\eta},
 \label{eq:p5-auxiliary-density}
\end{equation}
where $F_*$ is the joint minimum in
Proposition~\ref{prop:p5-joint-native-radius} and $\rho_\Lambda\geq0$.
The subtraction is independent of the fields and uses the common vacuum
normalization of Ref.~\cite{Guo2026SpectralGravity}. The constants $g>0$
and $Z_E>0$ retain the amplitude convention of the complete condensate.
For a retained degenerate pair with reference squared mass $m_0^2>0$
put
\[
 M=e^{-3\eta},\qquad B=m_0^2e^{-\eta},\qquad
 \mathsf X=-\frac12g^{\mu\nu}
             \partial_\mu\varphi^{\mathsf T}\partial_\nu\varphi,
 \qquad \varphi=(\varphi_1,\varphi_2)^{\mathsf T}.
\]
The pair can be taken from two real component directions of the
existing scalar doublet carrier with the constant internal harmonic
and vanishing fixed internal connection. This invariant internal
eigenspace has $A=m_0^2I$ and
$m_0^2=3\mu^2/(4L_{\rm ref}^2)$. The retained macroscopic fields are
$g_{\mu\nu}$ and $e$ together with $\varphi_1$ and $\varphi_2$ and the auxiliary radius.
Their action is
\begin{equation}
 S_{\rm ar}=\int\!\sqrt{-g}\,d^4x\left[
 \frac{\MP^2}{2}R-\frac{Z_E}{2}(\nabla e)^2
 +M\mathsf X-\frac B2|\varphi|^2-U(e,\eta)\right].
 \label{eq:p5-auxiliary-common-action}
\end{equation}
The macroscopic Einstein term has the inherited positive normalization.
The radius has an algebraic equation while $e$ and $\varphi$ carry the
temporal response. All variations including metric variation are
taken in this one action. In this initial-value construction $H_{\rm ini}$
denotes the expansion rate at $t=0$. The Hamiltonian constraint fixes
it from the chosen occupied state and its physical density
normalization.

\begin{proposition}[Occupied common background and physical propagation]
\label{prop:p5-auxiliary-native-closure}
For the complete unweighted conformal potential the action
\eqref{eq:p5-auxiliary-common-action} has a nonempty family of expanding
spatially flat FLRW solutions with nonzero matter phase charge.
There is a time interval on which the radius is a smooth generated
function of the retained fields and their first derivatives. The
Einstein constraints propagate and the three physical scalar kinetic
coefficients and squared characteristic speeds are positive.
The two tensor polarizations have the Einstein kinetic normalization
and unit characteristic speed.
\end{proposition}
\begin{proof}
Eliminating the radius requires control of its stationary equation
both at fixed field derivatives and at fixed canonical data. These
derivatives also determine the radius contribution to the physical
kinetic form.
The radius equation and its derivative at fixed matter derivatives are
\begin{align}
 \mathcal C&=U_{,\eta}+3M\mathsf X
                         -\frac B2|\varphi|^2=0,\notag\\
 \mathfrak h&=\mathcal C_{,\eta}
       =U_{,\eta\eta}-9M\mathsf X+\frac B2|\varphi|^2 .
 \label{eq:p5-auxiliary-source-constraint}
\end{align}
Thus the source is determined together with the matter and metric.
Where $\mathfrak h>0$ the implicit-function theorem gives
$\eta=\Gamma[g,e,\varphi,\nabla\varphi]$. The substituted action
retains first derivatives only. Its source exchange cancels against
the $e,\eta$ sector by the common-action identity
\eqref{eq:p5-native-source-ward}.

To construct an occupied background choose
\begin{equation}
 \sqrt2L_c<L_0\leq2L_c,\qquad
 z_0=L_0^{-2},\quad z_c=L_c^{-2},\quad
 \eta_0=-\log(L_0/L_{\rm ref}).
 \label{eq:p5-auxiliary-initial-radius}
\end{equation}
Let $y_0>0$ be the unique amplitude stationary point at $L_0$ and set
$u_0=gL_0^2y_0$ and $e_0=\mu\sqrt{y_0}$. Its equation is
$gL_0^2 I'(u_0)=8(z_c-z_0)$. The required occupied source density is
\begin{equation}
 \mathcal A_0=-U_{,\eta}(e_0,\eta_0)
      =16\mu^4y_0(z_c-2z_0)>0 .
 \label{eq:p5-auxiliary-occupied-density}
\end{equation}
At proper time $t=0$ take
\begin{equation}
 \begin{gathered}
 a=1,\quad \dot e=0,\quad
 \varphi=(r_0,0),\quad
 \dot\varphi=(0,\omega_0r_0),\\
 r_0^2=\frac{\mathcal A_0}{B_0},\quad
 \omega_0^2=\frac{B_0}{M_0},\quad
 H_{\rm ini}^2=\frac{U(e_0,\eta_0)+\mathcal A_0}{3\MP^2}.
 \end{gathered}
 \label{eq:p5-auxiliary-occupied-data}
\end{equation}
with $H_{\rm ini}>0$. The initial matter kinetic and potential densities are
both $\mathcal A_0/2$. Thus $\mathcal C=0$ exactly. The amplitude force
also vanishes initially. Since $U\geq\rho_\Lambda\geq0$ these data
satisfy the Hamiltonian constraint with positive $H_{\rm ini}$ and homogeneity
satisfies the momentum constraint. Direct differentiation of the same
potential gives the strict margin
\begin{equation}
 \mathfrak h_0
  =\mu^4\left[32y_0(4z_0-z_c)+4u_0^2I''(u_0)\right]>0 .
 \label{eq:p5-auxiliary-positive-margin}
\end{equation}
Indeed $z_c/4\leq z_0<z_c/2$ and $I''(u_0)>0$. The concrete choice
$L_0^2=3L_c^2$ has
$\mathfrak h_0=2\mathcal A_0+4\mu^4u_0^2I''(u_0)>2\mathcal A_0$.

With comoving momenta $p_e=a^3Z_E\dot e$ and
$p=a^3M\dot\varphi$ the common background satisfies the closed system
\begin{align}
 \dot e&=\frac{p_e}{a^3Z_E},&
 \dot p_e&=-a^3U_{,e},\notag\\
 \dot\varphi&=\frac{p}{a^3M},&
 \dot p&=-a^3B\varphi,\notag\\
 \dot a&=aH,&
 \dot H&=-\frac{Z_E\dot e^2+M|\dot\varphi|^2}{2\MP^2},\notag\\
 0&=U_{,\eta}+\frac{3|p|^2}{2a^6M}
                              -\frac B2|\varphi|^2 .
 \label{eq:p5-auxiliary-background-odes}
\end{align}
At fixed canonical data the last equation has derivative
\begin{equation}
 \mathfrak h_H=U_{,\eta\eta}+9\mathcal T+\mathcal V
             =\mathfrak h+18\mathcal T>0,\qquad
 \mathcal T=\frac{|p|^2}{2a^6M},\quad
 \mathcal V=\frac B2|\varphi|^2 .
 \label{eq:p5-auxiliary-hamiltonian-curvature}
\end{equation}
For fixed $g>0$ and canonical data with $a>0$ and
$y=(e/\mu)^2>0$, the radius equation has exactly one root
$\eta\in\mathbb R$. Put
$z=L^{-2}=L_{\rm ref}^{-2}e^{2\eta}$ and $u=gy/z$. The complete
potential in Proposition~\ref{prop:p5-joint-native-radius} gives
\[
 \mu^{-4}U_{,\eta}=16zy-2uI'(u),\qquad
 \mu^{-4}U_{,\eta\eta}
       =32zy+4uI'(u)+4u^2I''(u)>0.
\]
Writing $Q_2=\sum_j\beta_j/\gamma_j^2<\infty$, its spectral sum obeys
\[
 0\leq I'(u)\leq Q_2u,\qquad
 I'(u)\geq\frac{\beta_{j_0}}{2\gamma_{j_0}}
 \quad\text{for }u\geq\gamma_{j_0},
\]
where any term with $\beta_{j_0}>0$ may be used. At fixed canonical
data $\mathcal T$ is proportional to $e^{3\eta}$ and
$\mathcal V$ to $e^{-\eta}$, with nonnegative coefficients. Thus
$\mathcal C_H:=U_{,\eta}+3\mathcal T-\mathcal V$ is strictly
increasing, tends to $-\infty$ as $\eta\to-\infty$ and to $+\infty$
as $\eta\to+\infty$. The implicit-function theorem gives a smooth
radius function on this domain of canonical data. Substituting it
into \eqref{eq:p5-auxiliary-background-odes} gives a smooth
finite-dimensional system and hence a unique local solution.
The positivity of $\mathfrak h_H$ makes the canonical elimination
regular; $\mathfrak h$ enters the physical kinetic form.
For the occupied initial data, $H>0$ and $\mathfrak h>0$
persist on a nonzero interval by continuity. The energy density satisfies
\[
 \rho=\frac{Z_E}{2}\dot e^2+\mathcal T+\mathcal V+U,
 \qquad
 \dot\rho=-3H\bigl(Z_E\dot e^2+2\mathcal T\bigr).
\]
Consequently $d(3\MP^2H^2-\rho)/dt=0$. This proves propagation of
the background Hamiltonian constraint and the common source balance.
At the occupied initial state differentiation of the source equation
also gives its strictly positive initial compression rate
\begin{equation}
 \dot\eta_0=
 \frac{9H_{\rm ini}\mathcal A_0}{\mathfrak h_0+9\mathcal A_0}>0,
 \qquad \dot L_0=-L_0\dot\eta_0 .
 \label{eq:p5-auxiliary-initial-feedback}
\end{equation}
Here $\dot e_0=0$ and
$\varphi_0^{\mathsf T}\dot\varphi_0=p_0^{\mathsf T}\dot p_0=0$ and
so the explicit scale-factor derivative of $3\mathcal T$ is
$-18H_{\rm ini}\mathcal T_0$. Division by $\mathfrak h_H$ proves the formula.

The physical quadratic action follows by eliminating the auxiliary
radius and metric constraints. The radius momentum is zero. Its preservation gives the
algebraic source constraint. Their Poisson bracket is nonzero by
\eqref{eq:p5-auxiliary-hamiltonian-curvature}. Thus this pair removes
the radius phase-space pair. Around the homogeneous solution put
$q=(e,\varphi_1,\varphi_2)$ and $v=\dot\varphi$. At fixed lapse
the terms involving the radius perturbation and the field velocities
are
\[
 \frac12\dot Q^{\mathsf T}\mathcal G\dot Q
       +\delta\eta\,b^{\mathsf T}\dot Q
       -\frac{\mathfrak h}{2}(\delta\eta)^2,
 \qquad
 \mathcal G=\operatorname{diag}(Z_E,M,M),\quad
 b=(0,-3Mv_1,-3Mv_2)^{\mathsf T},
\]
up to terms with at most one derivative on the physical perturbations.
Eliminating $\delta\eta$ gives the positive rank-one contribution
$bb^{\mathsf T}/\mathfrak h$. There is no analogous spatial-gradient
term on a homogeneous background.

For nonzero spatial Fourier modes use spatially flat gauge. The
linear momentum constraint determines the lapse perturbation as
\begin{equation}
 \delta\mathcal N=
 \frac{Z_E\dot e\,Q_e+M v^{\mathsf T}Q_\varphi}
      {2\MP^2H} .
 \label{eq:p5-auxiliary-lapse-constraint}
\end{equation}
It contains no $\dot Q$. The scalar shift multiplies this constraint
in the quadratic action. Its terms cancel after substitution.
The remaining lapse terms affect the zero- and one-derivative
coefficients. Including both matter and Einstein variations gives
the complete scalar principal action
\begin{equation}
 \begin{gathered}
 S^{(2)}_{\rm sc}=\frac12\int a^3\,dt\,d^3x
 \left[\dot Q^{\mathsf T}\mathcal K\dot Q
 -\frac1{a^2}\partial_iQ^{\mathsf T}\mathcal G\partial_iQ
 +\text{lower-derivative terms}\right],\\
 \mathcal K=\mathcal G+\frac{bb^{\mathsf T}}{\mathfrak h} .
 \end{gathered}
 \label{eq:p5-auxiliary-physical-kinetic}
\end{equation}
This agrees with the general constrained scalar-field action in
Ref.~\cite[Eqs.~(41), (45)]{Langlois2008Kinetics}. Here the lapse
constraint and the auxiliary Schur complement have been evaluated
for the actual response coefficients. Both $\mathcal K$ and
$\mathcal G$ are positive definite. The generalized characteristic
eigenvalues are
\begin{equation}
 c_1^2=c_2^2=1,\qquad
 c_\parallel^2=
 \frac{\mathfrak h}{\mathfrak h+9M|v|^2}>0 .
 \label{eq:p5-auxiliary-characteristic-speeds}
\end{equation}
For $L_0^2=3L_c^2$ the initial nontrivial speed obeys
$c_\parallel^2>2/11$. The tensor action is
$\MP^2\int a^3[\dot\gamma_{ij}^2-a^{-2}(\partial_k\gamma_{ij})^2]
dt\,d^3x/8$ for transverse traceless $\gamma_{ij}$.
These are all the propagating principal blocks after the common
constraints have been removed.
\end{proof}

\paragraph{A certified interval of common evolution.}
Set $L_c^2=\pi$ and $g=1$ with $Z_E=2$ together with
$m_0^2/\mu^2=3/(8\pi)$ and $L_0^2=3\pi$ in
Proposition~\ref{prop:p5-auxiliary-native-closure}. For
$P=\MP/\mu\geq1$ and
$0\leq\rho_\Lambda/\mu^4\leq10^{-3}$ its occupied solution exists
throughout $0\leq\mu t\leq10^{-2}$. On this interval
\begin{equation}
 \frac{\mathfrak h}{\mu^4}>0.04769800,\qquad
 c_\parallel^2>0.19180880,\qquad
 0.23578722\leq\frac{d\vartheta}{d(\mu t)}
                         \leq0.34430776 .
 \label{eq:p5-auxiliary-finite-margins}
\end{equation}
The proof bounds the vector field and the algebraic source root on
a common phase-space box including the infinite spectral tail.
Use dimensionless fields $E=e/\mu$ and $\phi=\varphi/\mu$
with momenta $\widehat p_e=p_e/\mu^2$ and $\widehat p=p/\mu^2$ together with
$w=P H/\mu$ and $\alpha=e^\eta$. Dots in the following table
denote derivatives with respect to $\mu t$. Outward interval
evaluation gives the box and initial enclosures together with absolute derivative
bounds below. All bounds are uniform for $P\geq1$.
\begin{center}
\small
\begin{tabular}{llll}
\toprule
Coordinate & Box & Initial enclosure & $|\dot x|$ bound \\
\midrule
$a$ & $[0.99,1.01]$ & $1$ & $0.12120000$ \\
$E$ & $[0.101,0.105]$ & $[0.10249012,0.10352021]$ & $0.00206123$ \\
$\widehat p_e$ & $[-0.004,0.004]$ & $0$ & $0.11170501$ \\
$\phi_1$ & $[0.347,0.354]$ & $[0.34925525,0.35276545]$ & $0.00189878$ \\
$\phi_2$ & $[-0.004,0.004]$ & $0$ & $0.11709085$ \\
$\widehat p_1$ & $[-0.003,0.003]$ & $0$ & $0.05510890$ \\
$\widehat p_2$ & $[0.178,0.185]$ & $[0.18099847,0.18281760]$ & $0.00062270$ \\
$w$ & $[0.06,0.12]$ & $[0.07709848,0.10895794]$ & $0.01116963$ \\
\bottomrule
\end{tabular}
\end{center}
On this box the canonical source equation
$\mathcal C_H:=U_{,\eta}+3\mathcal T-\mathcal V=0$ has
\[
 \frac{\mathcal C_H(\alpha=0.79)}{\mu^4}<-0.00264330,
 \qquad
 \frac{\mathcal C_H(\alpha=0.85)}{\mu^4}>0.00401790 .
\]
Its positive derivative
$\partial_\alpha\mathcal C_H=\mathfrak h_H/\alpha$ gives a unique
source in $0.79<\alpha<0.85$. The spectral evaluations retain
$j=0,\ldots,N$ with $N=64$. In the notation of
\eqref{eq:p5-inherited-stationary-function} they use
\[
 0\leq\mathcal H(u)-\mathcal H_N(u)\leq r_Nu,\qquad
 0\leq\mathcal H'(u)-\mathcal H_N'(u)\leq r_N,
 \qquad r_N=\frac8{9(2N+1)}.
\]
These bounds follow from $\gamma_j\geq3(2j+1)^2/4$ and the
integral bound on $\sum_{j>N}(2j+1)^{-2}$. Monotonicity of
$\mathcal H$ and decrease of $\mathcal H'$ enclose the full response
on each argument interval. Eight distinct spectral endpoints
require 520 shell evaluations with the response and its derivative
evaluated together. They yield the displayed source brackets and
\eqref{eq:p5-auxiliary-finite-margins} on the entire
box~\cite{Guo2026GenesisFramework}. The initial roots obey
$u_0\in[0.099,0.101]$ and $u_*\in[0.10,0.12]$. The initial
energy enclosure uses the joint-minimum inequality
$I(u_*)\geq0$ and
$I(u_0)\leq\mathcal H'(0)u_0^2/2$ where
$\mathcal H'(0)=\pi(\pi+2)/8$.

For each coordinate the derivative bound multiplied by $10^{-2}$
is less than half the distance from its initial enclosure to the box
boundary. Integration of these bounds excludes a first exit before
$\mu t=10^{-2}$. The source root and its derivatives remain regular
there. Thus the ordinary differential equations continue to that time.
This proves the interval claim;
\eqref{eq:p5-auxiliary-physical-kinetic} and
\eqref{eq:p5-auxiliary-phase-charge} give its physical kinetic and
propagation bounds and its clock bounds.

The radius map has a finite-record construction on every compact
neighborhood with $0<h_-\leq\mathfrak h\leq h_+$. For fixed retained
data and a bracket containing its root the iteration
\[
 \eta_{n+1}=\eta_n-\frac{\mathcal C(\eta_n)}{h_+}
\]
is a contraction with factor at most $1-h_-/h_+$. It gives a geometric
error bound for the same source map $\Gamma$. Its domain and limit
are recorded together as in
Refs.~\cite{Guo2026SpectralArithmetic,Guo2026NumberDynamics}.

\paragraph{Conserved phase charge and the matter clock.}
In this auxiliary radius realization the common action depends on
rotational invariants of the retained real pair.  Its global phase rotation
gives the conserved charge
\begin{equation}
 \mathcal Q=a^3M(\varphi_1\dot\varphi_2-
                         \varphi_2\dot\varphi_1)
            =a^3Mr^2\dot\vartheta>0 .
 \label{eq:p5-auxiliary-phase-charge}
\end{equation}
The homogeneous matter equations are
$\partial_t(a^3M\dot\varphi_i)+a^3B\varphi_i=0$.
Multiplying the second by $\varphi_1$ and the first by $\varphi_2$ and
subtracting gives $\dot{\mathcal Q}=0$.
The initial data above have $\mathcal Q>0$. Thus the phase remains a
monotone clock on the regular solution interval.  Its rate
$\dot\vartheta=\mathcal Q/(a^3Mr^2)$ is determined by the conserved
charge, physical volume, compression and matter amplitude. The
evolution parameter orders the states; the matter phase supplies
the clock reading. Its proper time
readout is $ds=d\vartheta/\dot\vartheta=dt$ in the unit-lapse chart. Other positive clock charts give the same trajectory and rates by
\eqref{eq:p5-native-clock-reparametrization}. This realization
determines the matter history, occupied collective radius,
expanding metric and phase clock readout.
Its positive kinetic and characteristic forms hold on the interval
where $H>0$ and $\mathfrak h>0$. The density normalization and retained
fields and algebraic radius equation in
\eqref{eq:p5-auxiliary-common-action} specify this dynamical branch.
The production TT endpoint is determined by the sourced trace
defined in Section~\ref{sec:closure}.

\subsubsection{An effective realization with an independent foliation}
An independent foliation field $\chi$ gives a separate covariant
realization of the local constitutive law in
\eqref{eq:p5-feedback-action}. It specifies a macroscopic congruence. The matter clock $T$ describes the pressureless source trajectories.
The auxiliary-radius dynamics retain the field content of
\eqref{eq:p5-auxiliary-common-action} and the distance calculation
uses the Einstein--matter thermal background of
Section~\ref{sec:linprop}. For the foliation realization use signature $(-+++)$ with
$c=1$ and $M^2=\MP^2=(8\pi G)^{-1}$ and retain the constant endpoint
normalization $a_0$ of \eqref{eq:sigmaira0}. For a timelike gradient
of $\chi$ define
\begin{equation}
 n_\mu=-\frac{\partial_\mu\chi}
 {\sqrt{-g^{\alpha\beta}\partial_\alpha\chi\partial_\beta\chi}},
 \qquad A_\mu=n^\nu\nabla_\nu n_\mu,
 \qquad Y=A_\mu A^\mu .
 \label{eq:p5-clock-geometry}
\end{equation}
The function in the action is fixed by the same primitive
$\mathcal F$ as the local energy
\begin{align}
 \mathcal A(Y)&=2\mathcal F(\sqrt Y)-Y
 =a_0^2\left[y\sqrt{1+y^2}-\operatorname{arsinh}y-y^2\right],
 &y&=\frac{\sqrt Y}{a_0},
 \label{eq:p5-clock-invariant}\\
 S&=\frac{M^2}{2}\int d^4x\sqrt{-g}\,
       [R-2\Lambda-2\mathcal A(Y)]+S_{\rm m}.
 \label{eq:p5-common-covariant-action}
\end{align}
Here $\mathcal F(\sqrt Y)$ denotes its radial value and $S_{\rm m}$
contains the canonical radial field and pressureless source together with any
visible matter under consideration. The preferred-foliation action
and its covariant variation follow the construction of
Ref.~\cite{BlanchetMarsat2011}. Cosmological clock dynamics in this
class of theories are also discussed in
Ref.~\cite{BlanchetSkordis2024}. Equation~\eqref{eq:p5-clock-invariant}
specifies the full acceleration function from the present local
response. The independent foliation and its initial and boundary data
are the effective field content of this realization.
The static construction determines $\mathcal F$ and its spatial
response. The constrained dynamics of this field content provide a
comparison for generated variations retaining the same foliation sector.

\begin{proposition}[Local and cosmological limits of the foliation action]
\label{prop:p5-covariant-two-limits}
The action \eqref{eq:p5-common-covariant-action} has the following
properties.

\textnormal{(i)} Every smooth Einstein--matter solution admitting a
smooth geodesic foliation is a solution of the enlarged system with
that foliation. In a static metric with vanishing shift the choice
$\chi=t$ satisfies the clock equation and gives
\eqref{eq:endpointpoisson} in the leading weak-field
nonrelativistic limit.

\textnormal{(ii)} Consider a spatially flat expanding or contracting
dust--vacuum background with
\[
 3M^2H^2=\rho+\rho_\Lambda,\qquad
 -2M^2\dot H=\rho,\qquad \dot\rho=-3H\rho,
\]
where $\rho>0$ and $H\ne0$, $\rho_\Lambda$ is constant and the radial
field is at a stationary condensate with $m_\delta^2>0$. For each
nonzero Fourier mode $\kappa=|k|/a$ the constrained quadratic scalar
action consists of the Einstein radial--cold-source action and one
additional mode with positive kinetic coefficient and conserved
conjugate momentum. The effective cold density perturbation obeys
the Einstein cold-source equations for every value of this momentum.
The tensor quadratic action is the Einstein tensor action.
\end{proposition}

\begin{proof}
The identities $\mathcal A(0)=0$ and
\[
 \mathcal A_Y=\mu(\sqrt Y/a_0)-1,
 \qquad
 \frac{\partial\mathcal A}{\partial A_\mu}
       =2\mathcal A_Y A^\mu
\]
show that the added action and its first variation vanish when
$A_\mu=0$. The Einstein--matter equations and the clock equation
therefore hold on every smooth geodesic-foliation branch. This is a
local statement on a domain where the foliation remains smooth.

With
$K_{\mu\nu}=(g_\mu{}^\alpha+n_\mu n^\alpha)\nabla_\alpha n_\nu$
the clock equation can be written as
\begin{equation}
 \nabla_\mu\left[n^\mu\nabla_\nu
             (\mathcal A_Y A^\nu)\right]
 =\frac12 n^\mu\nabla_\mu\mathcal A
     +\mathcal A_Y A^\mu A^\nu K_{\mu\nu}.
 \label{eq:p5-clock-eom}
\end{equation}
In the static foliation $A_i=D_i\log N$ and $K_{ij}=0$. All time
derivatives in \eqref{eq:p5-clock-eom} vanish. With $N=1+\Phi$ at
leading order the gravitational and source Lagrangian is
\[
 -M^2\left[|\nabla\Phi|^2+
                 \mathcal A(|\nabla\Phi|^2)\right]-\rho\Phi
 =-\frac{\mathcal F(\nabla\Phi)}{4\pi G}-\rho\Phi.
\]
Its variation is \eqref{eq:endpointpoisson}. The independent matter
variation gives the nonrelativistic geodesic equation appearing in
\eqref{eq:p5-feedback-system}.

For the cosmological background choose $\chi=t$ and $T=t$.
Both congruences are geodesic. Thus the background equations above
follow directly. In spatially flat gauge write
$N=1+n$ and $T=t+\pi$ and $\chi=t+\sigma$. To first order
$A_i=\partial_i(n-\dot\sigma)$. Consequently the added quadratic
Lagrangian divided by $a^3$ is
$C(\dot\sigma-n)^2/2$ where $C=2M^2\kappa^2$.
It contains no linear shift perturbation. The shift constraint of
the Einstein radial--cold-source action therefore remains
\[
 n=B\pi,\qquad B=\frac{\rho}{2M^2H}.
\]
Eliminating the dust auxiliary variable gives the first-order
scalar action apart from the canonical radial fluctuation $Q_r$
\begin{equation}
 S^{(2)}_{\rm c}=\int dt\,a^3
 \left[p(\dot\pi-B\pi)-\frac D2\pi^2
             +\frac C2(\dot\sigma-B\pi)^2\right],
 \qquad
 D=\rho\kappa^2+\frac{3\rho^2}{2M^2}>0.
 \label{eq:p5-clock-first-order}
\end{equation}
The Einstein radial and dust constraints used here are those of
the common normalized matter action~\cite{Guo2026SpectralGravity}.
Let $P=a^3p$ and
$Q_\chi=a^3C(\dot\sigma-B\pi)$. Since $\sigma$ is cyclic
$\dot Q_\chi=0$. The Hamiltonian and canonical one-form are
\begin{align}
 H^{(2)}_{\rm c}
 &=B(P+Q_\chi)\pi+\frac{a^3D}{2}\pi^2
                         +\frac{Q_\chi^2}{2a^3C},\\
 P\,d\pi+Q_\chi\,d\sigma
 &=R\,d\pi+Q_\chi\,d\zeta_\chi,
 &R&=P+Q_\chi,\qquad \zeta_\chi=\sigma-\pi .
 \label{eq:p5-clock-canonical-split}
\end{align}
This transformation is independent of time. Set $r=R/a^3$. Integration by parts in $R\dot\pi$ and elimination of $\pi$ give
\begin{equation}
 S^{(2)}_{\rm scal}=\int dt\,a^3\left\{
 \frac12\dot Q_r^2-\frac12(\kappa^2+m_\delta^2)Q_r^2
 +\frac{[\dot r+(3H+B)r]^2}{2D}
 +\frac C2\dot\zeta_\chi^2\right\}.
 \label{eq:p5-clock-positive-reduction}
\end{equation}
The kinetic coefficients $1,D^{-1},C$ are all positive on the
stated domain. Lapse variation identifies the foliation density as
$\delta\rho_\chi=Q_\chi/a^3$. Its linear momentum density and
spatial stress vanish. Thus $r=p+Q_\chi/a^3$ is the effective cold
density perturbation. Its equation and its metric constraints are
the Einstein cold-source equations. Matching the initial $r$ and
velocity therefore matches the cosmological scalar metric response.
The branch $Q_\chi=0$ is invariant. The extra equation is
\[
 a^3C\dot\zeta_\chi=Q_\chi,\qquad
 \zeta_\chi(t)=\zeta_\chi(t_i)+\frac{Q_\chi}{2M^2|k|^2}
                         \int_{t_i}^t\frac{dt'}{a(t')} .
\]
Finally $A_i$ has only a scalar perturbation at first order. The
added term contributes no tensor quadratic term.
\end{proof}

The additional scalar has zero spatial restoring term at quadratic
order and hence $c_s^2=0$. Its conserved momentum partitions the total
cold density between the matter and foliation sectors. The usual
gravitational growth of the cold density is retained. This result
establishes the kinetic signs and linear evolution on the specified
dust--vacuum background.

The static and cosmological reductions use different macroscopic
foliations with their respective constraints eliminated. In the static
sector with bounded $\nabla\phi$
\[
 \mathcal J(\varepsilon\nabla\phi)
 =\frac{\varepsilon|\varepsilon|}{a_0}
          |\nabla\phi|\nabla\phi+O(|\varepsilon|^4),
 \qquad D\mathcal J(0)=0.
\]
This quadratic leading flux is the static reduction of
\eqref{eq:p5-common-covariant-action}. The cosmological reduction
retains the independent clock fluctuation and gives
\eqref{eq:p5-clock-positive-reduction}. These are background-specific
reductions of the nonlinear field and foliation equations with the
initial and boundary data chosen in each branch.

\subsubsection{The independent foliation on an accelerated background}
The stationary branch of the independent-foliation action permits an
explicit reduction of its scalar velocity form. We use the leading slow-motion action of
Ref.~\cite{Flanagan2023} with the matter configuration held fixed when
taking this velocity Hessian. Put $\chi=t+\sigma$ and
$\Phi=\bar\Phi+\phi$. The acceleration invariant depends at this order on
$\Xi=\Phi-\dot\sigma+\tfrac12|\nabla\sigma|^2$. On a static background define
\begin{equation}
 H(x)=\mu(y)I+y\mu'(y)e\otimes e,\qquad
 y=\frac{|\nabla\bar\Phi|}{a_0},\qquad
 e=\frac{\nabla\bar\Phi}{|\nabla\bar\Phi|}.
 \label{eq:p5-accelerated-principal}
\end{equation}

\begin{proposition}[Reduced velocity form of the independent foliation]
\label{prop:p5-generated-clock-kinetic}
Let $\Omega$ be a bounded Lipschitz domain with fixed Dirichlet
perturbations and suppose $H$ is bounded and uniformly positive on
$\Omega$. Write $D=-\Delta$ and
$L=-\nabla\cdot H\nabla$ as maps from $H_0^1(\Omega)$ to its dual.
Equip $H_0^1(\Omega)$ with the gradient norm.
Eliminating the potential gives the scalar velocity operator
\begin{equation}
 K_{\rm red}=2M^2(DL^{-1}D-D).
 \label{eq:p5-clock-schur}
\end{equation}
For variations in this reduced foliation sector with the
other physical variables fixed and $\sigma=Bq$ the induced velocity
form is $K_{\rm gen}=B^\ast K_{\rm red}B$. A finite set of independent
directions $b_i\in H_0^1(\Omega)$ therefore has matrix
\begin{equation}
 (K_{\rm gen})_{ij}
 =2M^2\left\{
 \langle Db_i,L^{-1}Db_j\rangle-\langle b_i,Db_j\rangle
 \right\}.
 \label{eq:p5-clock-generated-matrix}
\end{equation}
These pairings use the Sobolev duality associated with the Dirichlet
forms. In particular $0<H\leq I$ implies $K_{\rm red}\geq0$.
\end{proposition}

\begin{proof}
The part of the quadratic Lagrangian containing the potential and
foliation velocity is
\[
 -M^2\left[
 \langle\phi,D\phi\rangle+
 \langle\phi-\dot\sigma,(L-D)(\phi-\dot\sigma)\rangle\right].
\]
The remaining terms contain no squared velocities of $\sigma$.
Potential variation gives
$L\phi=(L-D)\dot\sigma$. Uniform ellipticity supplies its unique
Dirichlet solution. Substitution gives
$M^2\langle\dot\sigma,(DL^{-1}D-D)\dot\sigma\rangle$.
Restricting $\dot\sigma$ to $B\dot q$ proves the matrix formula.
If $L\leq D$ in quadratic-form order then
$L^{-1}\geq D^{-1}$ on the dual space. This proves the last assertion.
\end{proof}

For a constant background and a nonzero wavevector put
$h=\hat k^{\mathsf T}H\hat k$. The reduced Lagrangian is
$M^2|k|^2(1-h)|\dot\sigma_k|^2/h$. For the RMS law
\begin{equation}
 h_\parallel=\frac{y(y^2+2)}{(1+y^2)^{3/2}},\qquad
 h_\parallel>1
 \quad\Longleftrightarrow\quad
 y>y_*:=\sqrt{\frac{\sqrt5-1}{2}}=0.786151377\ldots .
 \label{eq:p5-clock-acceleration-threshold}
\end{equation}
Thus the independent foliation has a negative longitudinal velocity
direction above this threshold in the static slow-motion branch.
Equation~\eqref{eq:p5-clock-schur} retains the spatially varying
background coefficients in this same approximation.

\paragraph{Radial specialization.}
For a smooth spherical static solution with regular perturbations at the
centre and fixed outer potential put
$\lambda(r)=\mu(y(r))+y(r)\mu'(y(r))$. For radial velocities with
gradient compactly supported in a region where $\lambda>0$
\begin{equation}
 L_{\rm kin}^{(2)}
 =4\pi M^2\int r^2\frac{1-\lambda(r)}{\lambda(r)}
                    |\partial_r\dot\sigma|^2\,dr .
 \label{eq:p5-clock-radial-form}
\end{equation}
Every open shell on which $\lambda>1$ supports a strictly negative
radial velocity direction. In the vacuum exterior of a positive spherical source of
mass $M_{\rm s}$ and support radius $R_{\rm s}$ such a shell exists
whenever $R_{\rm s}<r_*$ where
\[
 r_*=
 \left(\frac{y_*^2}{\sqrt{1+y_*^2}}\right)^{-1/2}
 \sqrt{\frac{GM_{\rm s}}{a_0}}
 =1.434632715\ldots\sqrt{\frac{GM_{\rm s}}{a_0}} .
\]
Radial integration of the potential constraint using regularity at
the centre gives
$\phi'=(\lambda-1)\dot\sigma'/\lambda$. Substitution into the
quadratic action proves~\eqref{eq:p5-clock-radial-form}.
The potential perturbation is constant on any region where
$\lambda=0$ and the velocity gradient vanishes. Its additive constant
is fixed by the outer boundary. A nonconstant smooth bump supported
in a shell with $\lambda>1$ makes the integral
strictly negative. The exterior Gauss law reads
$GM_{\rm s}/(a_0r^2)=y^2/\sqrt{1+y^2}$. The right-hand side is a strictly increasing function
of $y>0$. Evaluation at $y_*$ gives the stated radius.

\paragraph{Constitutive restriction in the independent-foliation action.}
For this action the kinetic restriction can be expressed directly in
terms of the constitutive law independently of its detailed interpolation.
For the independent-foliation action
\eqref{eq:p5-common-covariant-action} nonnegative scalar velocity
energy in every constant accelerated background requires
\[
 0<\mu(y)\leq1,\qquad 0<(y\mu(y))'\leq1 .
\]
A differentiable response with $0<\mu(y)<1$ and
$1-\mu(y)=o(1/y)$ at large $y$ cannot satisfy these inequalities
for every $y>0$. For the RMS response the smallest coefficient
normalized by $2M^2D$ is
\begin{equation}
 \inf_{y,\hat k}\frac{1-h(y,\hat k)}{h(y,\hat k)}
 =\frac{3\sqrt6}{8}-1=-0.081441346\ldots ,
 \qquad y=\sqrt2,\quad \hat k=e .
 \label{eq:p5-clock-minimum-kinetic}
\end{equation}
The transverse and longitudinal eigenvalues of $H$ are $\mu$ and
$(y\mu)'$. Positive elliptic response and nonnegative reduced
foliation kinetic energy require both to lie in $(0,1]$. Put
$d(y)=y(1-\mu(y))$. The longitudinal condition gives $d'\geq0$
whereas the stated large-$y$ condition gives $d(y)\to0$ with
$d(y)>0$. This is a contradiction. For the RMS law
$h_\parallel'=(2-y^2)/(1+y^2)^{5/2}$. Thus its maximum is
$4\sqrt2/(3\sqrt3)$ at $y=\sqrt2$. It exceeds the transverse
eigenvalue. This proves~\eqref{eq:p5-clock-minimum-kinetic}.

Within an extension retaining this reduced foliation block the static
primitive fixes the coefficient of the spatial acceleration invariant. The independent foliation turns that same coefficient into a velocity
form after potential reduction. Additional generated fields contribute
through their common action and constraints. On a direction attaining
\eqref{eq:p5-clock-minimum-kinetic} its contribution in the same
normalization must be at least $1-3\sqrt6/8$ to obtain a nonnegative
form provided that direction remains in the reduced physical space
with the same normalization. The bound is necessary on that retained
direction.
The foliation kinetic function in Ref.~\cite{BlanchetSkordis2024}
illustrates additional dynamical data. Its coefficients and its
background and perturbation equations enter the common matching.

\paragraph{Finite-direction control of the independent foliation.}

For the independent-foliation diagnostic of
\eqref{eq:p5-clock-generated-matrix} the retained vectors $b_i$ are
independent and $G_{ij}=\langle b_i,Db_j\rangle$ is its positive
gradient metric. The corresponding matrix bound certifies this
restricted foliation form. Its potential-solve error can be estimated
directly. For $Lw=Db$ and a computed $\widehat w$ with
residual $r=Db-L\widehat w$ coercivity
$\langle u,Lu\rangle\geq\alpha\|\nabla u\|^2$ gives
\[
 \|w-\widehat w\|_{H_0^1}\leq
 \alpha^{-1}\|r\|_{H^{-1}}.
\]
This provides the potential-solve contribution for that response.
All its source columns use the same factorization of $L$.

On the same fixed finite space Gaussian transport acts on a
quadratic form by congruence.
Its multipliers $e^{-t\lambda_j}$ are nonzero. Thus it preserves the
inertia of that form. In the unrestricted Fourier algebra at $y=1$ the directions
$(1,\pm1,0)$ have $h=5/(4\sqrt2)<1$ whereas their sum direction has
$h=3/(2\sqrt2)>1$. The product of their cosine modes contains that
sum mode. Closure under these Fourier products therefore reaches
directions of both signs in this independent-foliation form. Its
cosmological reduction in Proposition~\ref{prop:p5-covariant-two-limits}
concerns the geodesic dust--vacuum background.

\paragraph{Finite mode verification.}
The finite calculation~\cite{Guo2026GenesisFramework} uses an
Einstein--de Sitter background in units $M^2=1$.
Here $a=t^{2/3}$ and $H=2/(3t)$ and $\rho=4/(3t^2)$ with $|k|=4$ and
$1\leq t\leq3$. For initial values
$(p_0,\pi_0,\sigma_0)=(-0.01,0.003,-0.002)$ at $t=1$ the exact
solutions are
\begin{align}
 \pi(t)&=\pi_0t,\\
 p(t)&=\frac{C_0}{t^3}
       -\frac{4\pi_0|k|^2}{5t^{4/3}}
       -\frac{8\pi_0/3+Q_\chi}{t^2},
 &C_0&=p_0+\frac{4\pi_0|k|^2}{5}
                         +\frac{8\pi_0}{3}+Q_\chi,\\
 \sigma(t)&=\sigma_0+\pi_0(t-1)
       +\frac{3Q_\chi}{2|k|^2}(t^{1/3}-1).
 \label{eq:p5-clock-eds-solution}
\end{align}
Integrations for $Q_\chi=0$ and $0.007$ are checked against these
expressions at 25 times. A third integration with
$(p_0,Q_\chi)=(-0.017,0.007)$ compares the same initial effective
cold density as the zero-charge solution. The sampled maximum errors
in $p$ and $\sigma$ across the three integrations are
$1.73\times10^{-12}$ and $1.44\times10^{-17}$ respectively.
For equal initial effective sources the sampled maximum differences
in $r$ and $n$ are $1.21\times10^{-12}$ and
$1.74\times10^{-18}$. The three integrations use a total of 720
right-hand-side evaluations and 795 background evaluations
including the 75 output samples. These are finite numerical checks of
\eqref{eq:p5-clock-first-order} complementary to the action and
constraint calculation above.

%% file: sections-en/declarations.tex
%% =============================================================================
%% Declarations (before references, per JHEP convention)
%% =============================================================================

\section*{Declarations}
\label{sec:declarations}

\hspace*{\parindent}Conflict of interest. The author declares no competing interests.

Data availability. The calculations use the
Coarse Graining Genesis Framework~\cite{Guo2026GenesisFramework}.
Appendix~\ref{app:repro} identifies the production calculation of 24 October 2026
and the archived fixed-background thermal evaluation~\cite{Guo2026ThermalCalculation}
with their source and data checksums, numerical conventions and comparison
inputs. The software repository is given in
Ref.~\cite{Guo2026GenesisFramework}. The article is deposited at
\url{https://doi.org/10.5281/zenodo.22067469}.

The Lean proof library~\cite{GuoLeanFoundations} provides
kernel checked declarations for the indexed clauses of the inherited
source and basic physical interfaces including source exchange and
centre trace cancellation, the charge lattice and the global group kernel.
It also checks covariance under common frame changes, compensation under
finite resolution changes, phase charge algebra and tensor identities
in the local five coefficient class.
The theorem index records the statement mappings and comparison classes;
the transitive axiom audit records their dependencies.
The article derives the covariant source work and total stress identity.
Physical calibration and the charged-lepton conversion $G_\ell$ have the
analytic or numerical status specified at their defining relations.